\documentclass[longauth]{aa} 
\pdfoutput=1
\usepackage[stable]{footmisc}
\usepackage{amsmath}
\usepackage{txfonts}
\usepackage{placeins}
\usepackage{natbib}
\bibpunct{(}{)}{;}{a}{}{,} %
\usepackage{graphicx}
\usepackage[breaklinks, colorlinks, citecolor=blue]{hyperref}
\usepackage{overpic}
\usepackage{booktabs}
\usepackage{longtable}
\usepackage{pdflscape}
\usepackage{fixltx2e}
\usepackage{rotating}

\usepackage[table,usenames,dvipsnames]{xcolor}

\def\setsymbol#1#2{\expandafter\def\csname #1\endcsname{#2}}
\def\getsymbol#1{\csname #1\endcsname}

\def\Planck{\textit{Planck}}

\newbox\tablebox    \newdimen\tablewidth
\def\leaderfil{\leaders\hbox to 5pt{\hss.\hss}\hfil}
\def\endPlancktable{\tablewidth=\columnwidth 
    $$\hss\copy\tablebox\hss$$
    \vskip-\lastskip\vskip -2pt}
\def\endPlancktablewide{\tablewidth=\textwidth 
    $$\hss\copy\tablebox\hss$$
    \vskip-\lastskip\vskip -2pt}
\def\tablenote#1 #2\par{\begingroup \parindent=0.8em
    \abovedisplayshortskip=0pt\belowdisplayshortskip=0pt
    \noindent
    $$\hss\vbox{\hsize\tablewidth \hangindent=\parindent \hangafter=1 \noindent
    \hbox to \parindent{$^#1$\hss}\strut#2\strut\par}\hss$$
    \endgroup}
\def\doubleline{\vskip 3pt\hrule \vskip 1.5pt \hrule \vskip 5pt}

\def\L2{\ifmmode L_2\else $L_2$\fi}

\def\DeltaT{\ifmmode \Delta T\else $\Delta T$\fi}
\def\deltat{\ifmmode \Delta t\else $\Delta t$\fi}
\def\fknee{\ifmmode f_{\rm knee}\else $f_{\rm knee}$\fi}
\def\Fmax{\ifmmode F_{\rm max}\else $F_{\rm max}$\fi}
\def\solar{\ifmmode{\rm M}_{\mathord\odot}\else${\rm M}_{\mathord\odot}$\fi}
\def\Msolar{\ifmmode{\rm M}_{\mathord\odot}\else${\rm M}_{\mathord\odot}$\fi}
\def\Lsolar{\ifmmode{\rm L}_{\mathord\odot}\else${\rm L}_{\mathord\odot}$\fi}
\def\inv{\ifmmode^{-1}\else$^{-1}$\fi}
\def\mo{\ifmmode^{-1}\else$^{-1}$\fi}
\def\sup#1{\ifmmode ^{\rm #1}\else $^{\rm #1}$\fi}
\def\expo#1{\ifmmode \times 10^{#1}\else $\times 10^{#1}$\fi}
\def\,{\thinspace}
\def\lsim{\mathrel{\raise .4ex\hbox{\rlap{$<$}\lower 1.2ex\hbox{$\sim$}}}}
\def\gsim{\mathrel{\raise .4ex\hbox{\rlap{$>$}\lower 1.2ex\hbox{$\sim$}}}}

\def\simprop{\mathrel{\raise .4ex\hbox{\rlap{$\propto$}\lower 1.2ex\hbox{$\sim$}}}}
\def\deg{\ifmmode^\circ\else$^\circ$\fi}
\def\pdeg{\ifmmode $\setbox0=\hbox{$^{\circ}$}\rlap{\hskip.11\wd0 .}$^{\circ}
          \else \setbox0=\hbox{$^{\circ}$}\rlap{\hskip.11\wd0 .}$^{\circ}$\fi}
\def\arcs{\ifmmode {^{\scriptstyle\prime\prime}}
          \else $^{\scriptstyle\prime\prime}$\fi}
\def\arcm{\ifmmode {^{\scriptstyle\prime}}
          \else $^{\scriptstyle\prime}$\fi}
\newdimen\sa  \newdimen\sb
\def\parcs{\sa=.07em \sb=.03em
     \ifmmode \hbox{\rlap{.}}^{\scriptstyle\prime\kern -\sb\prime}\hbox{\kern -\sa}
     \else \rlap{.}$^{\scriptstyle\prime\kern -\sb\prime}$\kern -\sa\fi}
\def\parcm{\sa=.08em \sb=.03em
     \ifmmode \hbox{\rlap{.}\kern\sa}^{\scriptstyle\prime}\hbox{\kern-\sb}
     \else \rlap{.}\kern\sa$^{\scriptstyle\prime}$\kern-\sb\fi}
\def\ra[#1 #2 #3.#4]{#1\sup{h}#2\sup{m}#3\sup{s}\llap.#4}
\def\dec[#1 #2 #3.#4]{#1\deg#2\arcm#3\arcs\llap.#4}
\def\deco[#1 #2 #3]{#1\deg#2\arcm#3\arcs}
\def\rra[#1 #2]{#1\sup{h}#2\sup{m}}

\def\dots{\relax\ifmmode \ldots\else $\ldots$\fi}
\def\WHzsr{\ifmmode $W\,Hz\mo\,sr\mo$\else W\,Hz\mo\,sr\mo\fi}
\def\mHz{\ifmmode $\,mHz$\else \,mHz\fi}
\def\GHz{\ifmmode $\,GHz$\else \,GHz\fi}
\def\mKs{\ifmmode $\,mK\,s$^{1/2}\else \,mK\,s$^{1/2}$\fi}
\def\muKs{\ifmmode \,\mu$K\,s$^{1/2}\else \,$\mu$K\,s$^{1/2}$\fi}
\def\muKRJs{\ifmmode \,\mu$K$_{\rm RJ}$\,s$^{1/2}\else \,$\mu$K$_{\rm RJ}$\,s$^{1/2}$\fi}
\def\muKHz{\ifmmode \,\mu$K\,Hz$^{-1/2}\else \,$\mu$K\,Hz$^{-1/2}$\fi}
\def\MJysr{\ifmmode \,$MJy\,sr\mo$\else \,MJy\,sr\mo\fi}
\def\MJysrmK{\ifmmode \,$MJy\,sr\mo$\,mK$_{\rm CMB}\mo\else \,MJy\,sr\mo\,mK$_{\rm CMB}\mo$\fi}
\def\microns{\ifmmode \,\mu$m$\else \,$\mu$m\fi}

\def\muK{\ifmmode \,\mu$K$\else \,$\mu$\hbox{K}\fi}
\def\microK{\ifmmode \,\mu$K$\else \,$\mu$\hbox{K}\fi}
\def\muW{\ifmmode \,\mu$W$\else \,$\mu$\hbox{W}\fi}
\def\kms{\ifmmode $\,km\,s$^{-1}\else \,km\,s$^{-1}$\fi}
\def\kmsMpc{\ifmmode $\,\kms\,Mpc\mo$\else \,\kms\,Mpc\mo\fi}
\providecommand{\sorthelp}[1]{}

\begin{document} 

\author{\small
Planck Collaboration: P.~A.~R.~Ade\inst{84}
\and
N.~Aghanim\inst{57}
\and
H.~D.~Aller\inst{6}
\and
M.~F.~Aller\inst{6}
\and
M.~Arnaud\inst{71}
\and
J.~Aumont\inst{57}
\and
C.~Baccigalupi\inst{82}
\and
A.~J.~Banday\inst{93, 11}
\and
R.~B.~Barreiro\inst{62}
\and
N.~Bartolo\inst{29, 63}
% ...
}
\institute{\small
APC, AstroParticule et Cosmologie, Universit\'{e} Paris Diderot, CNRS/IN2P3, CEA/lrfu, Observatoire de Paris, Sorbonne Paris Cit\'{e}, 10, rue Alice Domon et L\'{e}onie Duquet, 75205 Paris Cedex 13, France\goodbreak
\and
Aalto University Mets\"{a}hovi Radio Observatory and Dept of Radio Science and Engineering, P.O. Box 13000, FI-00076 AALTO, Finland\goodbreak
\and
Aalto University Mets\"{a}hovi Radio Observatory, P.O. Box 13000, FI-00076 AALTO, Finland\goodbreak
\and
African Institute for Mathematical Sciences, 6-8 Melrose Road, Muizenberg, Cape Town, South Africa\goodbreak
\and
Agenzia Spaziale Italiana Science Data Center, Via del Politecnico snc, 00133, Roma, Italy\goodbreak
\and
Astronomy Department, University of Michigan, 830 Dennison Building, 500 Church Street, Ann Arbor, Michigan 48109-1042, U.S.A.\goodbreak
\and
Astrophysics Group, Cavendish Laboratory, University of Cambridge, J J Thomson Avenue, Cambridge CB3 0HE, U.K.\goodbreak
\and
Astrophysics \& Cosmology Research Unit, School of Mathematics, Statistics \& Computer Science, University of KwaZulu-Natal, Westville Campus, Private Bag X54001, Durban 4000, South Africa\goodbreak
\and
Atacama Large Millimeter/submillimeter Array, ALMA Santiago Central Offices, Alonso de Cordova 3107, Vitacura, Casilla 763 0355, Santiago, Chile\goodbreak
\and
CITA, University of Toronto, 60 St. George St., Toronto, ON M5S 3H8, Canada\goodbreak
\and
CNRS, IRAP, 9 Av. colonel Roche, BP 44346, F-31028 Toulouse cedex 4, France\goodbreak
\and
Cahill Center for Astronomy and Astrophysics, California Institute of Technology, Pasadena CA,  91125, USA\goodbreak
\and
California Institute of Technology, Pasadena, California, U.S.A.\goodbreak
\and
Centro de Estudios de F\'{i}sica del Cosmos de Arag\'{o}n (CEFCA), Plaza San Juan, 1, planta 2, E-44001, Teruel, Spain\goodbreak
\and
Computational Cosmology Center, Lawrence Berkeley National Laboratory, Berkeley, California, U.S.A.\goodbreak
\and
DSM/Irfu/SPP, CEA-Saclay, F-91191 Gif-sur-Yvette Cedex, France\goodbreak
\and
DTU Space, National Space Institute, Technical University of Denmark, Elektrovej 327, DK-2800 Kgs. Lyngby, Denmark\goodbreak
\and
D\'{e}partement de Physique Th\'{e}orique, Universit\'{e} de Gen\`{e}ve, 24, Quai E. Ansermet,1211 Gen\`{e}ve 4, Switzerland\goodbreak
\and
Departamento de Astrof\'{i}sica, Universidad de La Laguna (ULL), E-38206 La Laguna, Tenerife, Spain\goodbreak
\and
Departamento de F\'{\i}sica, Universidad de Oviedo, Avda. Calvo Sotelo s/n, Oviedo, Spain\goodbreak
\and
Department of Physics \& Astronomy, University of British Columbia, 6224 Agricultural Road, Vancouver, British Columbia, Canada\goodbreak
\and
Department of Physics and Astronomy, Dana and David Dornsife College of Letter, Arts and Sciences, University of Southern California, Los Angeles, CA 90089, U.S.A.\goodbreak
\and
Department of Physics and Astronomy, University College London, London WC1E 6BT, U.K.\goodbreak
\and
Department of Physics, Florida State University, Keen Physics Building, 77 Chieftan Way, Tallahassee, Florida, U.S.A.\goodbreak
\and
Department of Physics, Gustaf H\"{a}llstr\"{o}min katu 2a, University of Helsinki, Helsinki, Finland\goodbreak
\and
Department of Physics, Princeton University, Princeton, New Jersey, U.S.A.\goodbreak
\and
Department of Physics, Purdue University, 525 Northwestern Ave, West Lafayette, IN 47907, USA\goodbreak
\and
Department of Physics, University of California, Santa Barbara, California, U.S.A.\goodbreak
\and
Dipartimento di Fisica e Astronomia G. Galilei, Universit\`{a} degli Studi di Padova, via Marzolo 8, 35131 Padova, Italy\goodbreak
\and
Dipartimento di Fisica e Scienze della Terra, Universit\`{a} di Ferrara, Via Saragat 1, 44122 Ferrara, Italy\goodbreak
\and
Dipartimento di Fisica, Universit\`{a} La Sapienza, P. le A. Moro 2, Roma, Italy\goodbreak
\and
Dipartimento di Fisica, Universit\`{a} degli Studi di Milano, Via Celoria, 16, Milano, Italy\goodbreak
\and
Dipartimento di Matematica, Universit\`{a} di Roma Tor Vergata, Via della Ricerca Scientifica, 1, Roma, Italy\goodbreak
\and
Discovery Center, Niels Bohr Institute, Blegdamsvej 17, Copenhagen, Denmark\goodbreak
\and
European Southern Observatory, ESO Vitacura, Alonso de Cordova 3107, Vitacura, Casilla 19001, Santiago, Chile\goodbreak
\and
European Space Agency, ESAC, Planck Science Office, Camino bajo del Castillo, s/n, Urbanizaci\'{o}n Villafranca del Castillo, Villanueva de la Ca\~{n}ada, Madrid, Spain\goodbreak
\and
European Space Agency, ESTEC, Keplerlaan 1, 2201 AZ Noordwijk, The Netherlands\goodbreak
\and
Facolt\`{a} di Ingegneria, Universit\`{a} degli Studi e-Campus, Via Isimbardi 10, Novedrate (CO), 22060, Italy\goodbreak
\and
Finnish Centre for Astronomy with ESO (FINCA), University of Turku, V\"{a}is\"{a}l\"{a}ntie 20, FIN-21500, Piikki\"{o}, Finland\goodbreak
\and
Gran Sasso Science Institute, INFN, viale F. Crispi 7, 67100 L'Aquila, Italy\goodbreak
\and
HGSFP and University of Heidelberg, Theoretical Physics Department, Philosophenweg 16, 69120, Heidelberg, Germany\goodbreak
\and
Harvard-Smithsonian Center for Astrophysics, Cambridge, MA 02138 USA\goodbreak
\and
Haverford College Astronomy Department, 370 Lancaster Avenue, Haverford, Pennsylvania, U.S.A.\goodbreak
\and
Helsinki Institute of Physics, Gustaf H\"{a}llstr\"{o}min katu 2, University of Helsinki, Helsinki, Finland\goodbreak
\and
INAF - Osservatorio Astronomico di Padova, Vicolo dell'Osservatorio 5, Padova, Italy\goodbreak
\and
INAF - Osservatorio Astronomico di Roma, via di Frascati 33, Monte Porzio Catone, Italy\goodbreak
\and
INAF - Osservatorio Astronomico di Trieste, Via G.B. Tiepolo 11, Trieste, Italy\goodbreak
\and
INAF/IASF Bologna, Via Gobetti 101, Bologna, Italy\goodbreak
\and
INAF/IASF Milano, Via E. Bassini 15, Milano, Italy\goodbreak
\and
INFN, Sezione di Bologna, Via Irnerio 46, I-40126, Bologna, Italy\goodbreak
\and
INFN, Sezione di Roma 1, Universit\`{a} di Roma Sapienza, Piazzale Aldo Moro 2, 00185, Roma, Italy\goodbreak
\and
INFN, Sezione di Roma 2, Universit\`{a} di Roma Tor Vergata, Via della Ricerca Scientifica, 1, Roma, Italy\goodbreak
\and
ISDC, Department of Astronomy, University of Geneva, ch. d'Ecogia 16, 1290 Versoix, Switzerland\goodbreak
\and
Imperial College London, Astrophysics group, Blackett Laboratory, Prince Consort Road, London, SW7 2AZ, U.K.\goodbreak
\and
Infrared Processing and Analysis Center, California Institute of Technology, Pasadena, CA 91125, U.S.A.\goodbreak
\and
Institut Universitaire de France, 103, bd Saint-Michel, 75005, Paris, France\goodbreak
\and
Institut d'Astrophysique Spatiale, CNRS (UMR8617) Universit\'{e} Paris-Sud 11, B\^{a}timent 121, Orsay, France\goodbreak
\and
Institut d'Astrophysique de Paris, CNRS (UMR7095), 98 bis Boulevard Arago, F-75014, Paris, France\goodbreak
\and
Institute of Astronomy, University of Cambridge, Madingley Road, Cambridge CB3 0HA, U.K.\goodbreak
\and
Institute of Theoretical Astrophysics, University of Oslo, Blindern, Oslo, Norway\goodbreak
\and
Instituto de Astrof\'{\i}sica de Canarias, C/V\'{\i}a L\'{a}ctea s/n, La Laguna, Tenerife, Spain\goodbreak
\and
Instituto de F\'{\i}sica de Cantabria (CSIC-Universidad de Cantabria), Avda. de los Castros s/n, Santander, Spain\goodbreak
\and
Istituto Nazionale di Fisica Nucleare, Sezione di Padova, via Marzolo 8, I-35131 Padova, Italy\goodbreak
\and
Jet Propulsion Laboratory, California Institute of Technology, 4800 Oak Grove Drive, Pasadena, California, U.S.A.\goodbreak
\and
Jodrell Bank Centre for Astrophysics, Alan Turing Building, School of Physics and Astronomy, The University of Manchester, Oxford Road, Manchester, M13 9PL, U.K.\goodbreak
\and
Kavli Institute for Cosmological Physics, University of Chicago, Chicago, IL 60637, USA\goodbreak
\and
Kavli Institute for Cosmology Cambridge, Madingley Road, Cambridge, CB3 0HA, U.K.\goodbreak
\and
Kazan Federal University, 18 Kremlyovskaya St., Kazan, 420008, Russia\goodbreak
\and
LAL, Universit\'{e} Paris-Sud, CNRS/IN2P3, Orsay, France\goodbreak
\and
LERMA, CNRS, Observatoire de Paris, 61 Avenue de l'Observatoire, Paris, France\goodbreak
\and
Laboratoire AIM, IRFU/Service d'Astrophysique - CEA/DSM - CNRS - Universit\'{e} Paris Diderot, B\^{a}t. 709, CEA-Saclay, F-91191 Gif-sur-Yvette Cedex, France\goodbreak
\and
Laboratoire de Physique Subatomique et Cosmologie, Universit\'{e} Grenoble-Alpes, CNRS/IN2P3, 53, rue des Martyrs, 38026 Grenoble Cedex, France\goodbreak
\and
Laboratoire de Physique Th\'{e}orique, Universit\'{e} Paris-Sud 11 \& CNRS, B\^{a}timent 210, 91405 Orsay, France\goodbreak
\and
Lawrence Berkeley National Laboratory, Berkeley, California, U.S.A.\goodbreak
\and
Lebedev Physical Institute of the Russian Academy of Sciences, Astro Space Centre, 84/32 Profsoyuznaya st., Moscow, GSP-7, 117997, Russia\goodbreak
\and
Max-Planck-Institut f\"{u}r Astrophysik, Karl-Schwarzschild-Str. 1, 85741 Garching, Germany\goodbreak
\and
National Radio Astronomy Observatory, P.O. Box 0, Socorro, NM 87801, USA\goodbreak
\and
National University of Ireland, Department of Experimental Physics, Maynooth, Co. Kildare, Ireland\goodbreak
\and
Nicolaus Copernicus Astronomical Center, Bartycka 18, 00-716 Warsaw, Poland\goodbreak
\and
Niels Bohr Institute, Blegdamsvej 17, Copenhagen, Denmark\goodbreak
\and
Optical Science Laboratory, University College London, Gower Street, London, U.K.\goodbreak
\and
SISSA, Astrophysics Sector, via Bonomea 265, 34136, Trieste, Italy\goodbreak
\and
School of Mathematics and Physics, University of Tasmania, Private Bag 37, Hobart, Australia, TAS 7001\goodbreak
\and
School of Physics and Astronomy, Cardiff University, Queens Buildings, The Parade, Cardiff, CF24 3AA, U.K.\goodbreak
\and
Sorbonne Universit\'{e}-UPMC, UMR7095, Institut d'Astrophysique de Paris, 98 bis Boulevard Arago, F-75014, Paris, France\goodbreak
\and
Space Research Institute (IKI), Russian Academy of Sciences, Profsoyuznaya Str, 84/32, Moscow, 117997, Russia\goodbreak
\and
Space Science Institute, 4750 Walnut Street, Suite 205, Boulder, CO 80301, U.S.A.\goodbreak
\and
Space Sciences Laboratory, University of California, Berkeley, California, U.S.A.\goodbreak
\and
Special Astrophysical Observatory, Russian Academy of Sciences, Nizhnij Arkhyz, Zelenchukskiy region, Karachai-Cherkessian Republic, 369167, Russia\goodbreak
\and
Sub-Department of Astrophysics, University of Oxford, Keble Road, Oxford OX1 3RH, U.K.\goodbreak
\and
Tuorla Observatory, Department of Physics and Astronomy, University of Turku, V\"ais\"al\"antie 20, FIN-21500, Piikki\"o, Finland\goodbreak
\and
UPMC Univ Paris 06, UMR7095, 98 bis Boulevard Arago, F-75014, Paris, France\goodbreak
\and
Universit\'{e} de Toulouse, UPS-OMP, IRAP, F-31028 Toulouse cedex 4, France\goodbreak
% ...
}

   \title{\Planck\ intermediate results. XLV. Radio spectra of northern extragalactic radio sources}

   \date{}


  \abstract{Continuum spectra covering centimetre to submillimetre wavelengths are presented for a northern sample of 104~extragalactic radio sources, mainly active galactic nuclei, based on four-epoch \Planck\ data. The nine \Planck\ frequencies, from 30 to 857\,GHz, are complemented by a set of simultaneous ground-based radio observations between 1.1 and 37\,GHz. The single-survey \Planck\ data confirm that the flattest high-frequency radio spectral indices are close to zero, indicating that the original accelerated electron energy spectrum is much harder than commonly thought, with power-law index around 1.5 instead of the canonical 2.5. The radio spectra peak at high frequencies and exhibit a variety of shapes. For a small set of low-$z$ sources, we find a spectral upturn at high frequencies, indicating the presence of intrinsic cold dust.  Variability can generally be approximated by achromatic variations, while sources with clear signatures of evolving shocks appear to be limited to the strongest outbursts.}


\titlerunning{Planck spectra of extragalactic radio sources}

\authorrunning{Planck Collaboration}

   \maketitle
%

\section{Introduction}

In radio-loud active galactic nuclei (AGN), jets of relativistic matter emerge symmetrically from the core. 
Non-thermal radiation produced in the jets, rather than thermal emission from the accretion disk, dominates the spectral energy distributions (SEDs) of these kinds of sources. The SEDs typically consist of two broad-band bumps, the one at lower frequencies associated with synchrotron radiation, and the other at higher frequencies with inverse Compton (IC) radiation. There are usually several emission components simultaneously occurring in the jet, in addition to the quiescent jet. The ``shock-in-jet'' model \citep{marscher85, valtaoja92_moniIII} describes the flow of plasma with shocks that locally enhance the emission.

In \citet{aatrokoski11}, hereafter Paper I, we used data from the Planck Early Release Compact Source Catalogue (ERCSC)\footnote{\Planck\ (\url{http://www.esa.int/Planck}) is a project of the European Space Agency (ESA) with instruments provided by two scientific consortia funded by ESA member states and led by Principal Investigators from France and Italy, telescope reflectors provided through a collaboration between ESA and a scientific consortium led and funded by Denmark. and additional contributions from NASA (USA).} to plot the radio spectra and spectral energy distributions for a sample of 104 radio-bright, northern AGN. 
Data have also been acquired by a large collaboration of observatories from radio to gamma-ray frequencies. The near-simultaneous spectra were determined with good accuracy. 

In Paper I, we reported that the majority of high-radio-frequency spectral indices were surprisingly flat for synchrotron sources. The canonical optically-thin synchrotron spectral index is $-0.7$, while for many sources the ERCSC data gave a spectral index between $-0.2$ and $-0.4$. The distribution was compatible with the electron energy spectral index $s=1.5$, which is clearly harder than the assumed $s=2.5$. 

In this paper, we study the spectral index distribution with new single-epoch data from four \Planck\ all-sky Surveys (the definitions of which are given in \citealt{planck2013-p01}.) Throughout the paper, we adopt the convention $S_\nu \propto \nu^{\alpha}$, where $S_\nu$ is the flux density and $\alpha$ is the spectral index.  The errors of numerical values given with a ``$\pm$'' correspond to one standard deviation.

\section{Sample}   
\label{section_sample}

The complete sample presented in this paper consists of 104 northern and equatorial radio-loud AGN. It includes all AGN with declination $\geq-10^{\circ}$ that have a measured average radio flux density at 37\,GHz exceeding 1\,Jy. Most of the sample sources have been monitored at Mets\"ahovi Radio Observatory for many years, and the brightest sources have been observed for up to 30 years. This sample forms the core of Mets\"ahovi's \Planck-related observing programme, in which we aimed to observe the sources as simultaneously as possible and at least within two weeks. We did this by using the \Planck\ On-Flight Forecaster tool \citep[{\tt POFF},][]{massardi10} to predict when the sources were visible at each of the satellite's frequencies. We also provided the information to our multifrequency collaborators. In addition to this sample of bright sources, large samples of fainter sources were observed at Mets\"ahovi in support 
of \Planck\ \citep[e.g.,][]{torniainen05, nieppola07}.

The sample can be divided into subclasses as follows: 40 high-polarized quasars (HPQ); 14 low-polarized quasars (LPQ); 24 BL Lacertae objects (BLO); 17 quasi-stellar objects (QSO); eight radio galaxies (GAL); and one unclassified source (J184915+67064). We classify highly-polarized quasars as objects that have a measured optical polarization $\geq3\%$ in the literature, while low-polarized quasars have a measured polarization $<3\%$. Most of the quasars have no polarization information so they could be either HPQs or LPQs. Radio galaxies are non-quasar AGN. 

The full sample is listed in Table~\ref{sample}, presented at the end of the paper. Columns 1 and 2 give the B1950 name for the source and an alias. The coordinates of the sources are given in Cols. 4 and 5, and Col. 6 lists the redshift.

\section{Data}

We use data within the frequency range 1.1--857\,GHz. \Planck\ frequencies and the radio observatories that participated in simultaneous multifrequency campaigns with \Planck\ are listed in Table~\ref{observatories}. Supporting observations were taken within two weeks of the \Planck\ scans. Archival radio data were retrieved from the database maintained by Mets\"ahovi Radio Observatory.

For non-\Planck\ data we imposed a signal-to-noise (S/N) constraint of 4, while for \Planck\ data we applied the S/N threshold given in table~1 of \citet{Planck_compcat2013}.

\setcounter{table}{1}
\begin{table}[ht!]
\caption{Participating observatories and their observing frequencies.}
\label{observatories}
\vskip -6mm
\footnotesize
\setbox\tablebox=\vbox{
 \newdimen\digitwidth
 \setbox0=\hbox{\rm 0}
 \digitwidth=\wd0
 \catcode`*=\active
 \def*{\kern\digitwidth}
  \newdimen\dpwidth
  \setbox0=\hbox{.}
  \dpwidth=\wd0
  \catcode`!=\active
  \def!{\kern\dpwidth}
\halign{\hbox to 3.5cm{#\leaderfil}\tabskip 1.5em&
    #\hfil \tabskip 0em\cr
\noalign{\doubleline}
\omit\hfil Observatory\hfil&\omit\hfil Frequencies [GHz]\hfil\cr
\noalign{\vskip 3pt\hrule\vskip 4pt}
\Planck\ LFI& 30, 44, 70\cr
\Planck\ HFI& 100, 143, 217, 353, 545, 857\cr
Mets\"ahovi, Finland& 37\cr
OVRO, USA& 15\cr
RATAN-600, Russia& 1.1, 2.3, 4.8, 7.7, 11.2, 21.7\cr
UMRAO, USA& 4.8, 8.0, 14.5\cr
\noalign{\vskip 4pt\hrule}}}
\endPlancktable
\end{table}

\subsection{\Planck\ data}
\label{planckdata}

Details of the characteristics of the \Planck\ Low Frequency Instrument (LFI) and High Frequency Instrument (HFI) data and processing pipelines are to be found in \citet{planck2014-a03} and \citet{planck2014-a08}, respectively. \Planck\ observed the whole sky twice in one year \citep{planck2013-p03}, therefore each source was typically observed every six months.  The visibility period was usually several days, but depended on frequency and ecliptic latitude.  We consider two weeks as a single pointing.  Given the scanning strategy of the satellite, sources close to the ecliptic poles were observed more often and over a period of up to several weeks. For a small subset of sources, therefore, the \Planck\ flux densities used in this paper are averages of several pointings over the source's visibility period in one Survey. The longest visibility period is approximately two months (seven sources). The start and end times of the visibility periods for all sources are shown in Table~\ref{obstime}, and the number of pointings within each Survey, for each frequency for sources with multiple pointings, is shown in Table~\ref{pointings} (both tables are published at the end of the paper.). These are all calculated with {\tt POFF}. 
 
The \Planck\ flux densities used in this paper have been extracted from the full mission maps from the 2015 data release using the Mexican Hat Wavelet~2 source detection and flux density estimation pipelines in the \Planck\ LFI and HFI Data Processing Centres. For LFI, data detection pipeline photometry (DETFLUX) was used, while for HFI, aperture photometry (APERFLUX) was used. The calibration of \Planck\ is based on the dipole signal, and is consistent at approximately the 0.2\% level \citep{planck2014-a01}. The systematic uncertainties of the absolute flux density scale are under 1\% for the seven lowest frequencies and under 7\% for the two highest. The overall uncertainties of the flux densities vary between 30 and 270\,mJy, depending on frequency. See the Second \Planck\ Catalogue of Compact Sources \citep{planck2014-a35} and the references therein for further details.

\subsection{Radio and submillimetre data}

Six-frequency broadband radio spectra for 29 sources were obtained with the RATAN-600 radio telescope in transit mode by observing simultaneously at 1.1, 2.3, 4.8, 7.7, 11.2, and 21.7\,GHz 
\citep{parijskij93}. The parameters of the receivers are listed in Table~\ref{ratan}, where $\lambda$ is the wavelength, $f_{\rm c}$ is the central frequency, $\Delta f$ is the bandwidth, 
$\Delta F$ is the flux density detection limit per beam, and BW is the beam width (full width at half-maximum in right ascension). The detection limit for the RATAN single sector is 
approximately 8\,mJy (over a 
3\,s integration) under good conditions at the frequency of 4.8\,GHz and at an average antenna elevation of $42^{\circ}$. Data were reduced using the RATAN standard software 
FADPS (Flexible Astronomical Data Processing System) reduction package \citep{verkhodanov97b}. The flux density measurement procedure is described by \citet{mingaliev01,mingaliev12}. We 
use the data acquisition and control system for all continuum radiometers, as described by \citet{tsybulev11}. The following flux density calibrators were applied to obtain the 
calibration coefficients in the scale by \citet{baars77}: 3C~48, 3C~147, 3C~161, 3C~286, 3C~295, 3C~309.1, and NGC~7027. The measurements of some of the calibrators were corrected for 
angular size and linear polarization following the data from \citet{ott94} and \citet{tabara80}. In addition, the traditional RATAN flux density calibrators J0237$-$23, 3C~138, 
J1154$-$35, and J1347$+$12 were used. The total error in the flux density includes the uncertainty of the RATAN calibration curve and the error in the antenna temperature measurement. The systematic uncertainty of the absolute flux density scale (3--10{\%} at different RATAN frequencies) is also included in the flux density error.  

\setcounter{table}{4}
\begin{table}[ht!]
\caption{Parameters of the RATAN-600 antenna and radiometers.}
\label{ratan}
\vskip -6mm
\footnotesize
\setbox\tablebox=\vbox{
 \newdimen\digitwidth
 \setbox0=\hbox{\rm 0}
 \digitwidth=\wd0
 \catcode`*=\active
 \def*{\kern\digitwidth}
  \newdimen\signwidth
  \setbox0=\hbox{+}
  \signwidth=\wd0
  \catcode`!=\active
  \def!{\kern\signwidth}
\halign{\hbox to 1.8cm{#\leaderfil}\tabskip 2.0em&
    \hfil#\hfil&
    \hfil#\hfil\tabskip 0.5em&
    \hfil#\hfil&
    \hfil#\hfil\tabskip 0em\cr
\noalign{\doubleline}
\omit\hfil $\lambda$\hfil&$f_{\rm c}$&$\Delta f$&$\Delta F$&BW\cr
\omit\hfil [cm]\hfil&[GHz]&[GHz]&[mJy\,beam\mo]&[arcsec]\cr
\noalign{\vskip 3pt\hrule\vskip 4pt}
1.4&  21.7& 2.5*& *70& *11\cr
2.7&  11.2& 1.4*& *20& *16\cr
3.9&  *7.7& 1.0*& *25& *22\cr
6.3&  *4.8& 0.9*& **8& *36\cr
13.0& *2.3& 0.4*& *30& *80\cr
31.2& *1.1& 0.12& 160& 170\cr
\noalign{\vskip 4pt\hrule\vskip 5pt}}}
\endPlancktable
\tablefoot{Col.~1~-- wavelength; Col.~2~-- central frequency; Col.~3~-- bandwidth; Col.~4~-- flux density detection limit per beam; Col.~5~-- beam width.}
\end{table}

Centimetre-band observations were obtained with the University of Michigan 26-m paraboloid. This telescope was equipped with transistor-based radiometers operating at three primary frequencies centred at 4.8, 8.0, and 14.5\,GHz, with bandwidths of 0.68, 0.79, and 1.68\,GHz, respectively. Dual-horn feed systems were used at 8.0 and 14.5\,GHz, while at 4.8\,GHz a single-horn mode-switching receiver was used. Each observation comprised a series of 8 to 16 individual measurements obtained over a 25 to 45 minute time period using an on--off observing procedure at 4.8\,GHz and an on--on technique  (switching the target source between two feed horns closely spaced on the sky) at 8.0 and 14.5\,GHz. Drift scans were made across strong sources to verify the pointing correction curves, and observations of nearby calibrators selected from a grid were obtained every 1--2 hours to correct for temporal changes in the antenna aperture efficiency. The adopted flux density scale was based on \citet{baars77}, and used Cas A as the primary standard. A systematic uncertainty of 4\% in the flux density scale is included in the error bars.

The 15\,GHz observations were carried out as part of a high-cadence gamma-ray blazar monitoring programme using the Owens Valley Radio Observatory (OVRO) 40-m telescope \citep{richards11}. This programme, which commenced in late 2007, now includes about 1800 sources, each observed with a nominal twice-per-week cadence. The OVRO 40-m uses off-axis dual-beam optics and a cryogenic high electron mobility transistor (HEMT) low-noise amplifier, with a 15.0\,GHz centre frequency and 3\,GHz bandwidth. The two sky beams are Dicke-switched using the off-source beam as a reference, and the source is alternated between the two beams in an on--on fashion to remove atmospheric and ground contamination. 
Calibration is achieved using a temperature-stable diode noise source to remove receiver gain drifts, and the flux density scale is derived from observations of 3C~286 
assuming the \citet{baars77} value of 3.44\,Jy at 15.0\,GHz. The systematic uncertainty of about 5\% in the flux density scale is included in the error bars. Complete details of the reduction and calibration procedure can be found in \citet{richards11}.

The 37\,GHz observations were made with the 13.7-m Mets\"ahovi radio telescope using a 1\,GHz bandwidth, dual-beam receiver centred at 
36.8\,GHz. The observations were on-on observations, alternating the source and the sky in each feed horn. The integration time used to obtain each flux density data point typically ranged from 1200 to 1400\,s. The detection limit of the telescope at 37\,GHz is approximately 0.2\,Jy under optimal conditions. Data points with ${\rm S/N} < 4$ were handled as non-detections. The flux density scale was based on \citet{baars77}, and was set by observations of DR~21. Sources NGC~7027, 3C~274, and 3C~84 were used as secondary calibrators. A detailed description of the data reduction and analysis is given in \citet{terasranta98}. The error estimate in the flux density includes the contribution from the measurement rms and the uncertainty of the absolute calibration.

The flux density scale based on \citet{baars77} used in the aforementioned observations agrees reasonably well with the new scale proposed by \citet{perleybutler13}. The ratio of the two scales at the frequencies used is very close to 1 \citep[see table 13 in][]{perleybutler13}, and the differences are small compared to the measurement errors.

\section{Statistics}

\subsection{Spectral indices}

\label{spectralindices}

The near-simultaneous radio spectra of the sources in the sample for each observing epoch are shown in Figs.~\ref{0003-066_spectra}--\ref{2353+816_spectra}, presented at the end of the paper. Historical data, showing the range of variability for each source, are also included in the figures. In Paper I the radio to submillimetre spectra were fitted with two power laws to characterize the lower and higher frequency (LF and HF, respectively) part of the spectrum. As a better approximation of the spectra, 
\citet{LeonTavares2012} fitted them using a broken power-law model,  
%
%
\begin{equation}
	S(\nu) \propto \begin{cases}
		\nu^{\alpha_{\rm LF}} & \quad \text{for $\nu \leq \nu_{\rm break}$,}\\
		\nu_{\rm break}^{(\alpha_{\rm LF} - \alpha_{\rm HF})} \, \nu^{\alpha_{\rm HF}} & \quad \text{for $\nu > \nu_{\rm break}$,}
	\end{cases}
	\label{eq1}
\end{equation}
where $\alpha_{\rm LF}$ and $\alpha_{\rm HF}$ are the spectral indices for the LF and HF parts of the spectrum, and $\nu_{\rm break}$ is the break frequency. We apply the same model to the spectra of 104 sources obtained from the four \Planck\ Surveys used in this paper. In order to characterize the spectra better, we consider only those sources observed at five or more frequencies during one Survey. The goodness of fit was tested based on a $\chi^2$ distribution at 95\% confidence level. Thus, we obtained acceptable estimates for 62, 60, 58, and 51 sources in Surveys 1, 2, 3, and 4, respectively.

From the fits we have calculated the LF and HF spectral indices. The break frequencies that separate the low and high frequencies have been determined from the fit. All parameters for all sample sources are available in Table~\ref{fit_par}, presented at the end of the paper. Columns 1 and 2 give the source name and class. The spectral indices and break frequencies for each of the four Surveys are given in Columns 3, 4, 5, and 6.
Figures~\ref{hf_index}, \ref{lf_index}, and \ref{brk_freq} show the distributions of the LF and HF spectral indices, and the break frequencies, respectively, for all four epochs. Some sources show a contribution of dust in their high frequency spectra, suggested by an upturn at the highest or two highest \Planck\ frequencies in visual inspection (see also Sect.~\ref{risinghfspectra}). Unfortunately the large error bars at the highest frequencies make quantifying the significance of the spectral upturns difficult.
While for some sources the dust imprint seems 
unmistakable (see Table~\ref{dusty_sources}), up to a quarter of our sources may be suspected of dust contamination. For example, the spectrum of the first source in our sample, 0003$-$066, flattens at the two highest frequencies, and the flux during the four epochs is constant within the error bars, consistent with a thermal origin.
We have separated the evidently ``dusty'' sources in Table~\ref{dusty_sources} from others in the relevant calculations and figures. 

\begin{figure}
	\centering
	\includegraphics[width=\hsize]{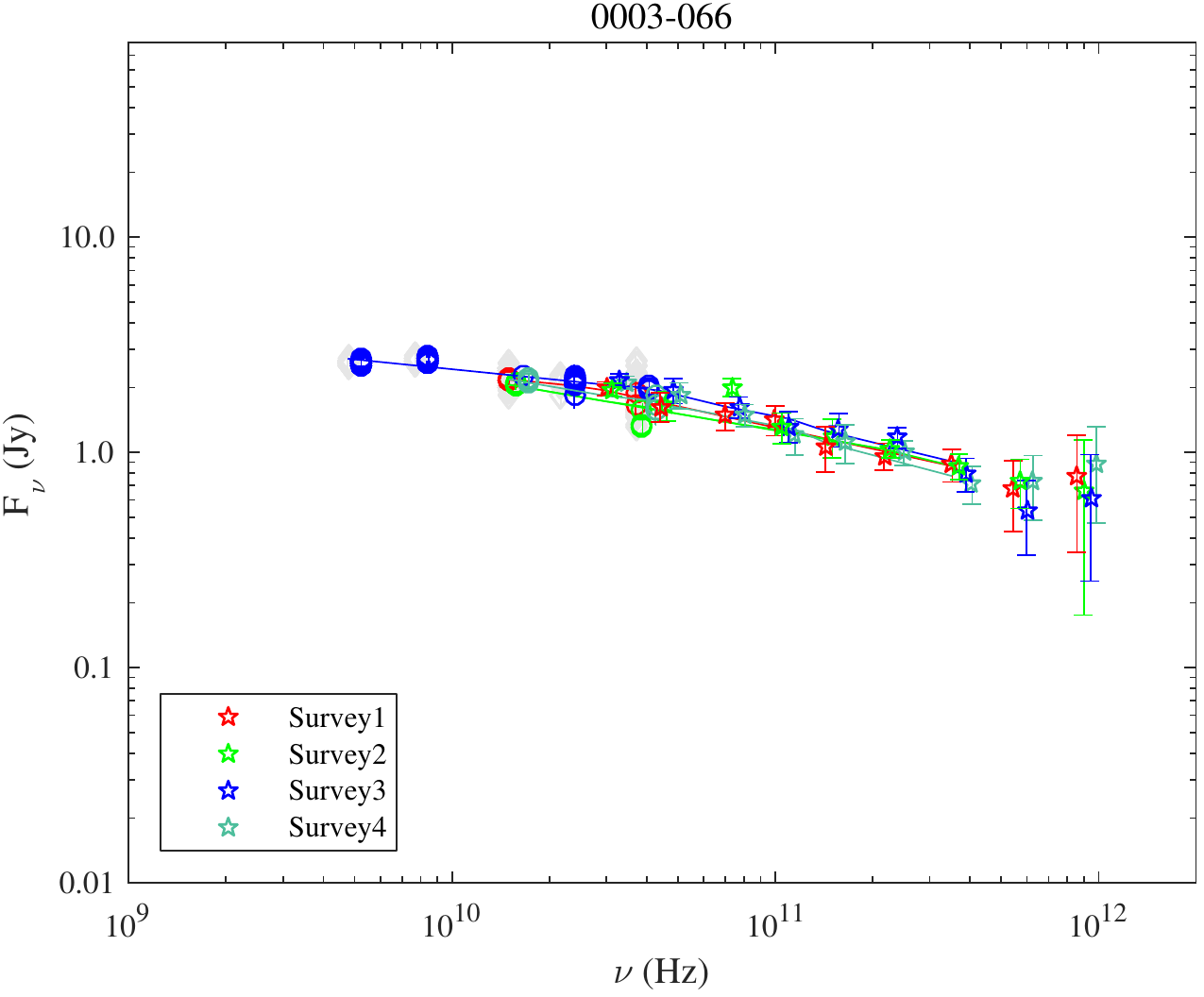}
	\caption{Radio spectrum of 0003$-$066: coloured stars, \Planck\ data from four Surveys; coloured circles, data simultaneous to the \Planck\ observations; grey circles, historical data; solid lines, broken power-law fits. The entire set of radio spectra for all 104 sources is shown in Figs.~\ref{0003-066_spectra}--\ref{2353+816_spectra} at the end of the paper.}
	\label{0003-066-spectra}
\end{figure}

The average spectral index and break frequency values, and their standard deviations of the distribution for sources with no dust and the dusty sources are 
shown in Figs.~\ref{hf_index} and \ref{brk_freq}. However, with the present data it is not possible to exclude with certainty the alternative non-thermal explanation, nor can we differentiate between Galactic
foreground cirrus contamination and possible intrinsic cold dust. The issue is discussed in more detail in Sect.~\ref{risinghfspectra}.
 Table~\ref{spec_par} lists the average spectral indices and break frequencies for each AGN subclass. The dusty sources have been excluded from this table. As a consequence, the number of GAL sources  was reduced to only four, and they were therefore left out of the table. The average low and high frequency indices of all Surveys of the GAL class are 0.337 and $-0.617$, respectively, while the average break frequency is 33.4\,GHz.

The break frequencies of the spectra in Table~\ref{spec_par} (see also Fig.~\ref{brk_freq} for the distribution) are remarkably high, tens of gigahertz. Practically all of the sources show strong variability, at least in the historical long-term flux density data, so none of them adhere to the original definition of a Gigahertz peaked-spectrum (GPS) source \citep[][and references therein]{odea98}. For many of them the turnover frequency also changes from epoch to epoch, due to the various phases of activity.

The low frequency spectral indices are close to zero, as expected. The high frequency indices are approximately $-0.5$. The abundance of flat high frequency radio spectra we reported in Paper I can also be seen in these distributions. On average 42~\% of the calculated high frequency indices are flatter than $-0.5$, while the spectral index of the canonical optically thin synchrotron spectrum is $-0.7$. 

In Paper I we argued that the most likely explanation for the flat high frequency spectra is a quite hard intrinsic electron energy spectrum with a power-law index well below the canonical value of approximately 2.5. As Fig.~\ref{hf_index} shows, the spectral index distributions do not range from $-0.7$ to $-1.2$, as would be expected from the canonical electron energy spectrum with consequent spectral steepening by energy losses; rather, the range of observed high frequency indices is from around $-0.2$ to $-0.7$, which would be expected from an electron energy spectrum index of approximately 1.5.

\begin{figure*}
   \centering
   \includegraphics[width=\hsize]{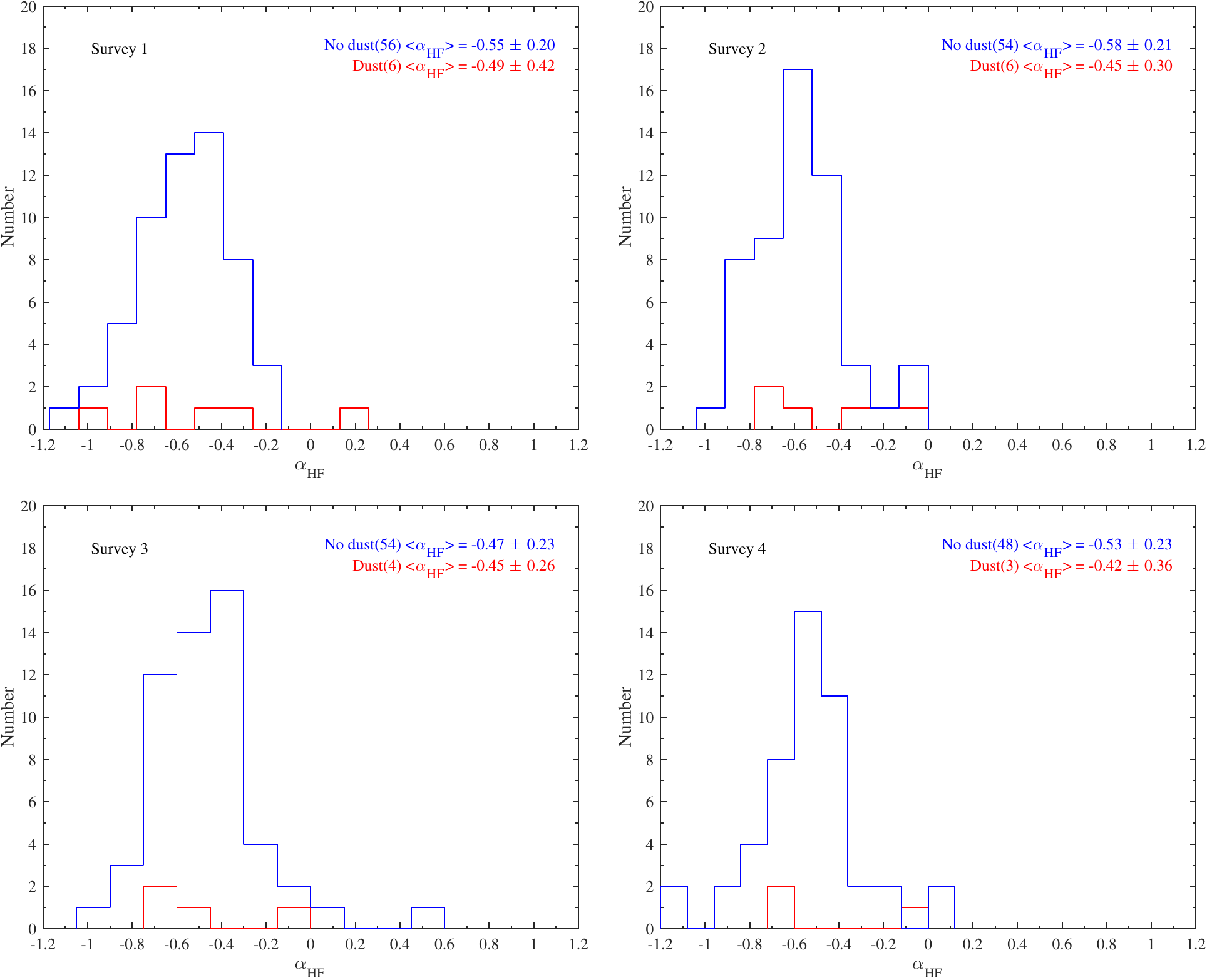}
      \caption{Distributions of the high frequency spectral indices for four \Planck\ Surveys: red, dusty sources; blue, sources with no dust (see Sect.~\ref{spectralindices} for details).}
         \label{hf_index}
   \end{figure*}

\begin{figure*}
   \centering
   \includegraphics[width=\hsize]{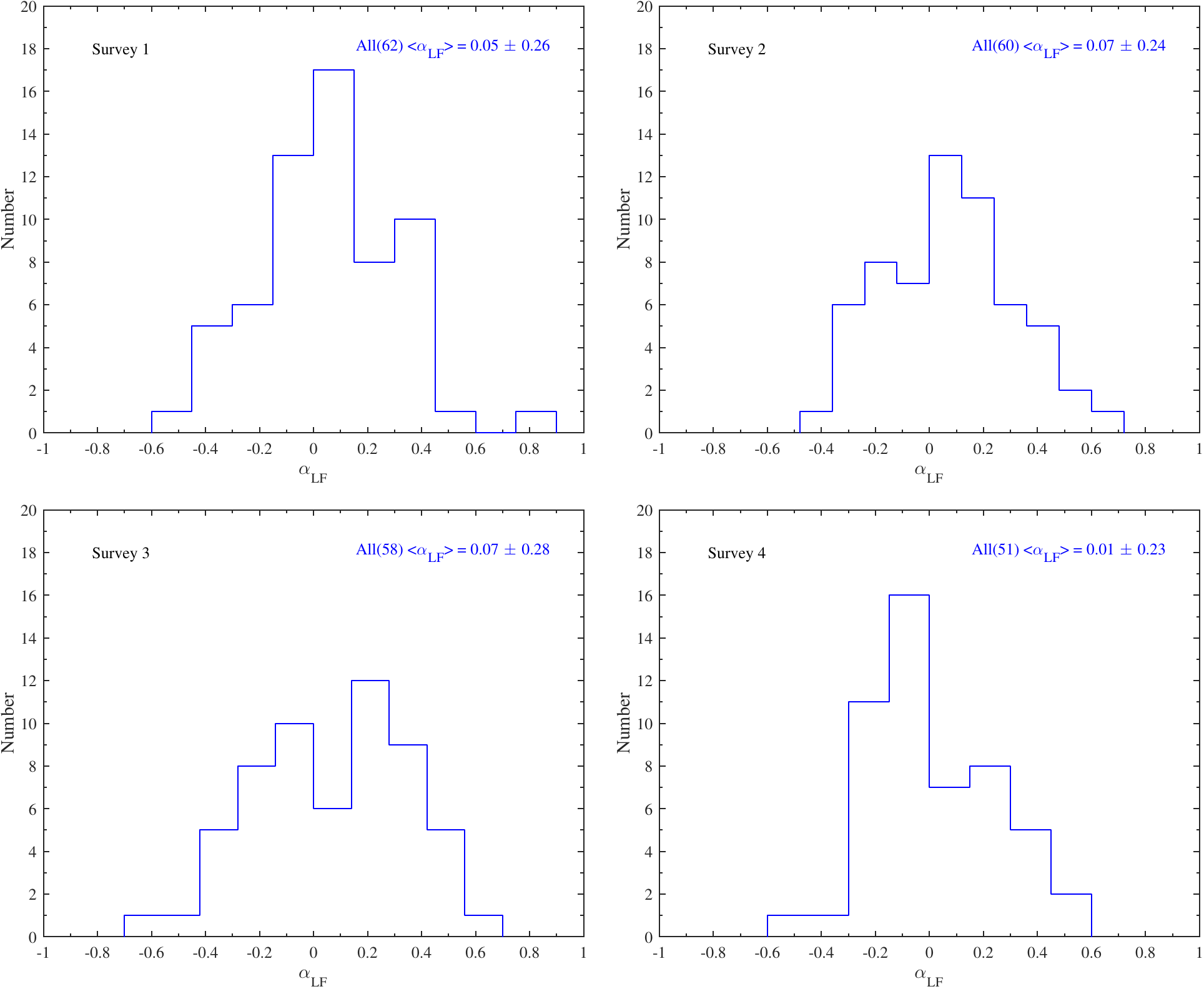}
      \caption{Distributions of the low frequency spectral indices for four \Planck\ Surveys.}
         \label{lf_index}
   \end{figure*}

\begin{figure*}
   \centering
   \includegraphics[width=\hsize]{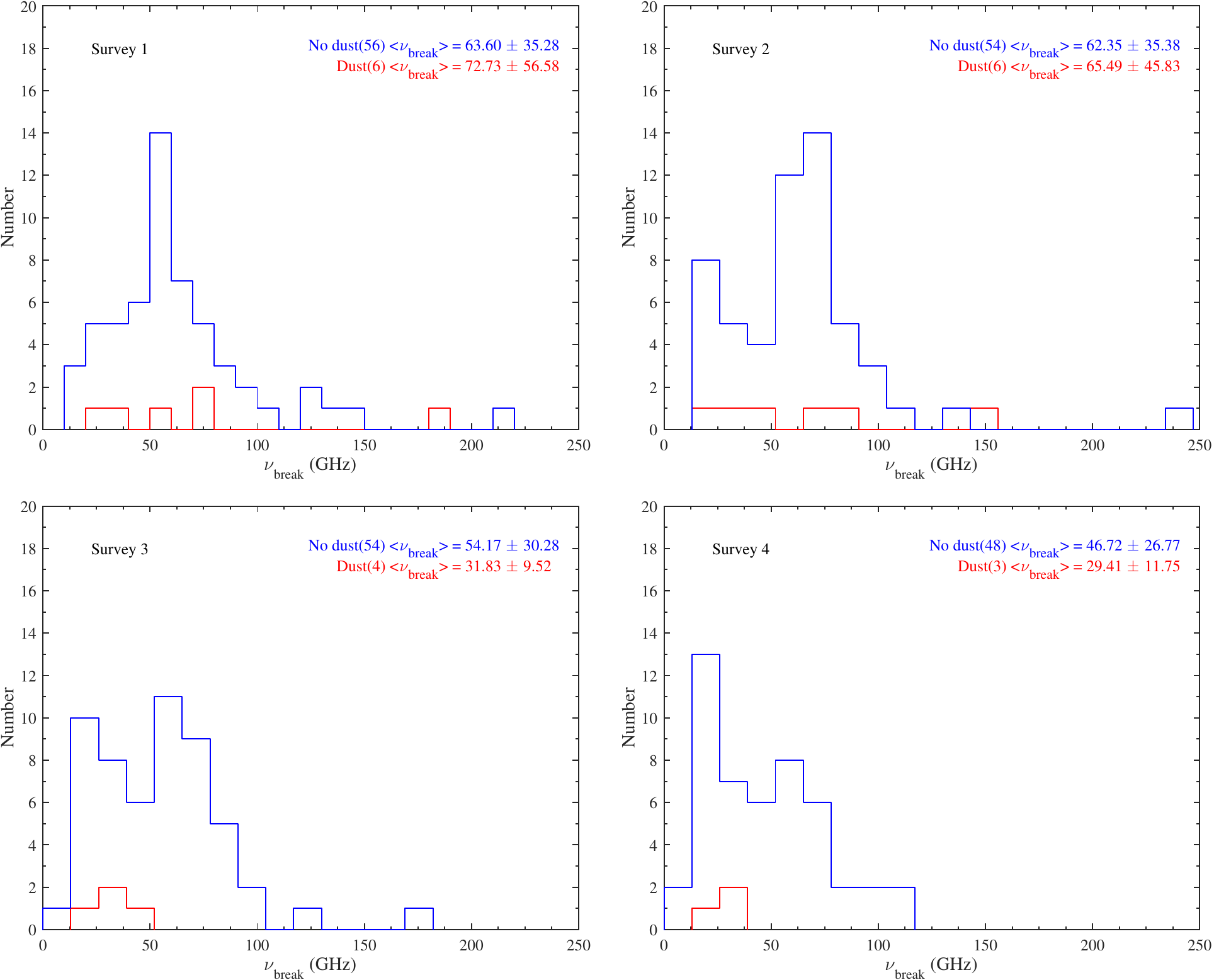}
      \caption{Distributions of the break frequencies for four \Planck\ Surveys: red, dusty sources; blue, sources with no dust (see Sect.~\ref{spectralindices} for details). }
         \label{brk_freq}
   \end{figure*}

\begin{figure*}
   \centering
   \includegraphics[width=\hsize]{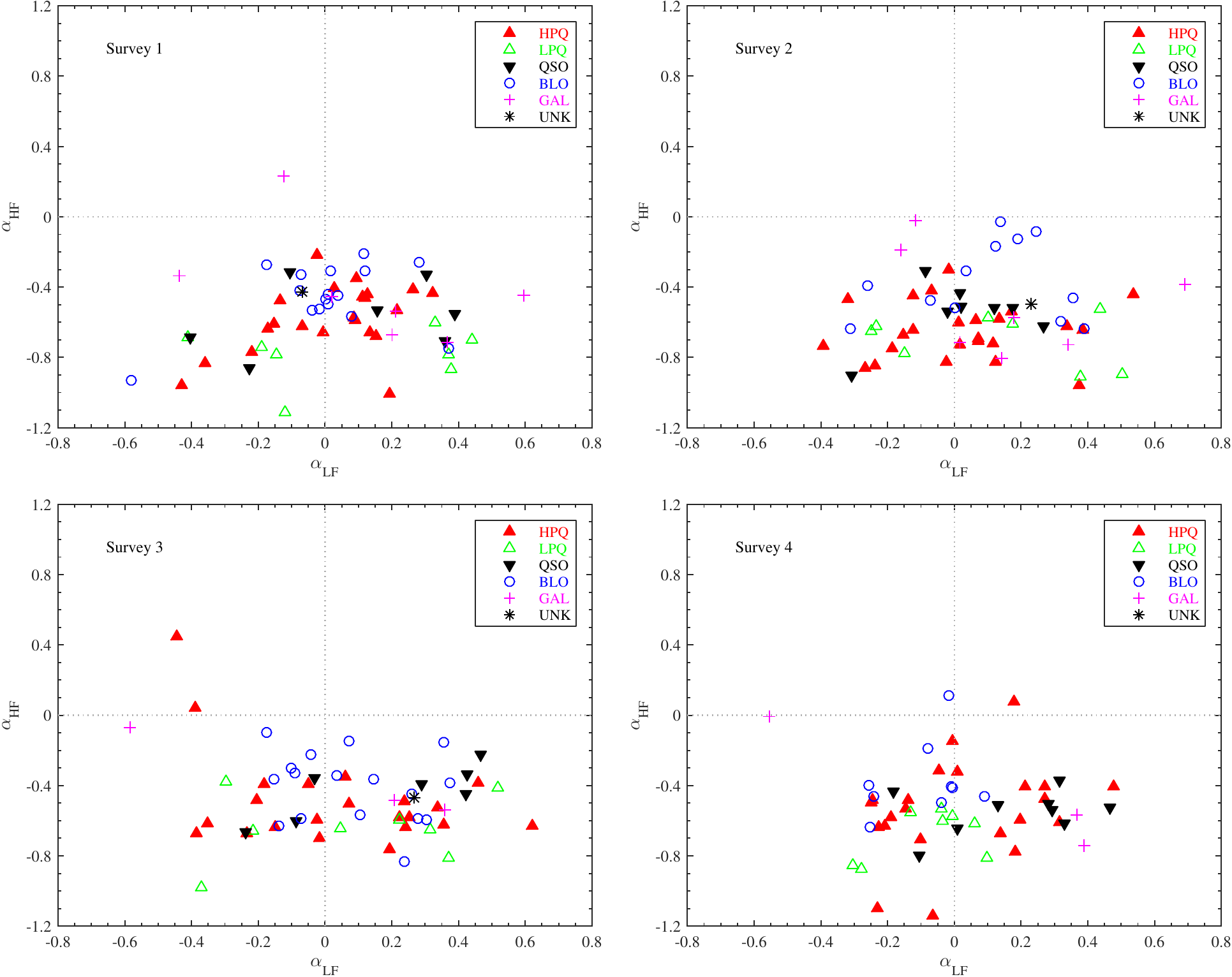}
      \caption{Correlations of the LF and HF spectral indices for four \Planck\ Surveys. The four quadrants signify different spectral shapes: inverted (upper right); peaking (lower right); steep (lower left); and upturn (upper left).}
         \label{spec_corr}
   \end{figure*}

\subsection{Differences between AGN types}

There are significant statistical differences between the spectral properties of the AGN subclasses in Table~\ref{spec_par}. While the distributions of the low frequency spectral index are compatible, the high frequency indices of BLOs and LPQs for all four epochs are drawn from different distributions, according to the Kruskal-Wallis analysis of variances ($P=0.001$--$0.008$ between Surveys). Also, for Survey~2, the HF indices of BLOs and HPQs are from different distributions ($P=0.001$). As can be seen in Table~\ref{spec_par}, the high frequency radio spectra of LPQs are clearly steeper than those of BLOs, indicating a possible difference in the energies of the electron population creating the spectra. A similar result was found in \citet{hovatta14}. There are no significant differences in the break frequencies, although they vary significantly between sources, which increases the uncertainty of the average values in Table~\ref{spec_par}.

The spectral shapes are further illustrated in Fig.~\ref{spec_corr}, which shows the radio colour plots, i.e., the correlation of the two spectral indices. The plot area can be divided into quadrants, which represent different spectral shapes: inverted (upper right); peaking (lower right); steep (lower left); and upturn (upper left). The sample sources are concentrated in the low-$\alpha_{\rm HF}$ part of the plots, as can be expected when the high frequency radio index is calculated up to the submillimetre domain. The low frequency indices are scattered around the zero line, including both rising and setting spectra. There are a few sources that have a rising high frequency spectrum, as can be seen in Fig.~\ref{spec_corr}. The galaxy appearing in the ``upturn'' quadrant in Survey~1 is 0238$-$084. Its spectrum has little variability between surveys, and truly appears to have an upturn in Fig.~\ref{0238-084_spectra}. Indeed, the core of this source has been found to be heavily obscured \citep{vermeulen03, kadler04}. The HPQ in the ``inverted'' quadrant in Survey~4 is 0736+017. The source passes through the ``peaking'' and ``upturn'' phase in Surveys 1 and 3. Thus, the later inverted spectra are likely to be caused by passing synchrotron blobs. The same is true for HPQ 0906+430, which showed upturn shapes in Survey~3; it exhibits a negative $\alpha_{\rm HF}$ in other surveys, although not very steep, generally above $-0.5$.

\setcounter{table}{7}
\begin{table*}
\caption{Average spectral indices at low, high, and break frequencies, excluding the dusty sources. \textit{N} is the total number of sources in each class.}
\label{spec_par}    
\vskip -6mm
\footnotesize
\setbox\tablebox=\vbox{
 \newdimen\digitwidth
 \setbox0=\hbox{\rm 0}
 \digitwidth=\wd0
 \catcode`*=\active
 \def*{\kern\digitwidth}
  \newdimen\signwidth
  \setbox0=\hbox{+}
  \signwidth=\wd0
  \catcode`!=\active
  \def!{\kern\signwidth}
\halign{\hbox to 1.8cm{#\leaderfil}\tabskip 2.0em&
    \hfil$#$\hfil\tabskip 1.0em&
    \hfil$#$\hfil\tabskip 1.0em&
    \hfil$#$\hfil&
    \hfil$#$\hfil\tabskip 2.0em&
    \hfil$#$\hfil\tabskip 1.0em&
    \hfil$#$\hfil\tabskip 1.0em&
    \hfil$#$\hfil&
    \hfil$#$\hfil\tabskip 0em\cr
\noalign{\doubleline}
\omit&\multispan4\hfil Survey 1\hfil&\multispan4\hfil Survey 2\hfil\cr
\noalign{\vskip -3pt}
\omit&\multispan4\hrulefill&\multispan4\hrulefill\cr
\omit\hfil Class\hfil&N&\alpha_{\rm LF}&\alpha_{\rm HF}&\nu_{\rm break}&N&\alpha_{\rm LF}&\alpha_{\rm HF}&\nu_{\rm break}\cr
\noalign{\vskip 3pt\hrule\vskip 4pt}
  BLO & 16 & 0.005\pm0.002 & -0.454\pm0.011 & 81.5\pm1.8 & 12 & 0.096\pm0.005 & -0.369\pm0.015 & 61.3\pm0.5\cr
  HPQ & 22 & 0.008\pm0.003 & -0.560\pm0.014 & 61.2\pm1.4 & 24 & 0.030\pm0.003 & -0.647\pm0.008 & 65.2\pm0.7\cr
  LPQ & 8 & 0.081\pm0.010 & -0.784\pm0.014 & 54.1\pm2.4 & 8 & 0.120\pm0.017 & -0.695\pm0.014 & 48.8\pm1.6\cr
  QSO & 8 & 0.205\pm0.020 & -0.495\pm0.013 & 55.5\pm0.6 & 8 & 0.017\pm0.010 & -0.530\pm0.037 & 72.6\pm5.6\cr
\omit&\multispan4\hrulefill&\multispan4\hrulefill\cr
\omit&\multispan4\hfil Survey 3\hfil& \multispan4\hfil Survey 4\hfil\cr
\noalign{\vskip -3pt}
\omit&\multispan4\hrulefill&\multispan4\hrulefill\cr
  BLO & 17 & 0.082\pm0.005 & -0.409\pm0.013 & 57.2\pm0.6 & 9 & -0.090\pm0.004 & -0.371\pm0.009 & 37.2\pm0.2\cr
  HPQ & 22 & 0.000\pm0.002 & -0.481\pm0.005 & 61.2\pm0.2 & 22 & !0.027\pm0.003 & -0.535\pm0.009 & 55.0\pm0.7\cr
  LPQ & 8 & 0.074\pm0.007 & -0.640\pm0.020 & 42.0\pm0.7 & 8 & -0.080\pm0.006 & -0.679\pm0.012 & 43.2\pm0.5\cr
  QSO & 7 & 0.230\pm0.008 & -0.410\pm0.018 & 44.7\pm1.8 & 9 & !0.150\pm0.004 & -0.541\pm0.025 & 43.5\pm1.0\cr
\noalign{\vskip 4pt\hrule\vskip 5pt}}}
\endPlancktablewide
\end{table*}

\section{Behaviour of flaring sources}

\subsection{Shocked jet models}

Ever since the first models were put forward in the 1980s \citep{marscher85,hughes85}, the most favoured description for the variations in spectral shape and total flux density (TFD) in the radio regime has been in terms of growing and decaying shock-like disturbances in a relativistic jet. This behaviour, originally seen and modelled in 3C~273 \citep{marscher85}, has been found to describe quite well a number of sources and flares. In particular, the predicted motion of the shock turnover peak from higher to lower frequencies and the time delays have been found to be in agreement with the predictions of the shocked jet models \citep[e.g.,][]{valtaoja88,stevens95,hovatta07,hovatta08,nieppola09}. Detailed modelling of the University of Michigan TFD and polarization data at 4.8, 8.0, and 14.5\,GHz using a shocked jet model code has been quite successful \citep{hughes85,aller02,aller14}. Furthermore, it has also been found that the zero epochs of the VLBI components correspond to the times when the millimetre TFD starts to rise, and that the variable TFD flux agrees with the corresponding VLBI component flux \citep{savolainen02}. Finally, the similarity of the TFD flares seen in centimetre- and millimetre-wave monitoring programmes has been argued to be a strong indication that all the radio variations have the same physical origin in growing and decaying shock-like structures propagating down the radio jet.

\begin{figure}
   \centering
   \includegraphics[width=\hsize]{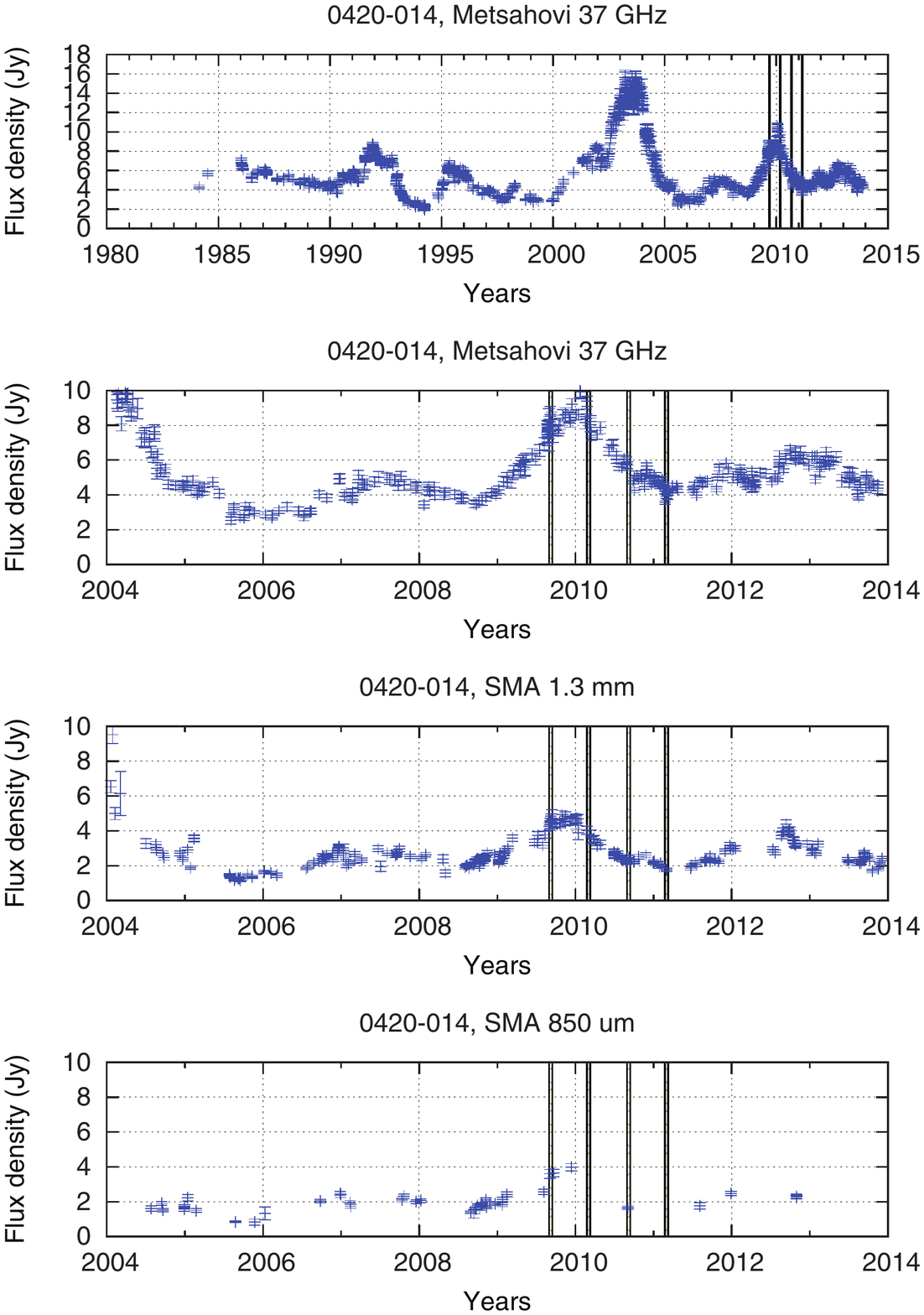}
      \caption{Flux density curves for 0420$-$014 at 37\,GHz (Mets\"ahovi), 1.3\,m, and 870\,$\mu$m (SMA). The top plot shows the long-term behaviour of the source and the three bottom plots show the period 2004--2014. The epochs of the four \Planck\ Surveys are marked with vertical lines.}
         \label{fluxcurves_0420}
   \end{figure}

\begin{figure}
   \centering
   \includegraphics[width=\hsize]{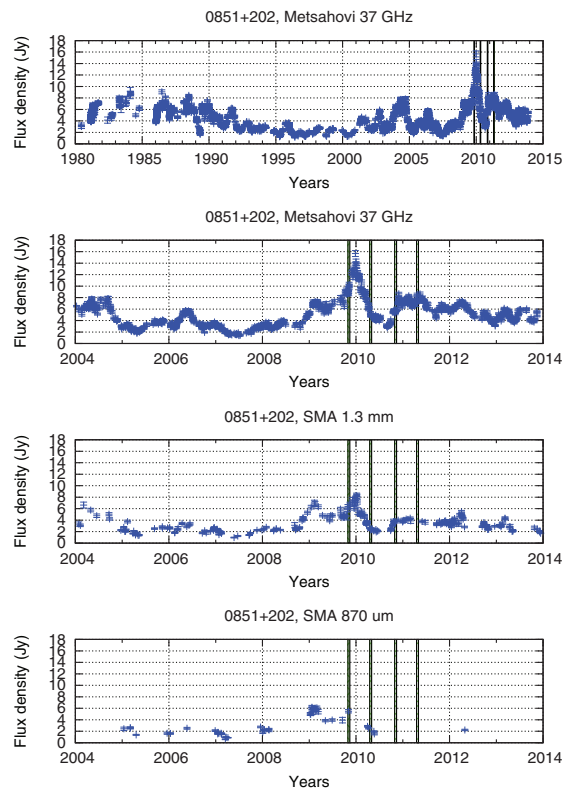}
      \caption{Flux density curves for 0851+202 at 37\,GHz (Mets\"ahovi), 1.3\,mm, and 870\,$\mu$m (SMA). See Fig.~\ref{fluxcurves_0420} for details. Some SMA data have been previously published in \citet{agudo11}. Some data come from a dedicated programme by Ann Wehrle.}
         \label{fluxcurves_0851}
   \end{figure}

\begin{figure}
   \centering
   \includegraphics[width=\hsize]{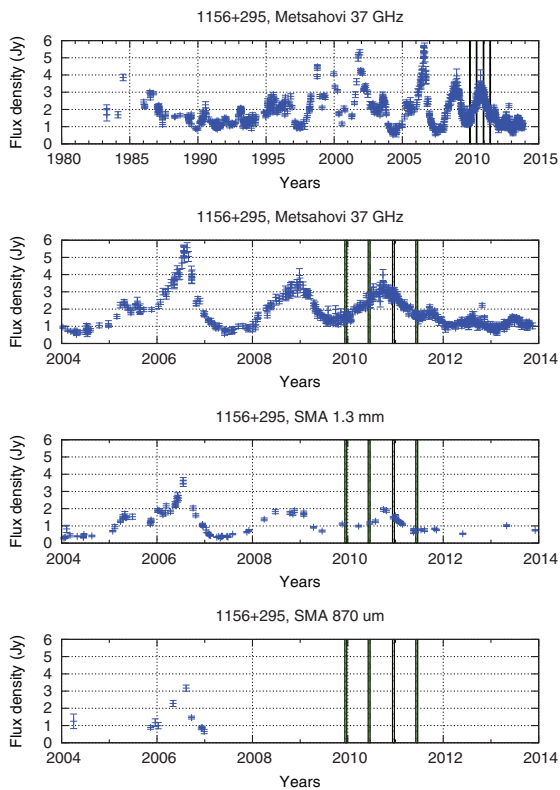}
      \caption{Flux density curves for 1156+295 at 37\,GHz (Mets\"ahovi), 1.3\,mm, and 870\,$\mu$m (SMA). See Fig.~\ref{fluxcurves_0420} for details. Some SMA data have been previously published in \citet{ramakrishnan14}.}
         \label{fluxcurves_1156}
   \end{figure}

\begin{figure}
   \centering
   \includegraphics[width=\hsize]{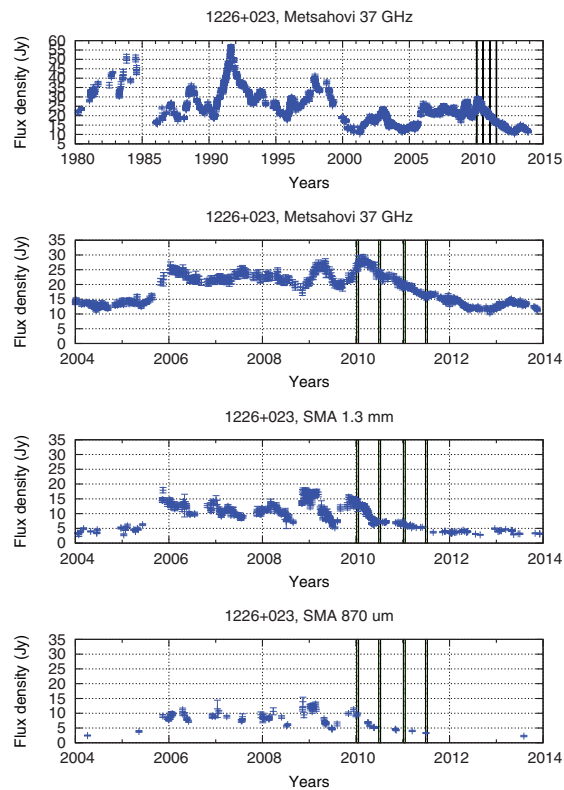}
      \caption{Flux density curves for 1226+023 at 37\,GHz (Mets\"ahovi), 1.3\,mm, and 870\,$\mu$m (SMA). See Fig.~\ref{fluxcurves_0420} for details. Some data come from a dedicated programme by Ann Wehrle.}
         \label{fluxcurves_1226}
   \end{figure}

\begin{figure}
   \centering
   \includegraphics[width=\hsize]{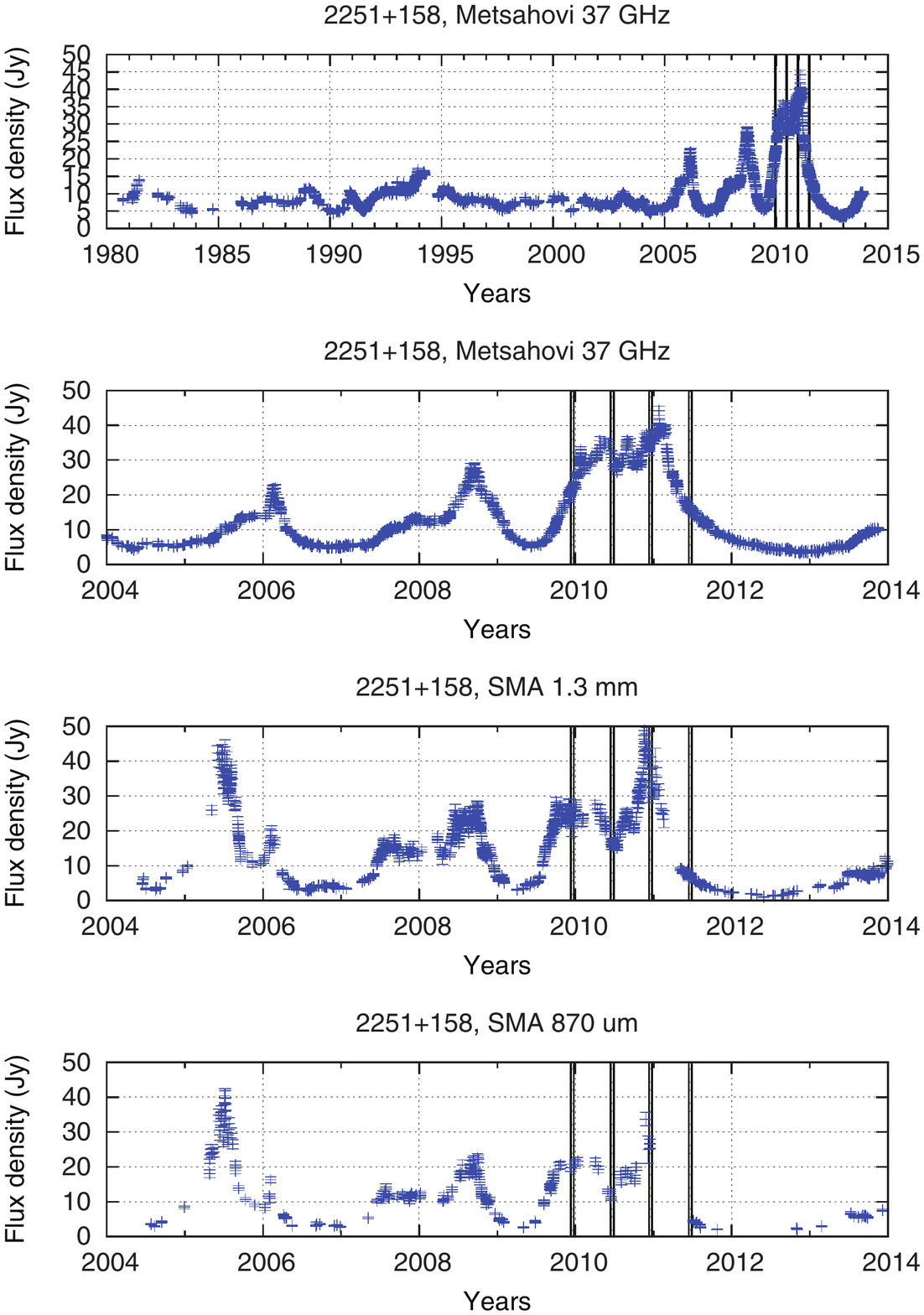}
      \caption{Flux density curves for 2251+158 at 37\,GHz (Mets\"ahovi), 1.3\,mm, and 870\,$\mu$m (SMA). See Fig.~\ref{fluxcurves_0420} for details.  A large part of the SMA data has been published; see \citet{raiteri08,raiteri11,villata09,jorstad10,jorstad13,pacciani10,vercellone10,vercellone11,ogle11,wehrle12}. Some data come from a dedicated programme by Ann Wehrle.}
         \label{fluxcurves_2251}
   \end{figure}

As is well known from AGN monitoring programmes, it is only rarely that we see a single, isolated radio flare without fine structure or superposition with other older and younger flares \citep[e.g.,][]{hovatta07}, as is also apparent in the discussion of the individual sources below. During a \Planck\ Survey snapshot, the observed radio spectrum is therefore virtually always a sum of several spectral components, including the contribution from the underlying radio jet spectrum. The sampling time of \Planck, approximately once in every six months, is typically too sparse to follow the evolution of a flare from the beginning to the end, even for those sources that were flaring during the \Planck\ mission (Figs.~\ref{fluxcurves_0420}--\ref{fluxcurves_2251}). For most of our sources, we therefore do not see clear evolving shock signatures. In  a few cases, the \Planck\ observations cover a strong radio flare seen in the Mets\"ahovi monitoring. In these cases we do see the expected new spectral component in the SED, with the spectral peak moving from higher to lower frequencies during subsequent survey epochs, as discussed below.

We limit our discussion here to the five sources with the most ``pure'' flaring occurring during the four \Planck\ Surveys, namely 0420$-$014 (OA~129), 0851+202 (OJ~287), 1156+295 (4C~29.45), 1226+023 (3C~273), and 2251+158 (3C~454.3), and discuss whether their behaviour is compatible with the shocked jet models. They all have a particularly well-sampled flaring event at 37\,GHz during the \Planck\ survey periods, and also have good multifrequency spectra for all four Surveys. In addition we have used 1.3\,mm (230\,GHz) and 870\,$\mu$m (345\,GHz) flux density data obtained at the Submillimeter Array (SMA) near the summit of Mauna Kea in Hawaii. The sources are included in an ongoing monitoring programme at the SMA to determine the flux densities of compact extragalactic radio sources that can be used as calibrators at millimetre wavelengths \citep{gurwell07}. Observations of available potential calibrators are from time to time observed for 3 to 5 minutes, and the measured source signal strength calibrated against known standards, typically Solar system objects (Titan, Uranus, Neptune, or Callisto). Data from this programme are updated regularly and are available at the SMA website.\footnote{\url{http://sma1.sma.hawaii.edu/callist/callist.html}}

The 37\,GHz flux curve from Mets\"ahovi and the 1.3\,mm and 870\,$\mu$m flux density curves from SMA for the five sources are shown in Figs.~\ref{fluxcurves_0420}--\ref{fluxcurves_2251}. The epochs of the four \Planck\ Surveys are marked with vertical lines. We analyse these sources in more detail in the following subsections.

\subsubsection{0420-014 (OA~129)} 

There is a clear, symmetric flare in the 37\,GHz flux curve, starting in late 2008 and lasting until early 2011 (Fig.~\ref{fluxcurves_0420}). The measured maximum flux density, reached in January 2010, was 10.5\,Jy. \Planck\ Surveys 1 and 2 occur at both sides of the peak, Survey~3 takes place during the flare decline, and at the time of Survey~4 the flux density has returned to normal.

This trend is quite clear throughout the multifrequency spectra in Fig.~\ref{0420-014_spectra}. Surveys 1 and 2 sample the highest flux densities in all frequencies, Survey~3 is slightly lower, and Survey~4 gives the faintest multifrequency spectrum. The spectral turnover can be discerned moving towards lower frequencies with decreasing flux density, and the optically thin spectrum steepens from epoch 1 to 4. Therefore we conclude that this source follows the Marscher \& Gear model well. See also \citet{aller14} for detailed modelling of this flare at 4.8, 8.0, and 14.5\,GHz.

\subsubsection{0851+202 (OJ~287)}

During the \Planck\ observing period, this source had its strongest observed flare at 37\,GHz, reaching  15.7\,Jy (Fig.~\ref{fluxcurves_0851}). Survey~1 occurred during its rise, and Survey~2 during its decay. Surveys 3 and 4 occurred at an active stage following the strong flare, with several quite intensive (roughly 8\,Jy) outbursts. Adhering to the shock model, the multifrequency spectrum of epoch 1 peaks at the highest frequencies and is the brightest. The steepening of the optically thin spectrum due to energy losses can be detected in the Survey~1 spectrum. The spectrum during Survey~2 features the flare decay stage, having a negative spectral index throughout. The spectra of epochs 3 and 4 are yet again brighter, and peak around or slightly below 37\,GHz. See \citet{aller14} for detailed modelling of this flare at 4.8, 8.0, and 14.5\,GHz.

\subsubsection{1156+295 (4C~29.45)}

1156+295 also has a well-sampled radio flare during the \Planck\ survey period (Fig.~\ref{fluxcurves_1156}). The first two Surveys sample the rising phase, and the last two Surveys occur in the decay stage. The flux density curves look quite smooth, but the multifrequency spectra reveal the presence of several flaring components, especially during Survey~1. The structure of the flare has been studied in great detail in \citet{ramakrishnan14}. During the active phase in 2007--2012, the flaring consists of four moving and one stationary component in the jet. The Survey~1 spectrum is the most complicated, and the spectrum becomes smoother towards Survey~4. See \citet{aller14} for detailed modelling of this flare at 4.8, 8.0, and 14.5\,GHz.

\subsubsection{1226+023 (3C~273)}
This source clearly shows the passing of a flare-inducing blob in its multifrequency spectra, particularly in Surveys 1 and 2. In Survey~1 there is a distinct bump in the spectrum, caused by the rising flare at 37\,GHz occurring at the same time (Fig.~\ref{fluxcurves_1226}). The 1.3\,mm flux density is already decreasing during Survey~1. In the Survey~2 spectrum, the bump can still be discerned below 100\,GHz, even though the flare is already decreasing at 37\,GHz as well. In Surveys~3 and 4, it has faded away, and the spectrum has returned to its steep form in the quiet stage. For a multiwavelength analysis of this source during the \Planck\ Surveys see, e.g., \citet{jorstad12}.

\subsubsection{2251+158 (3C~454.3)}

3C~454.3 was another source to undergo its strongest observed 37\,GHz flare during the \Planck\ Surveys (Fig.~\ref{fluxcurves_2251}). The flare had a complex structure, both at 37\,GHz and 1.3\,mm, and reached 44.2\,Jy at 37\,GHz in early 2011. Survey~1 sampled its rise, Surveys~2 and 3 its complex multicomponent peak, and Survey~4 the late decline stage. The multifrequency spectra showcase the essence of the Marscher \& Gear model. The spectrum peak moves to lower frequencies from Survey~1 to 2, and Survey~3 samples the highest flux densities close to the peak of the flare. Due to another synchrotron component, the spectrum peak is again at higher frequencies compared to Survey~2. In Survey~4, the flare is decreasing, the flux density has dropped, and the peak has moved to around 10\,GHz from  approximately 100\,GHz in Survey~1. For multiwavelength analyses of this source during the \Planck\ Surveys see also, for example, \citet{wehrle12} and \citet{jorstad13}.

\subsection{Alternative models for variability?}

Although the variability in our five example sources can be argued to be compatible with the shock behaviour predicted by the basic Marscher \& Gear model, it is worthwhile to discuss alternative models. In most other sources, the variability behaviour seems to be approximately achromatic, with the fluxes at all frequencies except the very lowest ones either rising and decreasing simultaneously. However, an earlier study \citep{angelakis12} with data over 4.5 years (including also millimetre wavebands) found that only eight out of 78 sources show achromatic variability.

Until now, the vast majority of AGN radio monitoring has been limited to relatively low frequencies, even though studies such as \citet{tornikoski93_southern,tornikoski00_southern} and \citet{hovatta08} also used data up to\,230\,GHz. Consequently, the comparisons with shocked jet models have also been done at low frequencies, and have mainly considered the strongest radio flares, those which also produce visible VLBI components emerging from the radio core.

A plausible working hypothesis is that the shocked jet description is a valid approximation for the strongest disturbances, which survive to emerge from the radio core and continue to propagate downstream as something that can be approximated with a single shocked portion of the jet. We note that the flaring epochs described above are, for each of the five sources, among the strongest seen in these sources during the 30+ years of Mets\"ahovi monitoring.

Multifrequency observations \citep[e.g.,][]{marscher08} and comparisons with VLBI data \citep[e.g.,][]{savolainen02} have demonstrated that the less extreme variations take place mainly within the radio core itself, or even upstream of it. The structure of the radio core is under debate, as are the processes that occur when a disturbance moving down the jet passes through it. One possibility is the turbulent extreme multi-zone model \citep[TEMZ;][]{wehrle12, marscher14}, in which a number of random cells are shocked by a disturbance passing through a standing conical shock, identified as the radio core. The observed flux density is the sum of radiation from the individual cells. Due to the stochastic nature of the TEMZ, the flux density variations can be essentially achromatic even over a very wide frequency range, as shown by model simulations (Fig. 3 in \citealp{marscher14} and priv. comm.). In \citet{wehrle12}, the TEMZ model was successfully applied to an outburst in~3C 454.3, during which the multifrequency variations, in particular from $10^{11}$ to $10^{15}$\,Hz, appeared to be nearly simultaneous. In many cases the TEMZ model may thus be a better overall description of the radio SED behaviour as seen by \Planck. For example, in 0716+714 the overall flux densities change by a factor of 3, and in 1253$-$055 by a factor of 2, but the shapes of the radio spectra remain unchanged within the observational error bars. This is very similar to what was seen in 3C~454.3 (see especially fig. 2 in \citealp{wehrle12}).

Geometric variability -- changes in the jet direction -- can naturally produce achromatic variations through changes in the Doppler boosting factor. Such models have been proposed to explain the radio variability in a number of sources \citep[e.g.,][]{raiteri12, chen13}. However, geometric models in general predict regularities (either periodicity or repetitive variability patterns) that have never been observed over longer timescales \citep{chen13}. A kink in the jet will produce a change in the overall flux density levels, but not spectral variability moving up and down.

Finally, achromatic variability can also arise naturally during shocks, at high frequencies (above the spectral turnover), where the source is optically thin and the fluxes therefore rise and fall in unison \citep{valtaoja92_moniIII}.

\section{Rising high frequency spectra}
\label{risinghfspectra}

Looking at our radio spectra, it is obvious that some sources show a clear upturn at the highest or two highest frequencies (see also Sect.~\ref{spectralindices}). The most likely reason for the upturns in most sources is contamination of the flux densities by Galactic cirrus clouds with temperatures from 15 to 25\,K. This interpretation is in accordance with the $\nu^{3}$ to $\nu^{4}$ spectral index (estimated by removing the synchrotron component extrapolated from lower frequencies), typical for Galactic cirrus clouds, and the estimated temperature for the sources where archival IRAS and other data impose upper limits to mid-infrared fluxes, restraining the range of possible temperatures and fluxes for cold thermal components.

As we noted in Paper I, even the most luminous known infrared-emitting AGN dust components would be completely swamped by the non-thermal radiation at the typical redshifts of our blazars. However, there remains the possibility that some blazar-type AGN could have extreme amounts of cold dust, detectable at the highest \Planck\ frequencies. In the commonly accepted evolutionary scenario, the starburst phase preceeds the AGN phase. Starburst galaxies typically contain copious amounts of dust, and in the AGN phase some blazars may consequently still have detectable dust components.  

We originally looked at the seven most promising candidates with clearly rising high frequency spectra. These sources, and their redshifts and optical classes, are listed in Table~\ref{dusty_sources}. The high-$z$ objects unfortunately lie in poor, confused foreground areas, based on IRAS and \textit{Herschel} data, and have cirrus flags in the Planck Catalogue of Compact Sources \citep{Planck_compcat2013}, with the possible exception of 2037+511. However, in sources with the lowest redshifts, detected previously in the mid-infrared, and located in unconfused areas, the highest \Planck\ frequency upturns indeed could be due to the coldest intrinsic dust component, around 20\,K.

For the three low-$z$ sources, 0238$-$084 (NGC~1052), 0415+379 (3C~111), and 0430+052 (3C~120), archival IRAS and other infrared data allow us to delineate the whole spectrum of this IR component. It can be roughly fitted with a single 15-K heated dust component (corresponding to log $\nu_{\rm peak} = 12.2$, or 200\microns). For all three sources, archival mid-infrared data indicate the presence of warmer dust components that cannot be due to Galactic cirrus clouds. Therefore, the roughly 20\,K components in these sources may indeed be intrinsic as well. This possibility requires further study. 
For example, \citet{malmrose11} detected dust using observations from the \textit{Spitzer Space Telescope} in two AGN also included in our sample, namely 4C~21.35 (1222+216) and CTA~102 (2230+114). In the first source we do not see an upturn, but in the second one a small upturn is visible. However, our \Planck\ data are from a considerably lower frequency range than \textit{Spitzer}'s.

\setcounter{table}{6}
\begin{table}
\caption{\label{dusty_sources}Sources with clearly rising high frequency spectra, along with their redshift, optical class, and figure number.}
\vskip -6mm
\footnotesize
\setbox\tablebox=\vbox{
 \newdimen\digitwidth
 \setbox0=\hbox{\rm 0}
 \digitwidth=\wd0
 \catcode`*=\active
 \def*{\kern\digitwidth}
  \newdimen\signwidth
  \setbox0=\hbox{+}
  \signwidth=\wd0
  \catcode`!=\active
  \def!{\kern\signwidth}
\halign{\hbox to 2.2cm{#\leaderfil}\tabskip 2.0em&
    \hfil#\hfil\tabskip=2.6em&
    #\hfil\tabskip=2.0em&
    \hfil#\hfil\tabskip 0em\cr
\noalign{\doubleline}
\omit\hfil Source\hfil&$z$&\llap{O}ptical class&Figure\cr
\noalign{\vskip 3pt\hrule\vskip 4pt}
0238$-$084& 0.005& Seyfert 2& \ref{0238-084_spectra}\cr
0333+321&   1.258& Quasar& \ref{0333+321_spectra}\cr
0415+379&   0.049& Seyfert 1& \ref{0415+379_spectra}\cr
0430+052&   0.033& Seyfert 1& \ref{0430+052_spectra}\cr
0446+112&   1.207& BL Lac& \ref{0446+112_spectra}\cr
1954+513&   1.223& Quasar& \ref{1954+513_spectra}\cr
2037+511&   1.686& Quasar& \ref{2037+511_spectra}\cr
\noalign{\vskip 4pt\hrule\vskip 5pt}}}
\endPlancktable
\end{table}

\section{Conclusions}

We have presented four-epoch radio-to-submillimetre spectra using simultaneous \Planck\ and auxiliary data for a complete flux-density-limited sample of 104 northern-to-equatorial bright extragalactic radio sources. Our main conclusions from this unique data set are the following.

\begin{enumerate}

\item The flattest high frequency radio spectra have indices close to zero, and most range between $-0.2$ and $-0.7$. This indicates that the original electron acceleration spectrum index is quite hard, 1.5 instead of the canonical 2.5.

\item Statistical differences between AGN subclasses are significant at high frequencies, particularly between BLOs and LPQs.

\item The radio spectra of these sources peak at remarkably high frequencies, tens of gigahertz. Most of them are highly variable, and for many the peak frequency also shifts with variability.

\item Although we see signatures of evolving shocks in the strongest radio flares, much of the high frequency variability may be better approximated by achromatic variations. Such variability is predicted by the recent TEMZ model, although other explanations are also possible.

\item We have found indications of intrinsic cold dust in low-$z$ AGN. A number of more distant AGN show an upturn at the high frequency end of their radio spectra that is most likely due to contamination from cold Galactic cirrus clouds, although we cannot exclude intrinsic dust contributions in some cases.

\end{enumerate}

A more comprehensive scientific analysis of this data set, with spectral energy distributions that include high frequency data (optical, UV, X-ray, and gamma-ray data) will be published in a subsequent paper.

\renewcommand{\thefigure}{1.\arabic{figure}}
\setcounter{figure}{0}

\onlfig{

\begin{figure*}
	\centering
	\begin{minipage}[b]{.47\textwidth}
	\includegraphics[width=\columnwidth]{0003-066_spectra}
	\caption{Radio spectrum of 0003$-$066: coloured stars, \Planck\ data from four Surveys; coloured circles, data simultaneous to the \Planck\ observations; grey circles, historical data and solid lines, broken power-law fits.}
	\label{0003-066_spectra}
	\end{minipage}\qquad
	\begin{minipage}[b]{.47\textwidth}
	\includegraphics[width=\columnwidth]{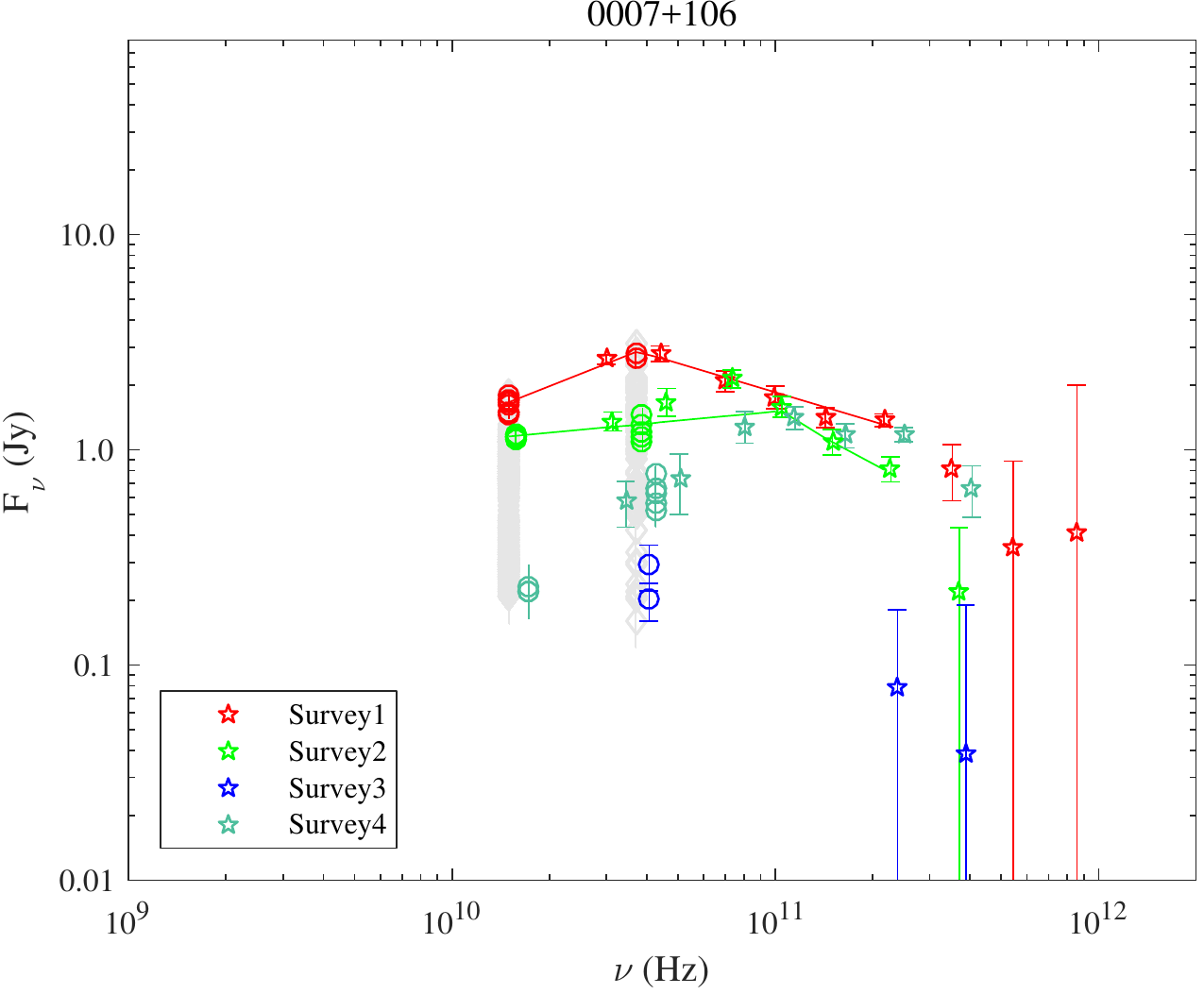}
	\caption{0007+106.}
	\label{0007+106_spectra}
	\end{minipage}
\end{figure*}

\begin{figure*}
	\centering
	\begin{minipage}[b]{.47\textwidth}
	\includegraphics[width=\columnwidth]{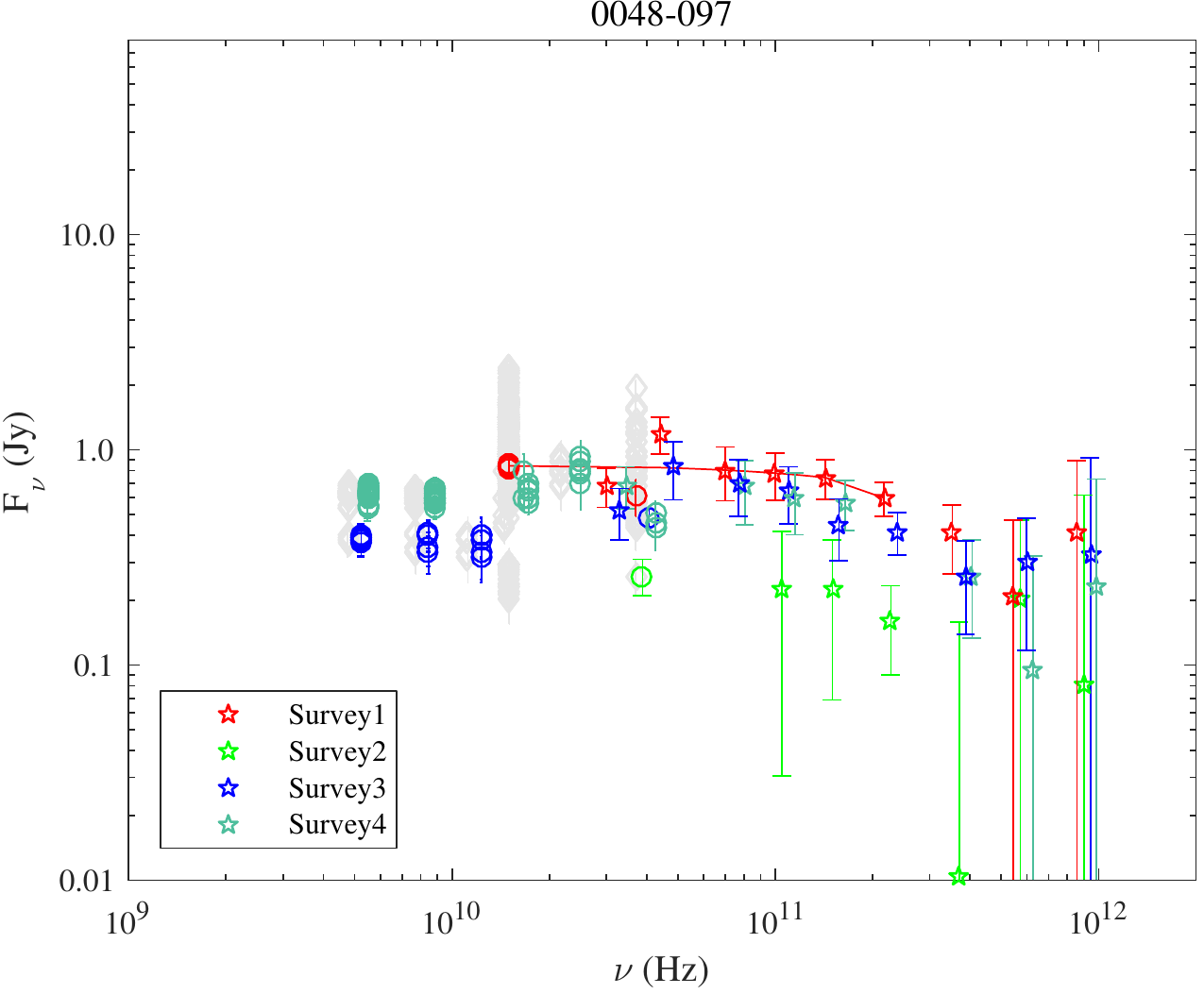}
	\caption{0048-097.}
	\label{0048-097_spectra}
	\end{minipage}\qquad
	\begin{minipage}[b]{.47\textwidth}
	\includegraphics[width=\columnwidth]{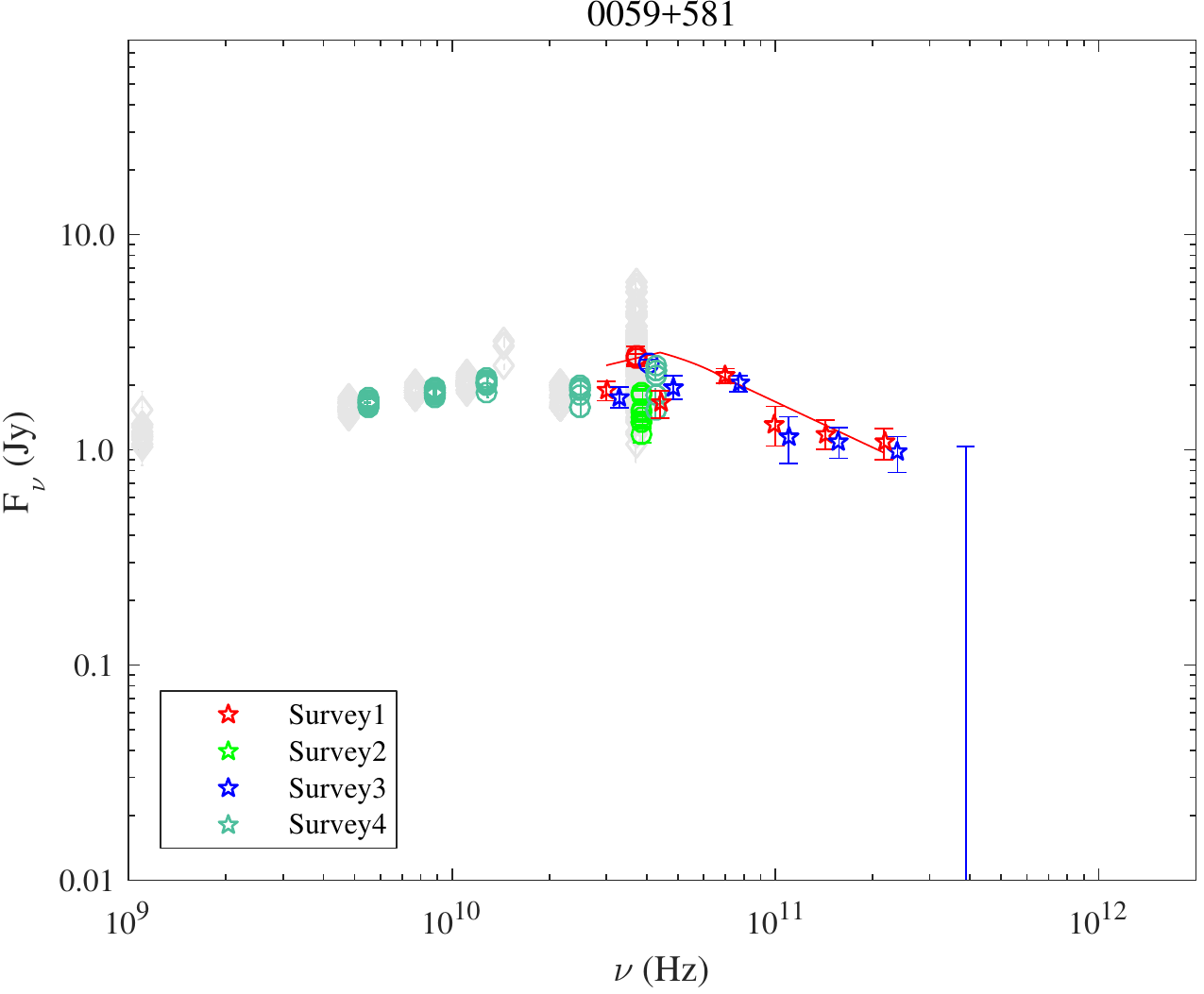}
	\caption{0059+581.}
	\label{0059+581_spectra}
	\end{minipage}
\end{figure*}

\begin{figure*}
	\centering
	\begin{minipage}[b]{.47\textwidth}
	\includegraphics[width=\columnwidth]{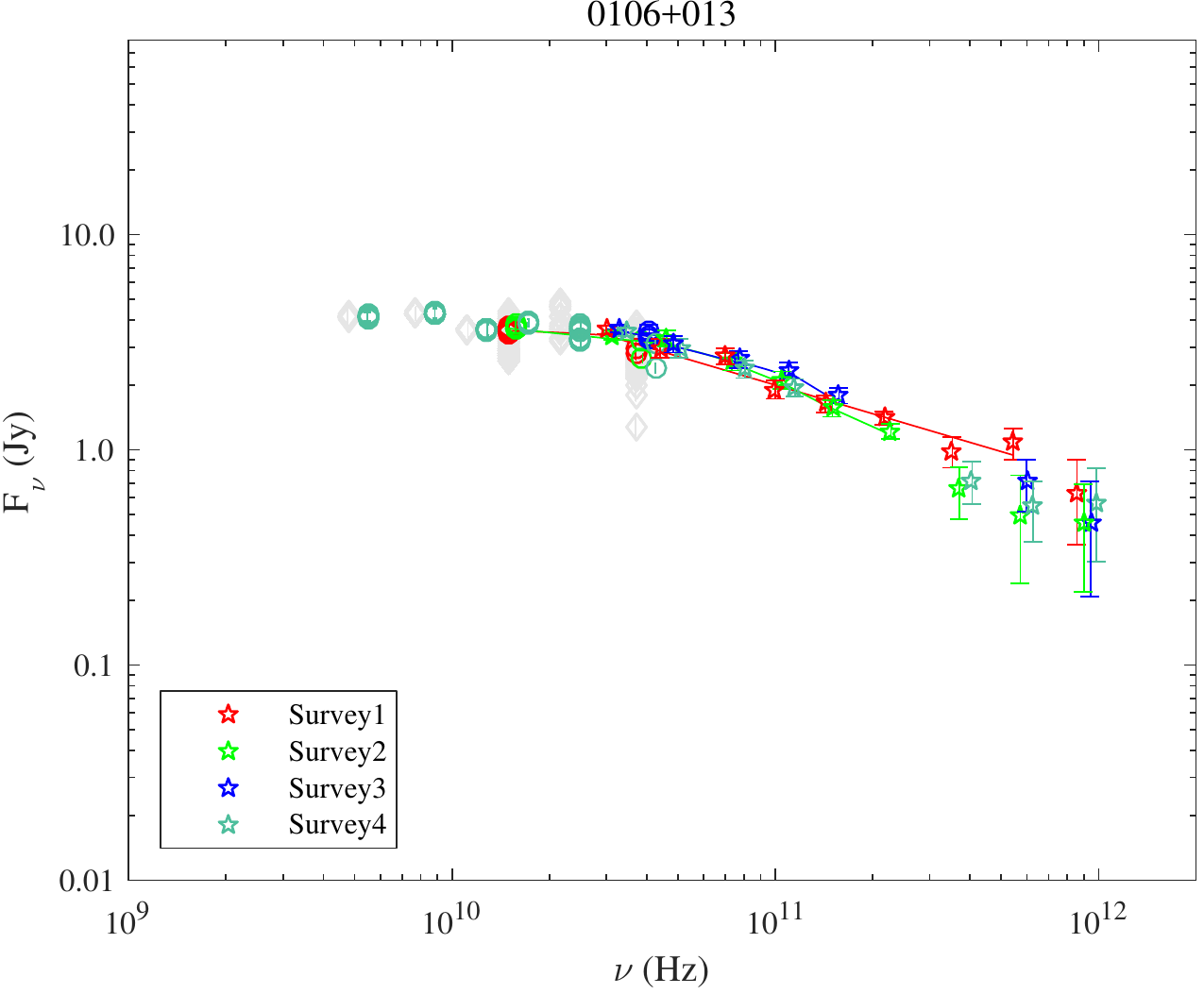}
	\caption{0106+013.}
	\label{0106+013_spectra}
	\end{minipage}\qquad
	\begin{minipage}[b]{.47\textwidth}
	\includegraphics[width=\columnwidth]{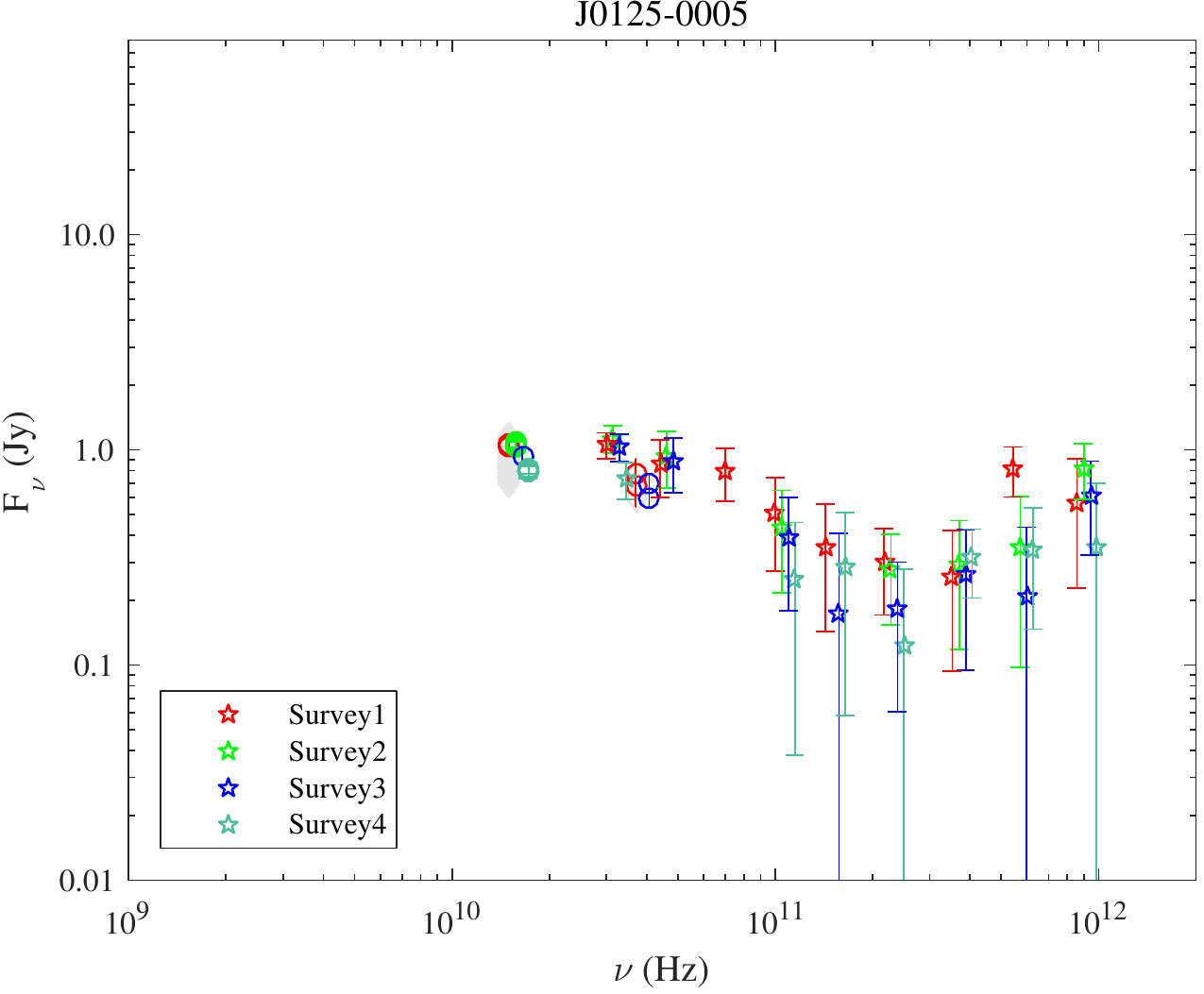}
	\caption{J0125$-$0005.}
	\label{J0125-0005_spectra}
	\end{minipage}
\end{figure*}

\clearpage

\begin{figure*}
	\centering
	\begin{minipage}[b]{.47\textwidth}
	\includegraphics[width=\columnwidth]{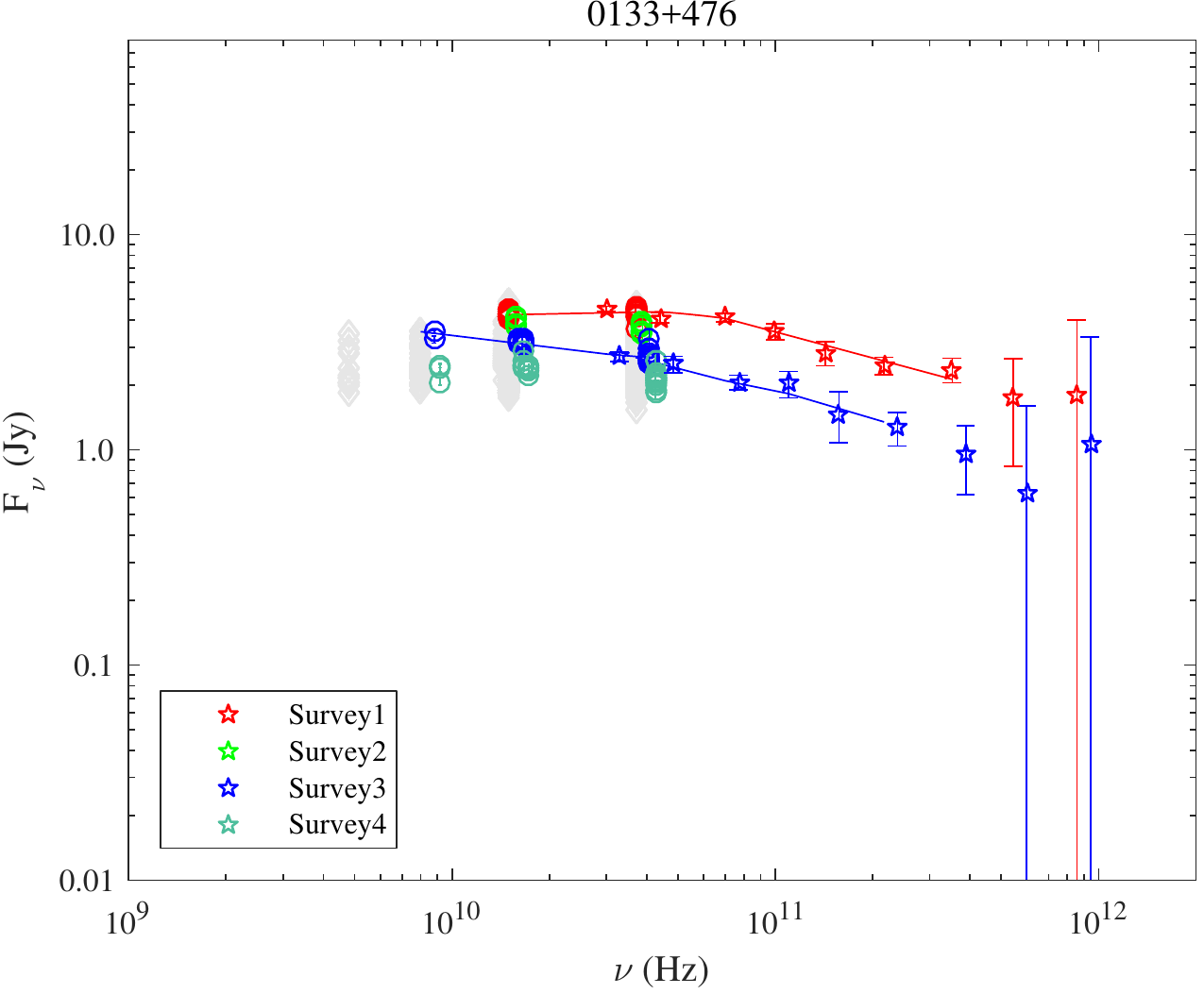}
	\caption{0133+476.}
	\label{0133+476_spectra}
	\end{minipage}\qquad
	\begin{minipage}[b]{.47\textwidth}
	\includegraphics[width=\columnwidth]{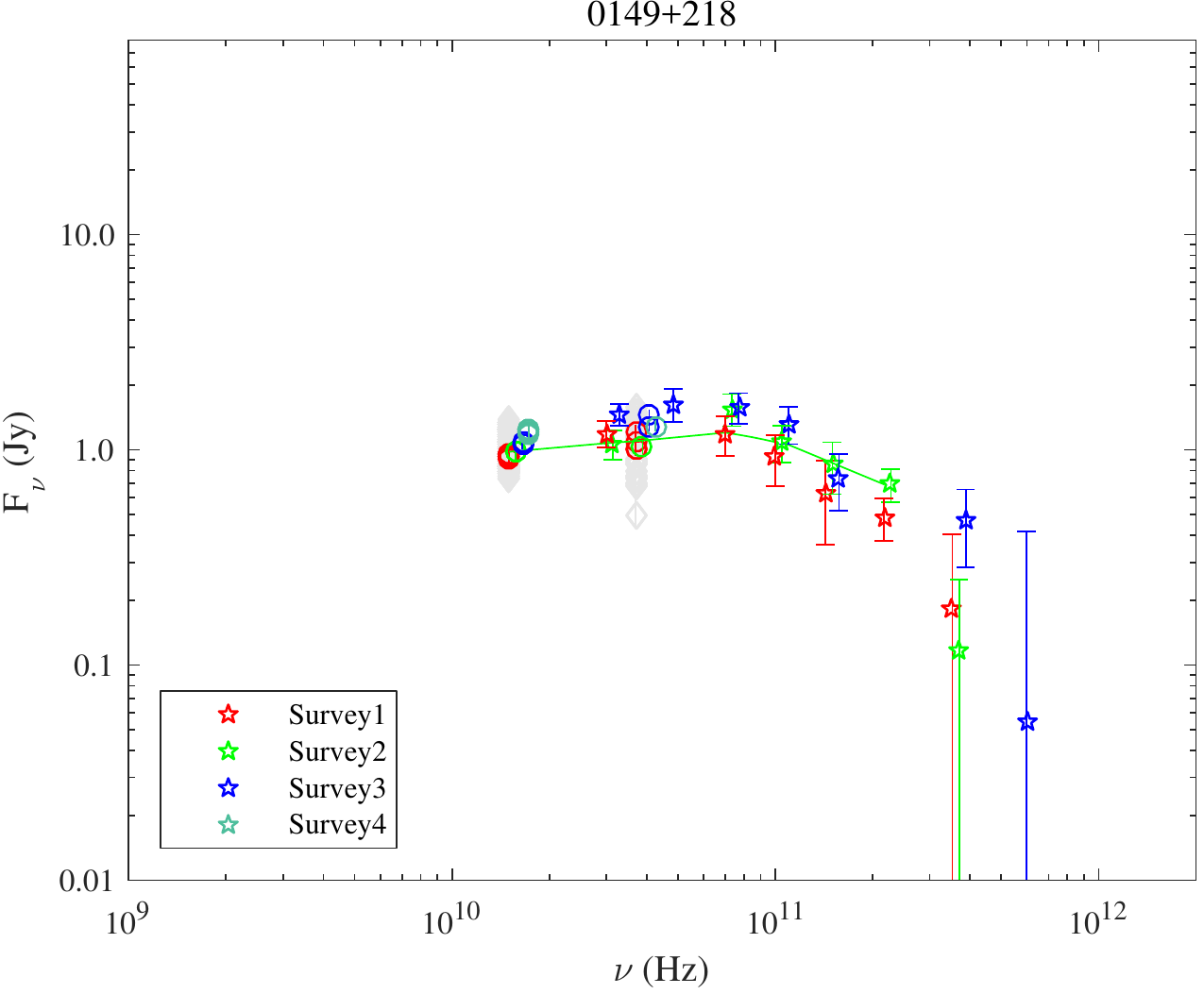}
	\caption{0149+218.}
	\label{0149+218_spectra}
	\end{minipage}
\end{figure*}

\begin{figure*}
	\centering
	\begin{minipage}[b]{.47\textwidth}
	\includegraphics[width=\columnwidth]{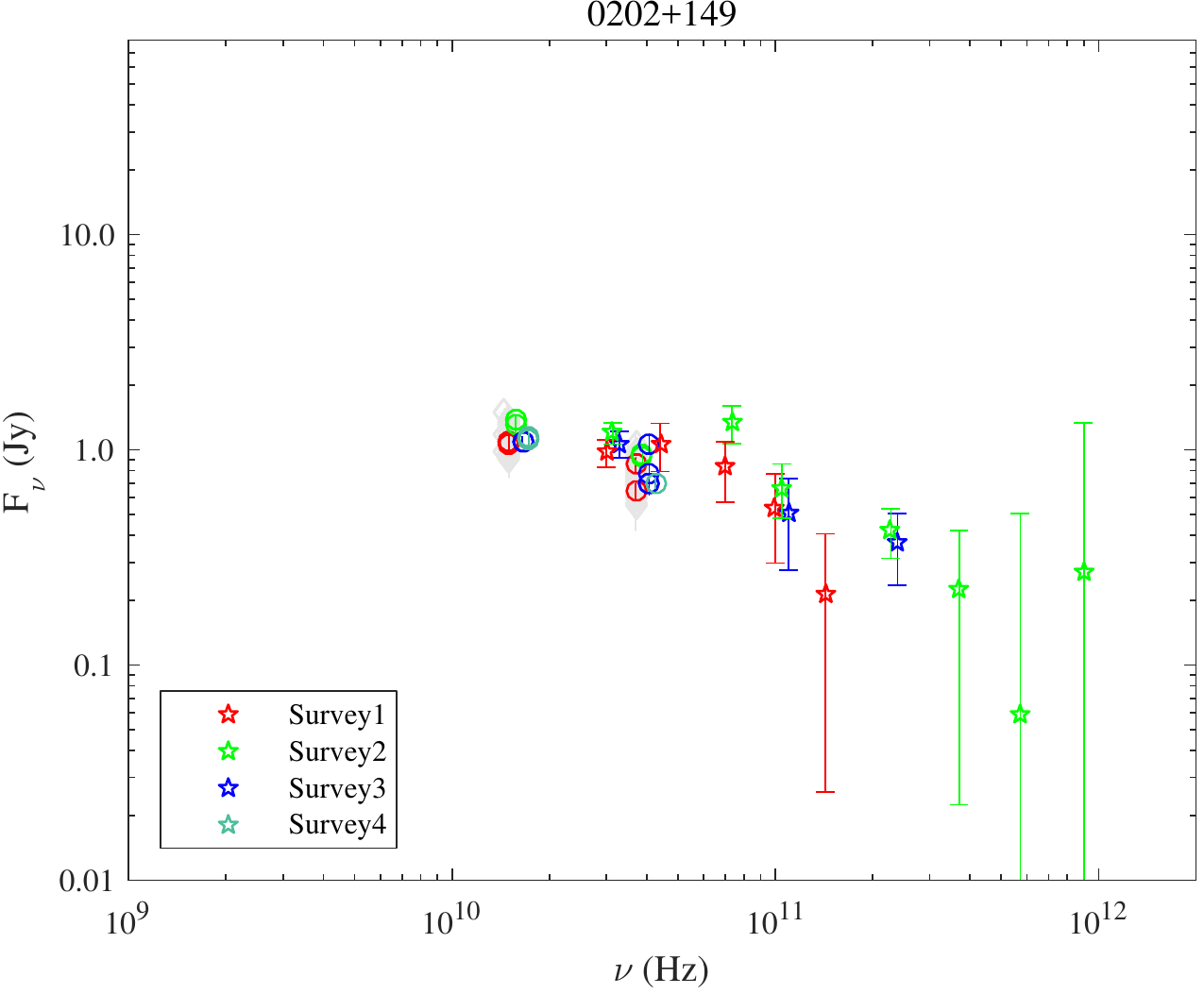}
	\caption{0202+149.}
	\label{0202+149_spectra}
	\end{minipage}\qquad
	\begin{minipage}[b]{.47\textwidth}
	\includegraphics[width=\columnwidth]{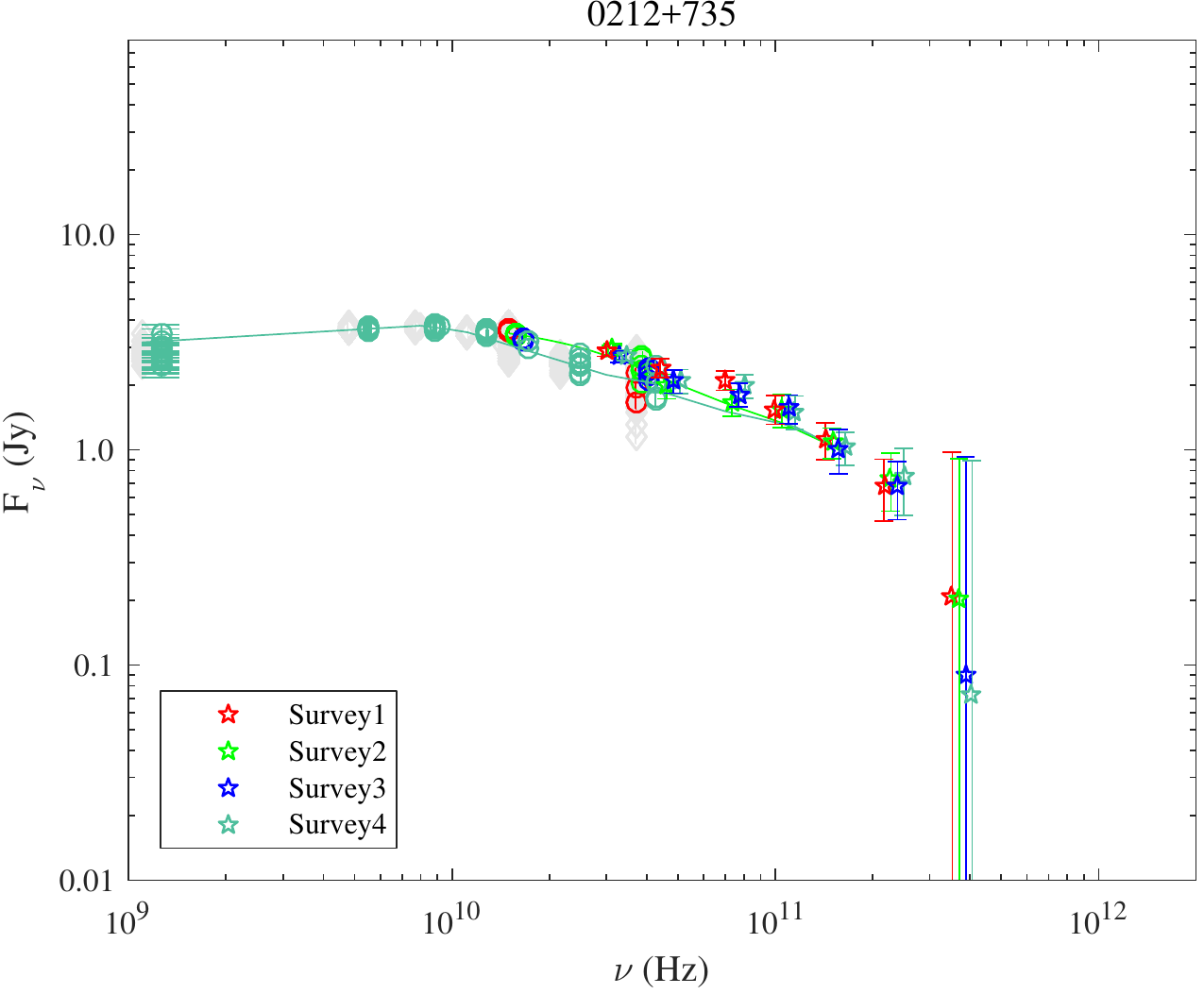}
	\caption{0212+735.}
	\label{0212+735_spectra}
	\end{minipage}
\end{figure*}

\begin{figure*}
	\centering
	\begin{minipage}[b]{.47\textwidth}
	\includegraphics[width=\columnwidth]{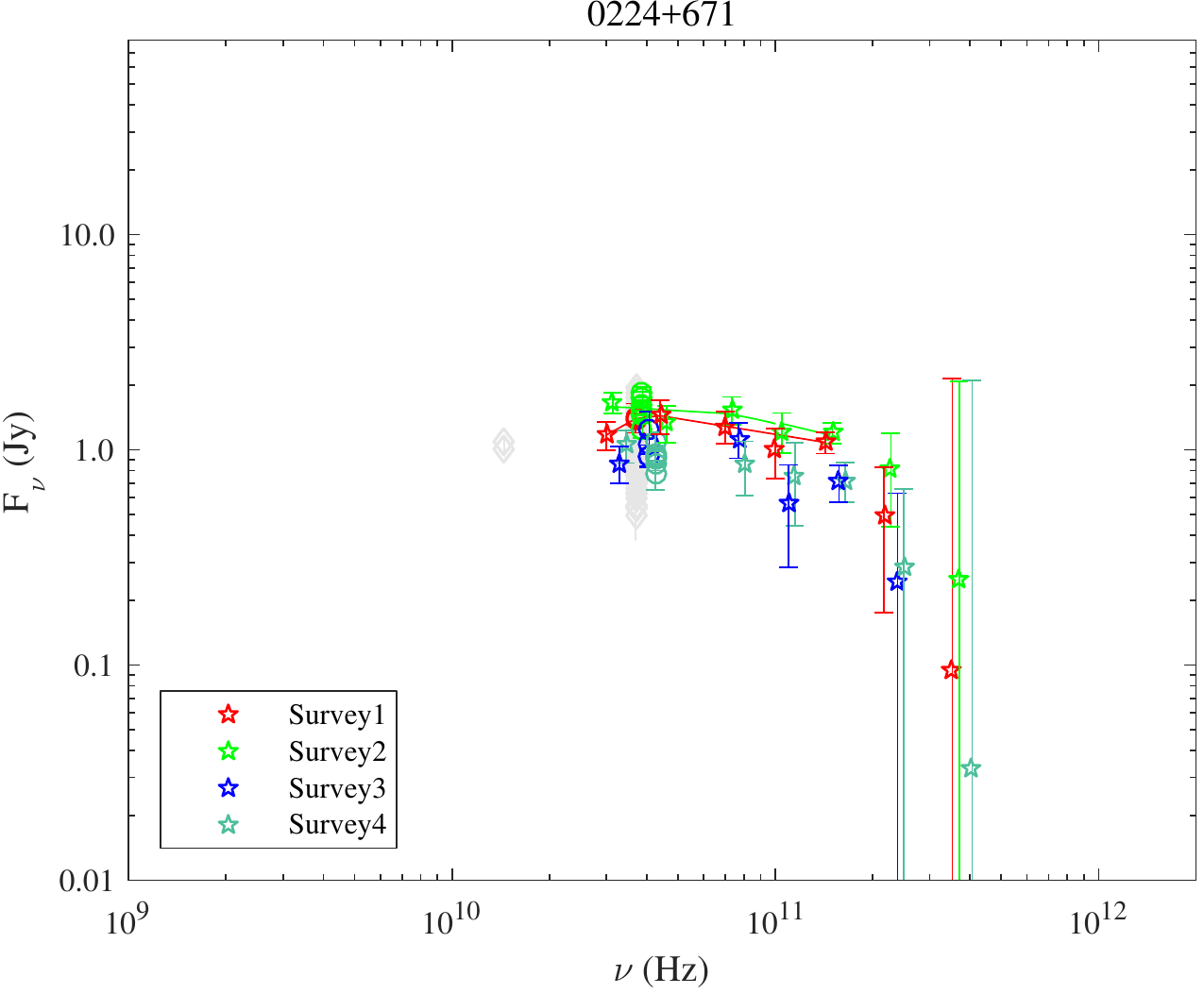}
	\caption{0224+671.}
	\label{0224+671_spectra}
	\end{minipage}\qquad
	\begin{minipage}[b]{.47\textwidth}
	\includegraphics[width=\columnwidth]{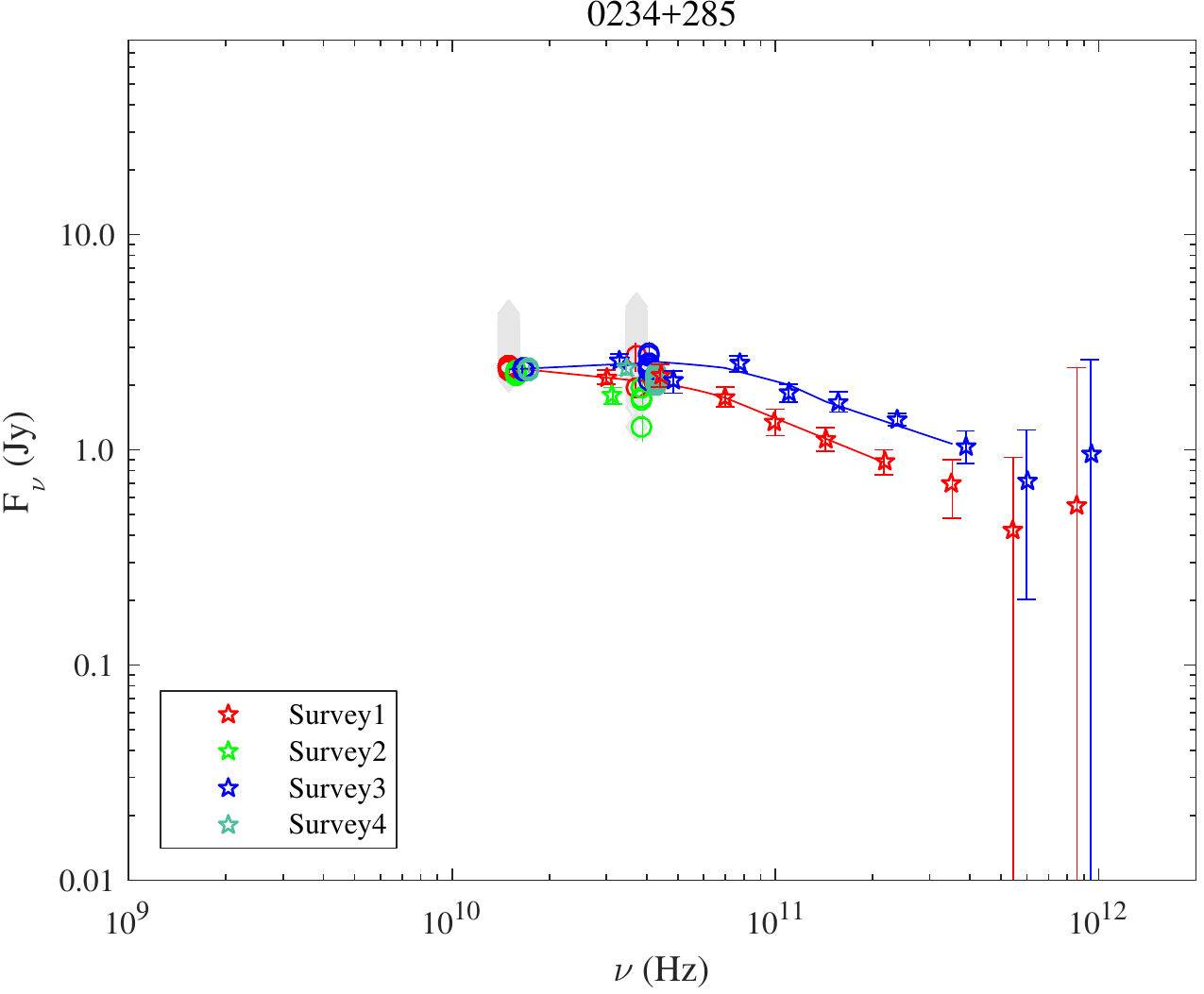}
	\caption{0234+285.}
	\label{0234+285_spectra}
	\end{minipage}
\end{figure*}

\clearpage

\begin{figure*}
	\centering
	\begin{minipage}[b]{.47\textwidth}
	\includegraphics[width=\columnwidth]{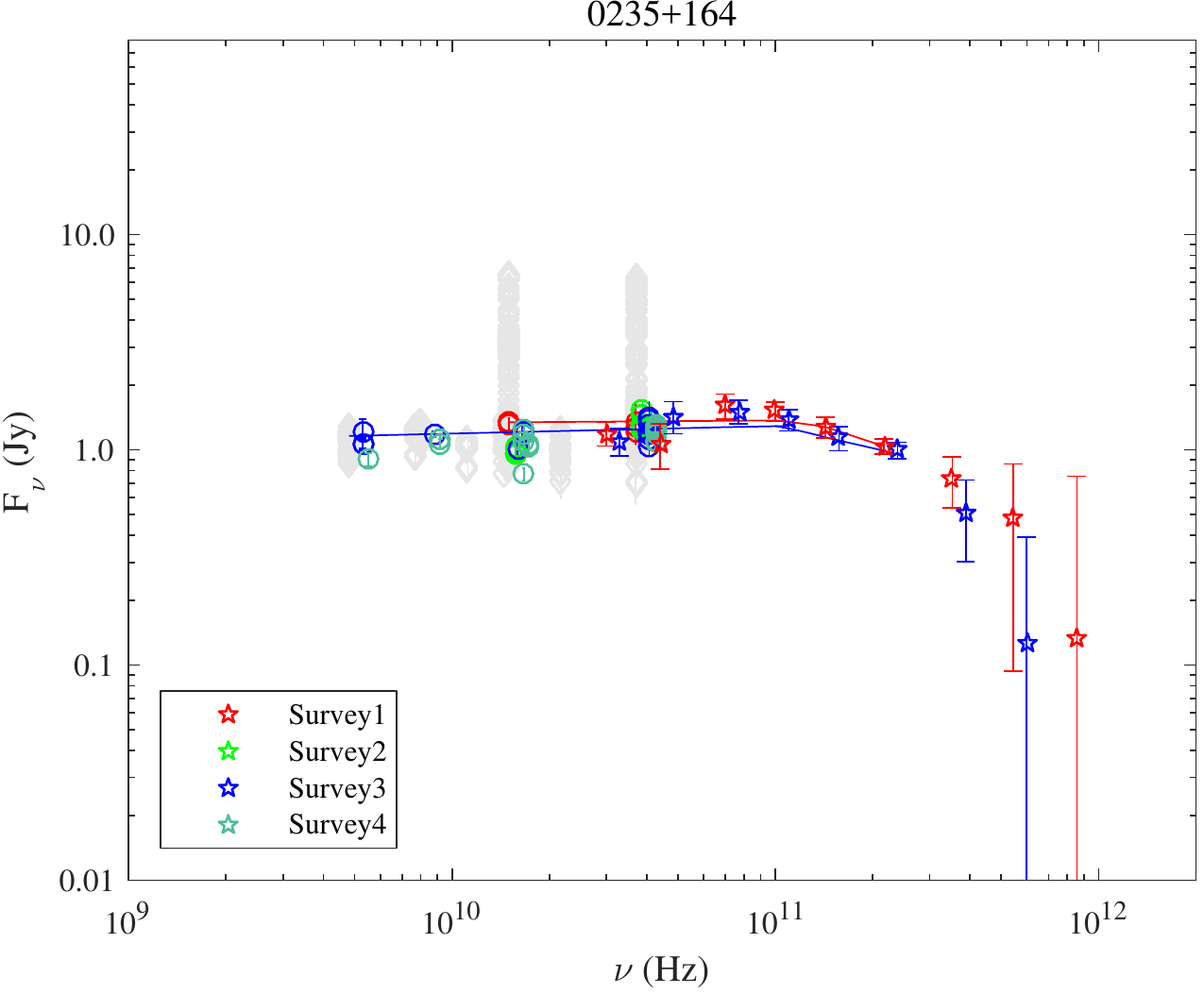}
	\caption{0235+164.}
	\label{0235+164_spectra}
	\end{minipage}\qquad
	\begin{minipage}[b]{.47\textwidth}
	\includegraphics[width=\columnwidth]{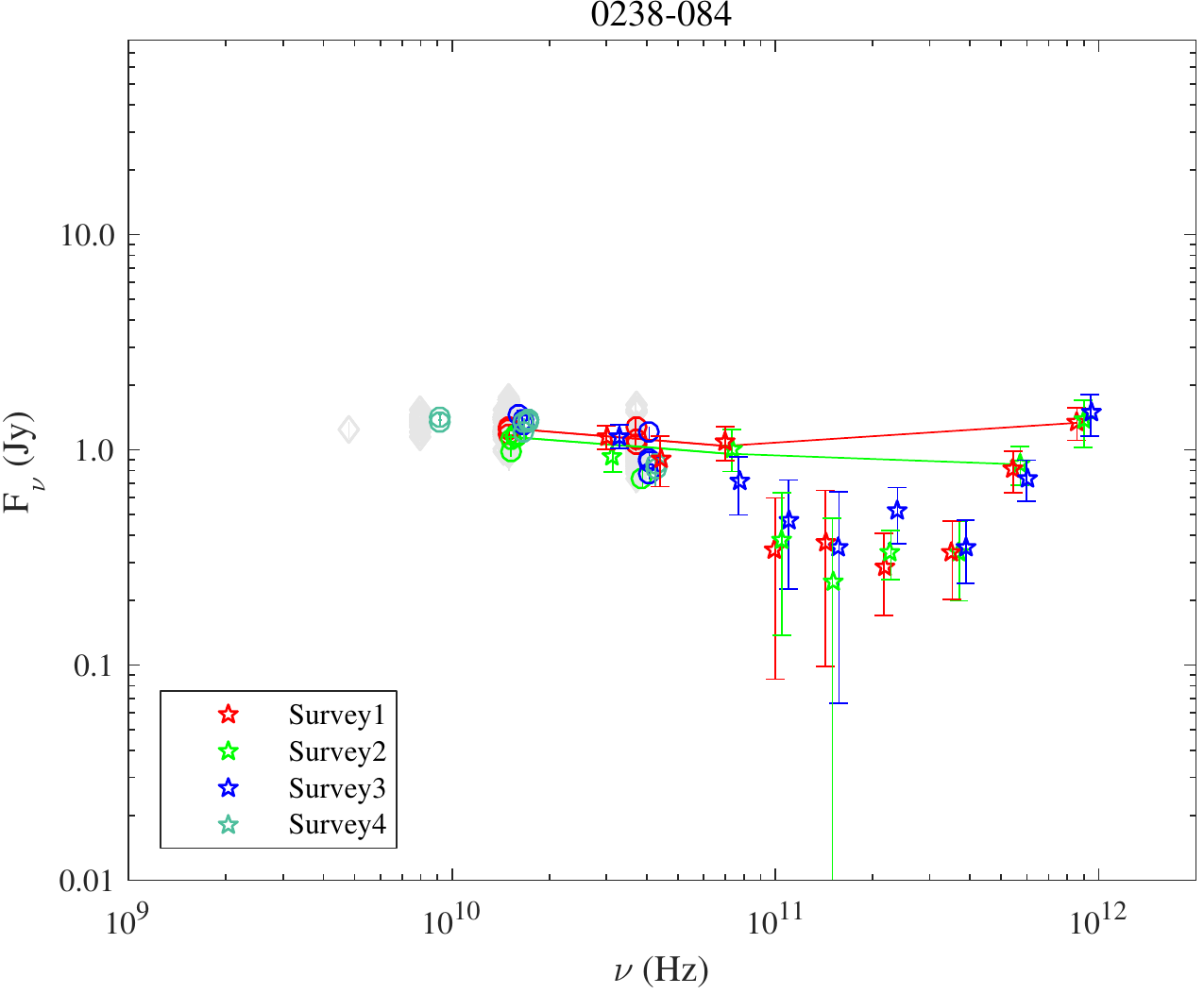}
	\caption{0238$-$084.}
	\label{0238-084_spectra}
	\end{minipage}
\end{figure*}

\begin{figure*}
	\centering
	\begin{minipage}[b]{.47\textwidth}
	\includegraphics[width=\columnwidth]{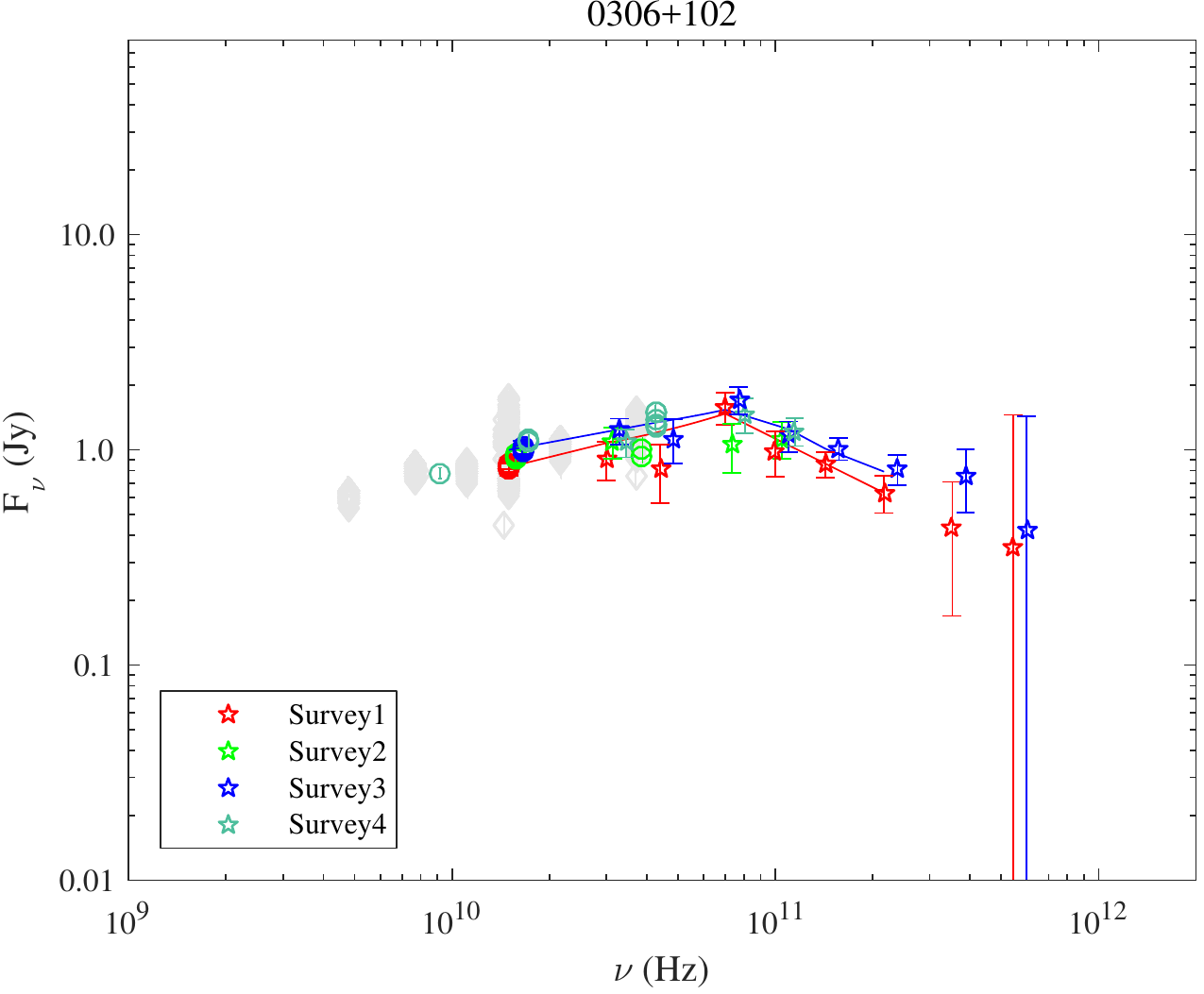}
	\caption{0306+102.}
	\label{0306+102_spectra}
	\end{minipage}\qquad
	\begin{minipage}[b]{.47\textwidth}
	\includegraphics[width=\columnwidth]{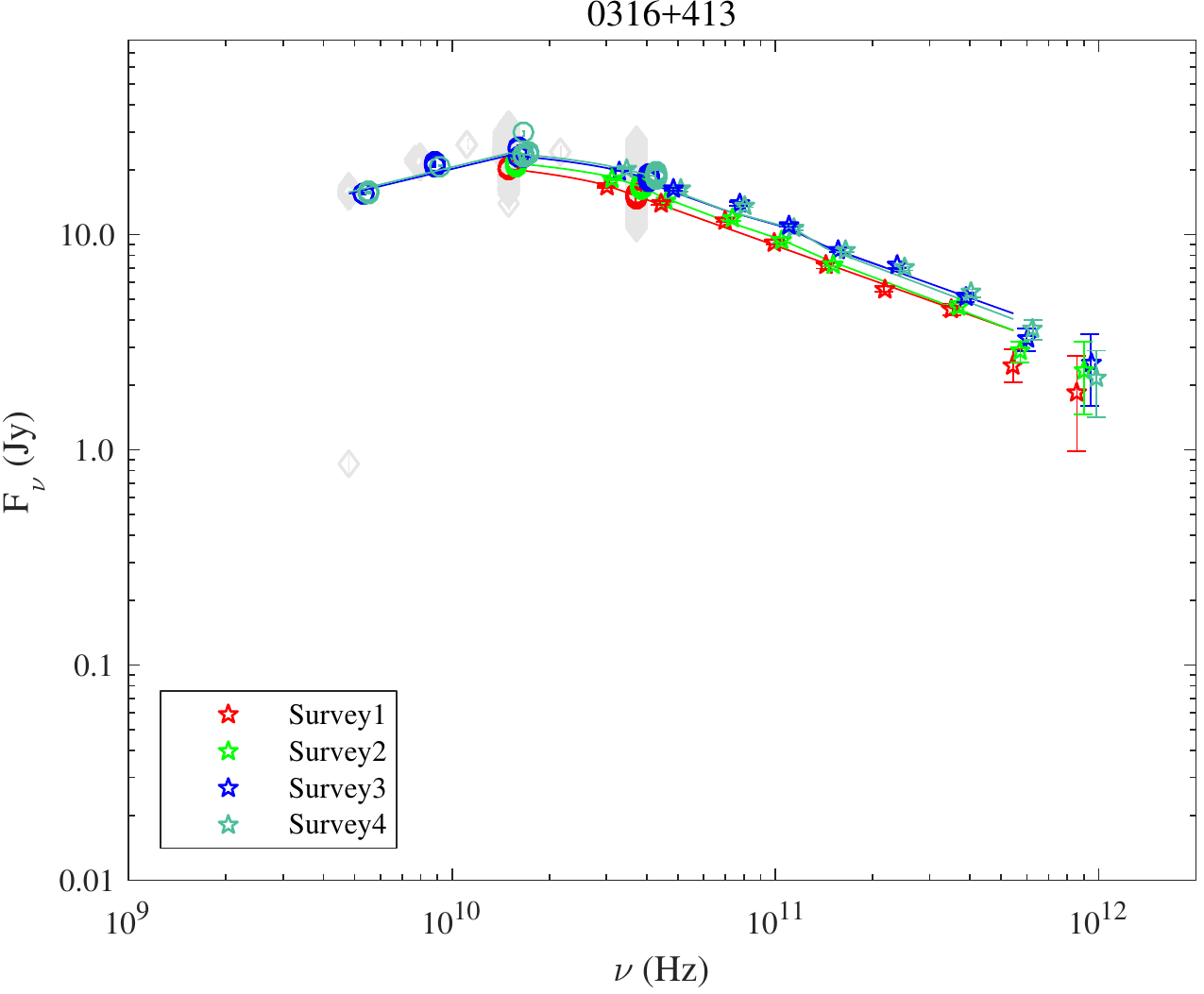}
	\caption{0316+413.}
	\label{0316+413_spectra}
	\end{minipage}
\end{figure*}

\begin{figure*}
	\centering
	\begin{minipage}[b]{.47\textwidth}
	\includegraphics[width=\columnwidth]{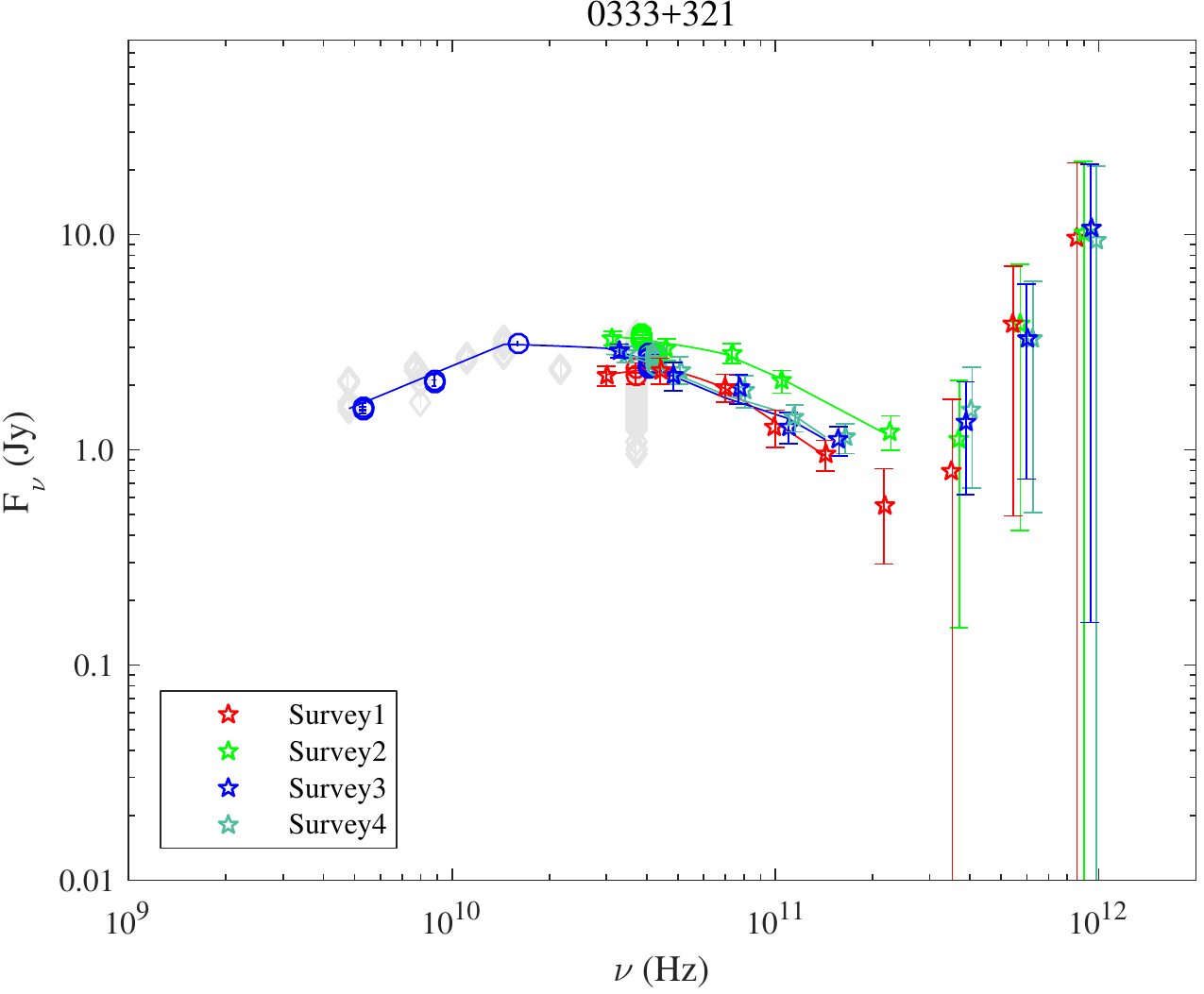}
	\caption{0333+321.}
	\label{0333+321_spectra}
	\end{minipage}\qquad
	\begin{minipage}[b]{.47\textwidth}
	\includegraphics[width=\columnwidth]{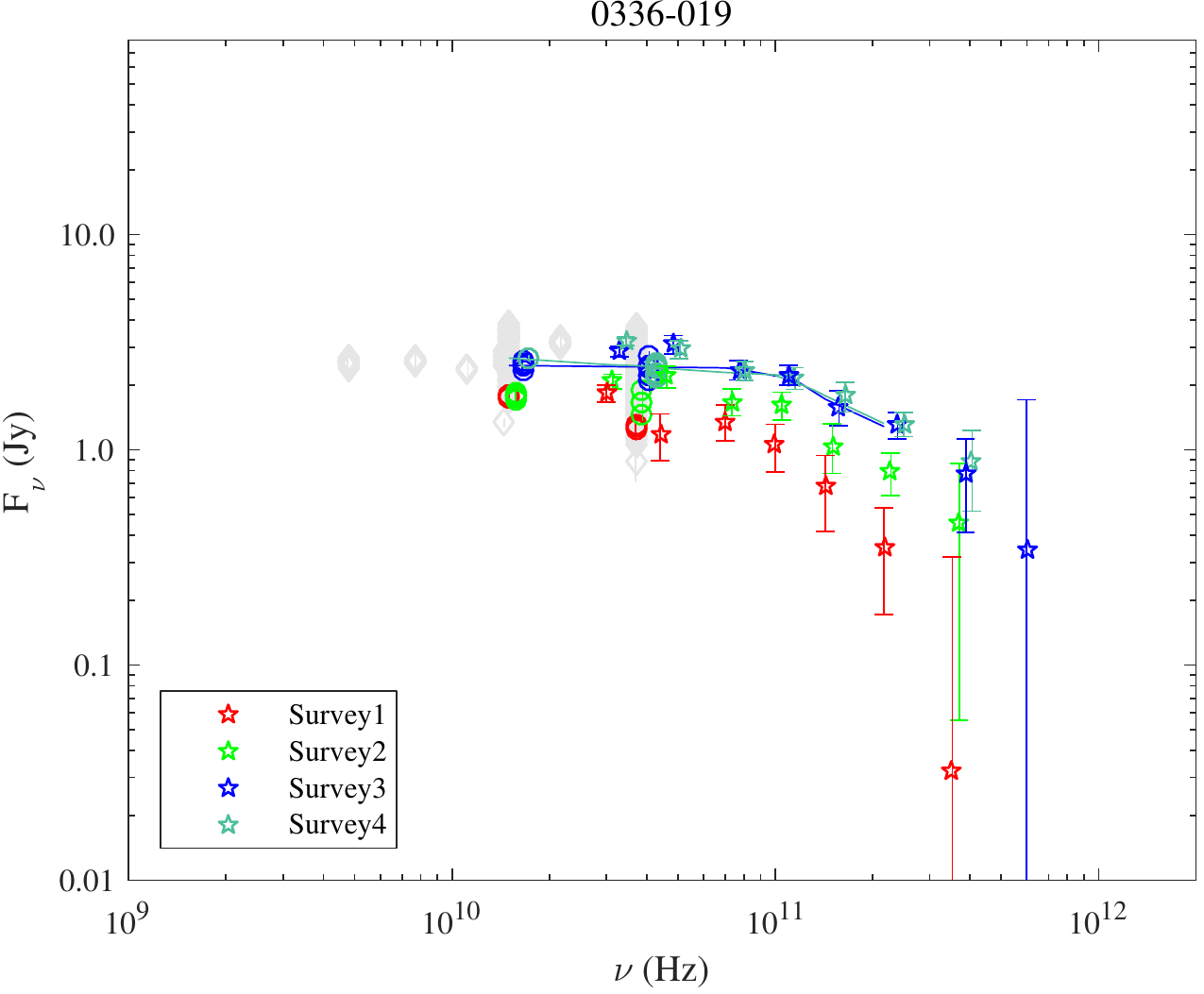}
	\caption{0336$-$019.}
	\label{0336-019_spectra}
	\end{minipage}
\end{figure*}

\clearpage

\begin{figure*}
	\centering
	\begin{minipage}[b]{.47\textwidth}
	\includegraphics[width=\columnwidth]{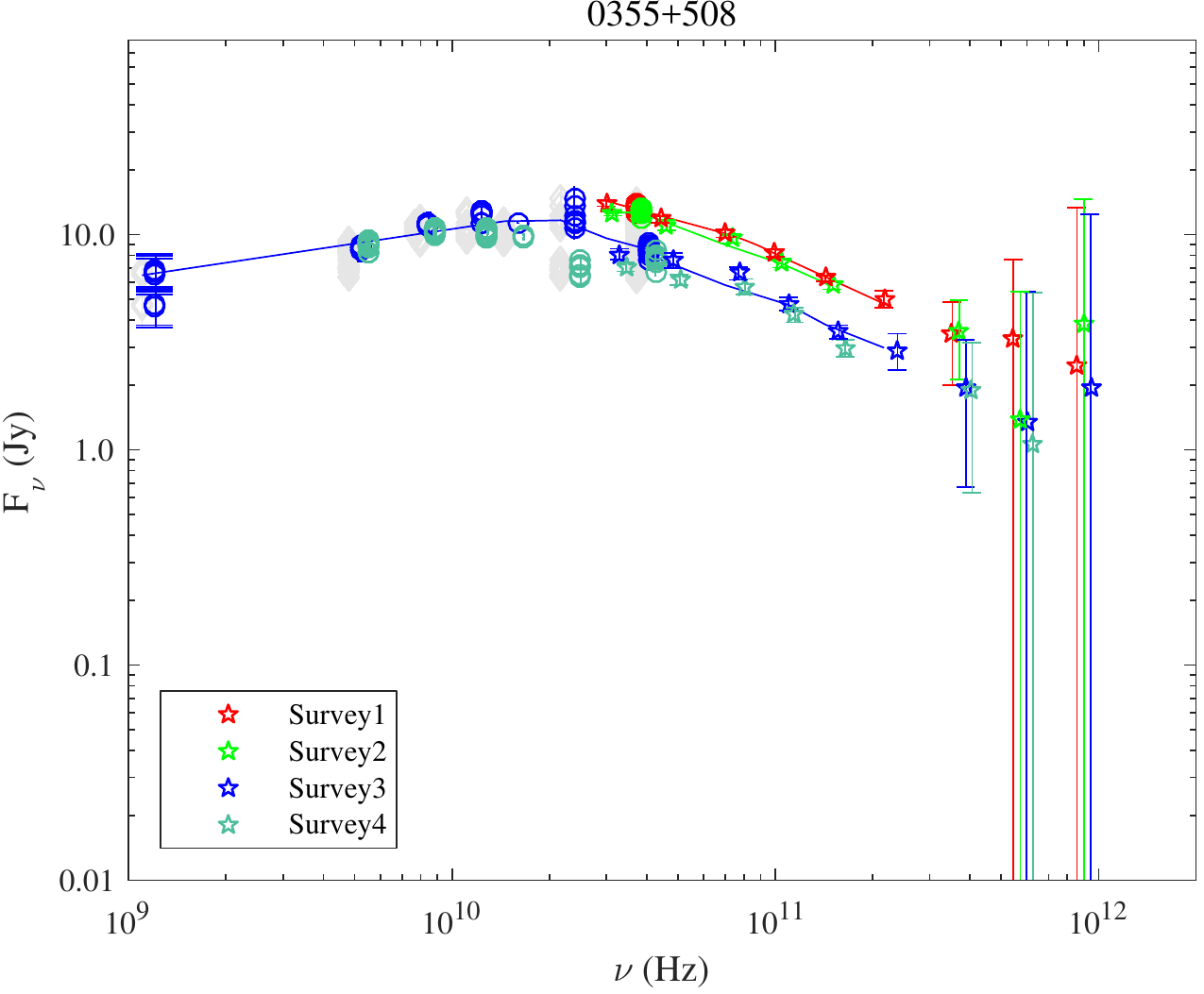}
	\caption{0355+508.}
	\label{0355+508_spectra}
	\end{minipage}\qquad
	\begin{minipage}[b]{.47\textwidth}
	\includegraphics[width=\columnwidth]{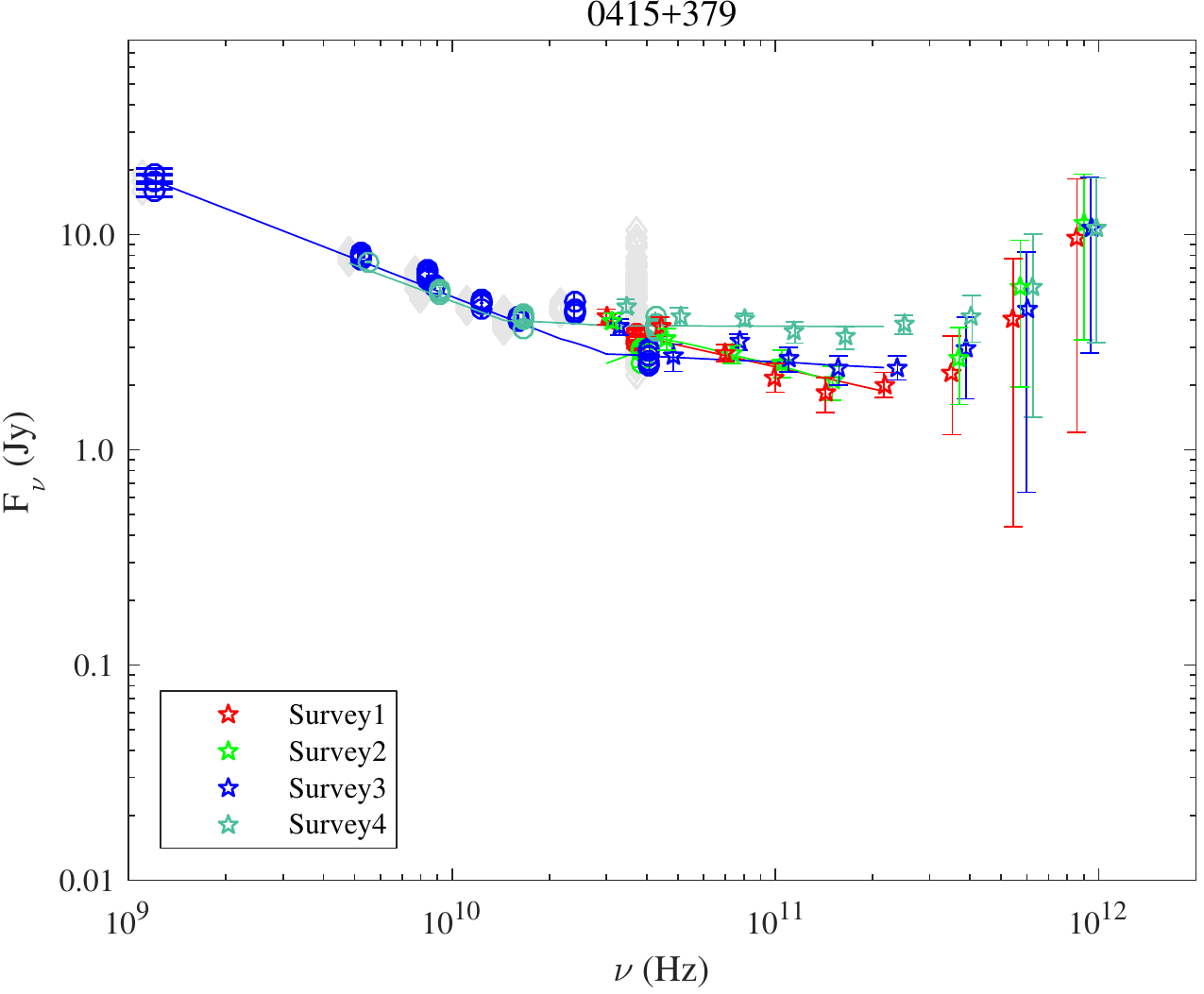}
	\caption{0415+379.}
	\label{0415+379_spectra}
	\end{minipage}
\end{figure*}

\begin{figure*}
	\centering
	\begin{minipage}[b]{.47\textwidth}
	\includegraphics[width=\columnwidth]{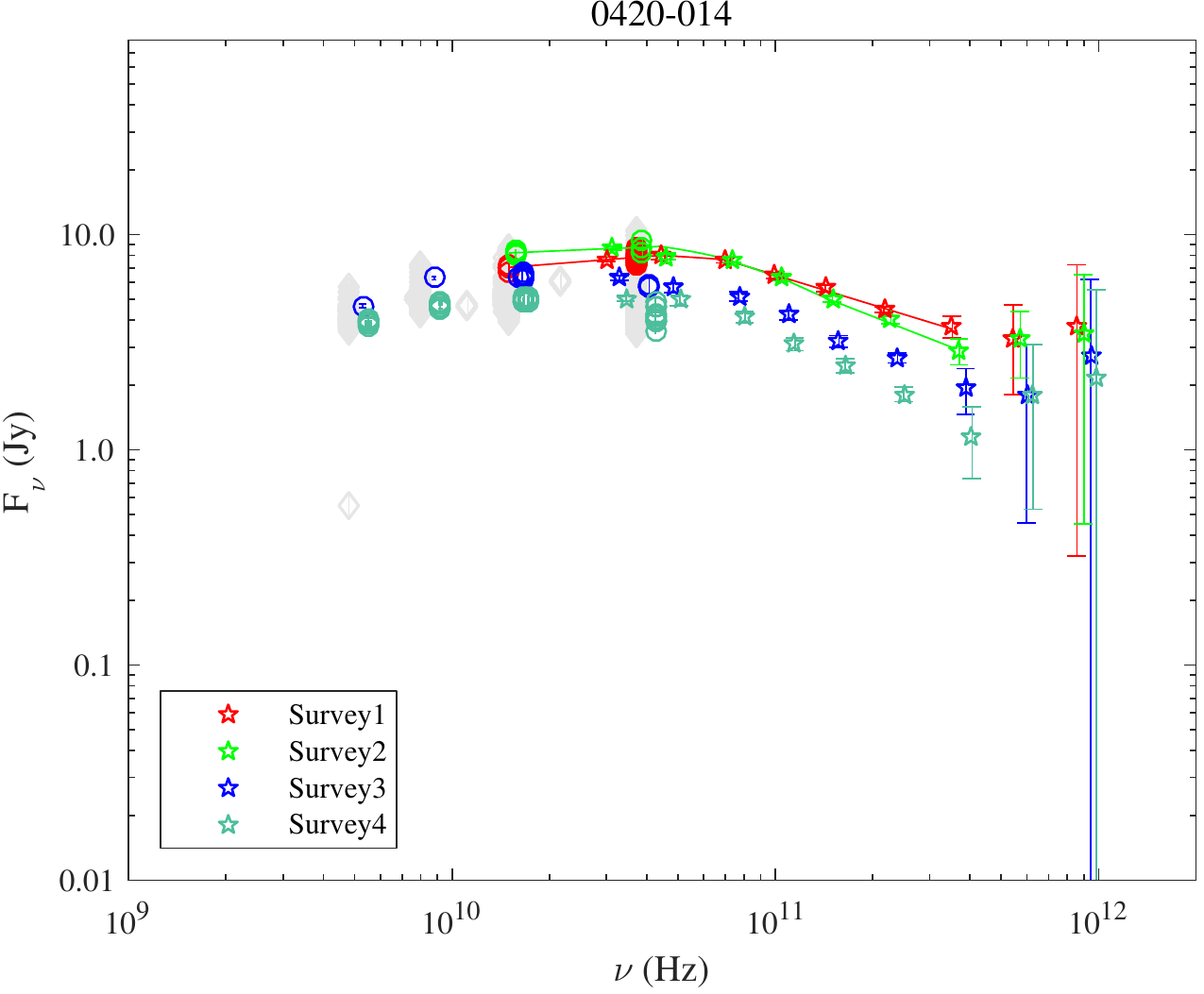}
	\caption{0420$-$014.}
	\label{0420-014_spectra}
	\end{minipage}\qquad
	\begin{minipage}[b]{.47\textwidth}
	\includegraphics[width=\columnwidth]{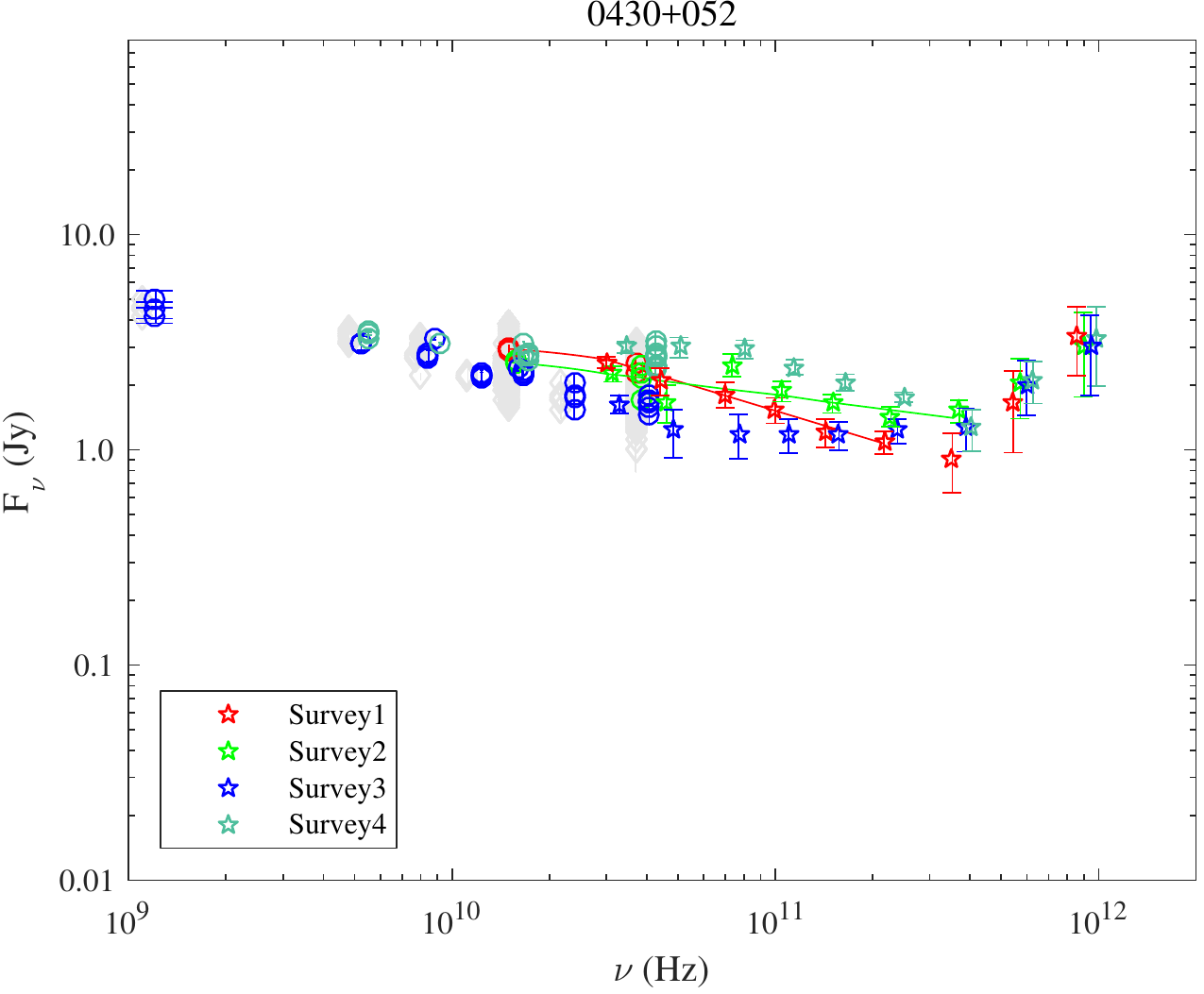}
	\caption{0430+052.}
	\label{0430+052_spectra}
	\end{minipage}
\end{figure*}

\begin{figure*}
	\centering
	\begin{minipage}[b]{.47\textwidth}
	\includegraphics[width=\columnwidth]{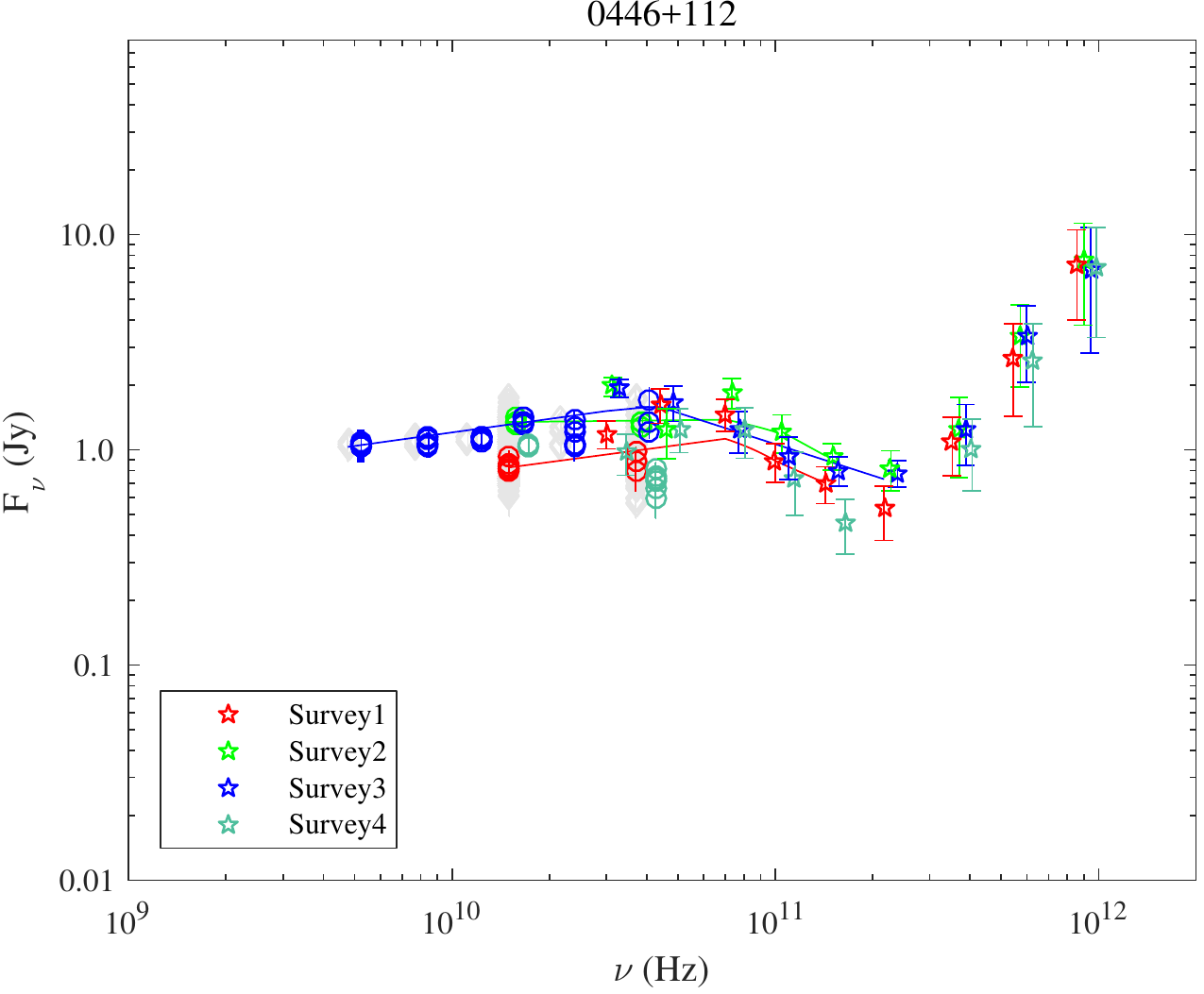}
	\caption{0446+112.}
	\label{0446+112_spectra}
	\end{minipage}\qquad
	\begin{minipage}[b]{.47\textwidth}
	\includegraphics[width=\columnwidth]{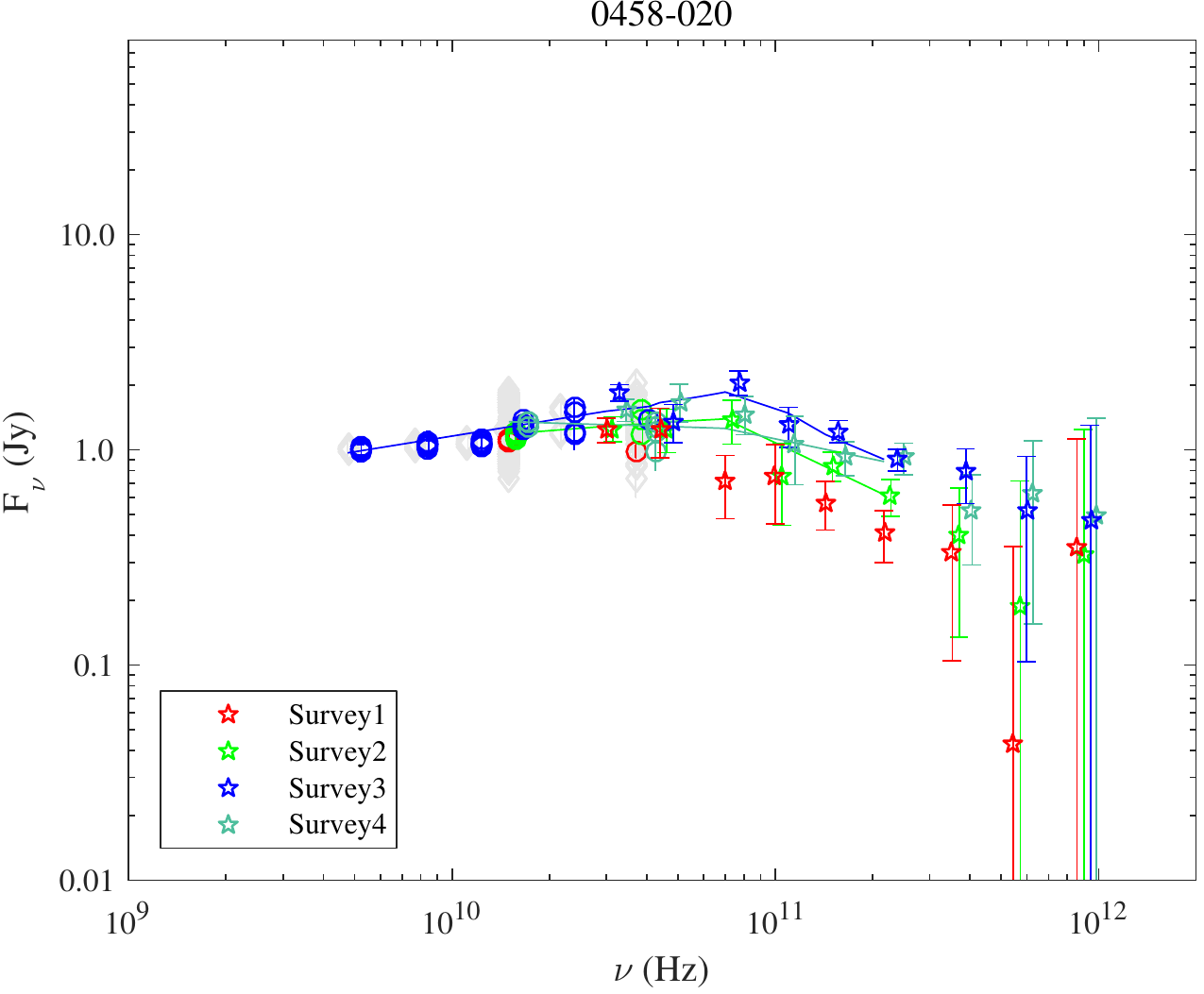}
	\caption{0458$-$020.}
	\label{0458-020_spectra}
	\end{minipage}
\end{figure*}

\clearpage

\begin{figure*}
	\centering
	\begin{minipage}[b]{.47\textwidth}
	\includegraphics[width=\columnwidth]{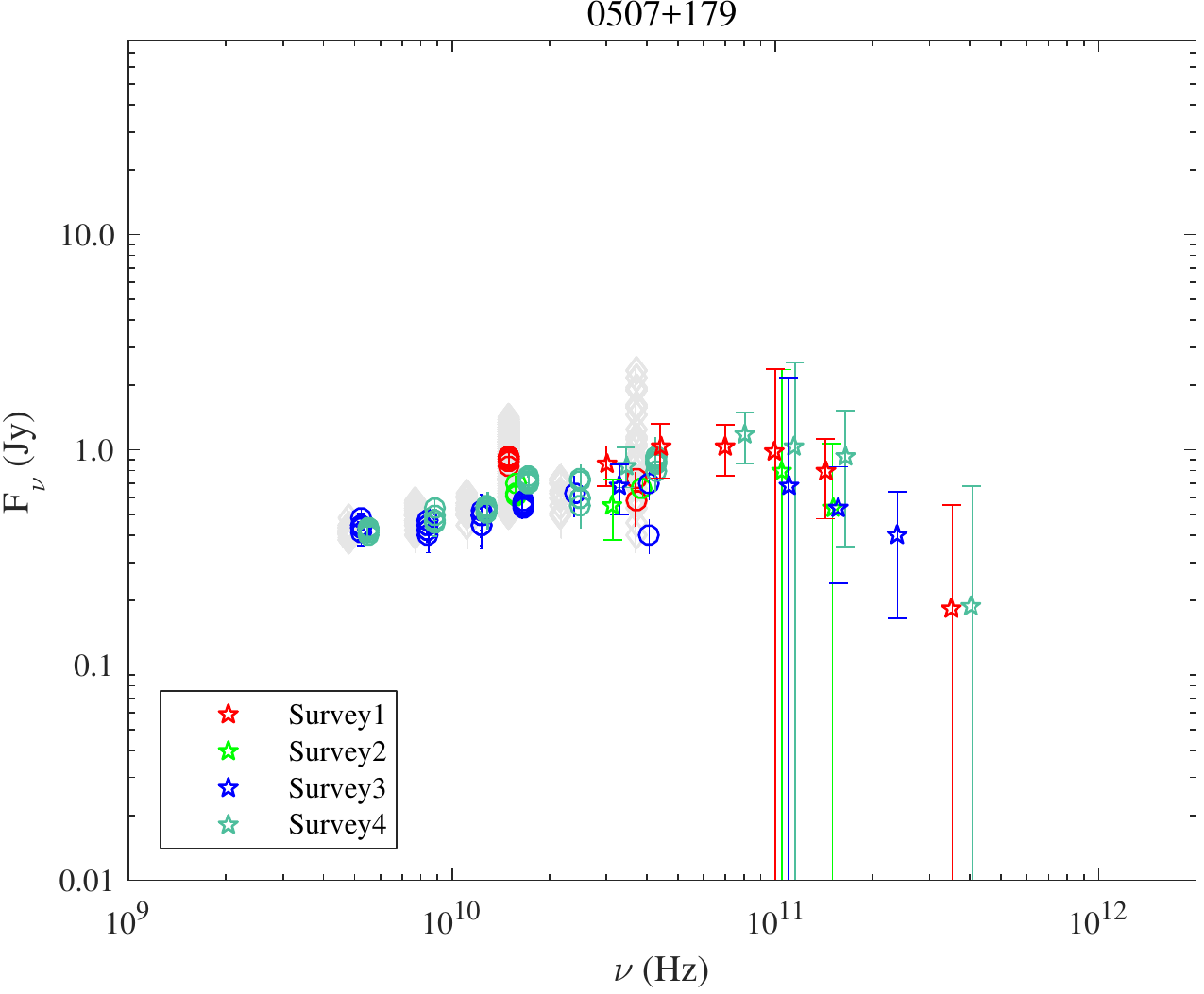}
	\caption{0507+179.}
	\label{0507+179_spectra}
	\end{minipage}\qquad
	\begin{minipage}[b]{.47\textwidth}
	\includegraphics[width=\columnwidth]{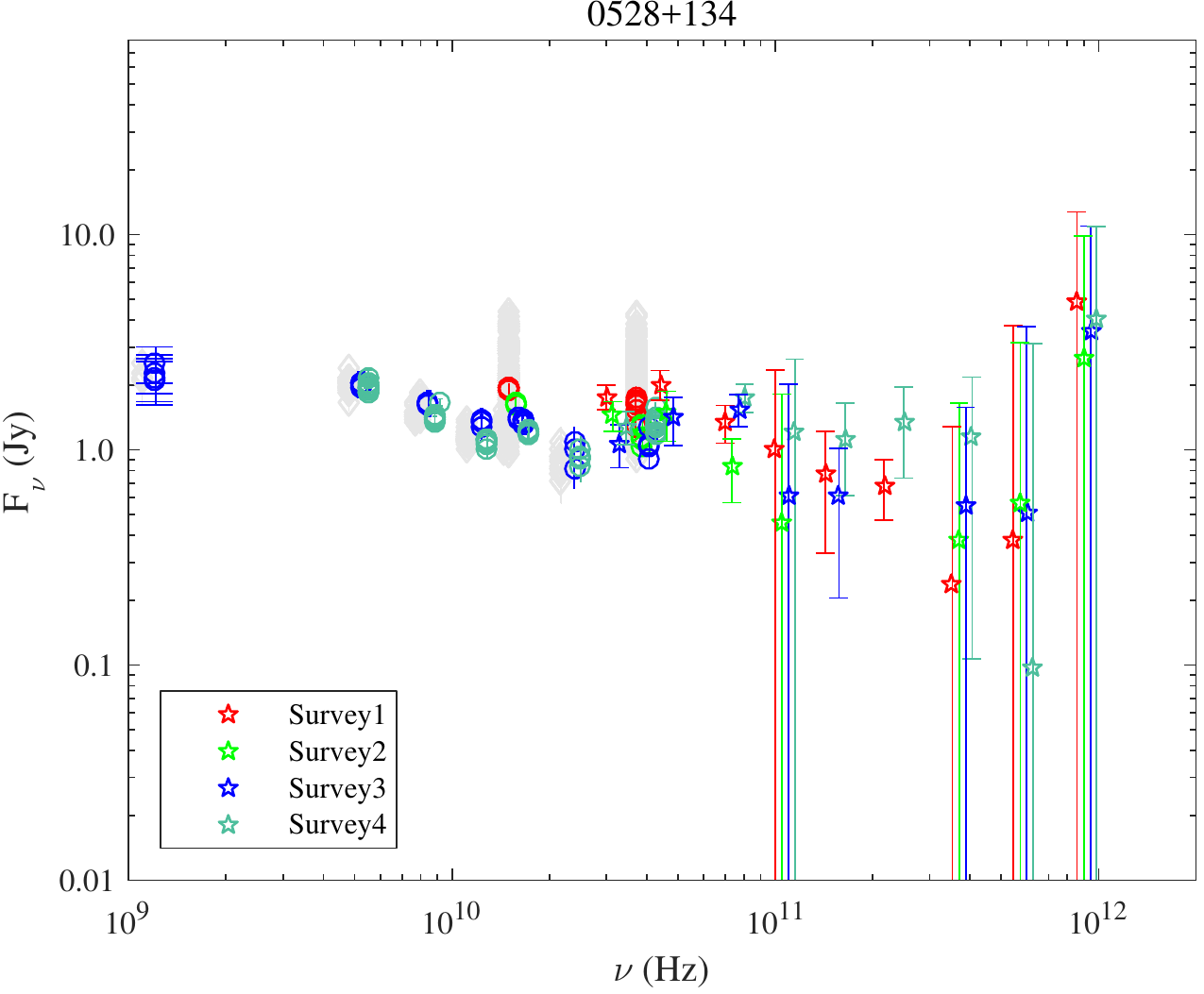}
	\caption{0528+134.}
	\label{0528+134_spectra}
	\end{minipage}
\end{figure*}

\begin{figure*}
	\centering
	\begin{minipage}[b]{.47\textwidth}
	\includegraphics[width=\columnwidth]{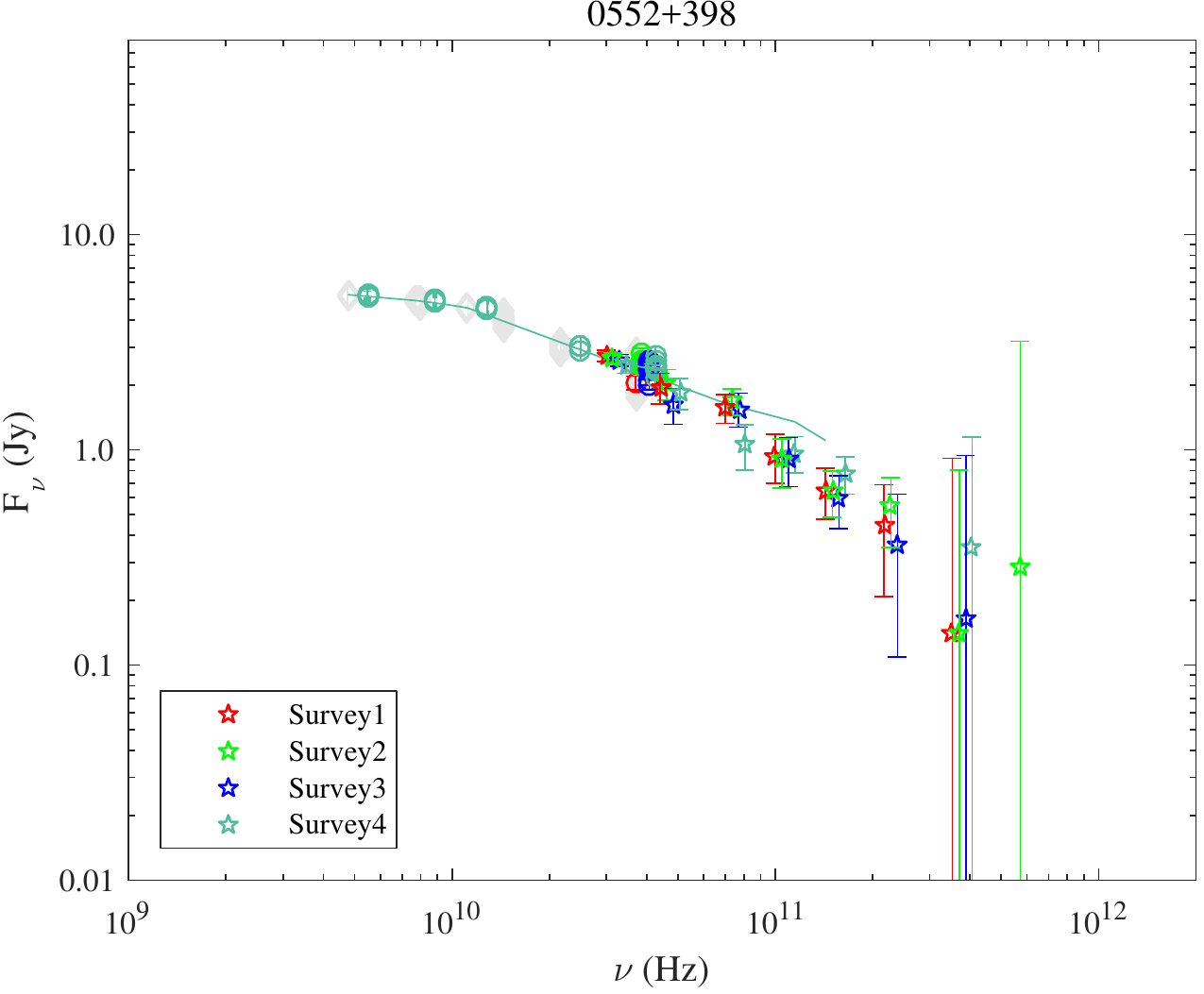}
	\caption{0552+398.}
	\label{0552+398_spectra}
	\end{minipage}\qquad
	\begin{minipage}[b]{.47\textwidth}
	\includegraphics[width=\columnwidth]{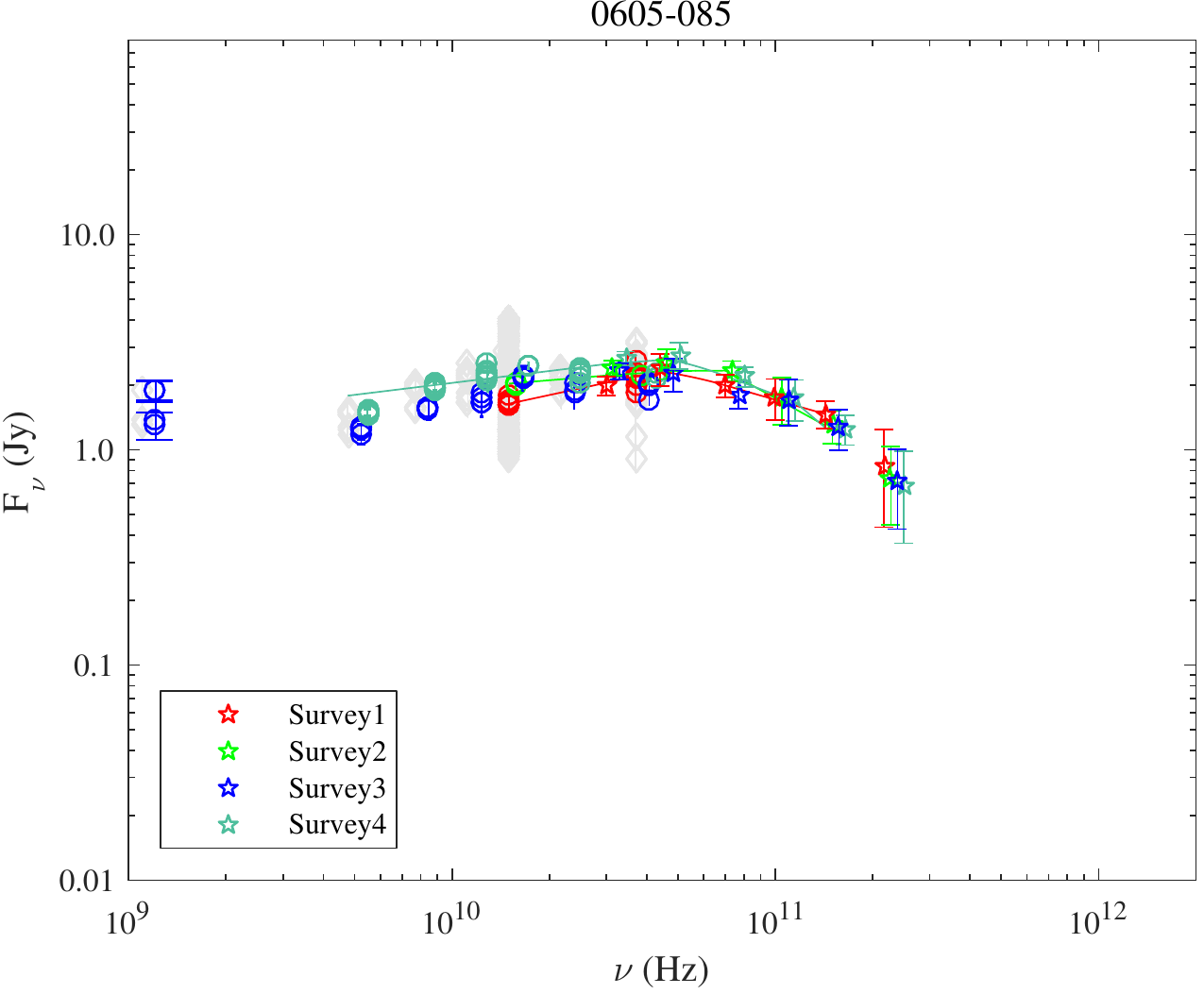}
	\caption{0605$-$085.}
	\label{0605-085_spectra}
	\end{minipage}
\end{figure*}

\begin{figure*}
	\centering
	\begin{minipage}[b]{.47\textwidth}
	\includegraphics[width=\columnwidth]{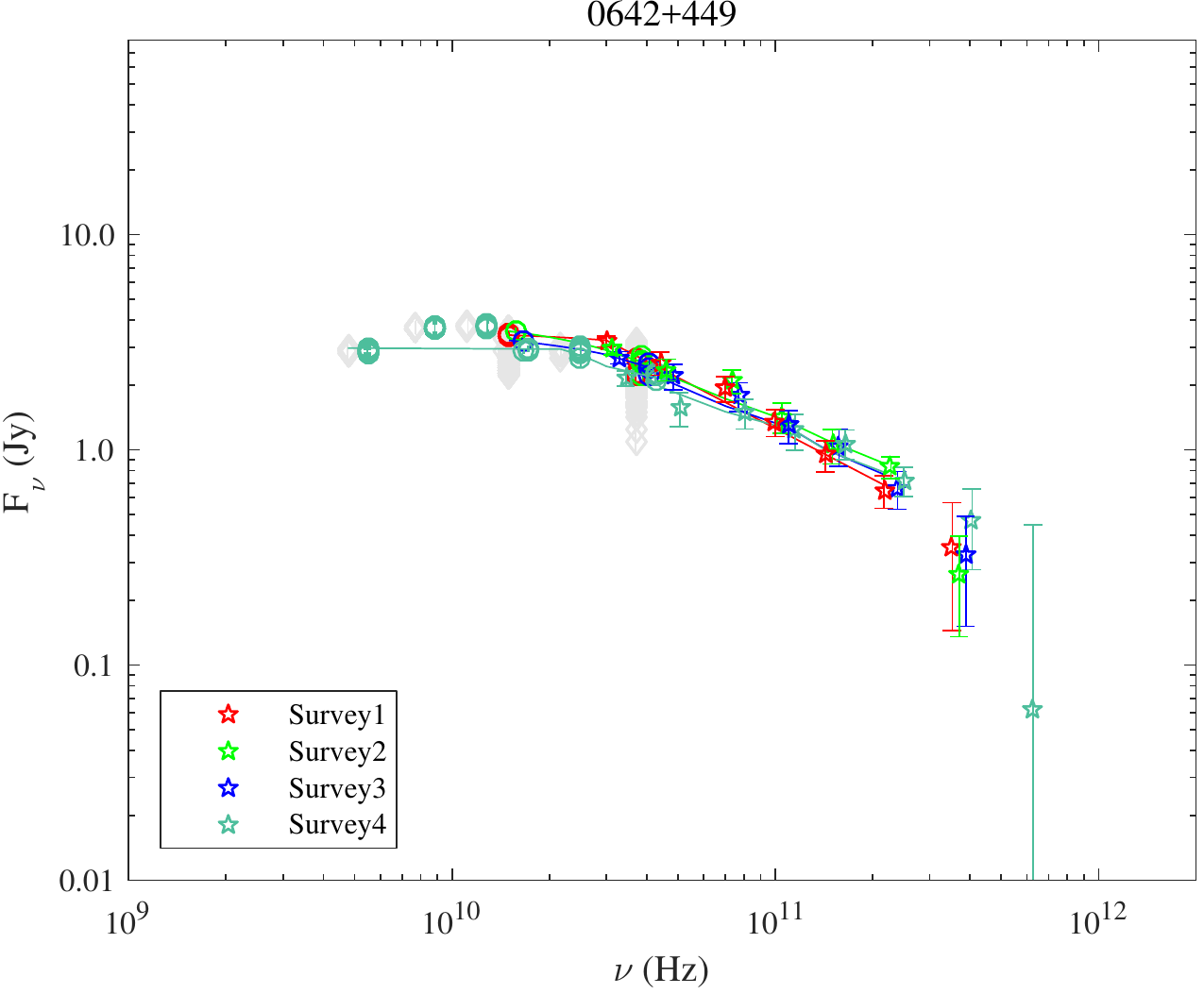}
	\caption{0642+449.}
	\label{0642+449_spectra}
	\end{minipage}\qquad
	\begin{minipage}[b]{.47\textwidth}
	\includegraphics[width=\columnwidth]{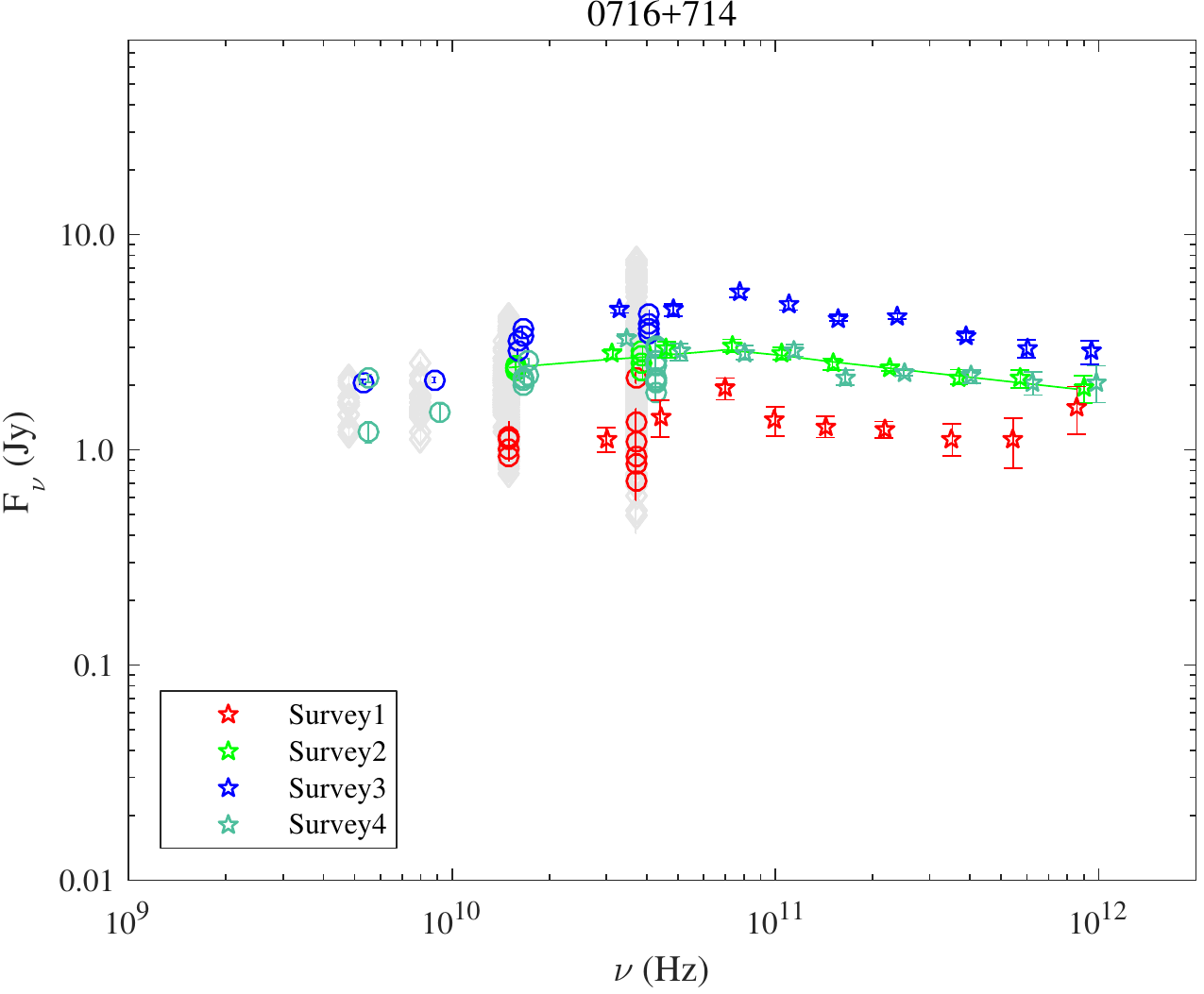}
	\caption{0716+714.}
	\label{0716+714_spectra}
	\end{minipage}
\end{figure*}

\clearpage

\begin{figure*}
	\centering
	\begin{minipage}[b]{.47\textwidth}
	\includegraphics[width=\columnwidth]{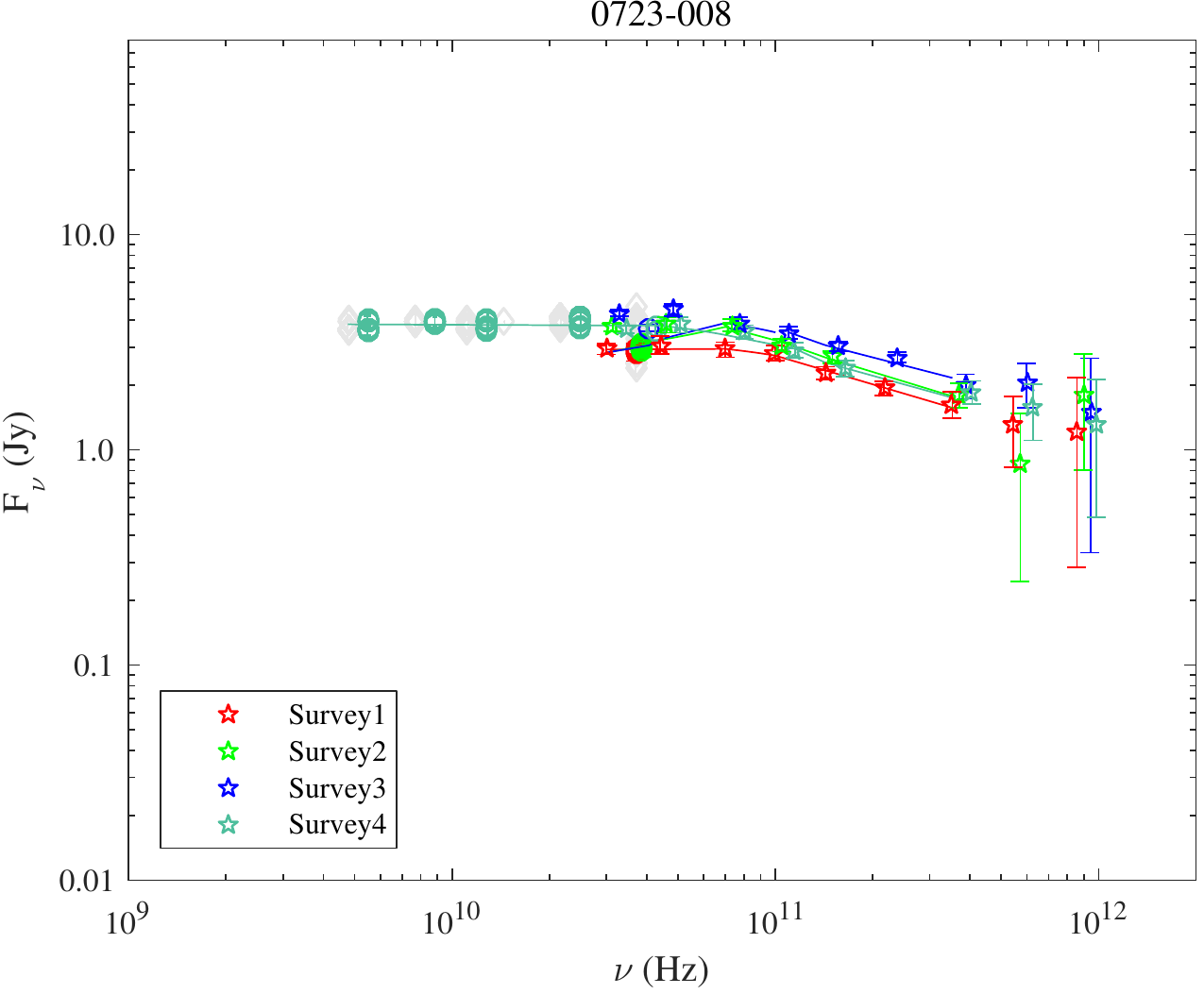}
	\caption{0723$-$008.}
	\label{0723-008_spectra}
	\end{minipage}\qquad
	\begin{minipage}[b]{.47\textwidth}
	\includegraphics[width=\columnwidth]{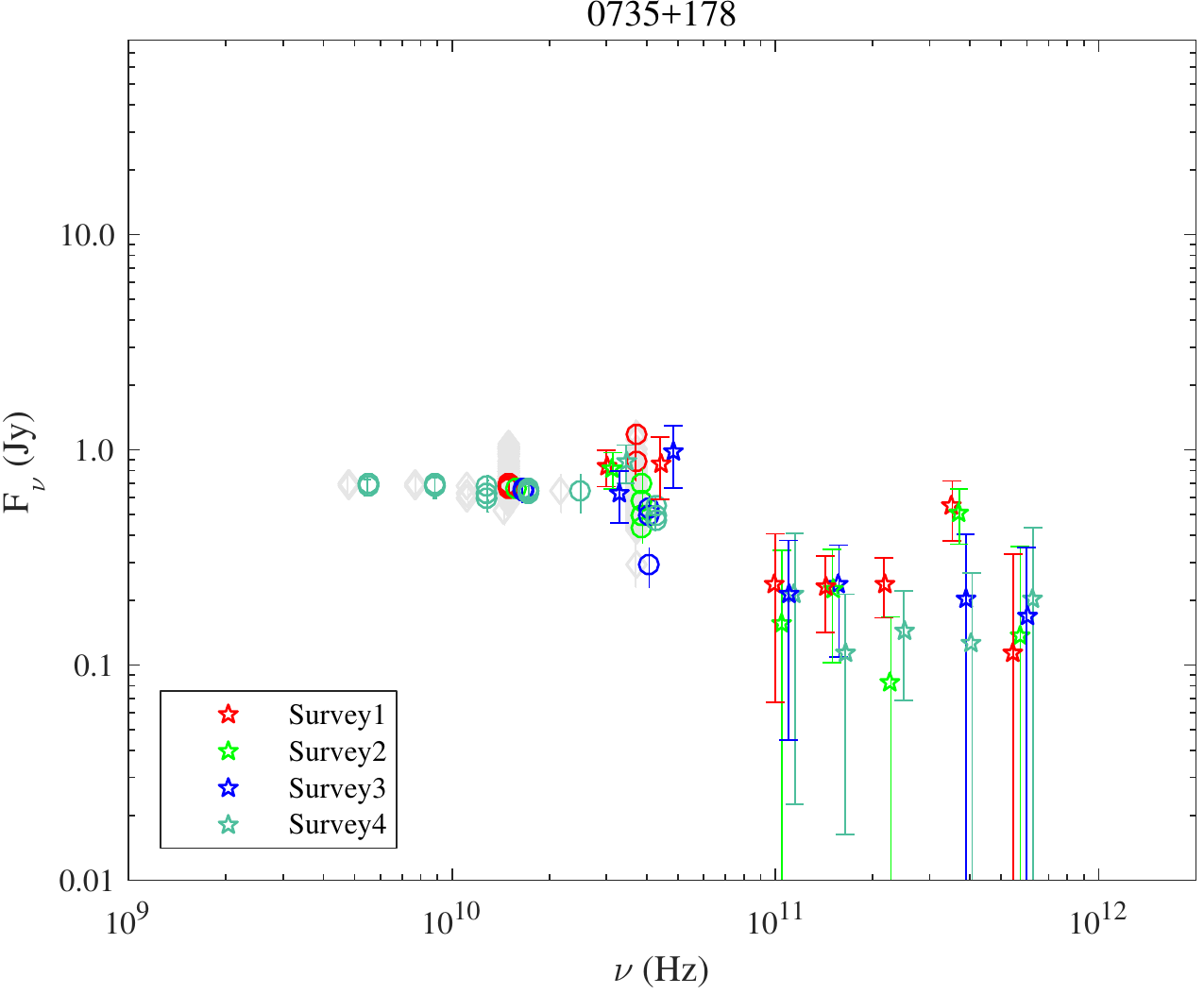}
	\caption{0735+178.}
	\label{0735+178_spectra}
	\end{minipage}
\end{figure*}

\begin{figure*}
	\centering
	\begin{minipage}[b]{.47\textwidth}
	\includegraphics[width=\columnwidth]{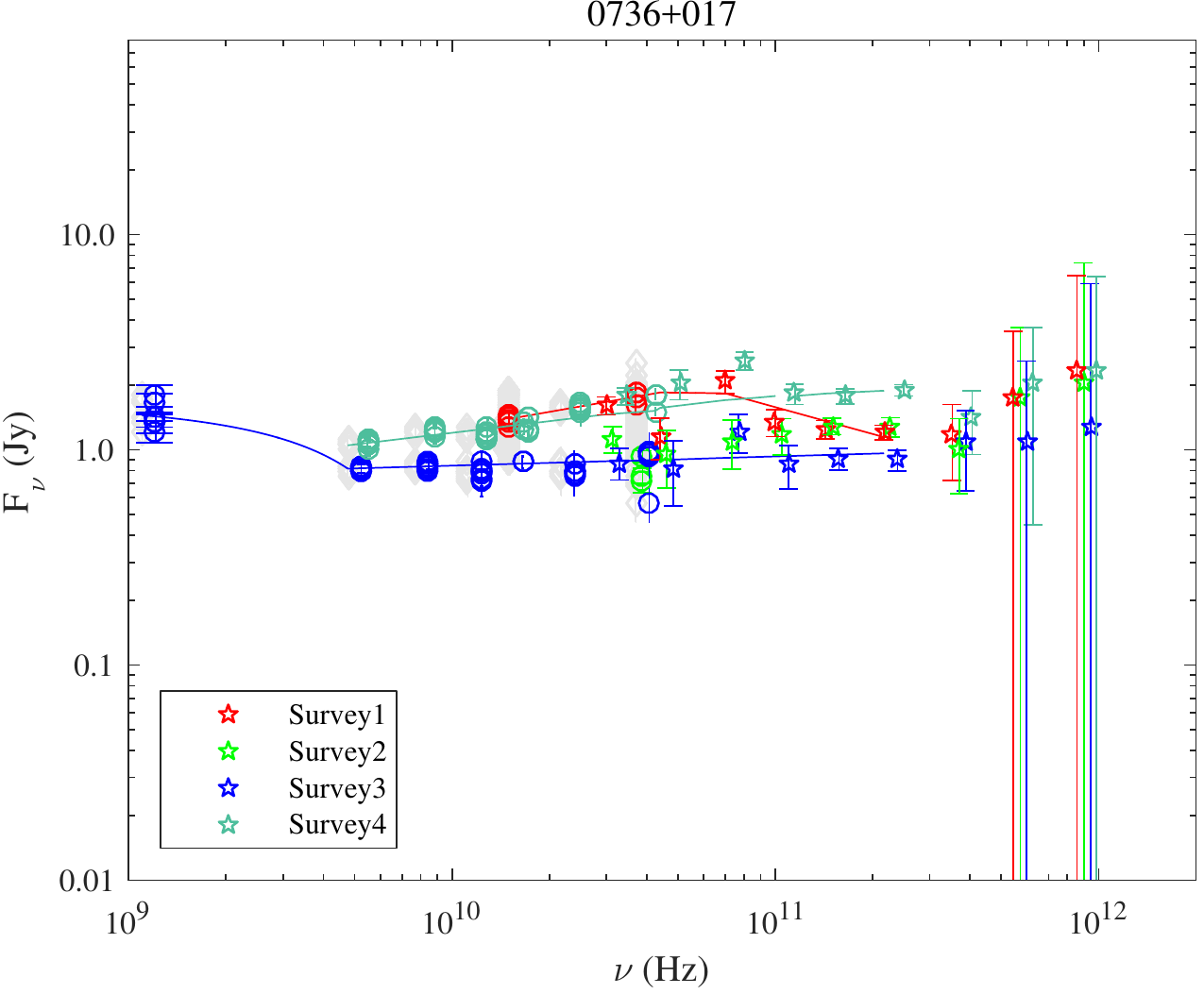}
	\caption{0736+017.}
	\label{0736+017_spectra}
	\end{minipage}\qquad
	\begin{minipage}[b]{.47\textwidth}
	\includegraphics[width=\columnwidth]{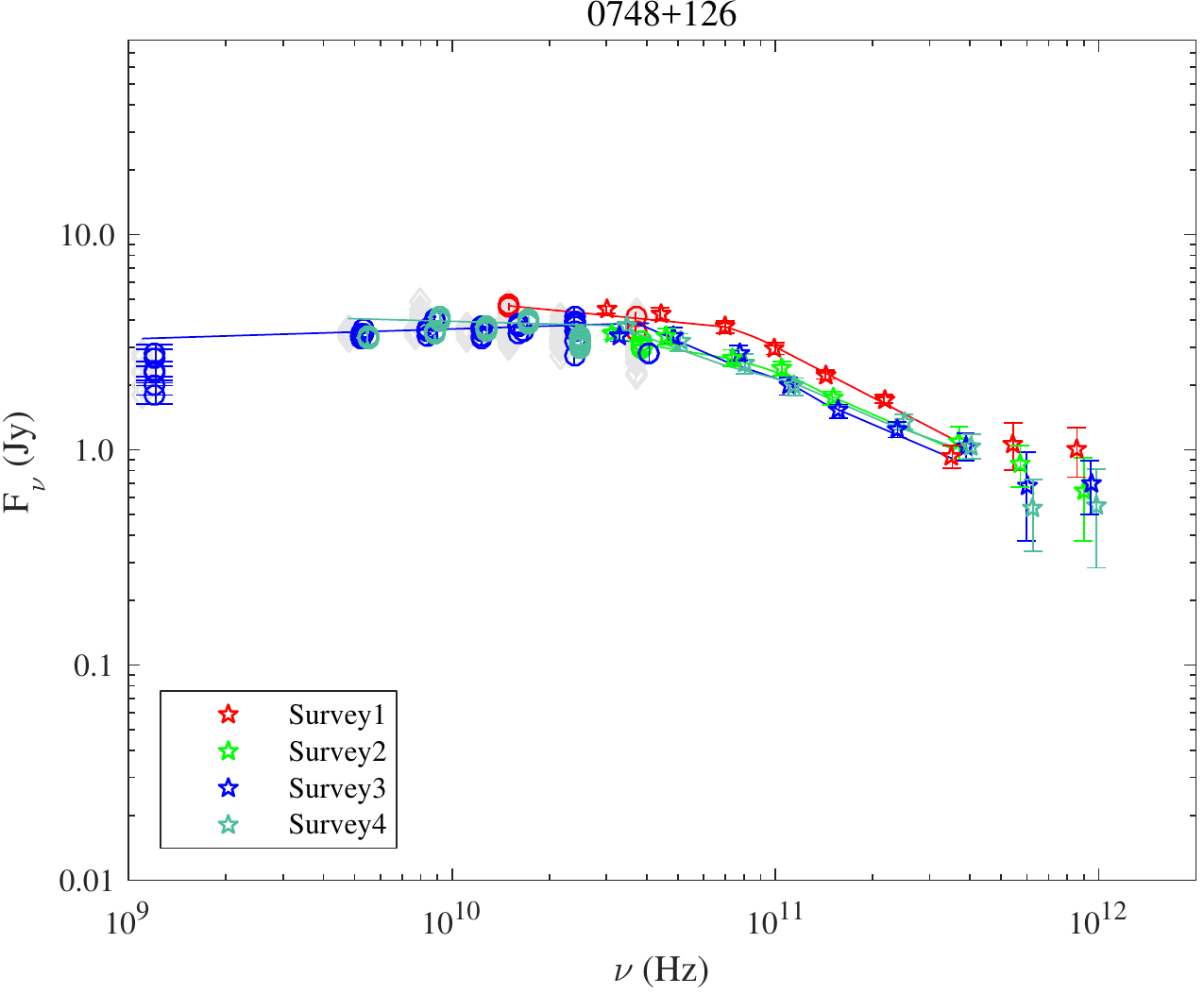}
	\caption{0748+126.}
	\label{0748+126_spectra}
	\end{minipage}
\end{figure*}

\begin{figure*}
	\centering
	\begin{minipage}[b]{.47\textwidth}
	\includegraphics[width=\columnwidth]{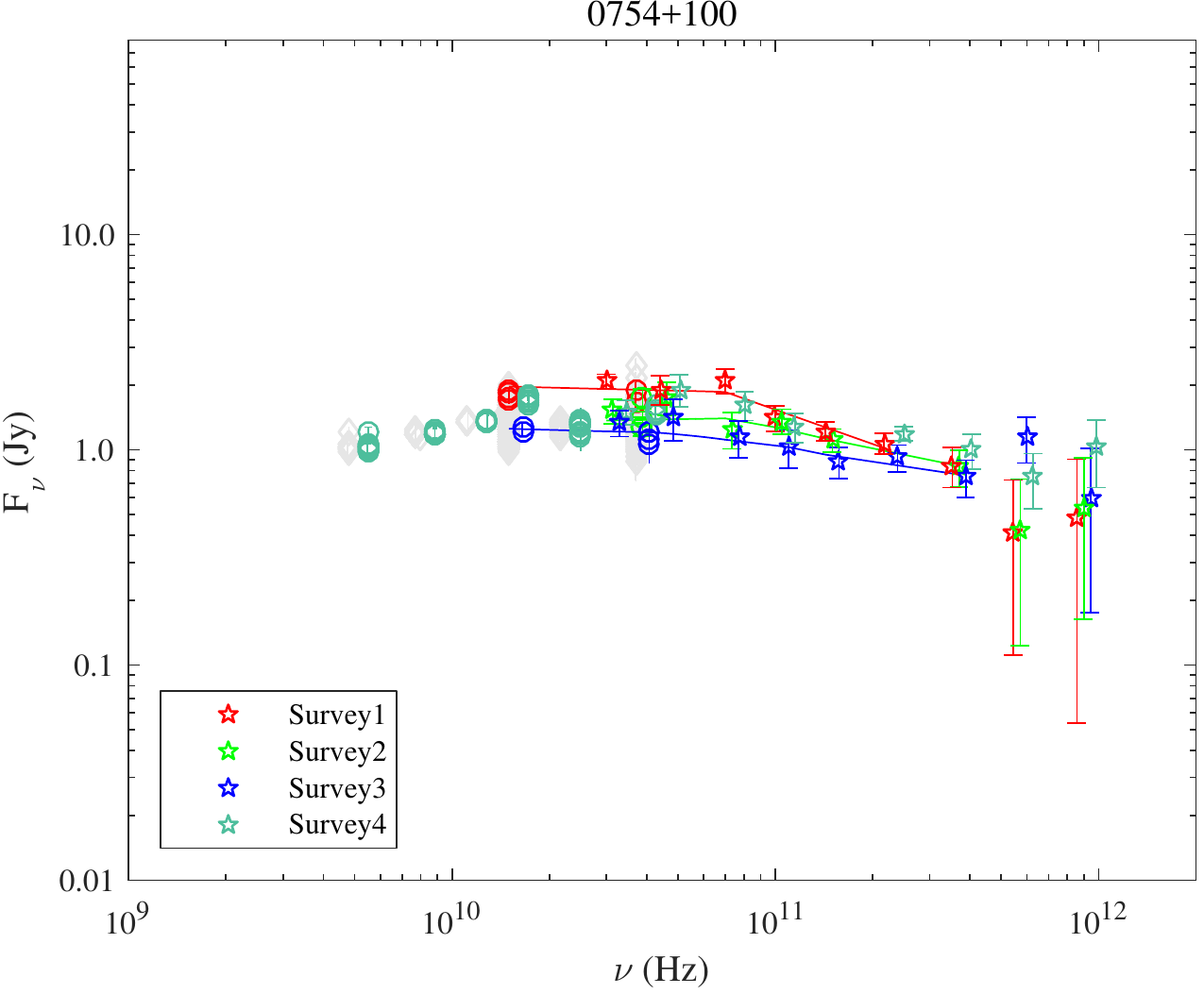}
	\caption{0754+100.}
	\label{0754+100_spectra}
	\end{minipage}\qquad
	\begin{minipage}[b]{.47\textwidth}
	\includegraphics[width=\columnwidth]{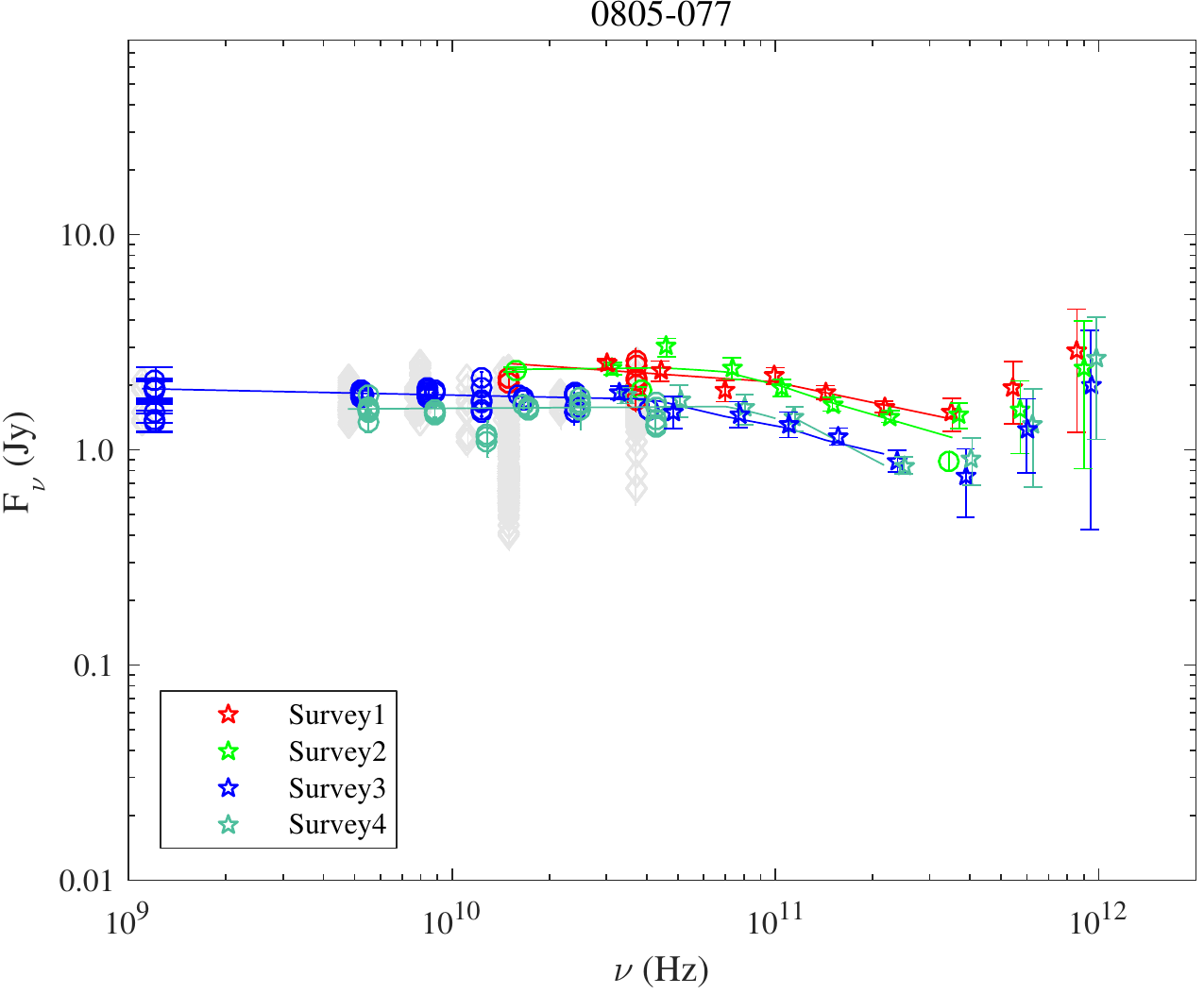}
	\caption{0805$-$077.}
	\label{0805-077_spectra}
	\end{minipage}
\end{figure*}

\clearpage

\begin{figure*}
	\centering
	\begin{minipage}[b]{.47\textwidth}
	\includegraphics[width=\columnwidth]{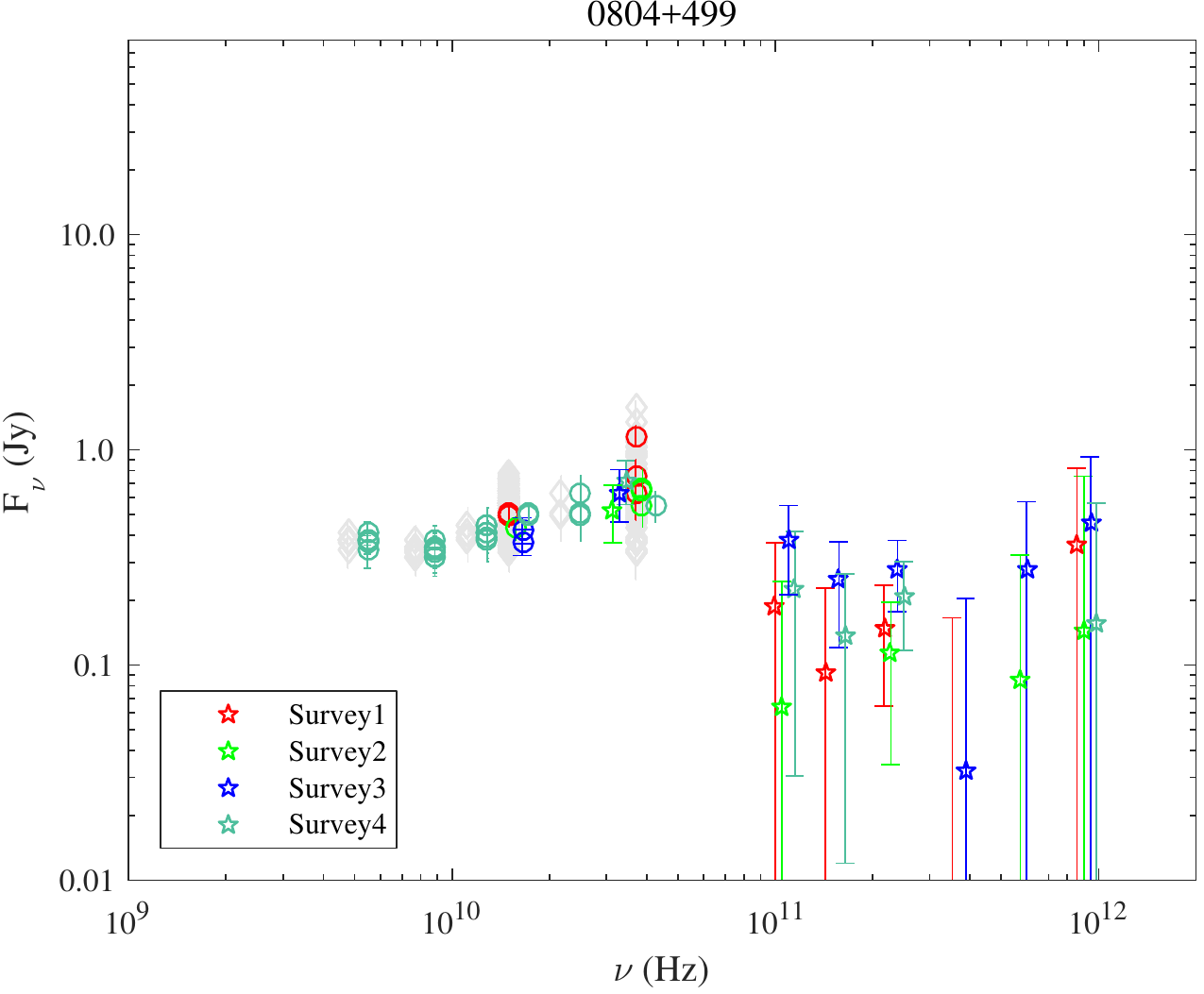}
	\caption{0804+499.}
	\label{0804+499_spectra}
	\end{minipage}\qquad
	\begin{minipage}[b]{.47\textwidth}
	\includegraphics[width=\columnwidth]{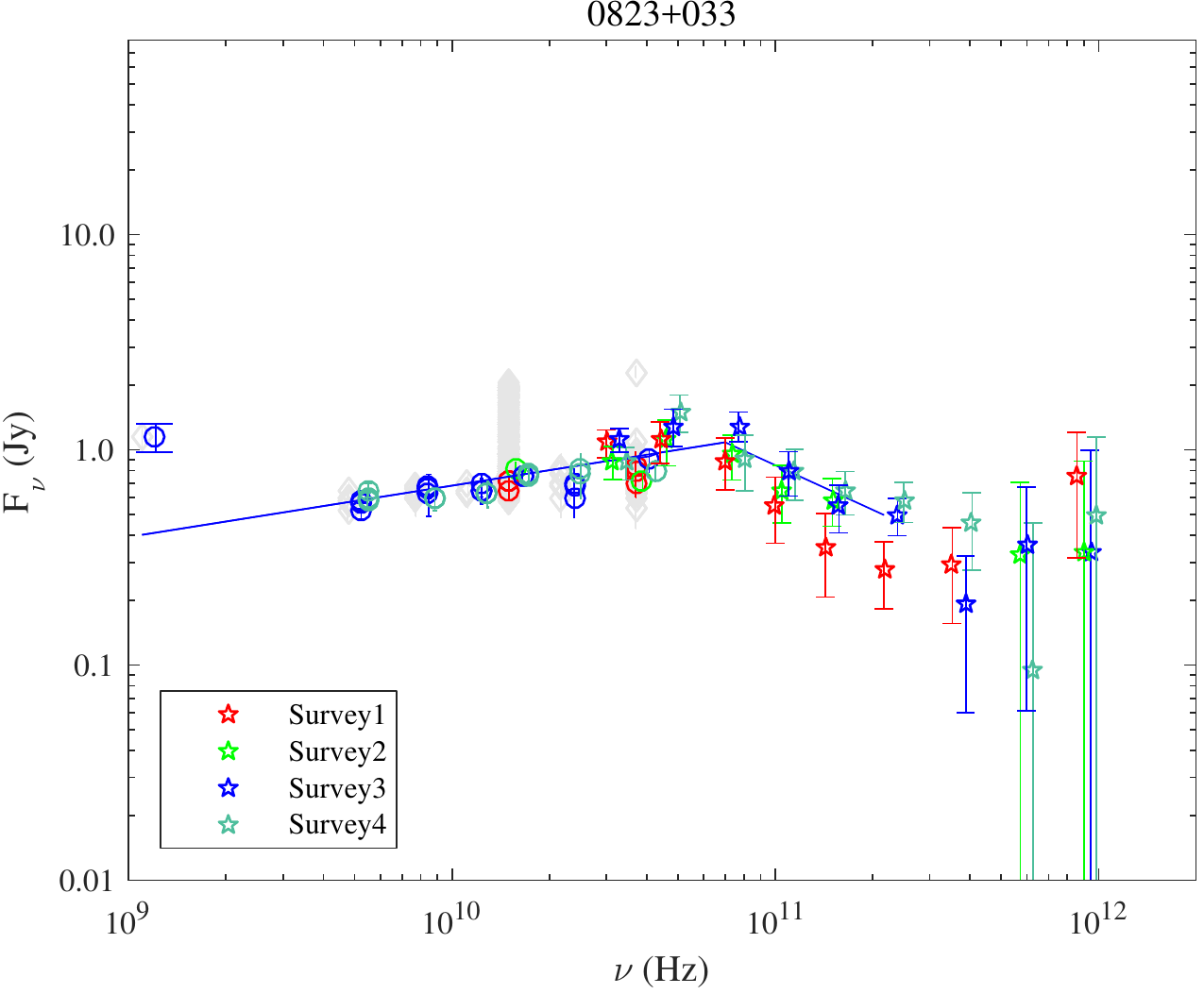}
	\caption{0823+033.}
	\label{0823+033_spectra}
	\end{minipage}
\end{figure*}

\begin{figure*}
	\centering
	\begin{minipage}[b]{.47\textwidth}
	\includegraphics[width=\columnwidth]{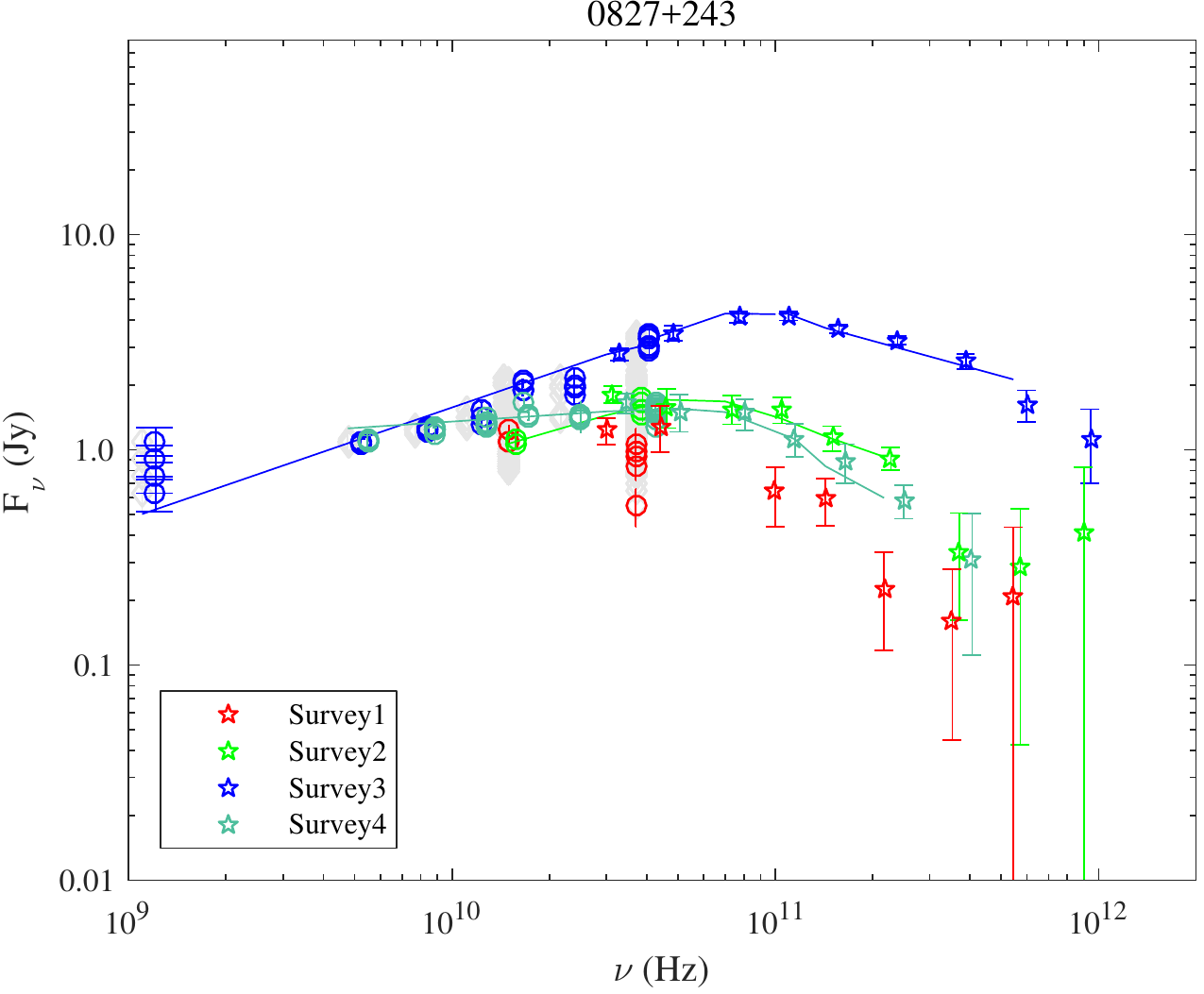}
	\caption{0827+243.}
	\label{0827+243_spectra}
	\end{minipage}\qquad
	\begin{minipage}[b]{.47\textwidth}
	\includegraphics[width=\columnwidth]{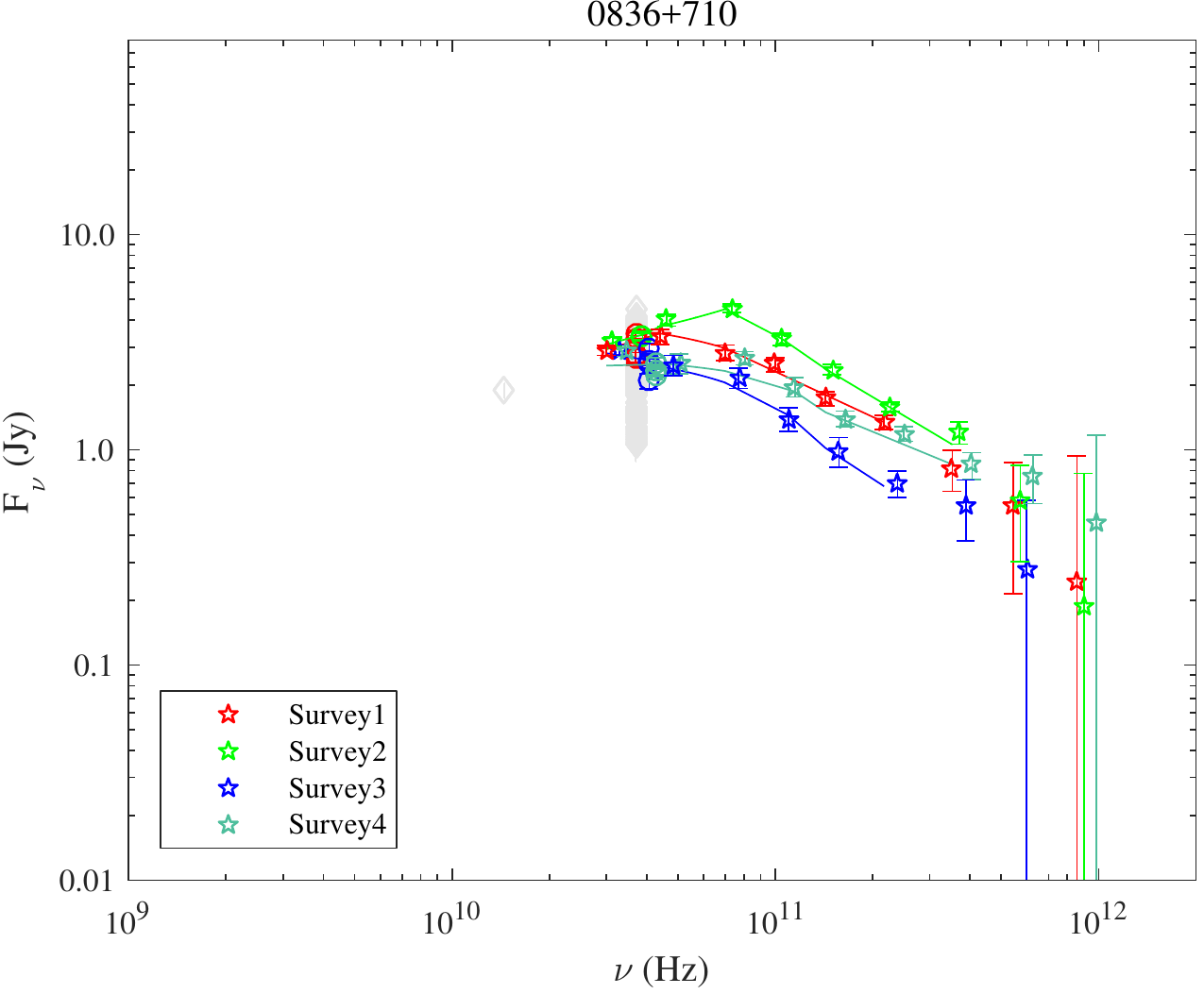}
	\caption{0836+710.}
	\label{0836+710_spectra}
	\end{minipage}
\end{figure*}

\begin{figure*}
	\centering
	\begin{minipage}[b]{.47\textwidth}
	\includegraphics[width=\columnwidth]{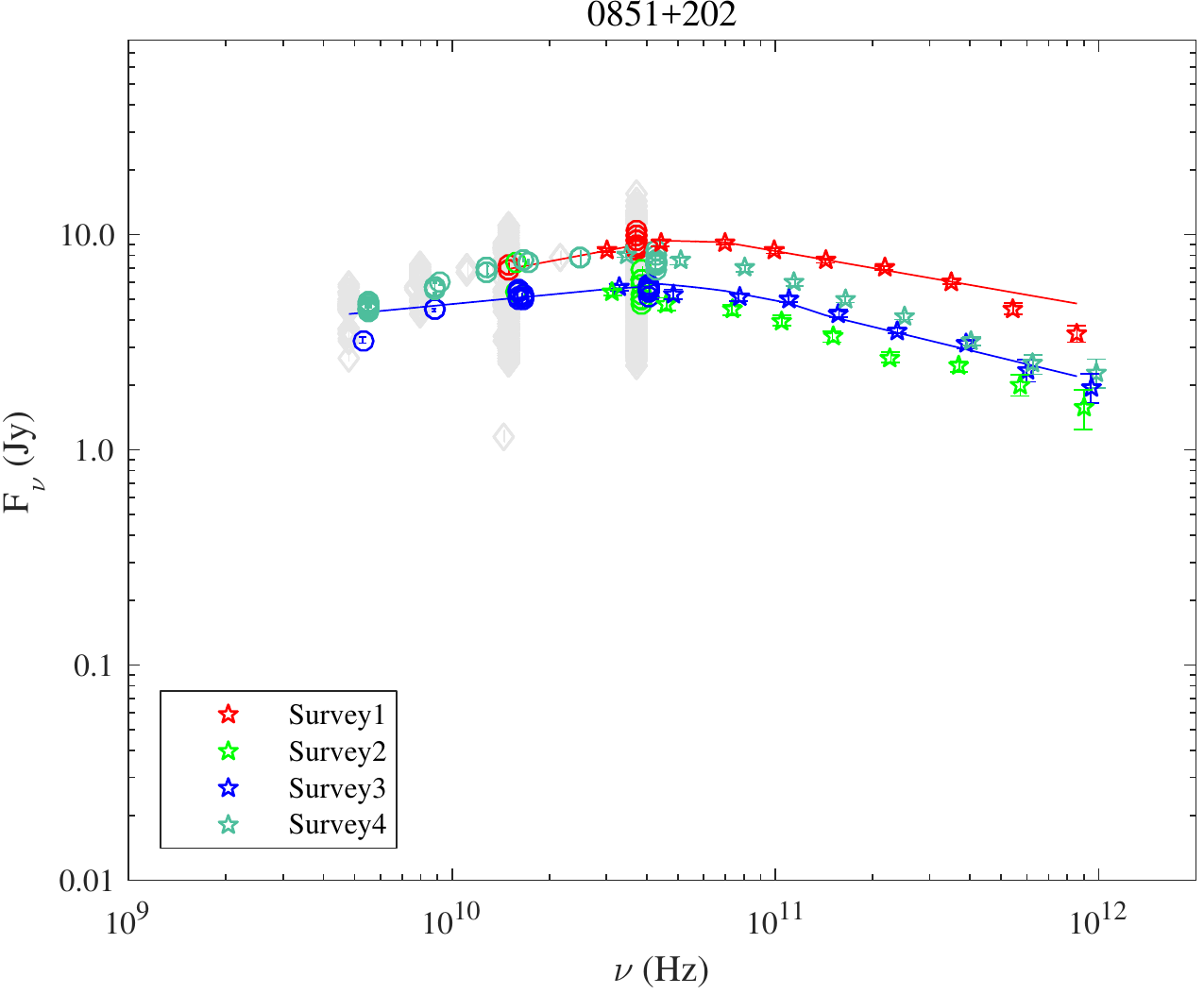}
	\caption{0851+202.}
	\label{0851+202_spectra}
	\end{minipage}\qquad
	\begin{minipage}[b]{.47\textwidth}
	\includegraphics[width=\columnwidth]{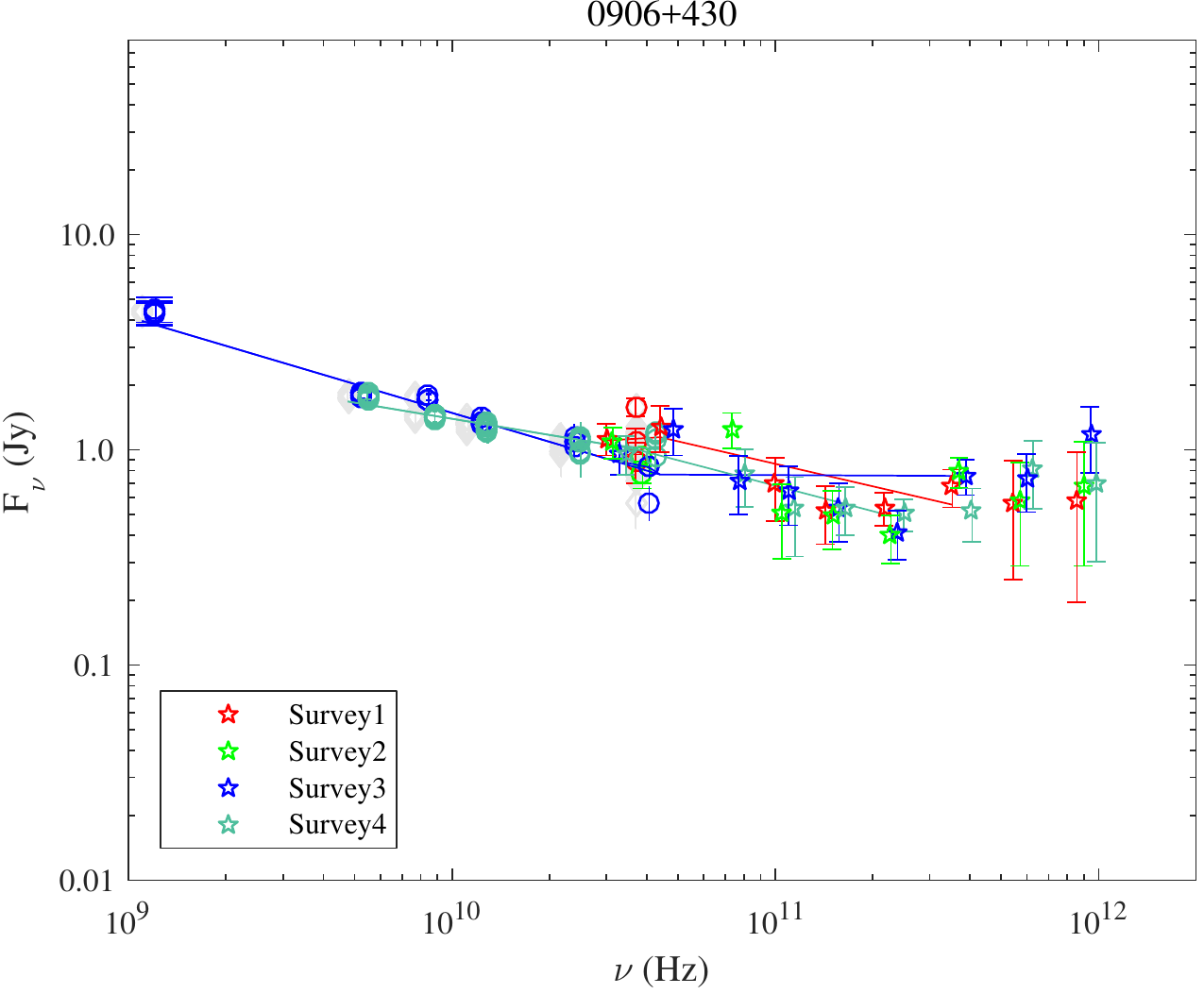}
	\caption{0906+430.}
	\label{0906+430_spectra}
	\end{minipage}
\end{figure*}

\clearpage

\begin{figure*}
	\centering
	\begin{minipage}[b]{.47\textwidth}
	\includegraphics[width=\columnwidth]{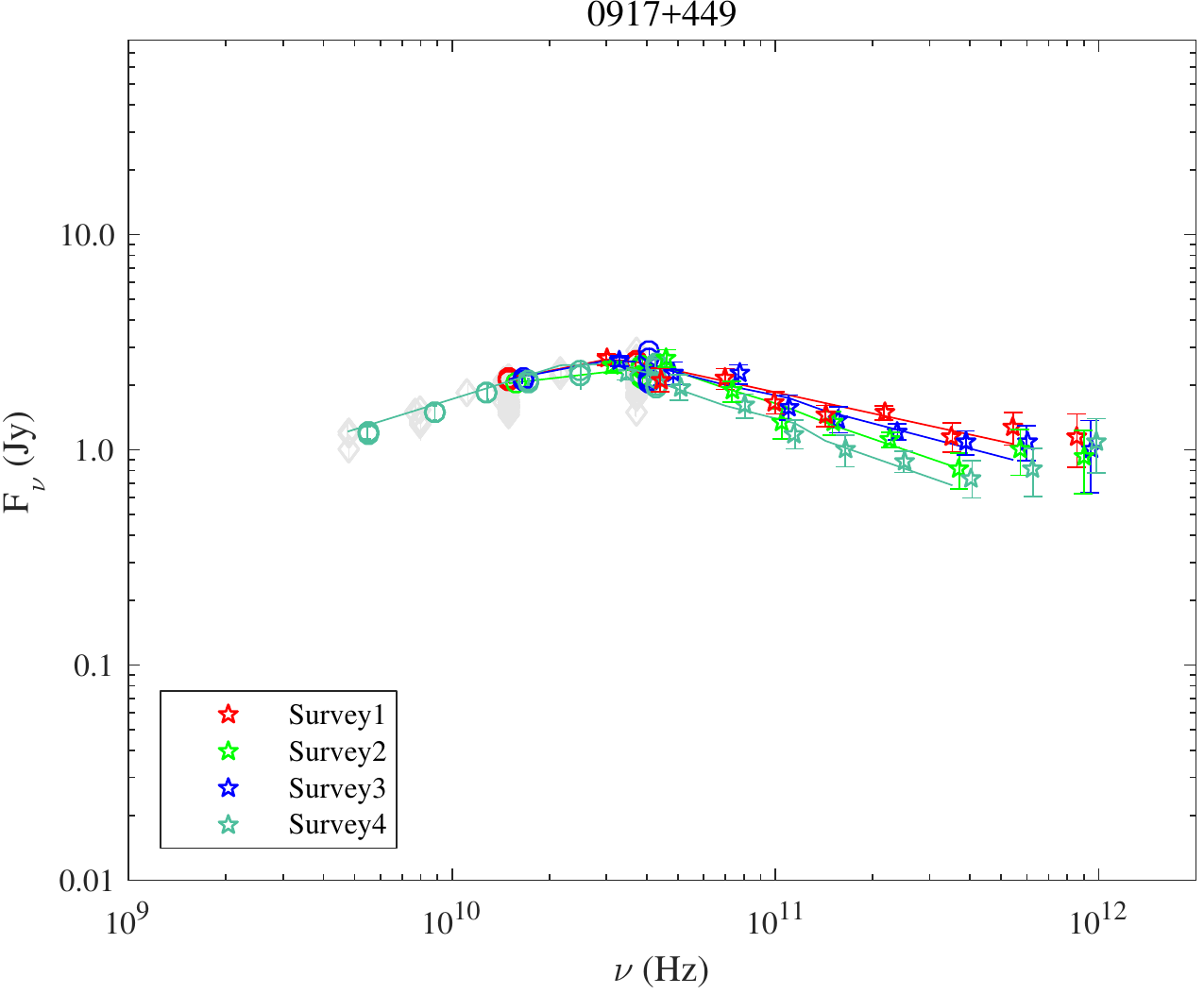}
	\caption{0917+449.}
	\label{0917+449_spectra}
	\end{minipage}\qquad
	\begin{minipage}[b]{.47\textwidth}
	\includegraphics[width=\columnwidth]{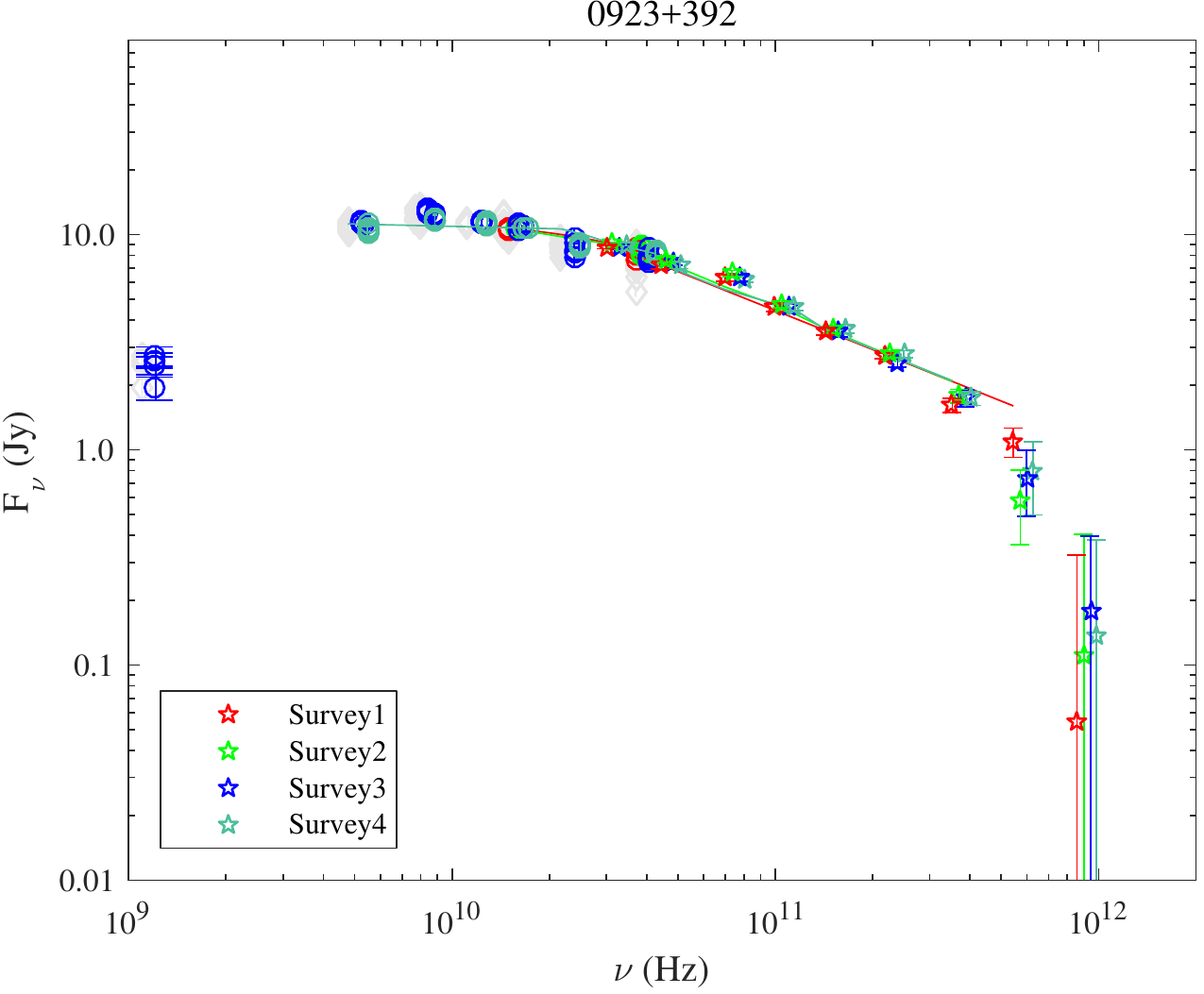}
	\caption{0923+392.}
	\label{0923+392_spectra}
	\end{minipage}
\end{figure*}

\begin{figure*}
	\centering
	\begin{minipage}[b]{.47\textwidth}
	\includegraphics[width=\columnwidth]{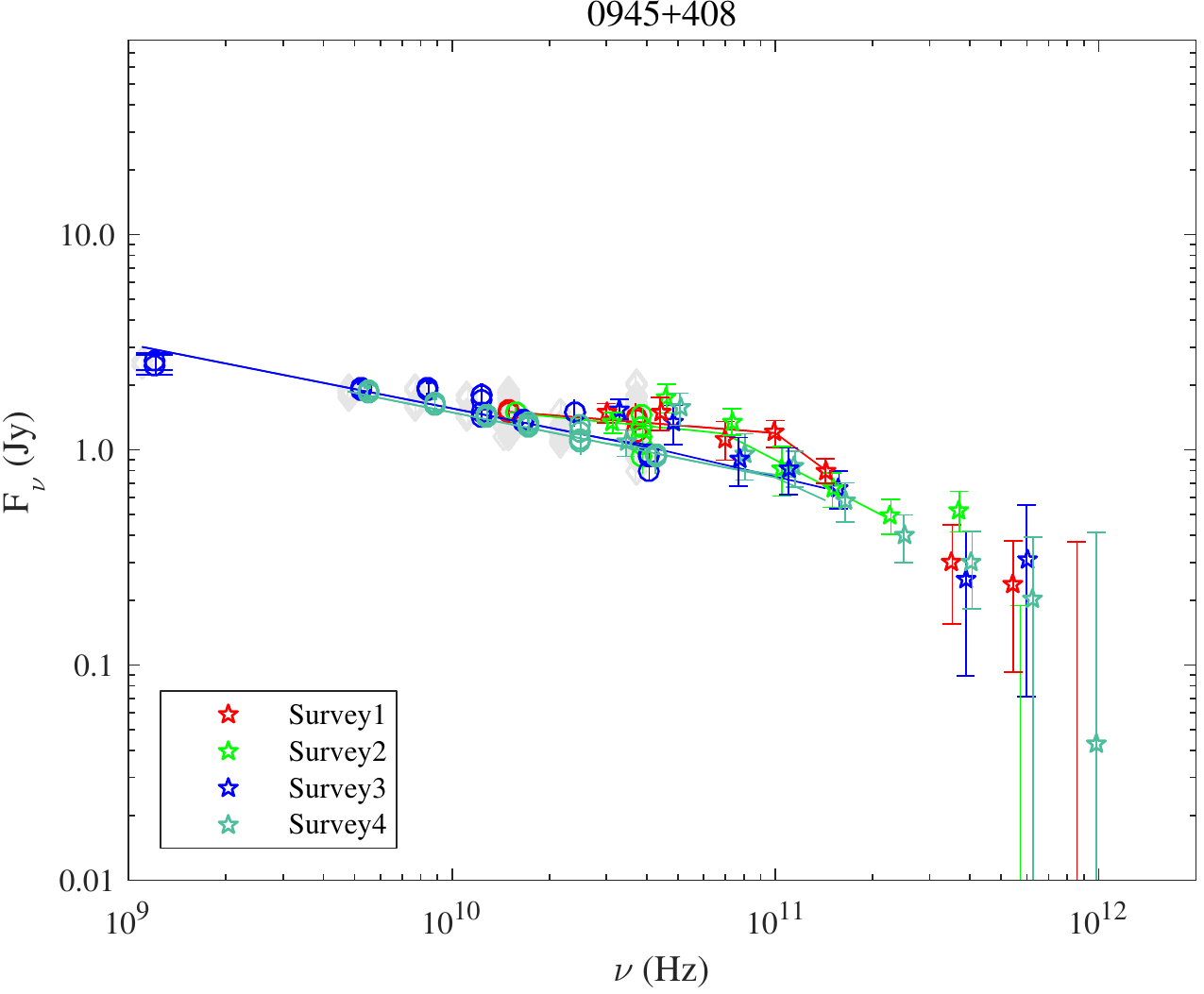}
	\caption{0945+408.}
	\label{0945+408_spectra}
	\end{minipage}\qquad
	\begin{minipage}[b]{.47\textwidth}
	\includegraphics[width=\columnwidth]{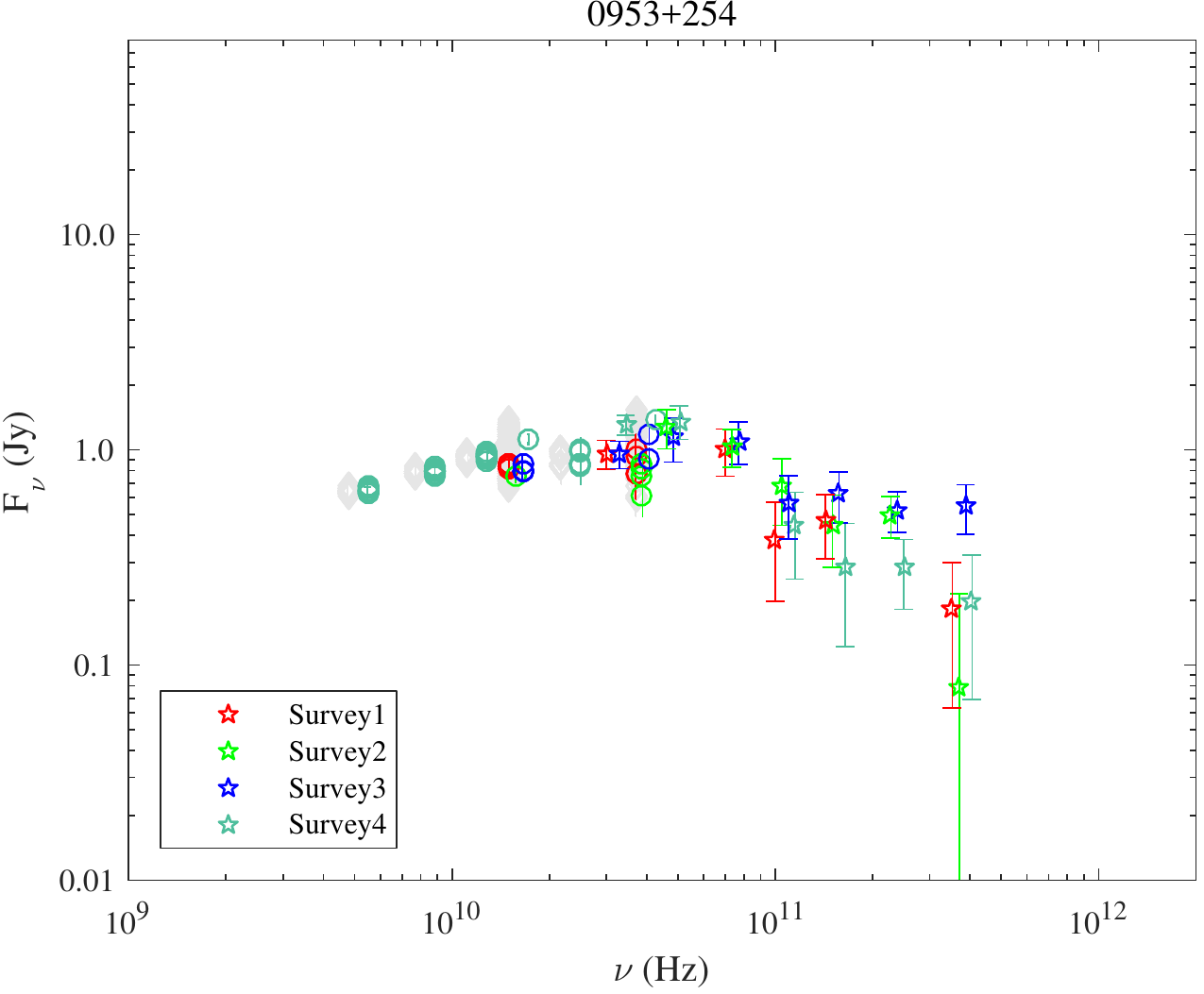}
	\caption{0953+254.}
	\label{0953+254_spectra}
	\end{minipage}
\end{figure*}

\begin{figure*}
	\centering
	\begin{minipage}[b]{.47\textwidth}
	\includegraphics[width=\columnwidth]{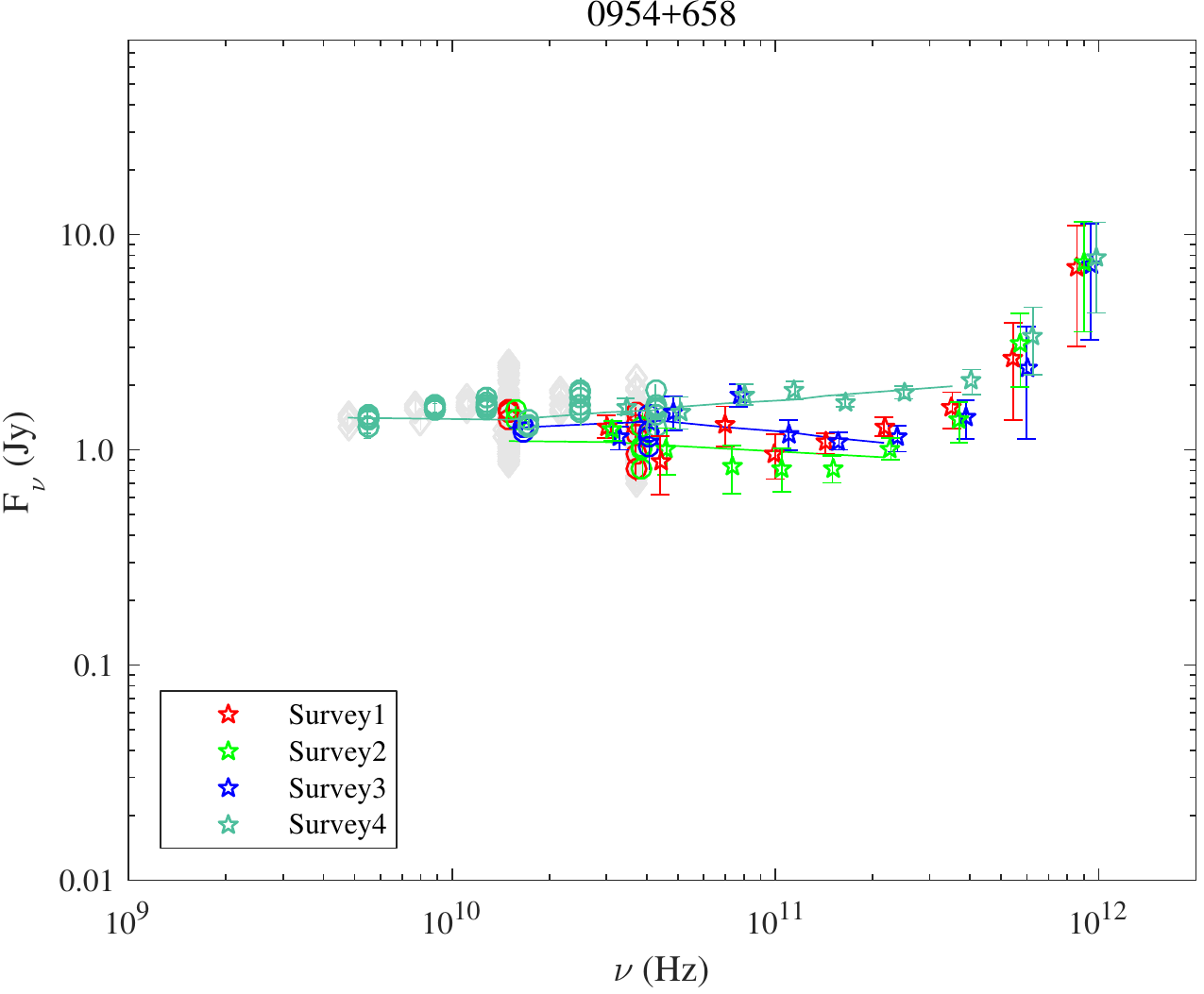}
	\caption{0954+658.}
	\label{0954+658_spectra}
	\end{minipage}\qquad
	\begin{minipage}[b]{.47\textwidth}
	\includegraphics[width=\columnwidth]{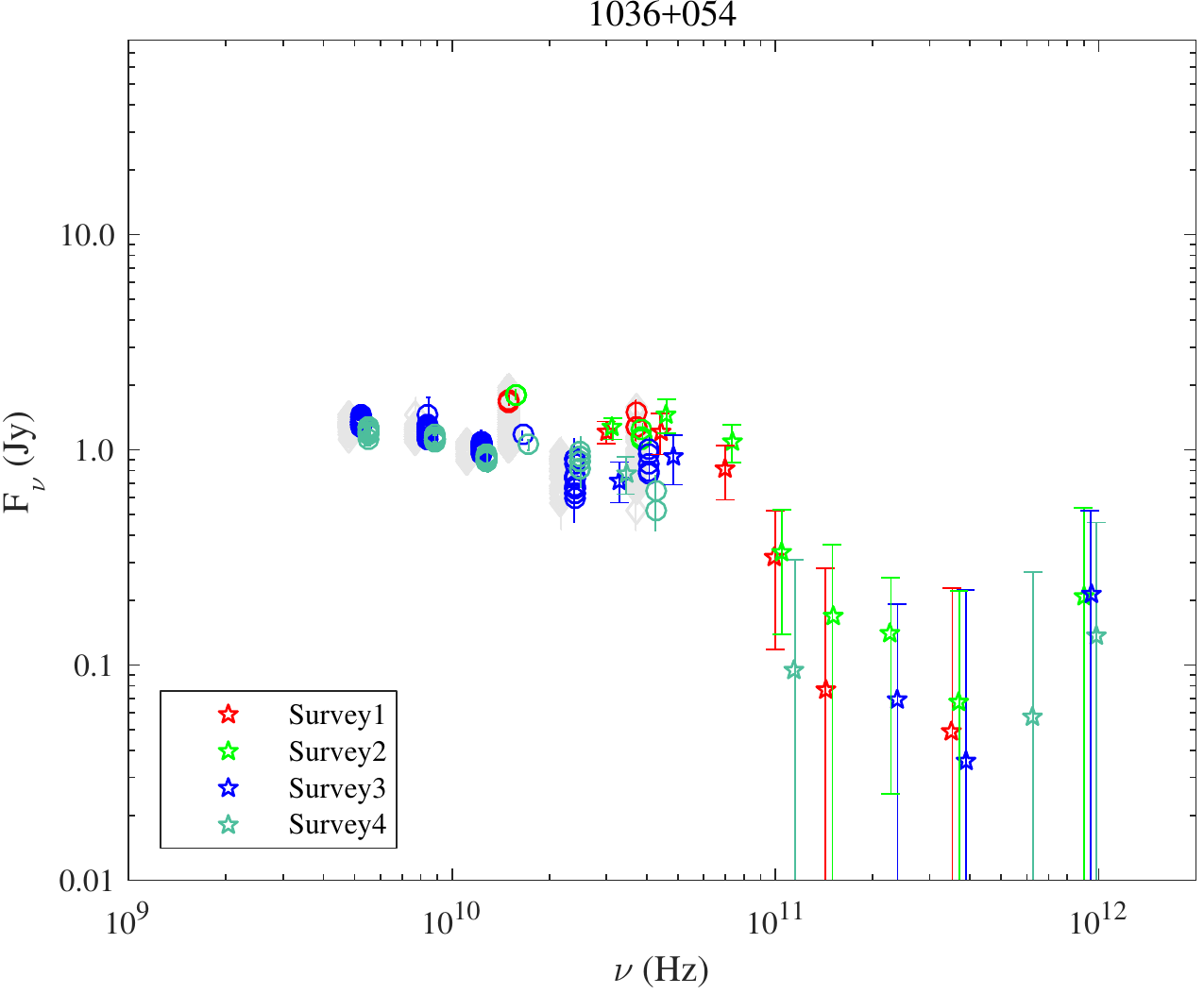}
	\caption{1036+054.}
	\label{1036+054_spectra}
	\end{minipage}
\end{figure*}

\clearpage

\begin{figure*}
	\centering
	\begin{minipage}[b]{.47\textwidth}
	\includegraphics[width=\columnwidth]{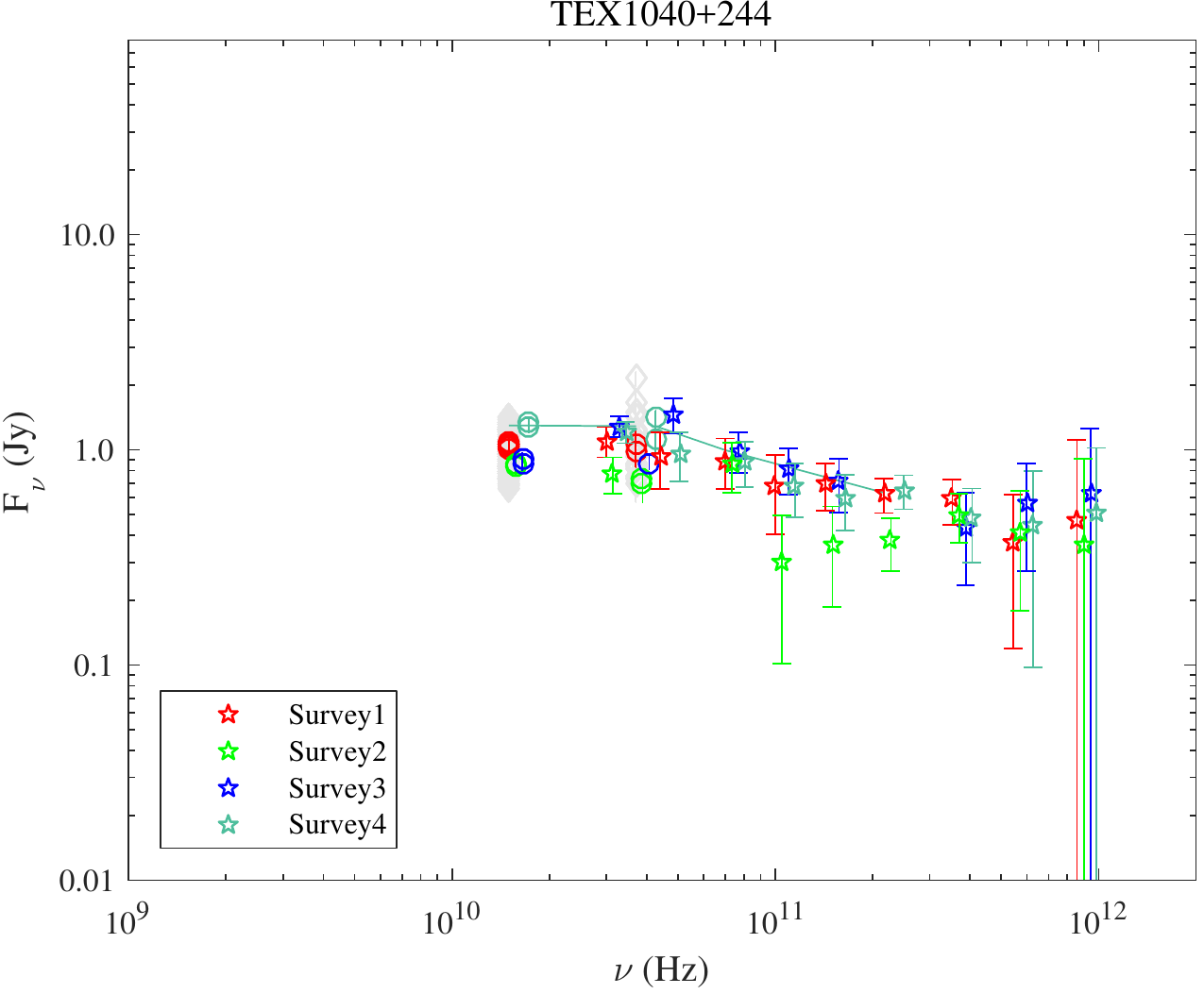}
	\caption{TEX1040+244.}
	\label{TEX1040+244_spectra}
	\end{minipage}\qquad
	\begin{minipage}[b]{.47\textwidth}
	\includegraphics[width=\columnwidth]{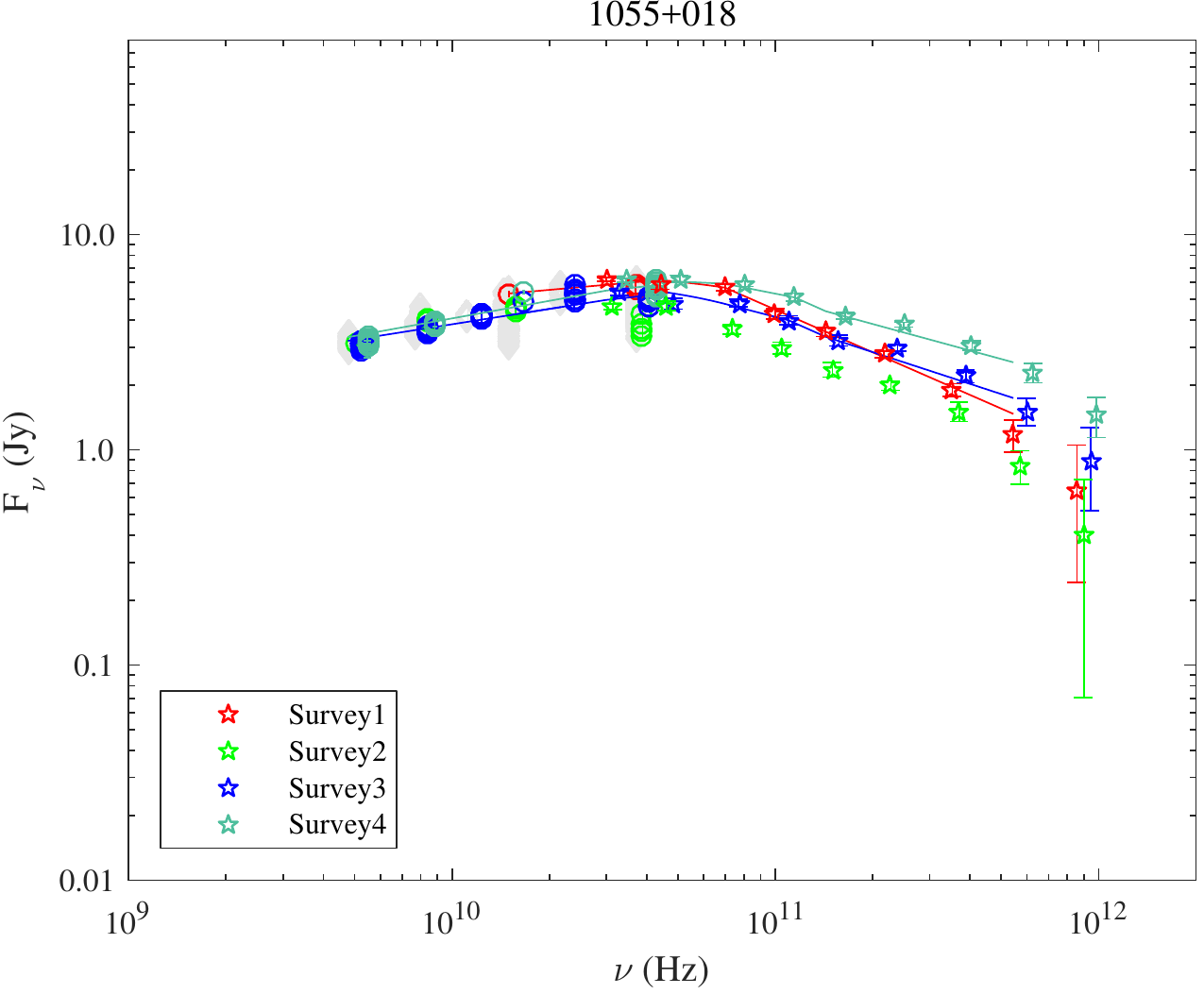}
	\caption{1055+018.}
	\label{1055+018_spectra}
	\end{minipage}
\end{figure*}

\begin{figure*}
	\centering
	\begin{minipage}[b]{.47\textwidth}
	\includegraphics[width=\columnwidth]{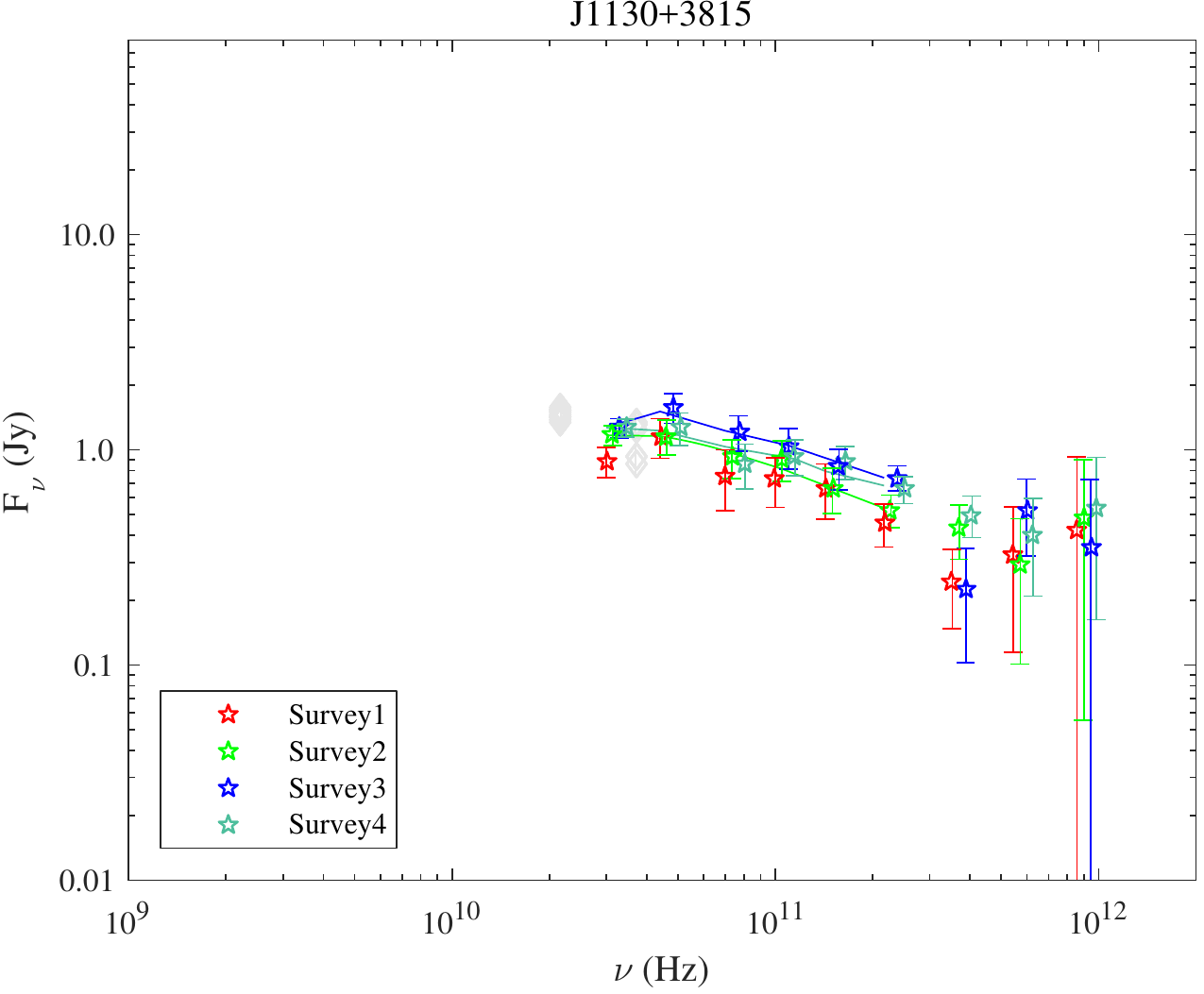}
	\caption{J1130+3815.}
	\label{J1130+3815_spectra}
	\end{minipage}\qquad
	\begin{minipage}[b]{.47\textwidth}
	\includegraphics[width=\columnwidth]{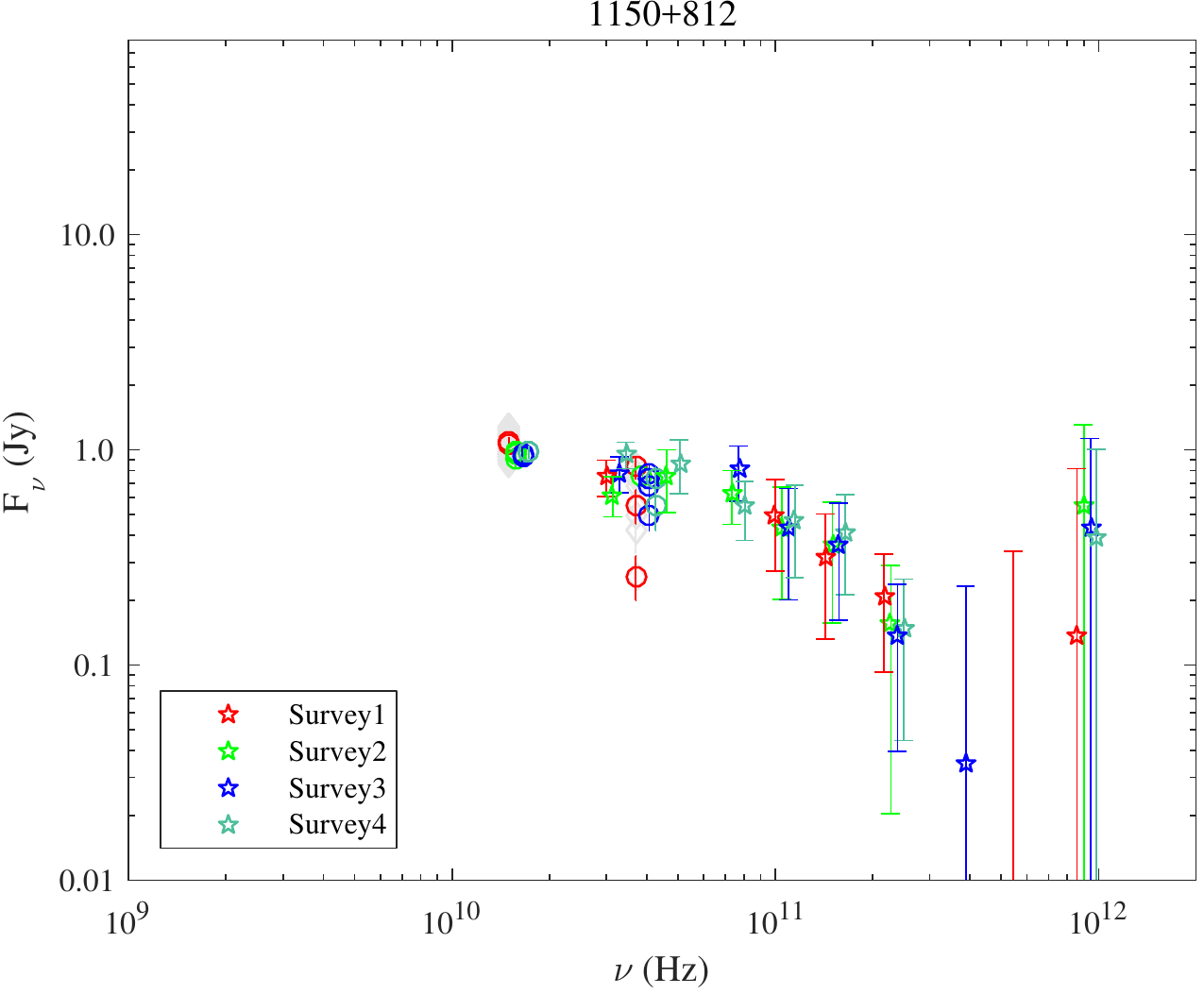}
	\caption{1150+812.}
	\label{1150+812_spectra}
	\end{minipage}
\end{figure*}

\begin{figure*}
	\centering
	\begin{minipage}[b]{.47\textwidth}
	\includegraphics[width=\columnwidth]{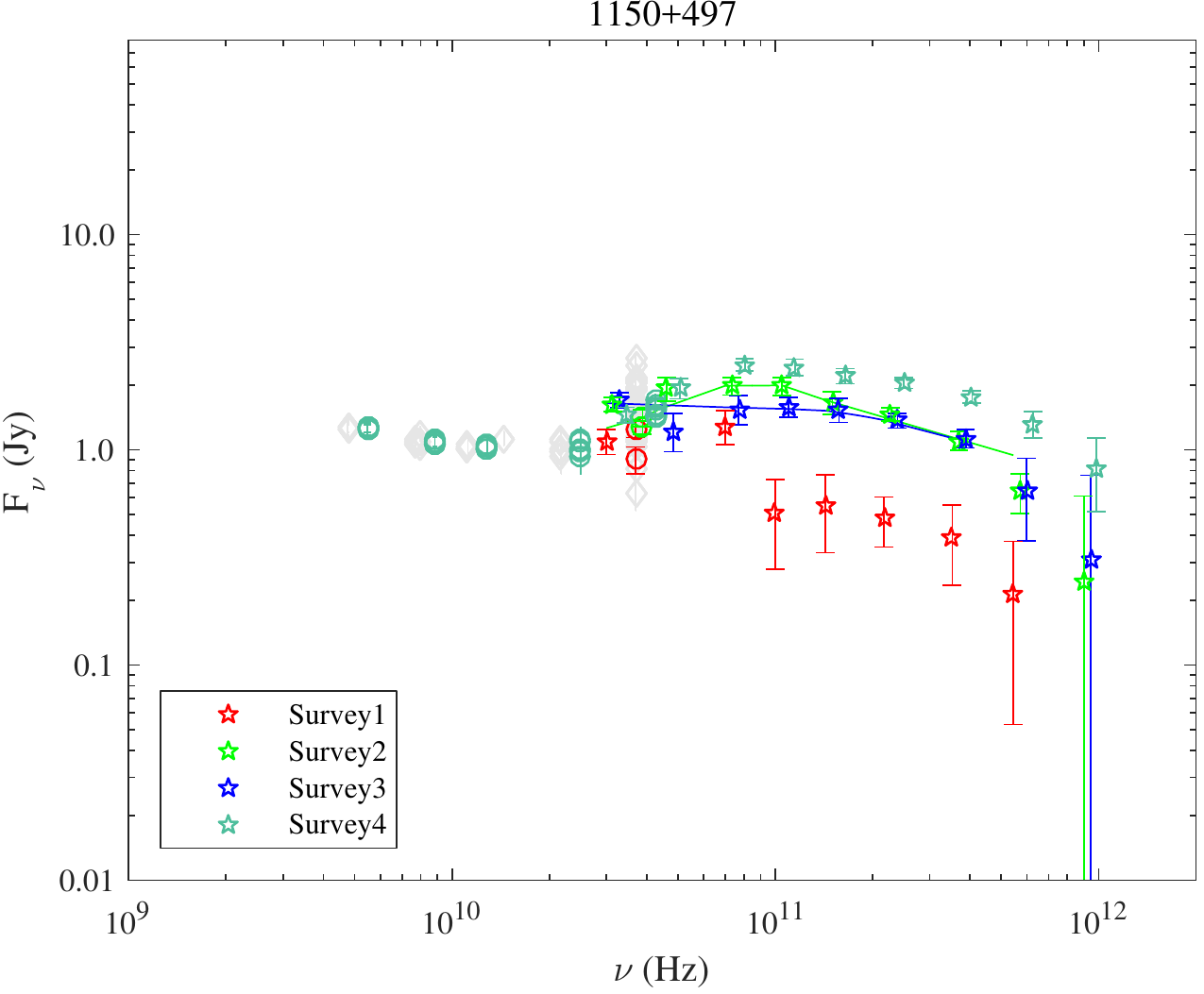}
	\caption{1150+497.}
	\label{1150+497_spectra}
	\end{minipage}\qquad
	\begin{minipage}[b]{.47\textwidth}
	\includegraphics[width=\columnwidth]{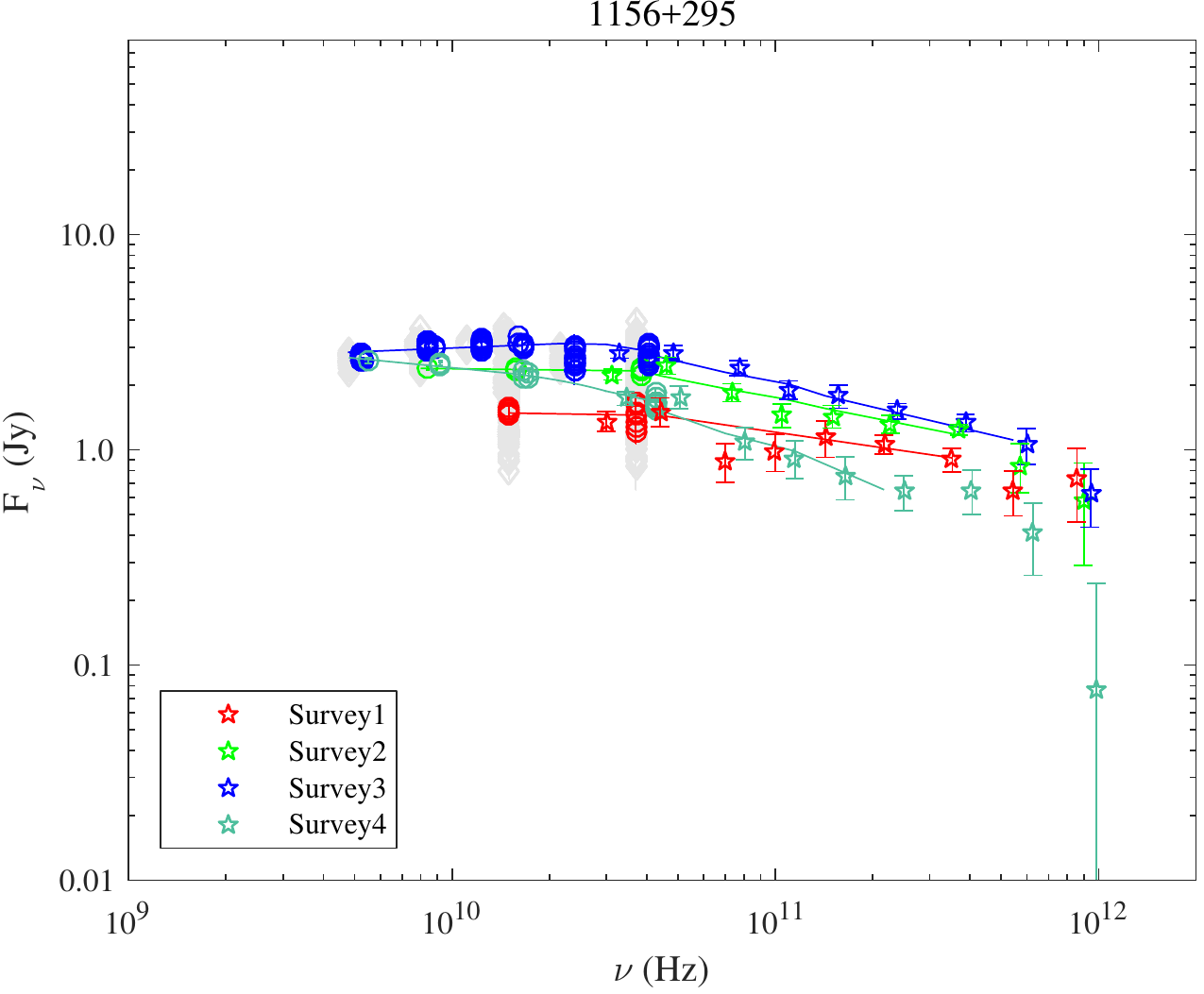}
	\caption{1156+295.}
	\label{1156+295_spectra}
	\end{minipage}
\end{figure*}

\clearpage

\begin{figure*}
	\centering
	\begin{minipage}[b]{.47\textwidth}
	\includegraphics[width=\columnwidth]{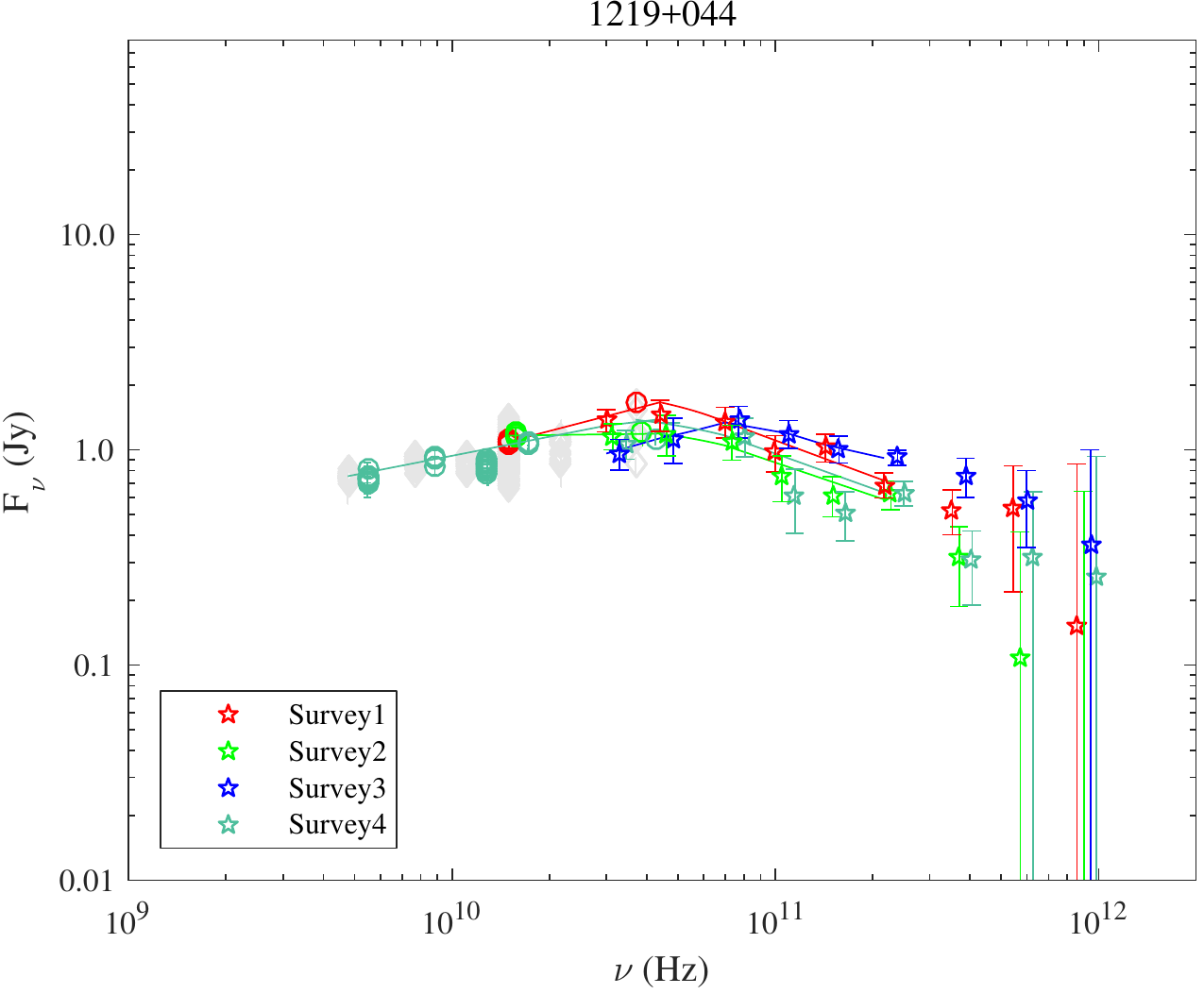}
	\caption{1219+044.}
	\label{1219+044_spectra}
	\end{minipage}\qquad
	\begin{minipage}[b]{.47\textwidth}
	\includegraphics[width=\columnwidth]{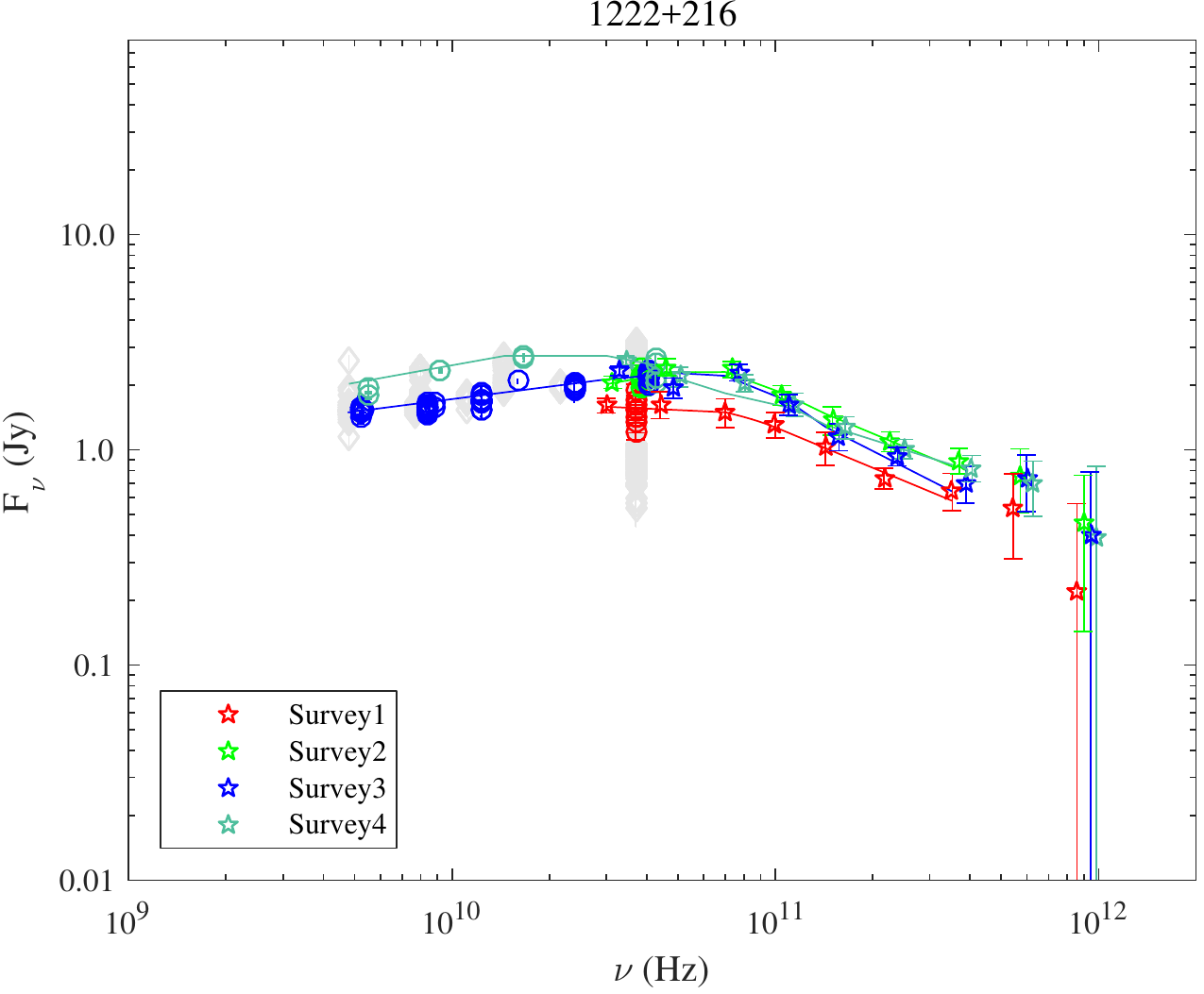}
	\caption{1222+216.}
	\label{1222+216_spectra}
	\end{minipage}
\end{figure*}

\begin{figure*}
	\centering
	\begin{minipage}[b]{.47\textwidth}
	\includegraphics[width=\columnwidth]{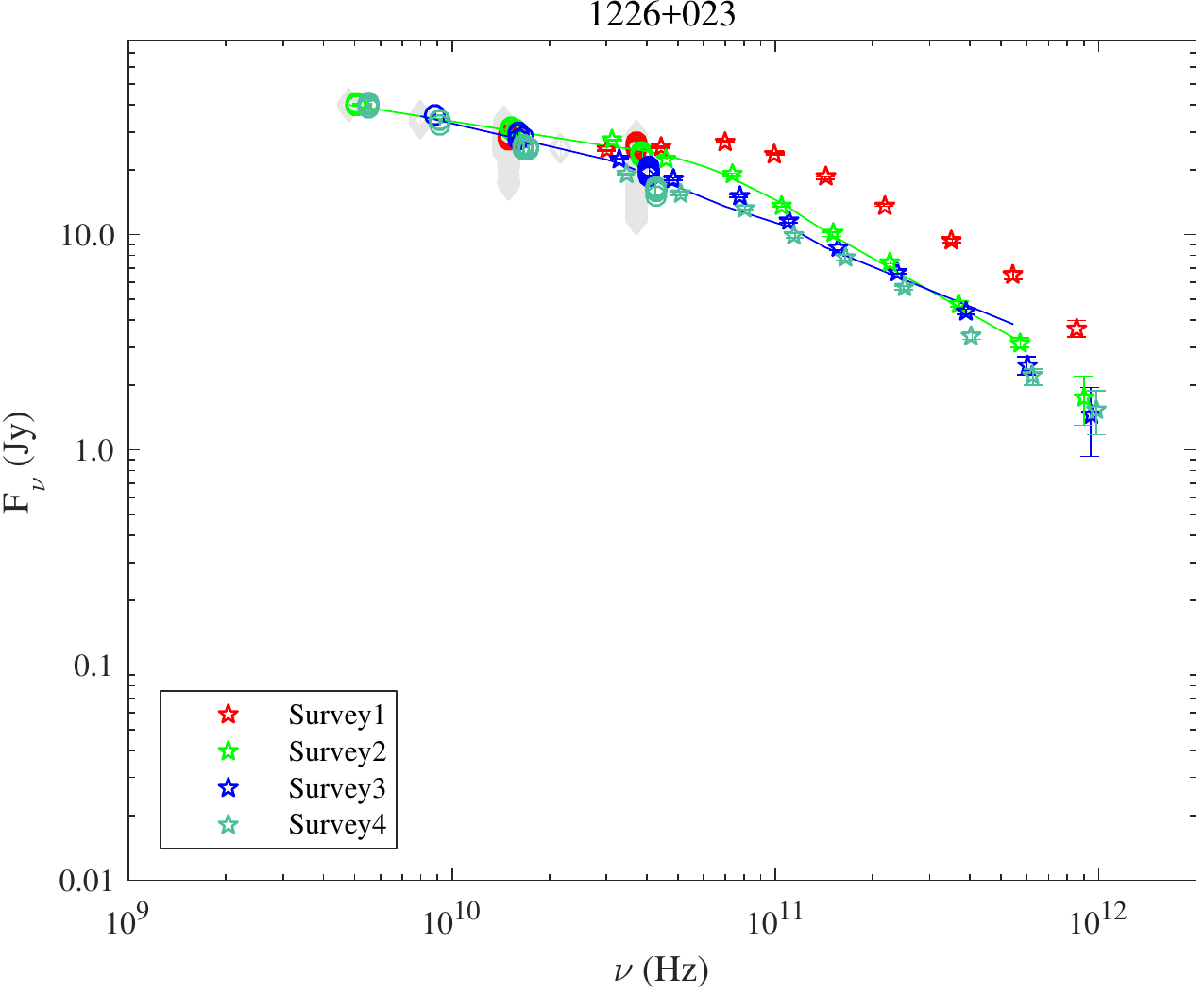}
	\caption{1226+023.}
	\label{1226+023_spectra}
	\end{minipage}\qquad
	\begin{minipage}[b]{.47\textwidth}
	\includegraphics[width=\columnwidth]{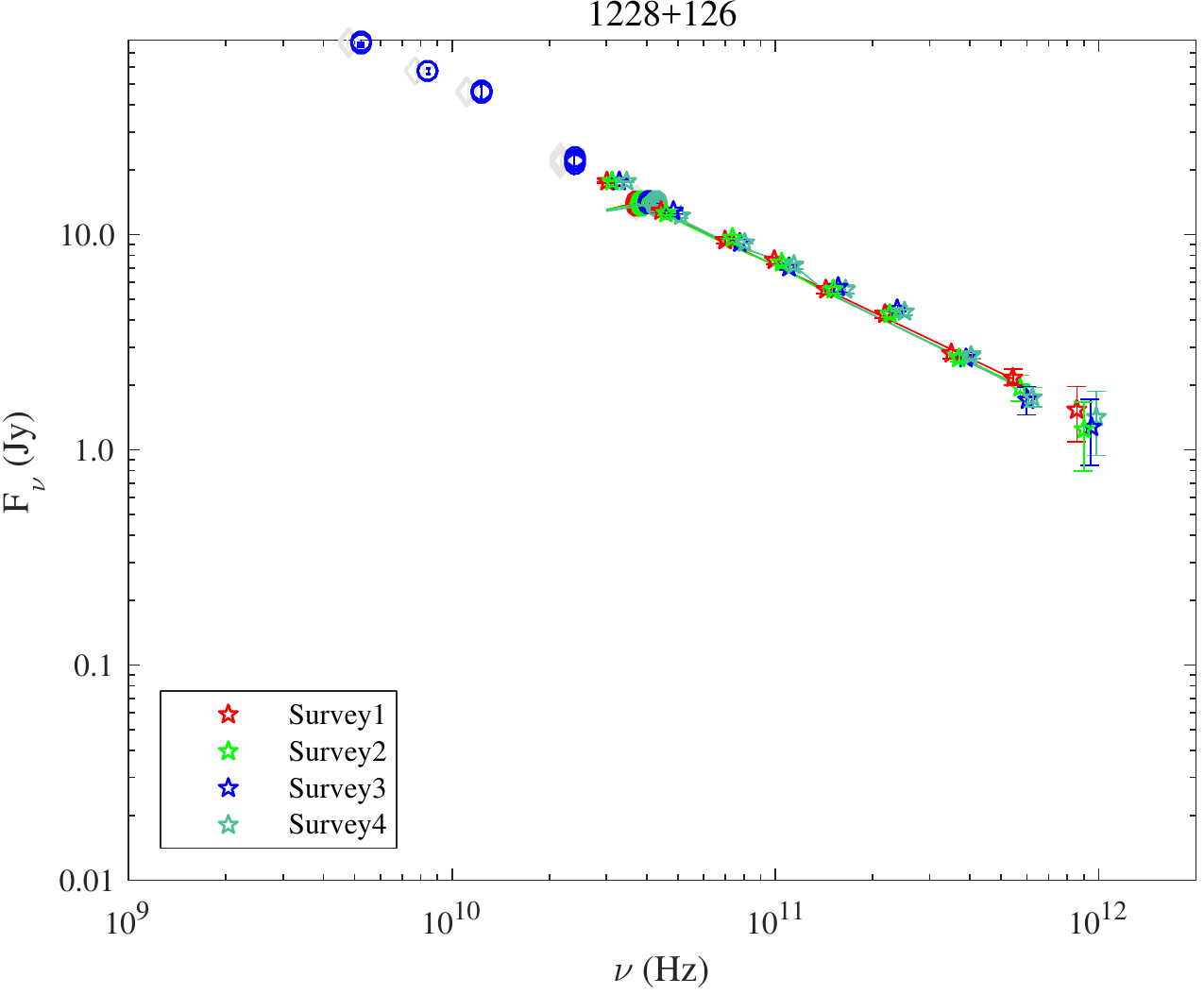}
	\caption{1228+126.}
	\label{1228+126_spectra}
	\end{minipage}
\end{figure*}

\begin{figure*}
	\centering
	\begin{minipage}[b]{.47\textwidth}
	\includegraphics[width=\columnwidth]{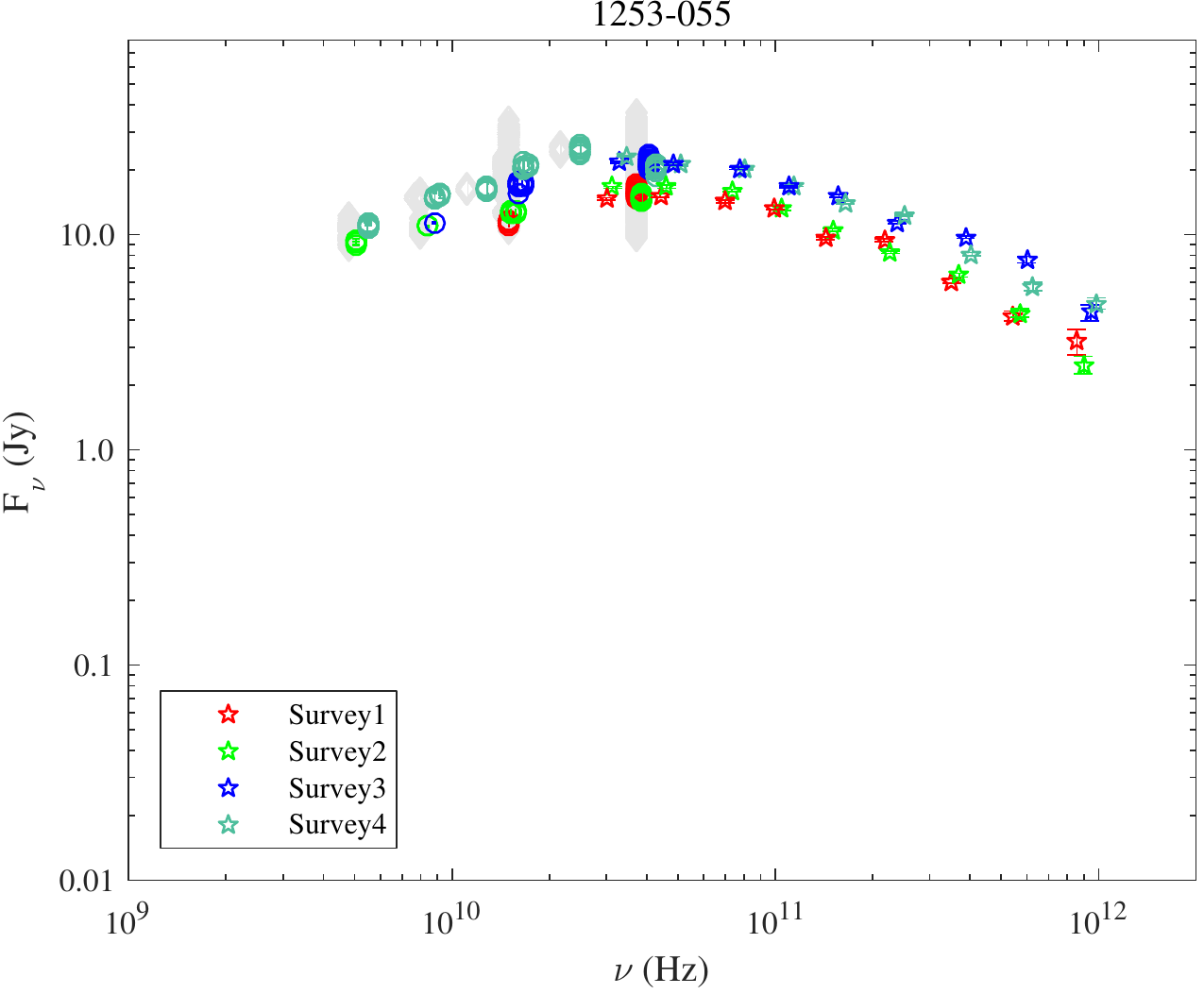}
	\caption{1253$-$055.}
	\label{1253-055_spectra}
	\end{minipage}\qquad
	\begin{minipage}[b]{.47\textwidth}
	\includegraphics[width=\columnwidth]{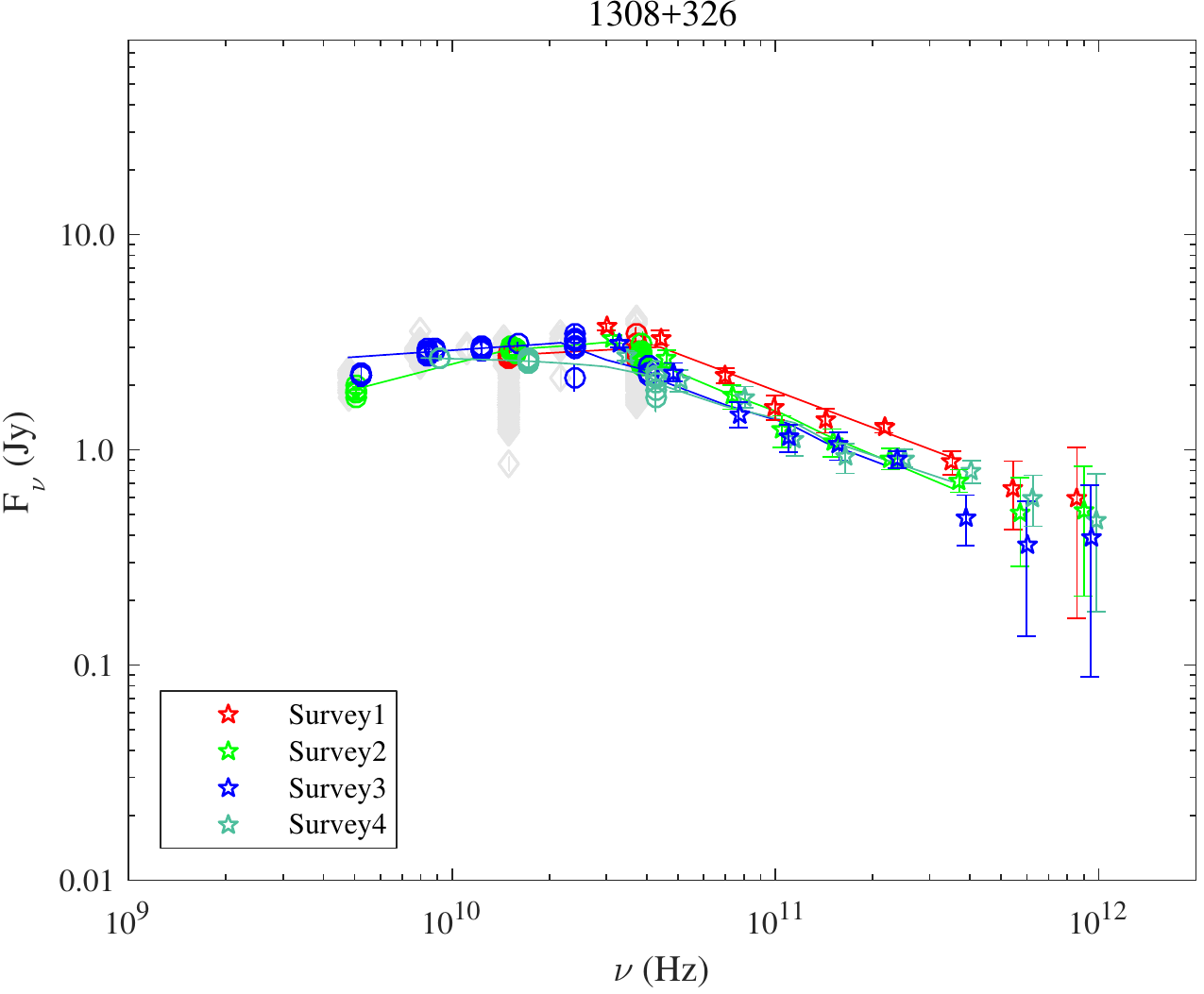}
	\caption{1308+326.}
	\label{1308+326_spectra}
	\end{minipage}
\end{figure*}

\clearpage

\begin{figure*}
	\centering
	\begin{minipage}[b]{.47\textwidth}
	\includegraphics[width=\columnwidth]{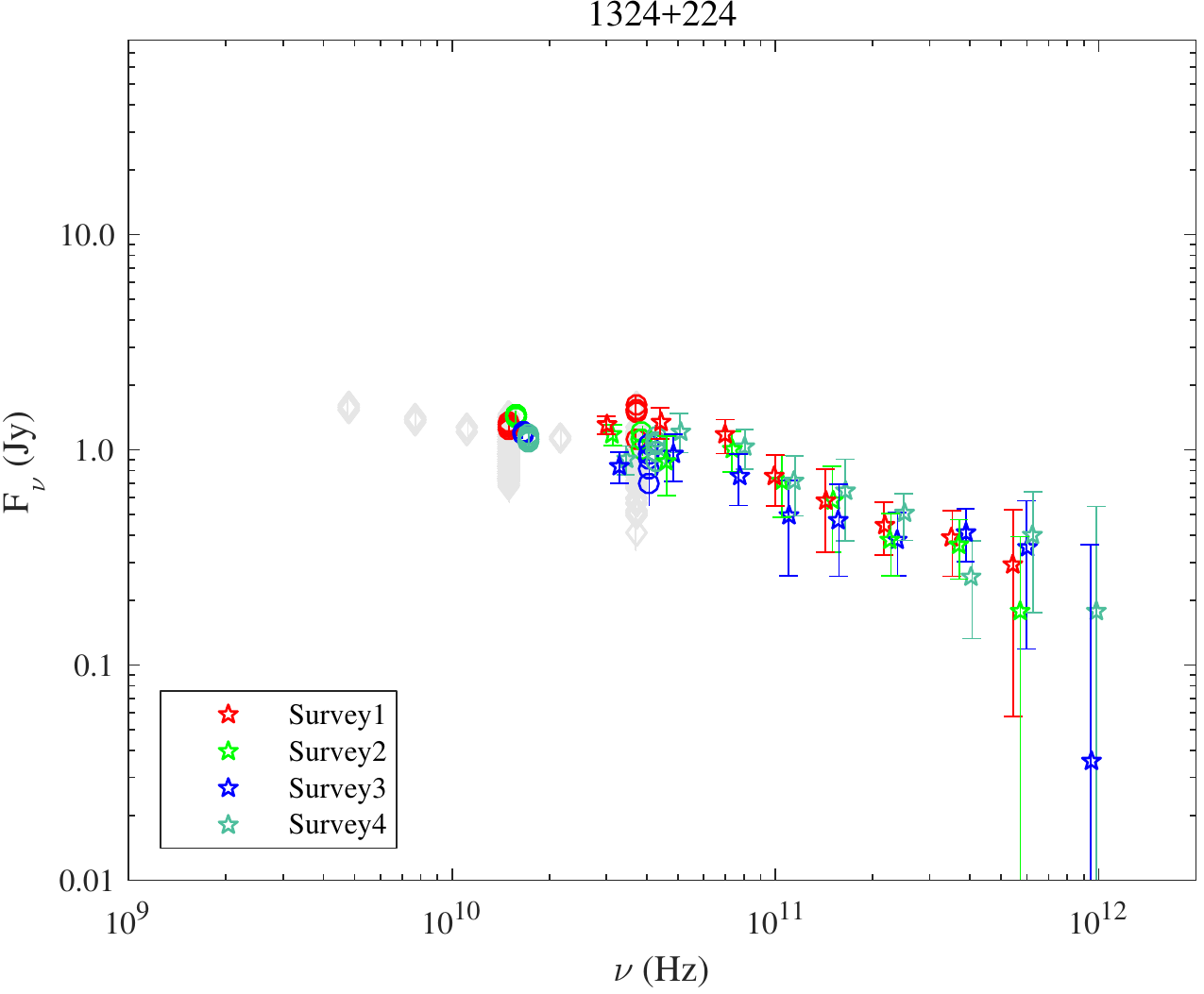}
	\caption{1324+224.}
	\label{1324+224_spectra}
	\end{minipage}\qquad
	\begin{minipage}[b]{.47\textwidth}
	\includegraphics[width=\columnwidth]{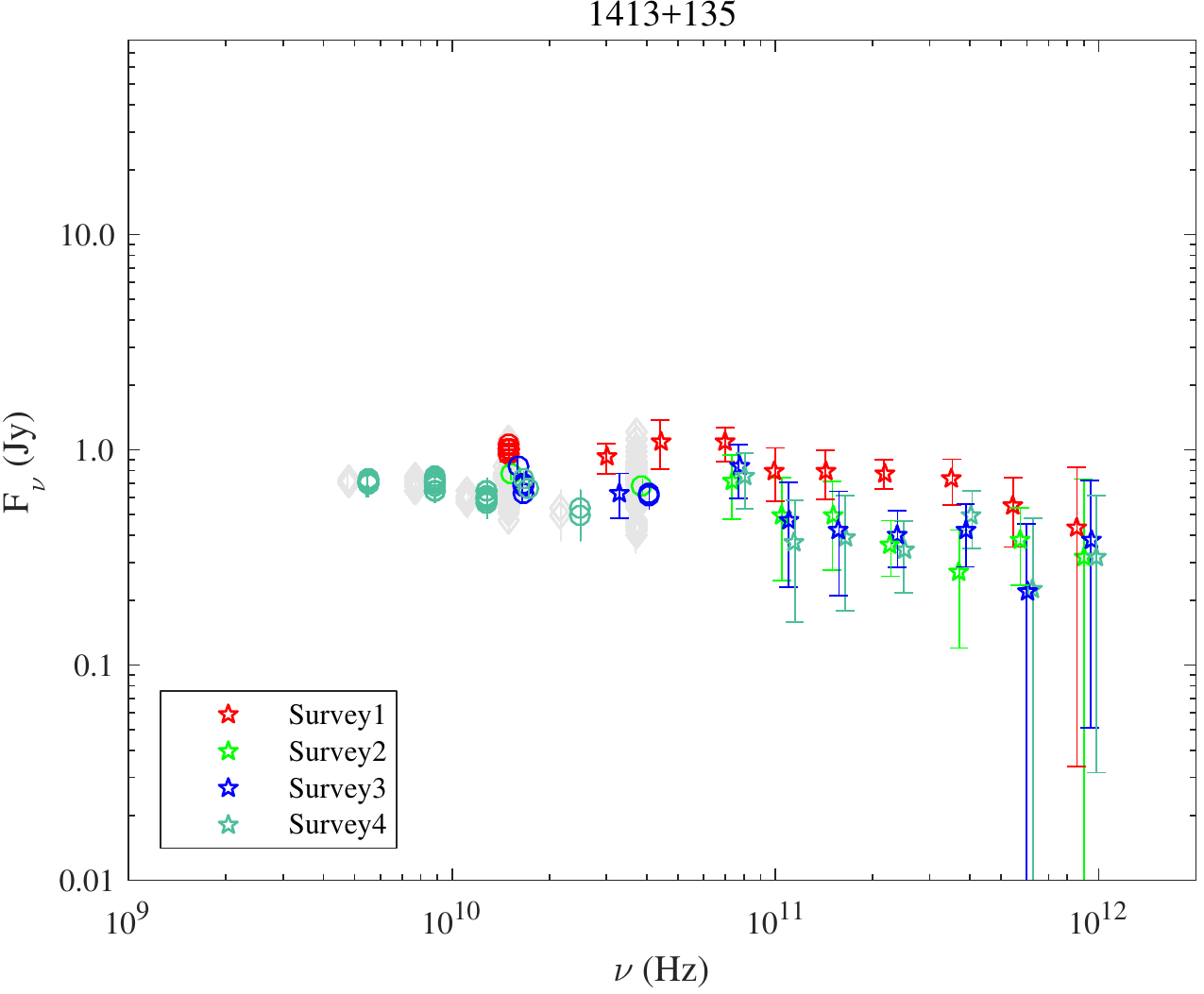}
	\caption{1413+135.}
	\label{1413+135_spectra}
	\end{minipage}
\end{figure*}

\begin{figure*}
	\centering
	\begin{minipage}[b]{.47\textwidth}
	\includegraphics[width=\columnwidth]{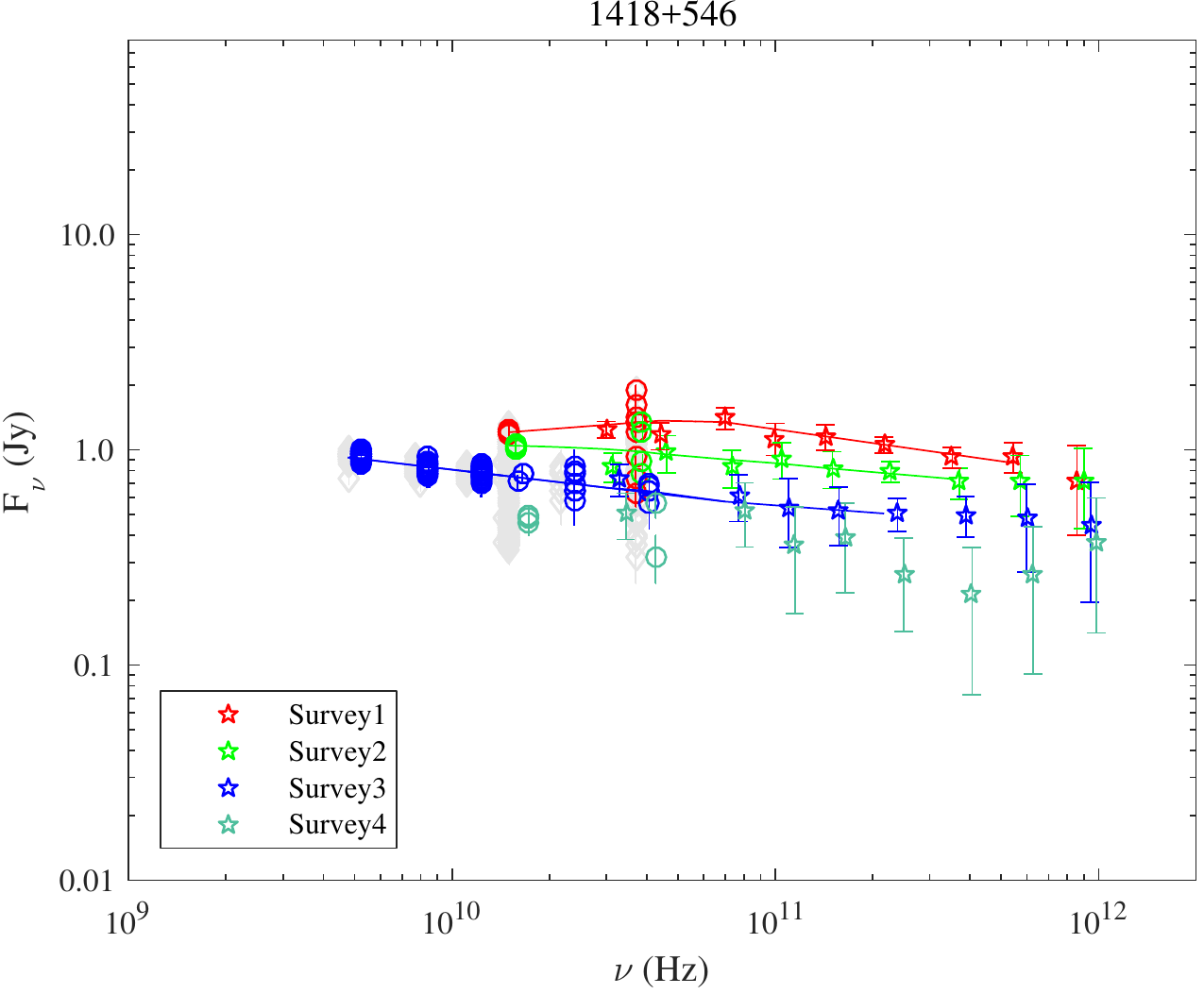}
	\caption{1418+546.}
	\label{1418+546_spectra}
	\end{minipage}\qquad
	\begin{minipage}[b]{.47\textwidth}
	\includegraphics[width=\columnwidth]{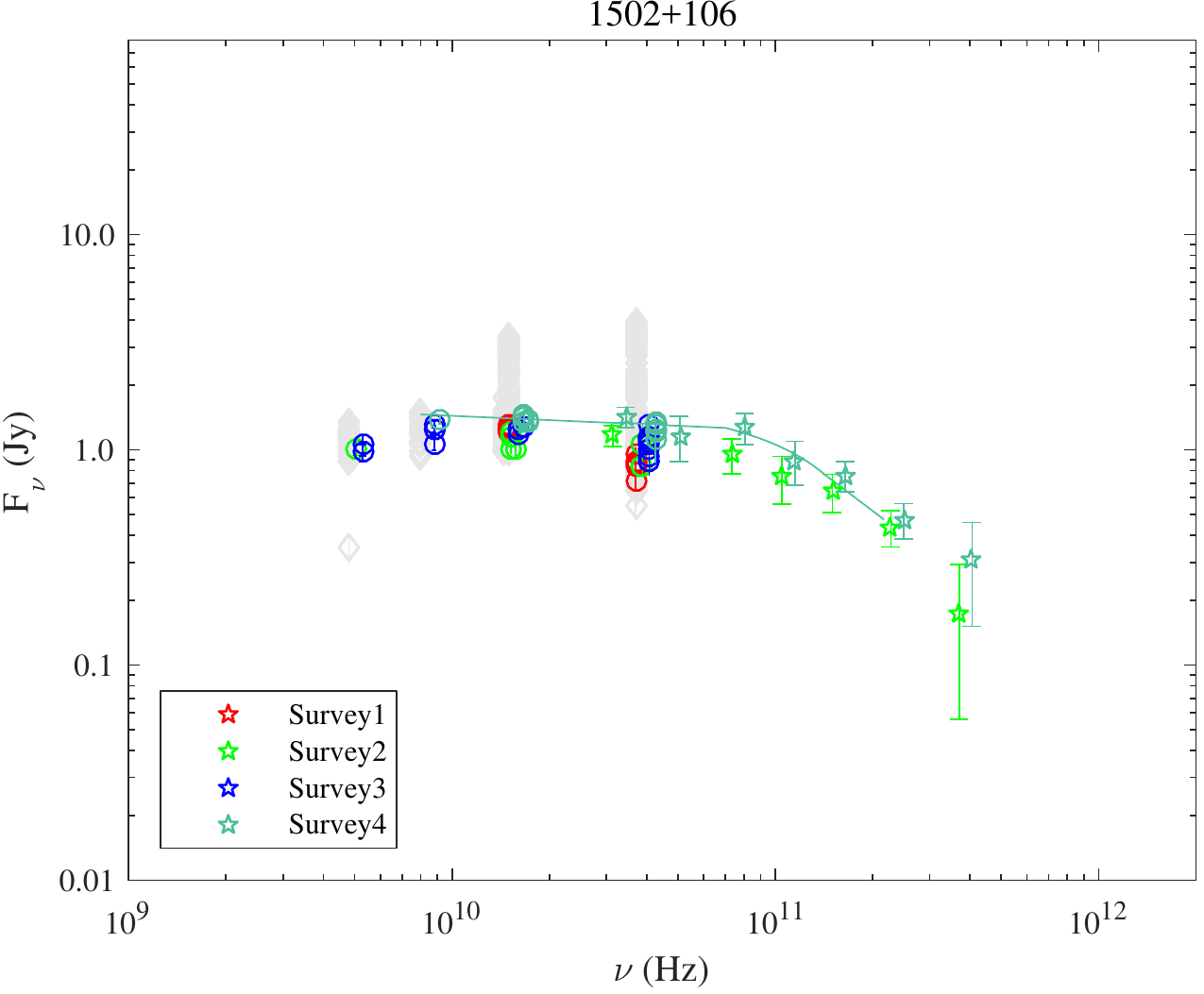}
	\caption{1502+106.}
	\label{1502+106_spectra}
	\end{minipage}
\end{figure*}

\begin{figure*}
	\centering
	\begin{minipage}[b]{.47\textwidth}
	\includegraphics[width=\columnwidth]{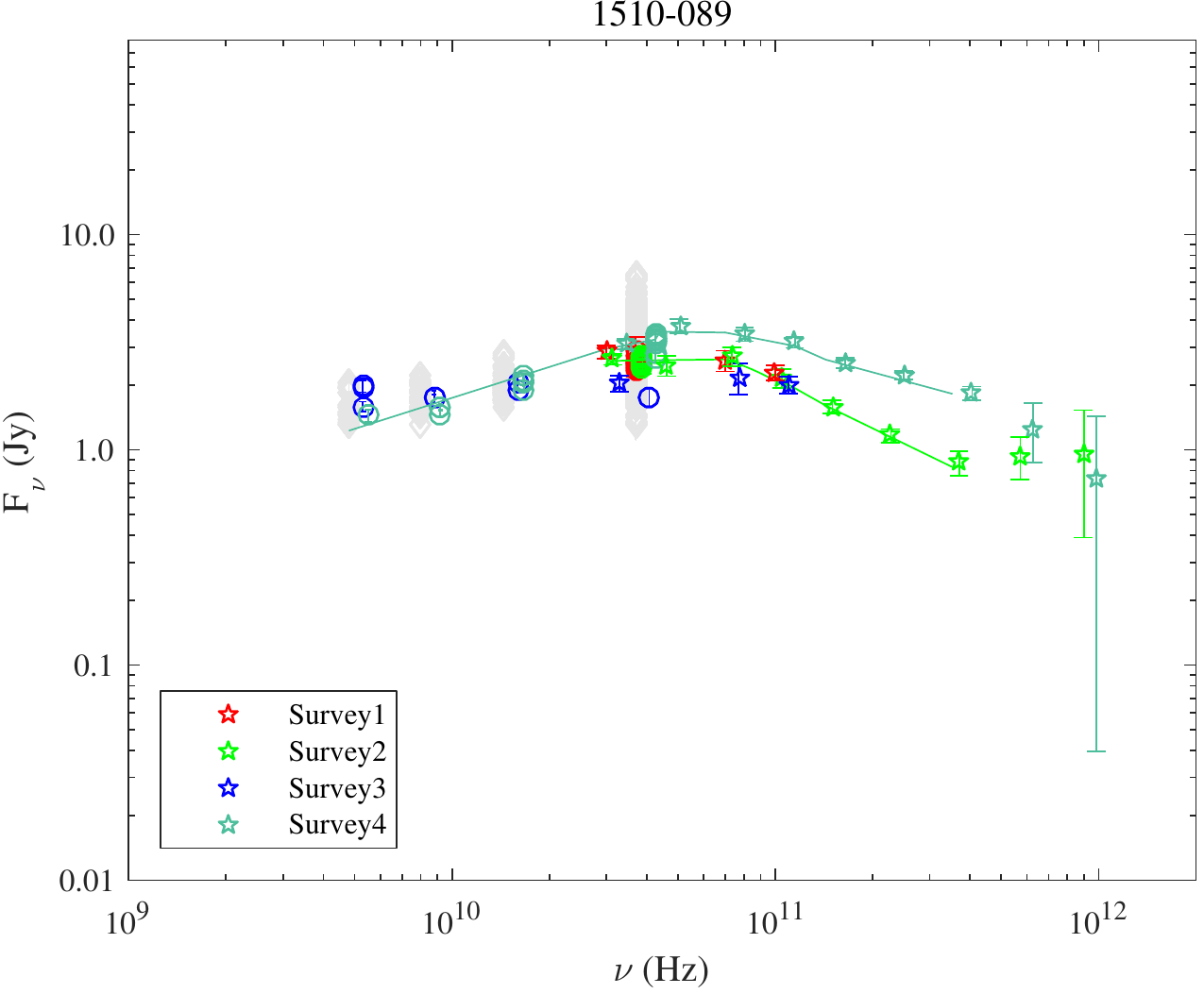}
	\caption{1510$-$089.}
	\label{1510-089_spectra}
	\end{minipage}\qquad
	\begin{minipage}[b]{.47\textwidth}
	\includegraphics[width=\columnwidth]{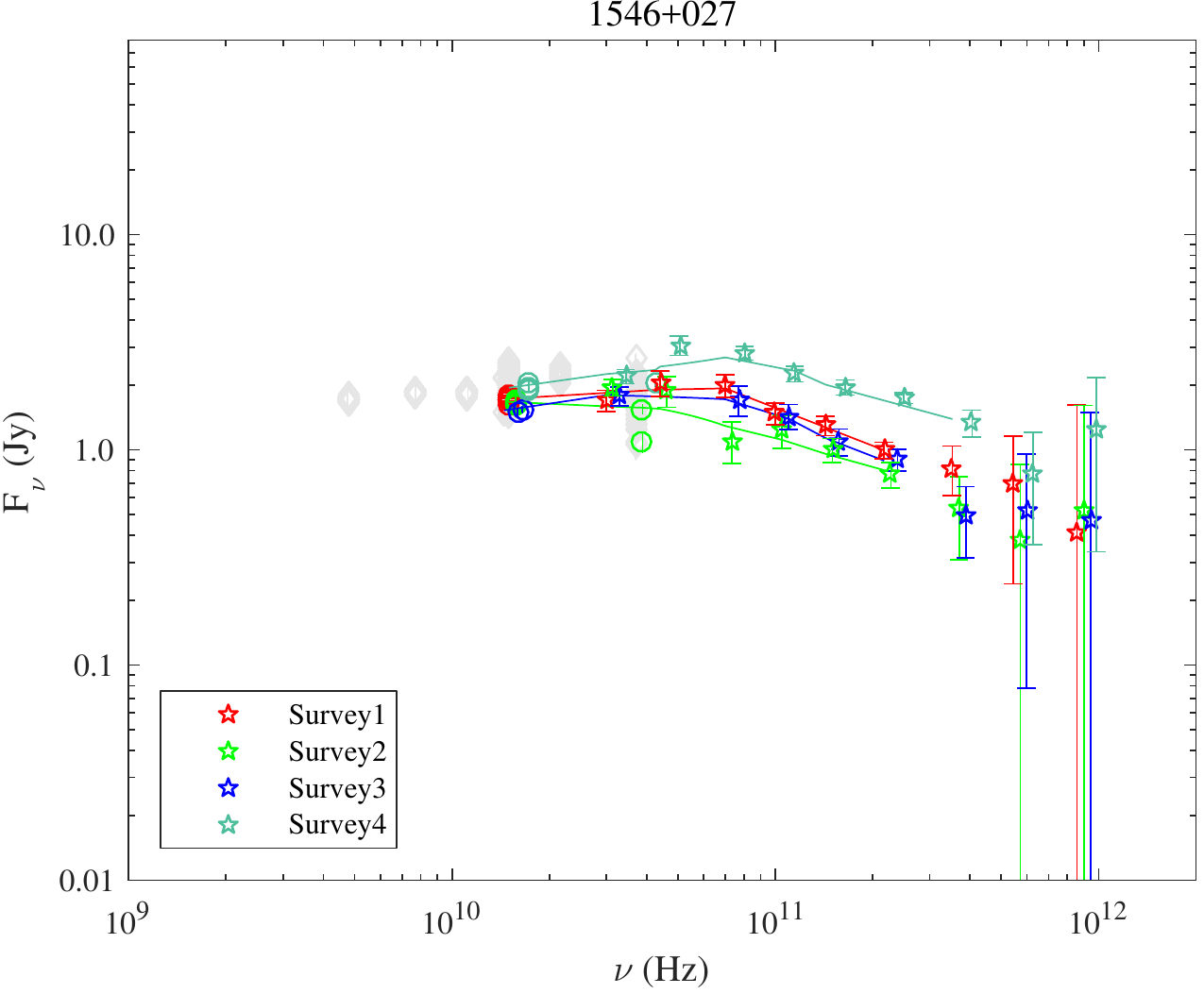}
	\caption{1546+027.}
	\label{1546+027_spectra}
	\end{minipage}
\end{figure*}

\clearpage

\begin{figure*}
	\centering
	\begin{minipage}[b]{.47\textwidth}
	\includegraphics[width=\columnwidth]{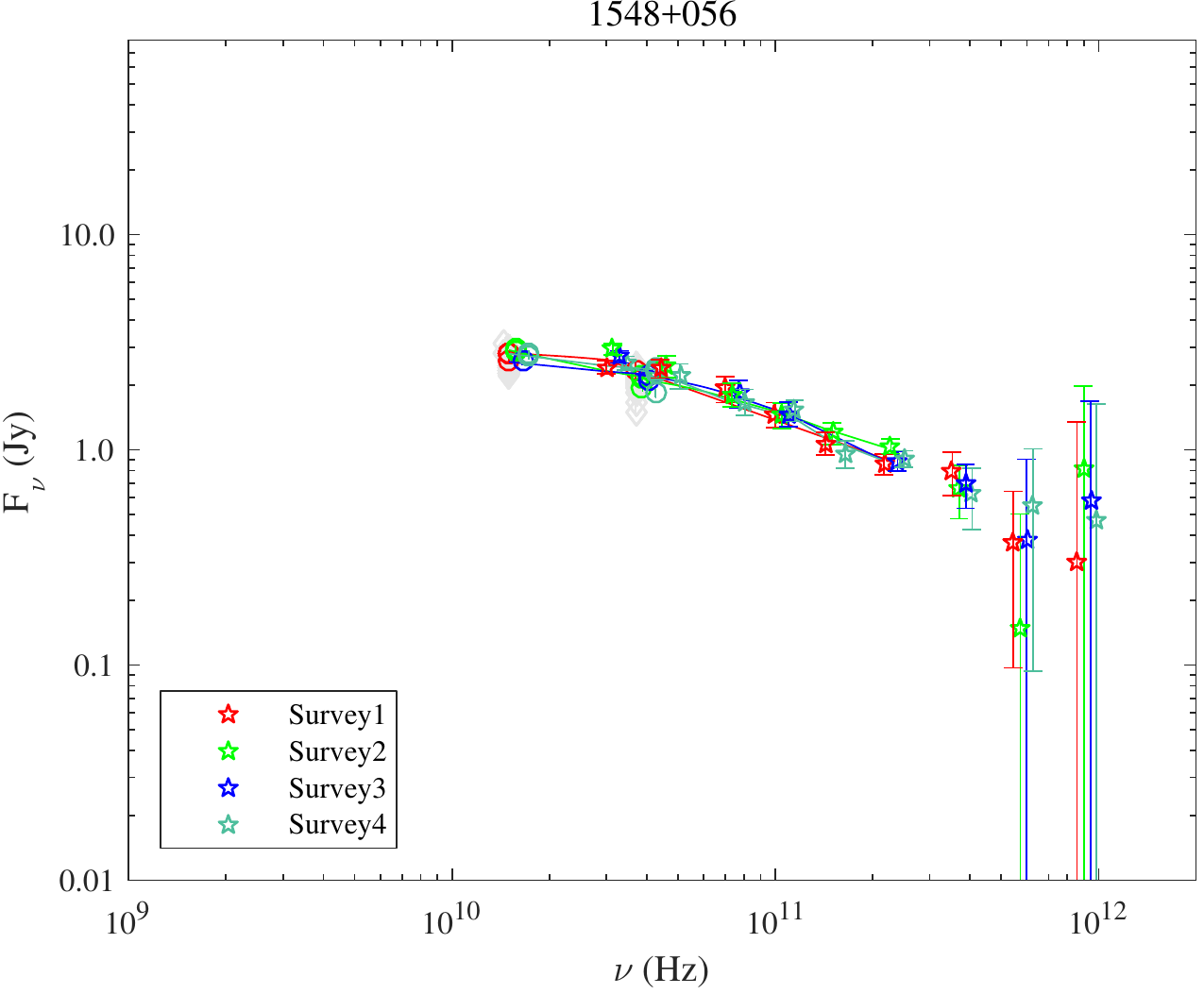}
	\caption{1548+056.}
	\label{1548+056_spectra}
	\end{minipage}\qquad
	\begin{minipage}[b]{.47\textwidth}
	\includegraphics[width=\columnwidth]{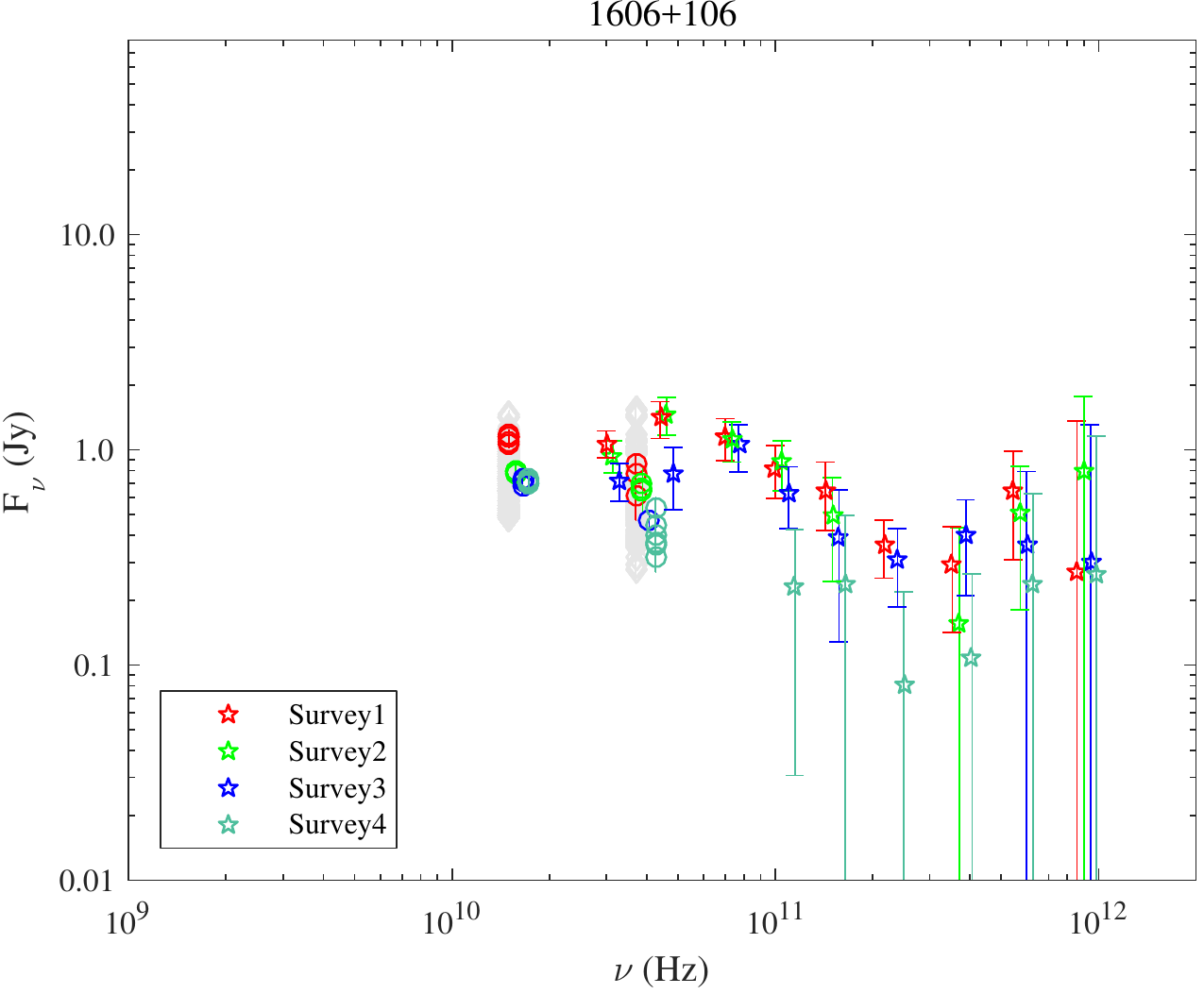}
	\caption{1606+106.}
	\label{1606+106_spectra}
	\end{minipage}
\end{figure*}

\begin{figure*}
	\centering
	\begin{minipage}[b]{.47\textwidth}
	\includegraphics[width=\columnwidth]{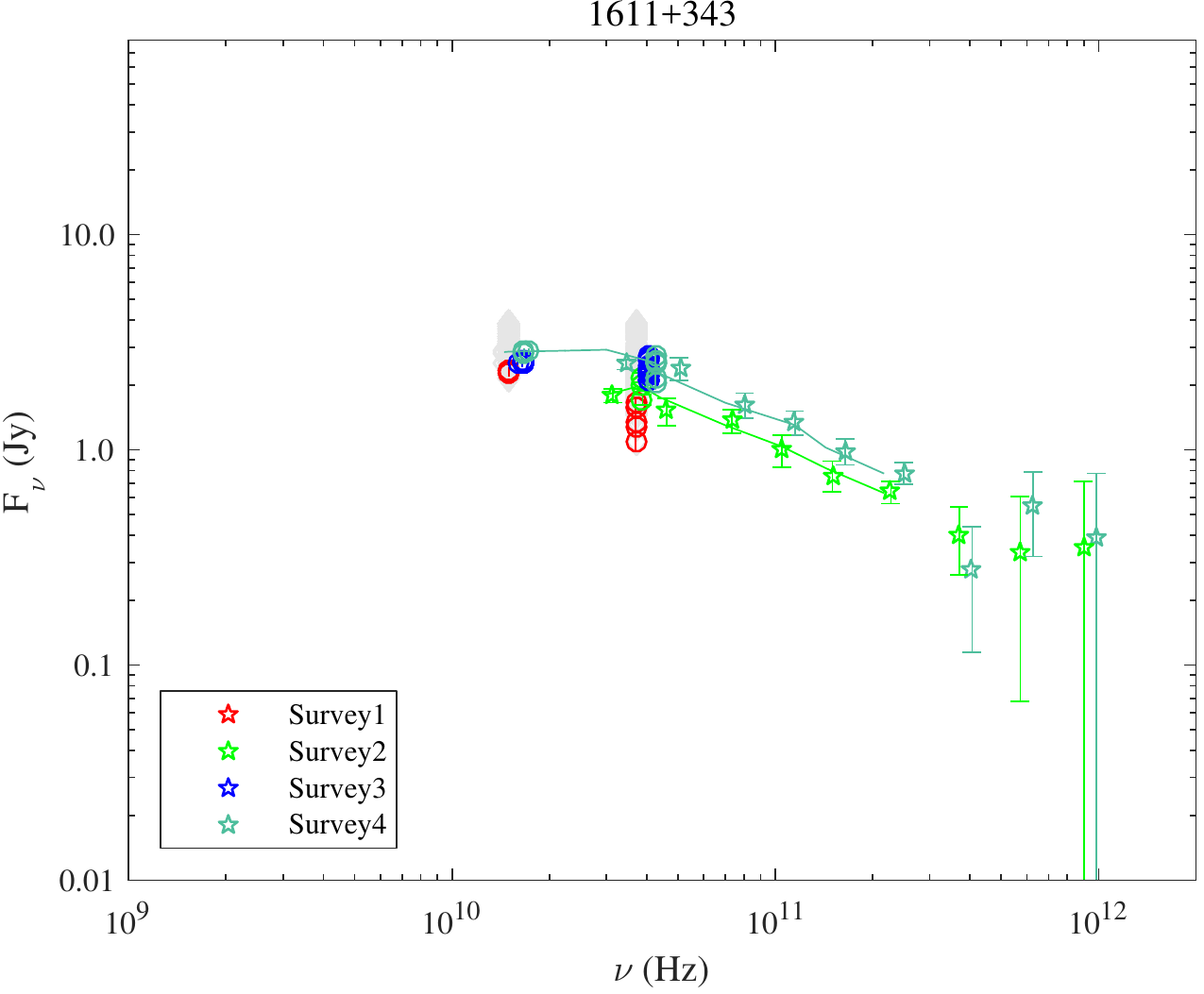}
	\caption{1611+343.}
	\label{1611+343_spectra}
	\end{minipage}\qquad
	\begin{minipage}[b]{.47\textwidth}
	\includegraphics[width=\columnwidth]{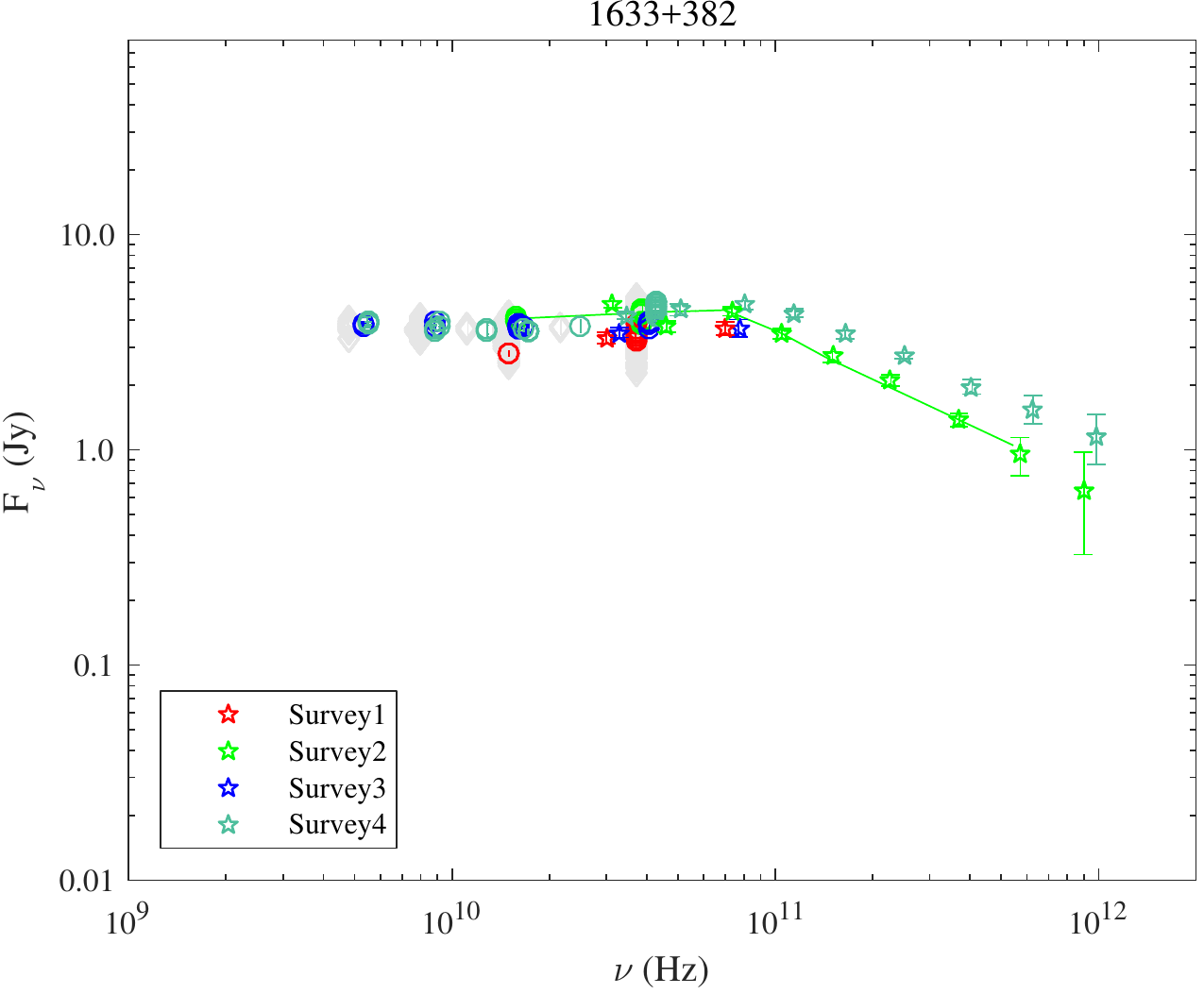}
	\caption{1633+382.}
	\label{1633+382_spectra}
	\end{minipage}
\end{figure*}

\begin{figure*}
	\centering
	\begin{minipage}[b]{.47\textwidth}
	\includegraphics[width=\columnwidth]{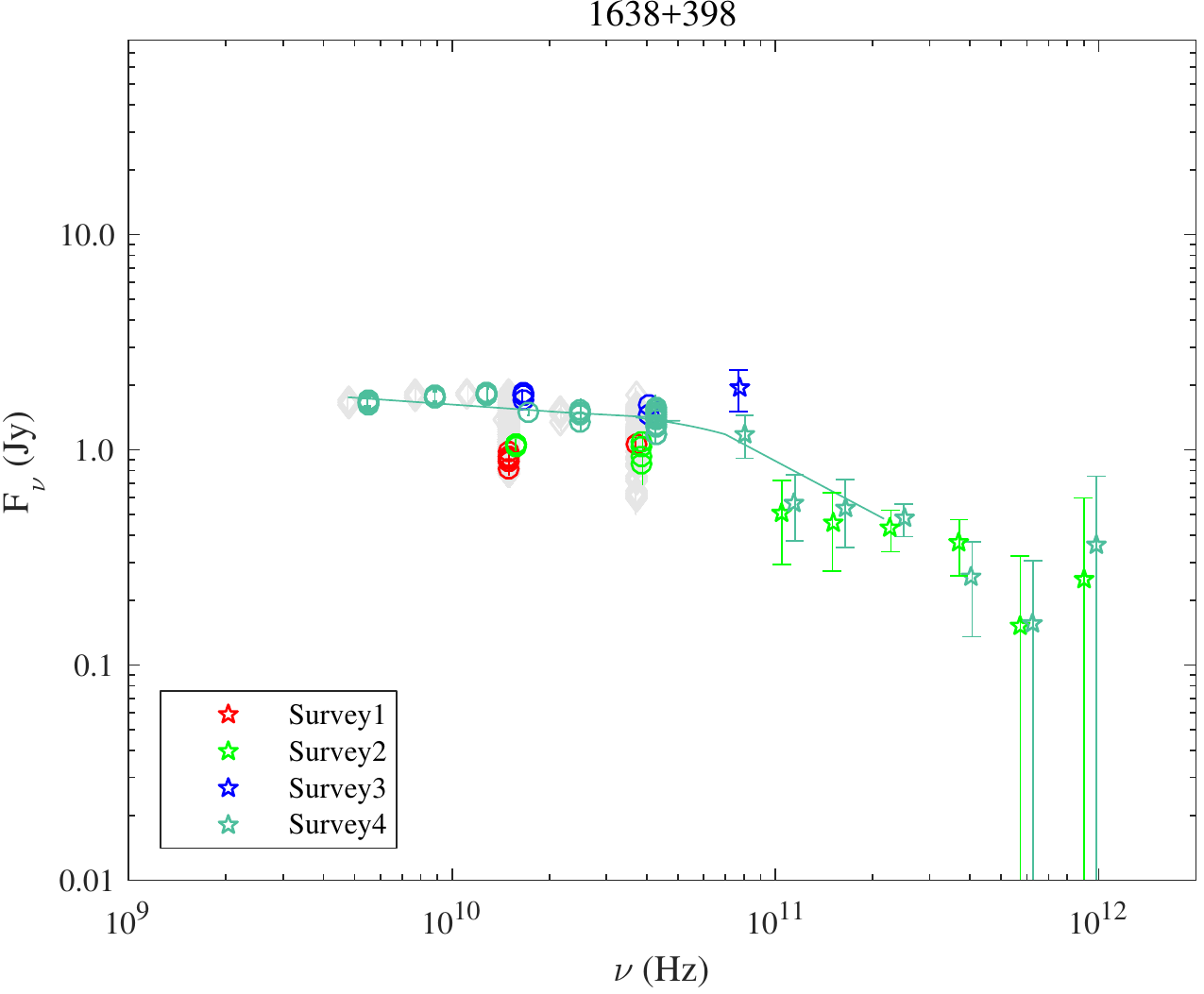}
	\caption{1638+398.}
	\label{1638+398_spectra}
	\end{minipage}\qquad
	\begin{minipage}[b]{.47\textwidth}
	\includegraphics[width=\columnwidth]{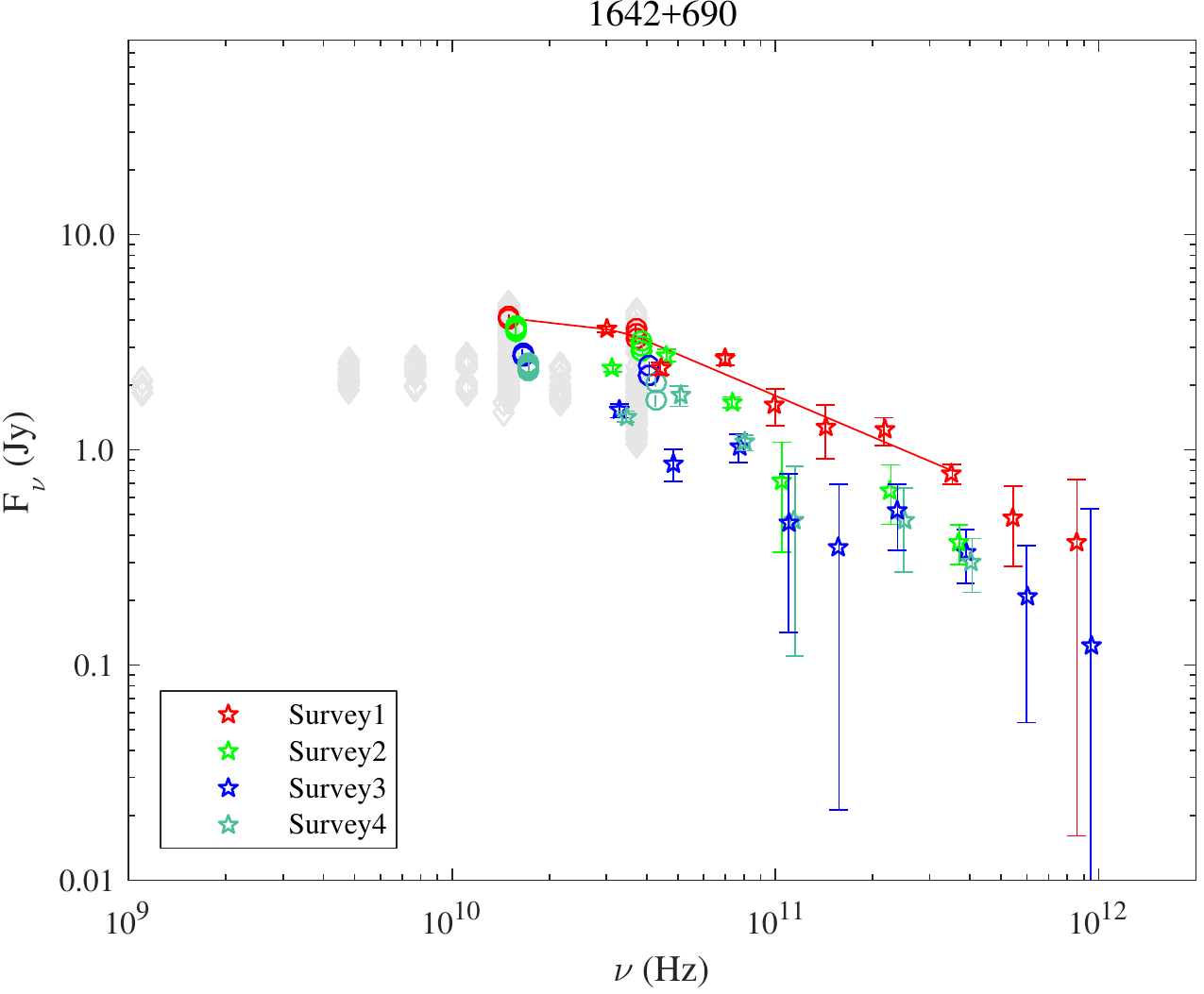}
	\caption{1642+690.}
	\label{1642+690_spectra}
	\end{minipage}
\end{figure*}

\clearpage

\begin{figure*}
	\centering
	\begin{minipage}[b]{.47\textwidth}
	\includegraphics[width=\columnwidth]{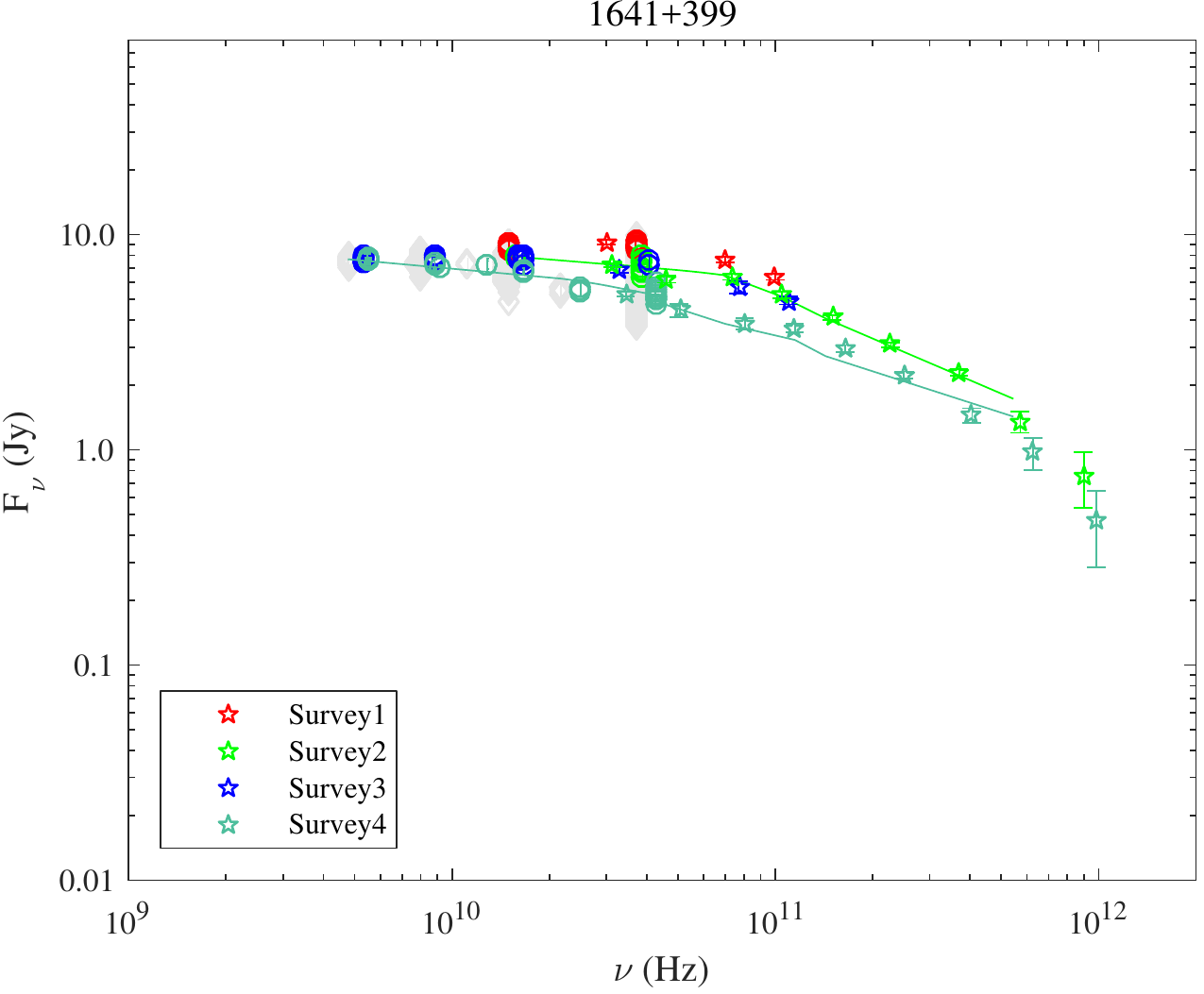}
	\caption{1641+399.}
	\label{1641+399_spectra}
	\end{minipage}\qquad
	\begin{minipage}[b]{.47\textwidth}
	\includegraphics[width=\columnwidth]{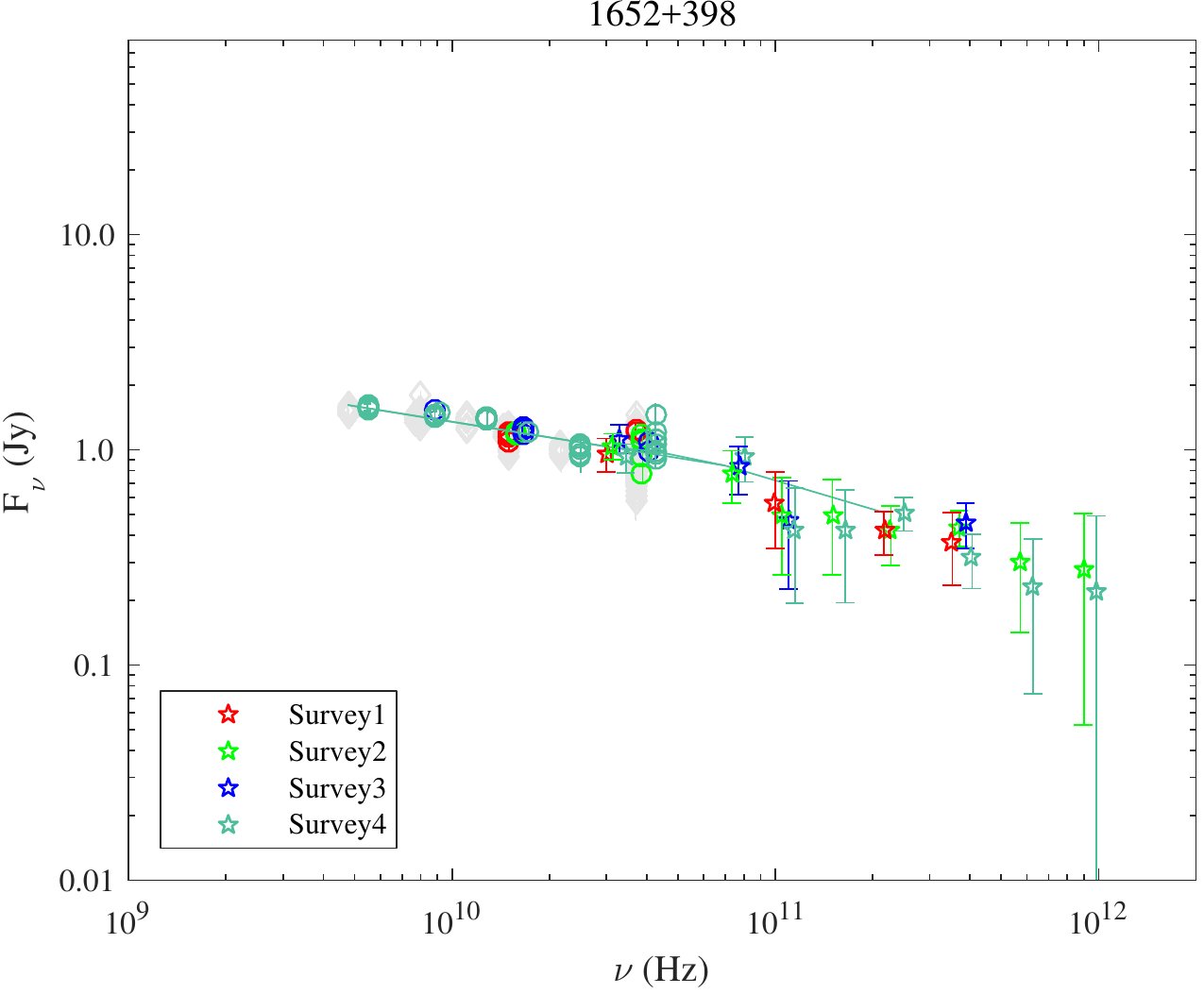}
	\caption{1652+398.}
	\label{1652+398_spectra}
	\end{minipage}
\end{figure*}

\begin{figure*}
	\centering
	\begin{minipage}[b]{.47\textwidth}
	\includegraphics[width=\columnwidth]{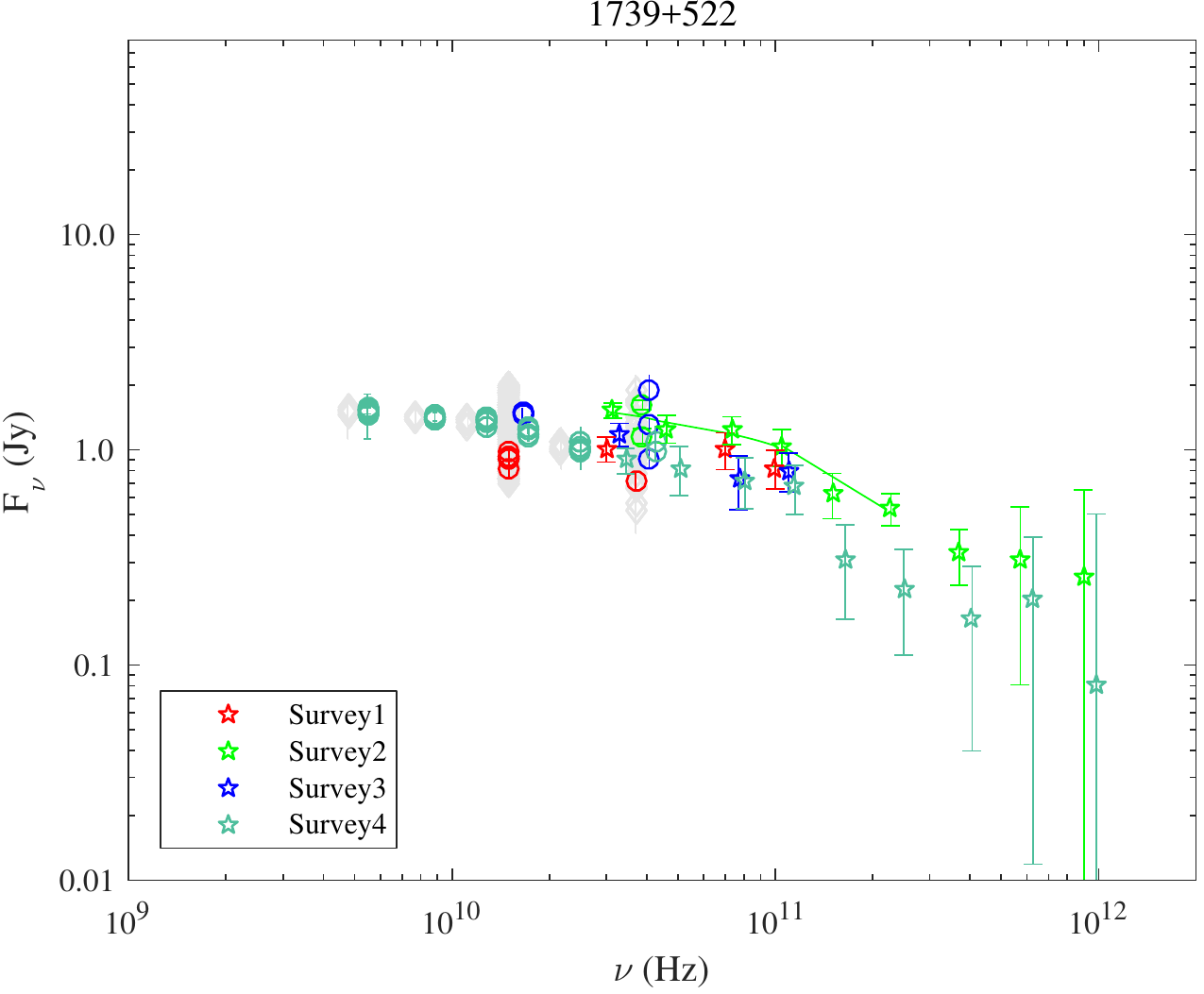}
	\caption{1739+522.}
	\label{1739+522_spectra}
	\end{minipage}\qquad
	\begin{minipage}[b]{.47\textwidth}
	\includegraphics[width=\columnwidth]{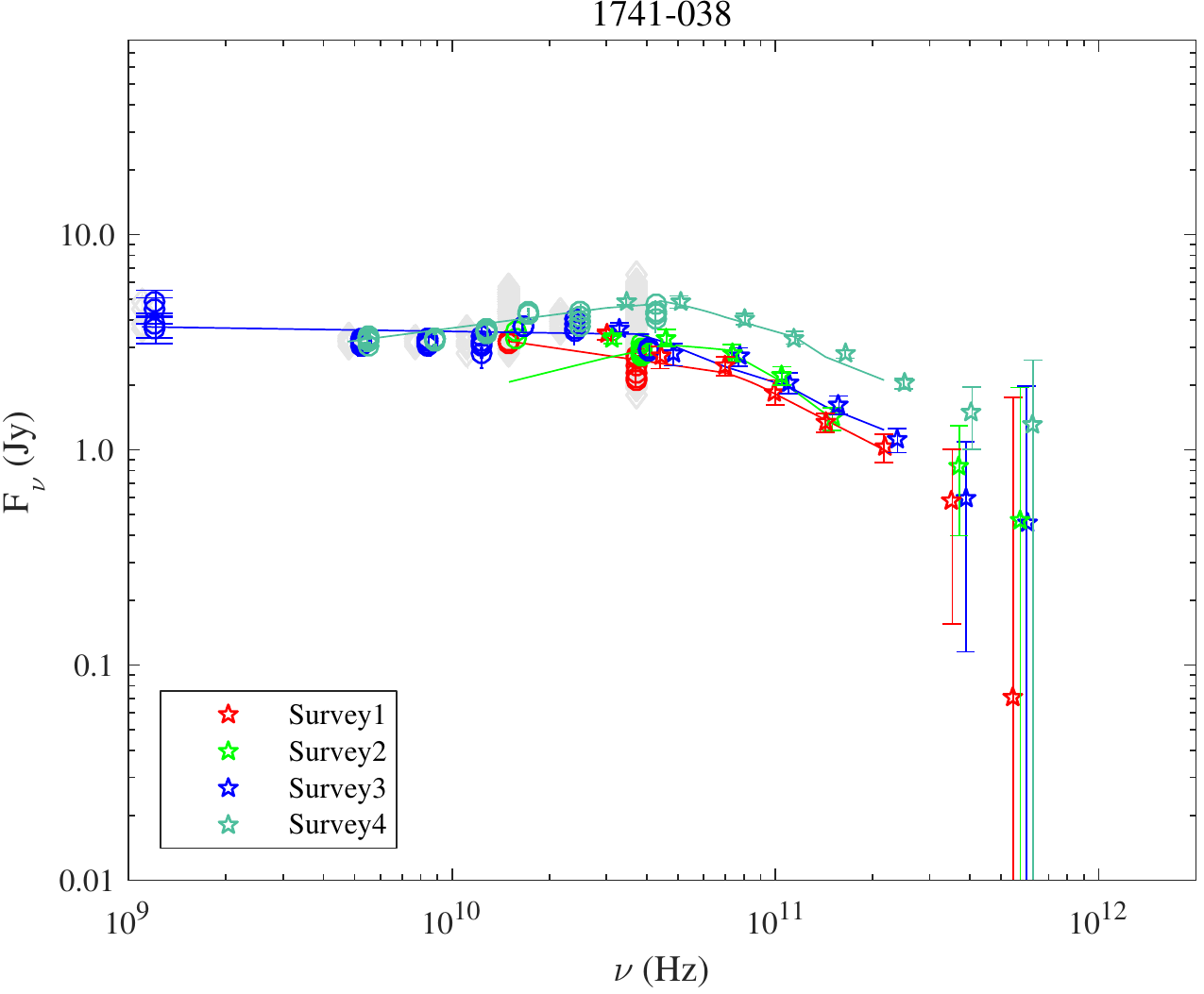}
	\caption{1741$-$038.}
	\label{1741-038_spectra}
	\end{minipage}
\end{figure*}

\begin{figure*}
	\centering
	\begin{minipage}[b]{.47\textwidth}
	\includegraphics[width=\columnwidth]{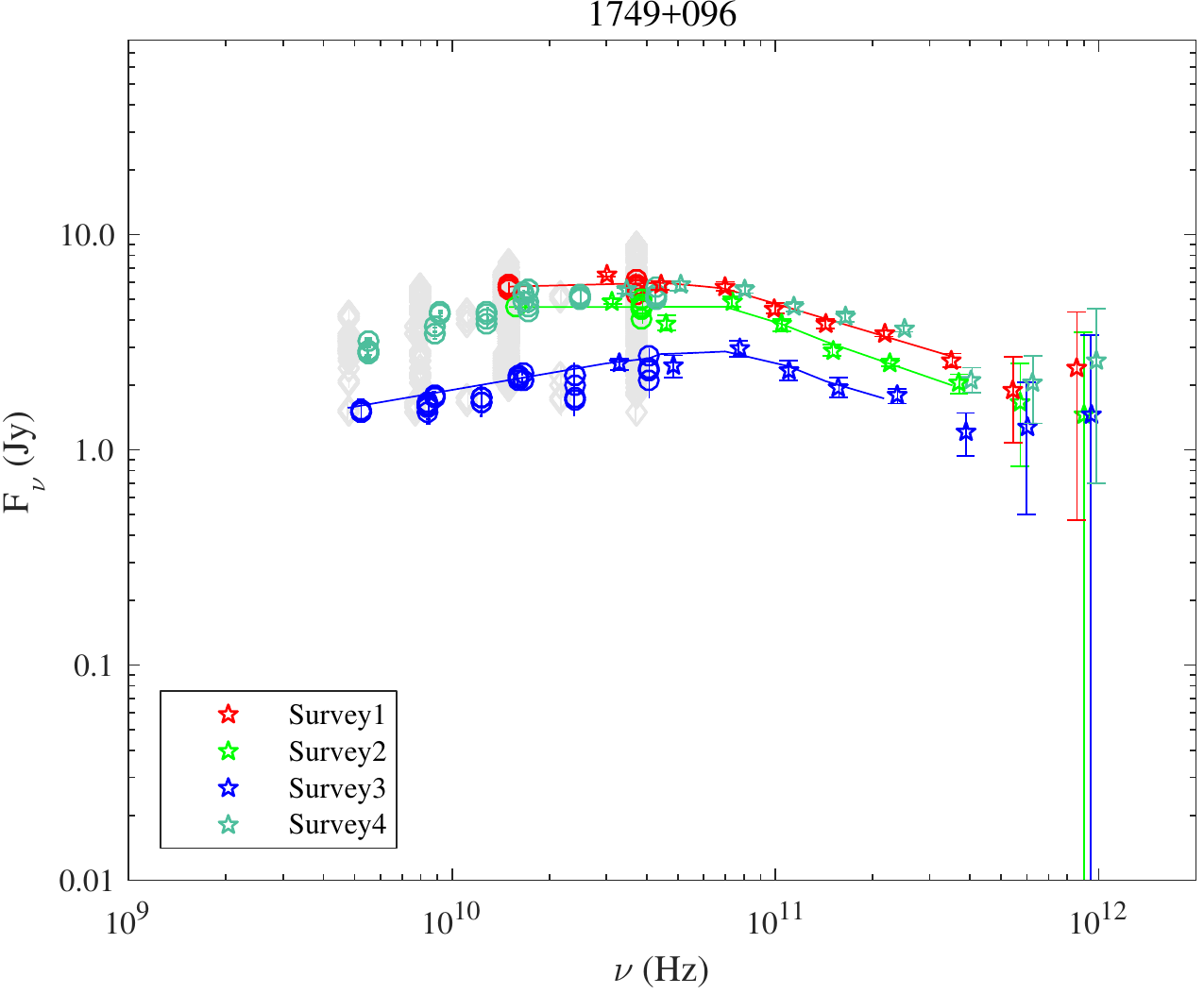}
	\caption{1749+096.}
	\label{1749+096_spectra}
	\end{minipage}\qquad
	\begin{minipage}[b]{.47\textwidth}
	\includegraphics[width=\columnwidth]{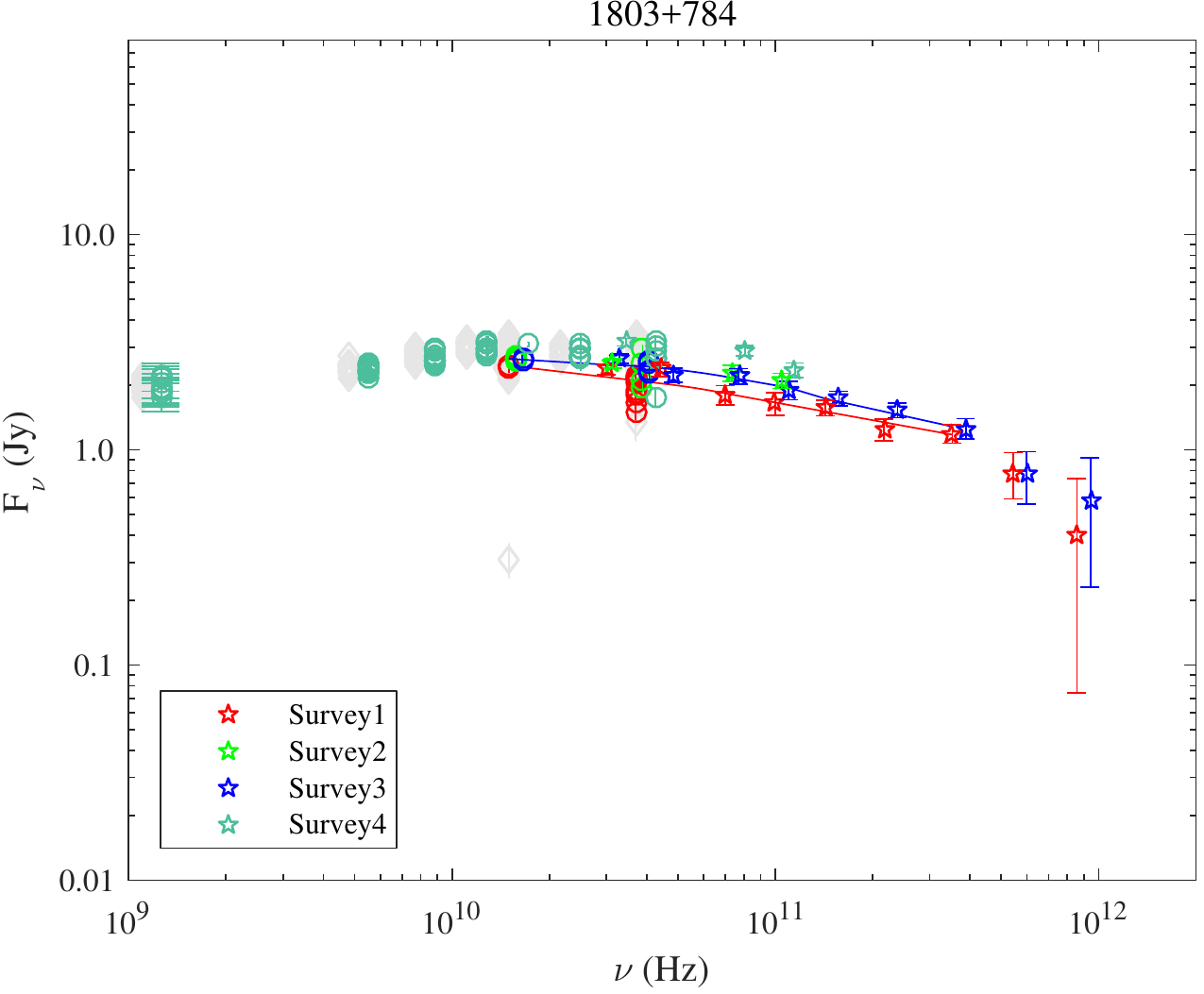}
	\caption{1803+784.}
	\label{1803+784_spectra}
	\end{minipage}
\end{figure*}

\clearpage

\begin{figure*}
	\centering
	\begin{minipage}[b]{.47\textwidth}
	\includegraphics[width=\columnwidth]{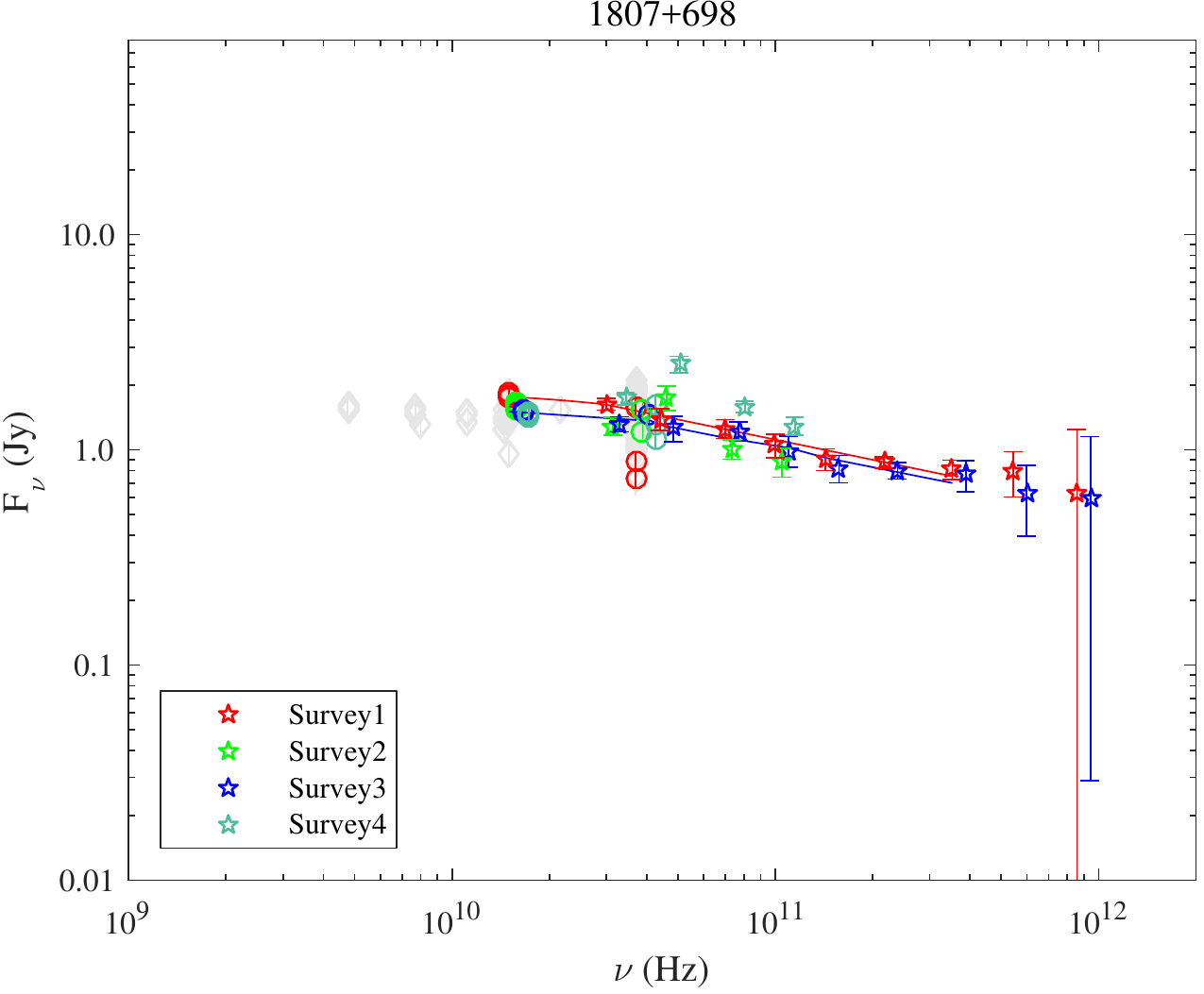}
	\caption{1807+698.}
	\label{1807+698_spectra}
	\end{minipage}\qquad
	\begin{minipage}[b]{.47\textwidth}
	\includegraphics[width=\columnwidth]{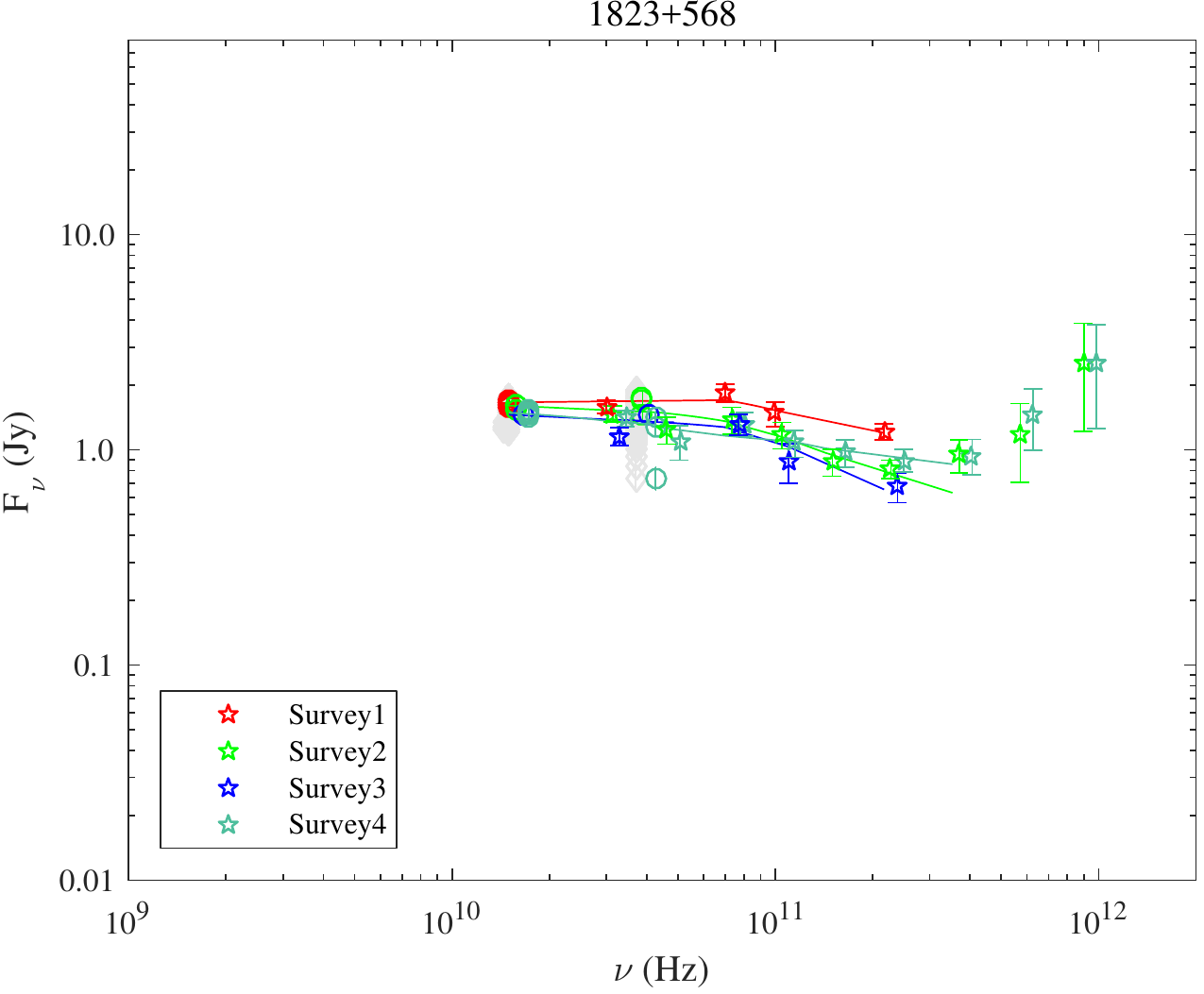}
	\caption{1823+568.}
	\label{1823+568_spectra}
	\end{minipage}
\end{figure*}

\begin{figure*}
	\centering
	\begin{minipage}[b]{.47\textwidth}
	\includegraphics[width=\columnwidth]{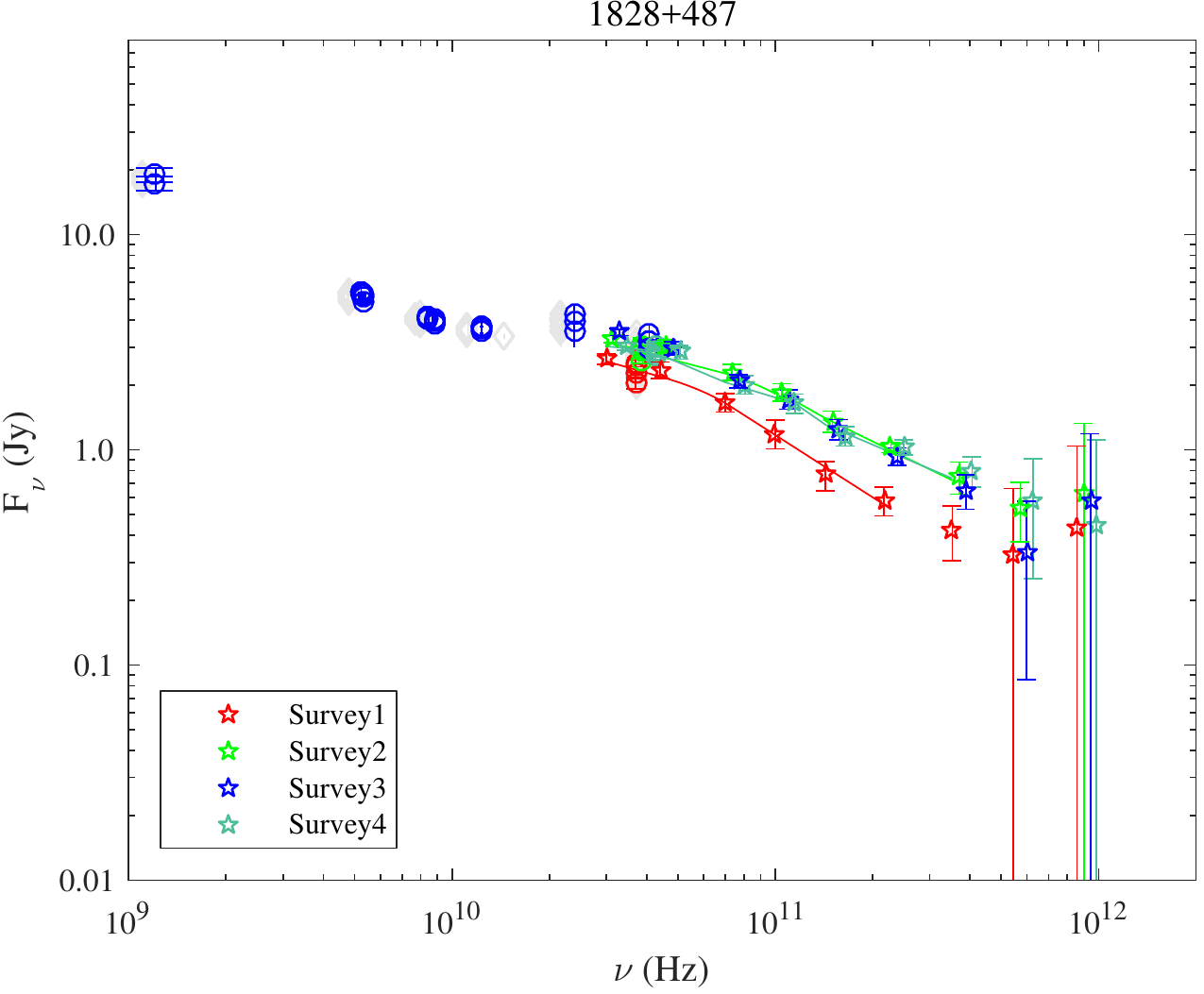}
	\caption{1828+487.}
	\label{1828+487_spectra}
	\end{minipage}\qquad
	\begin{minipage}[b]{.47\textwidth}
	\includegraphics[width=\columnwidth]{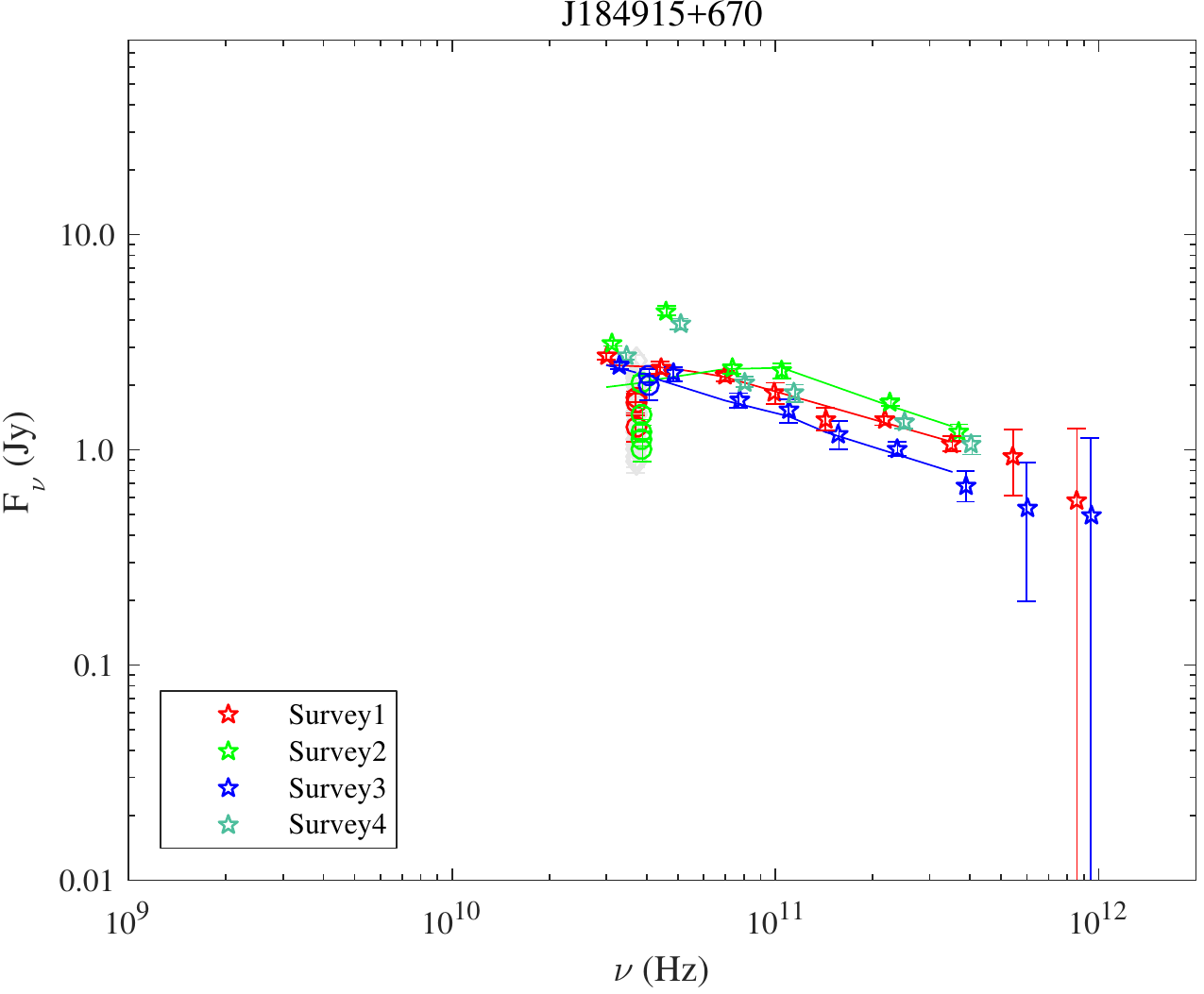}
	\caption{J184915+670.}
	\label{J184915+670_spectra}
	\end{minipage}
\end{figure*}

\begin{figure*}
	\centering
	\begin{minipage}[b]{.47\textwidth}
	\includegraphics[width=\columnwidth]{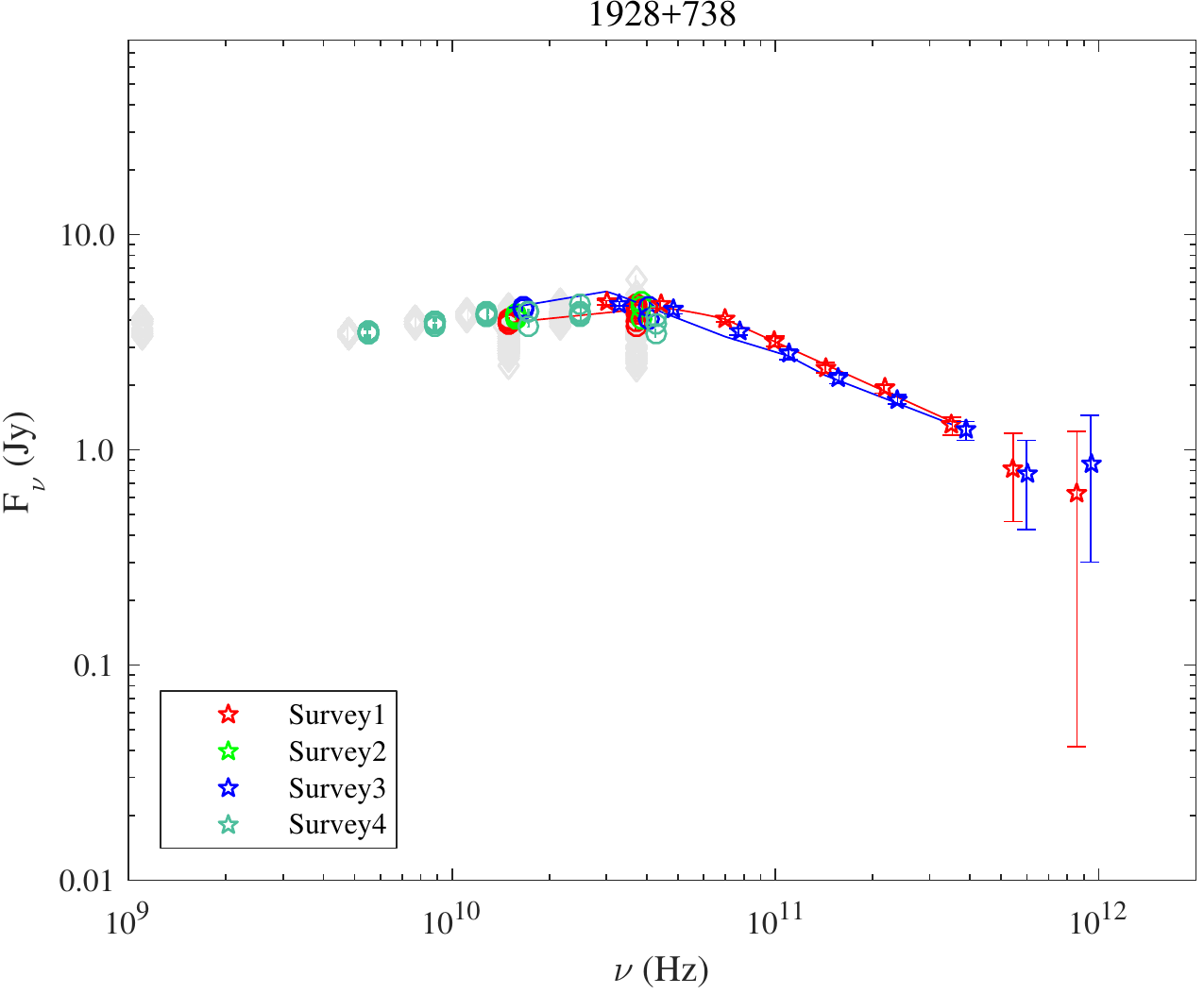}
	\caption{1928+738.}
	\label{1928+738_spectra}
	\end{minipage}\qquad
	\begin{minipage}[b]{.47\textwidth}
	\includegraphics[width=\columnwidth]{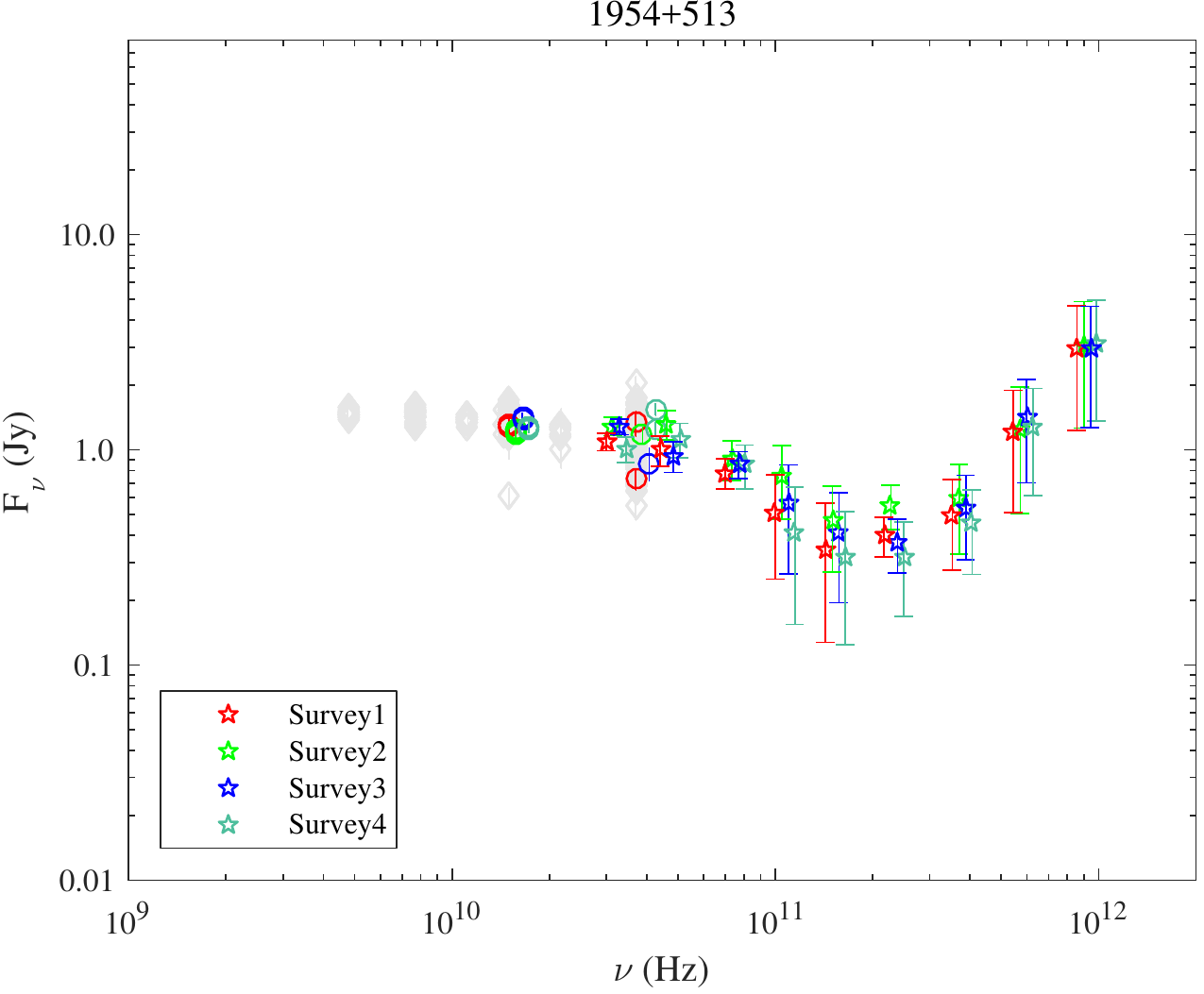}
	\caption{1954+513.}
	\label{1954+513_spectra}
	\end{minipage}
\end{figure*}

\clearpage

\begin{figure*}
	\centering
	\begin{minipage}[b]{.47\textwidth}
	\includegraphics[width=\columnwidth]{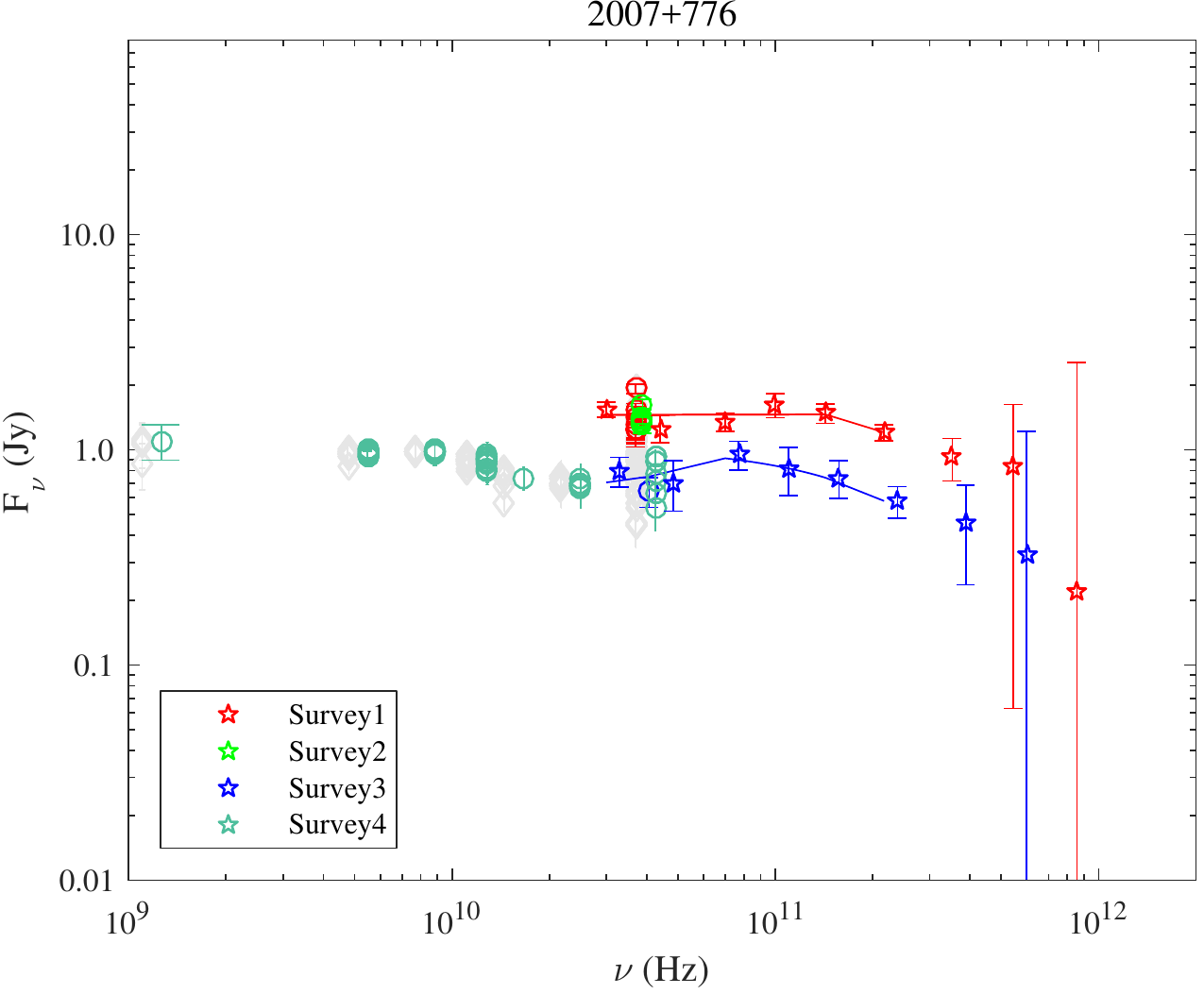}
	\caption{2007+776.}
	\label{2007+776_spectra}
	\end{minipage}\qquad
	\begin{minipage}[b]{.47\textwidth}
	\includegraphics[width=\columnwidth]{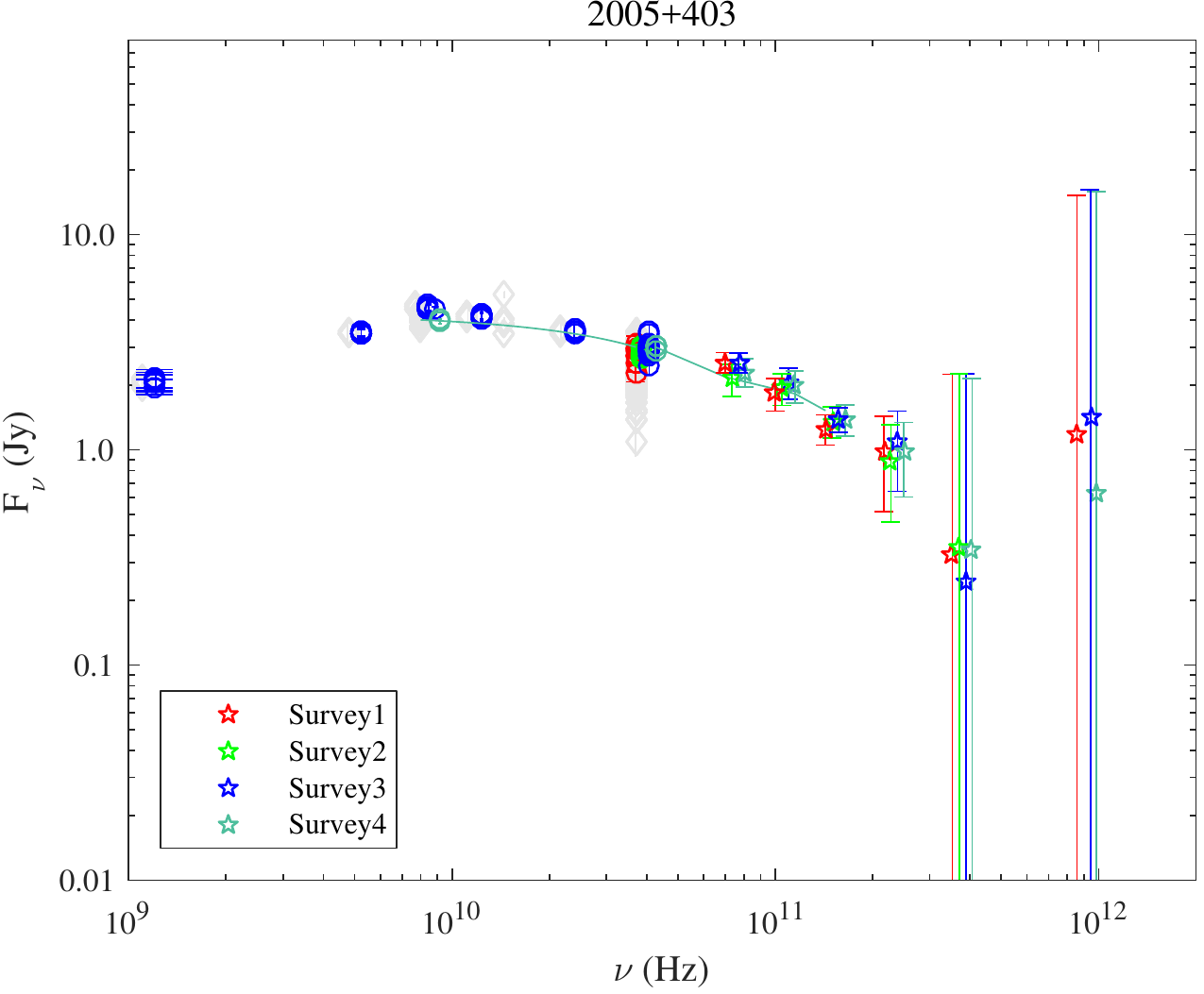}
	\caption{2005+403.}
	\label{2005+403_spectra}
	\end{minipage}
\end{figure*}

\begin{figure*}
	\centering
	\begin{minipage}[b]{.47\textwidth}
	\includegraphics[width=\columnwidth]{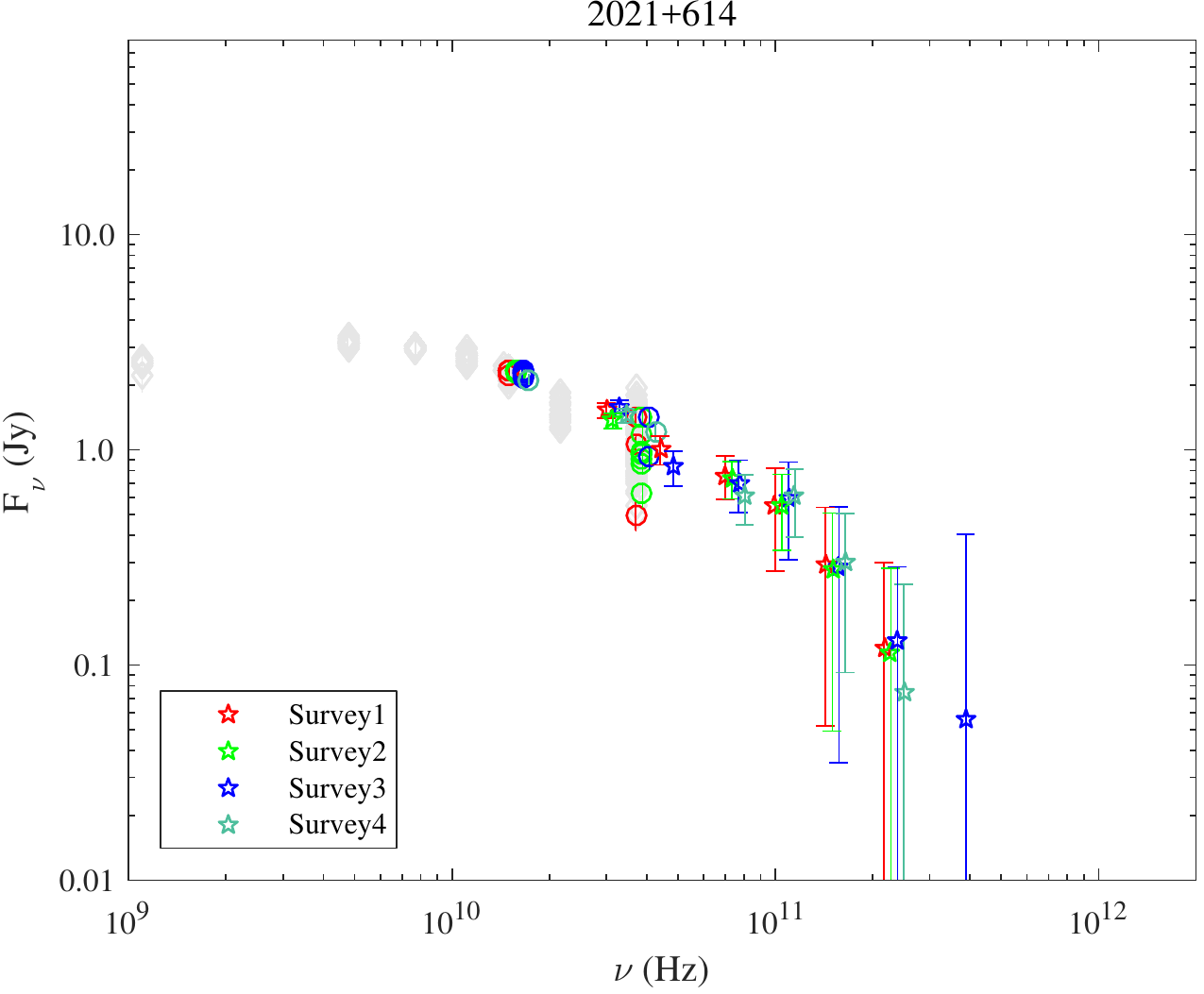}
	\caption{2021+614.}
	\label{2021+614_spectra}
	\end{minipage}\qquad
	\begin{minipage}[b]{.47\textwidth}
	\includegraphics[width=\columnwidth]{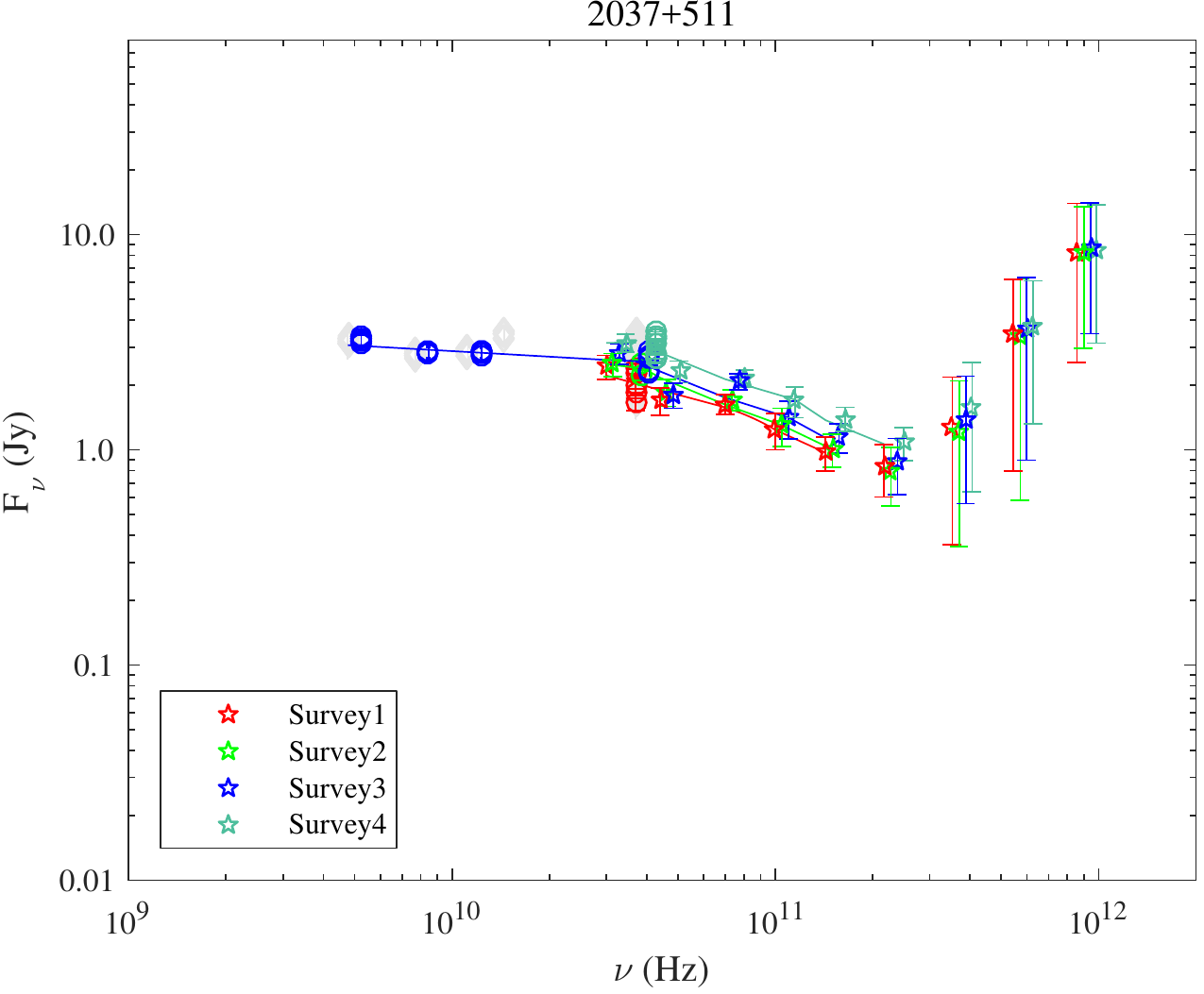}
	\caption{2037+511.}
	\label{2037+511_spectra}
	\end{minipage}
\end{figure*}

\begin{figure*}
	\centering
	\begin{minipage}[b]{.47\textwidth}
	\includegraphics[width=\columnwidth]{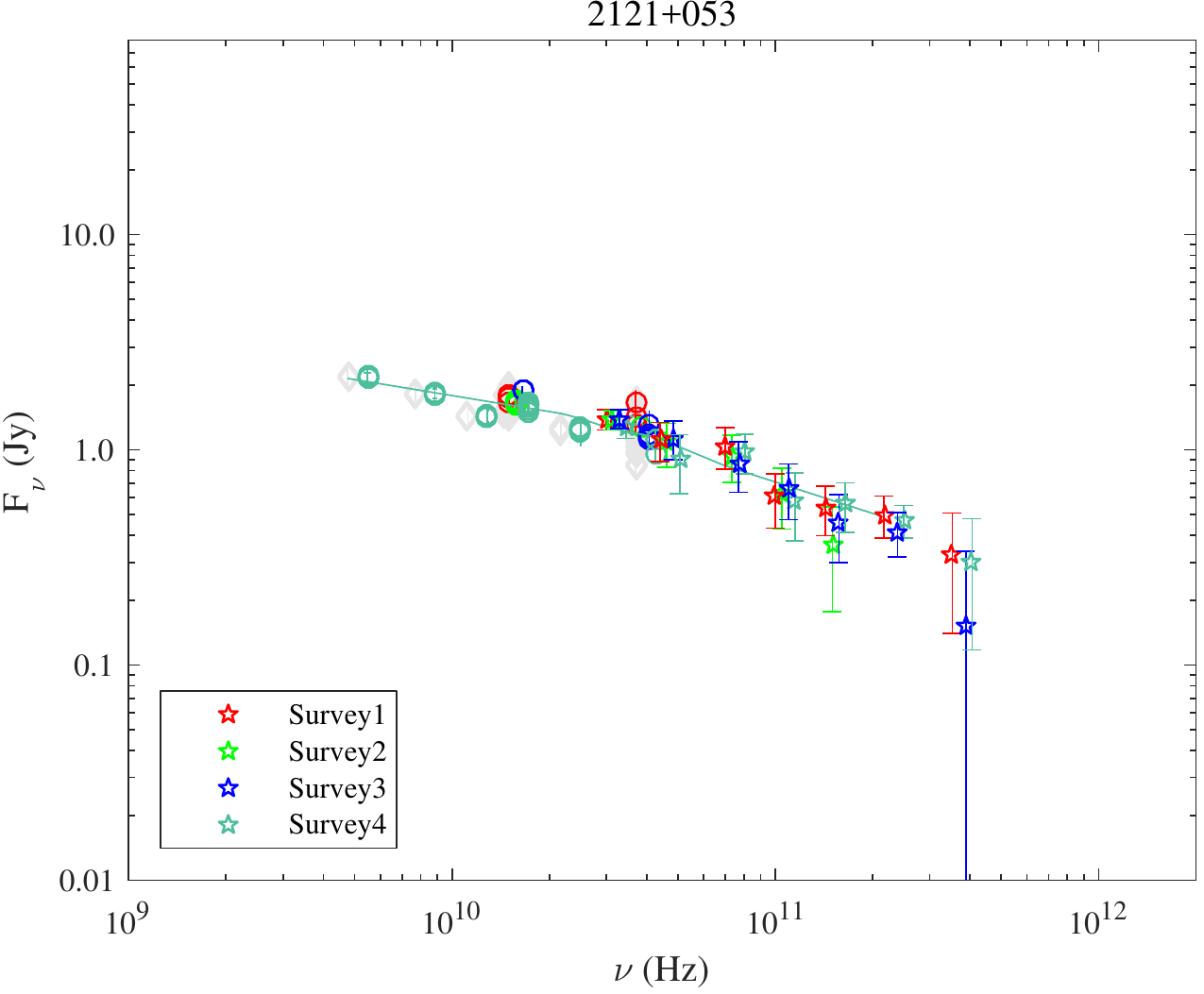}
	\caption{2121+053.}
	\label{2121+053_spectra}
	\end{minipage}\qquad
	\begin{minipage}[b]{.47\textwidth}
	\includegraphics[width=\columnwidth]{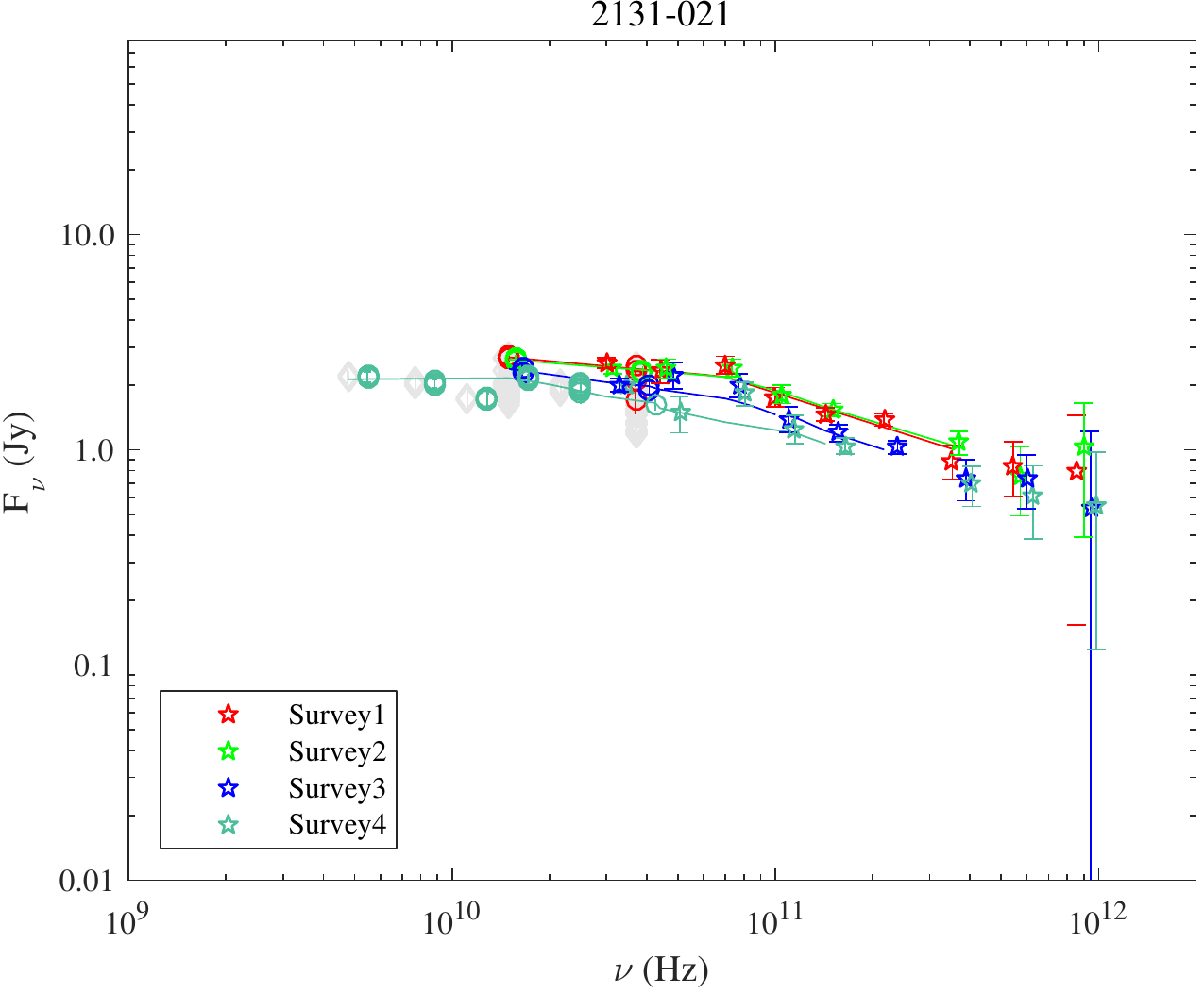}
	\caption{2131$-$021.}
	\label{2131-021_spectra}
	\end{minipage}
\end{figure*}

\clearpage

\begin{figure*}
	\centering
	\begin{minipage}[b]{.47\textwidth}
	\includegraphics[width=\columnwidth]{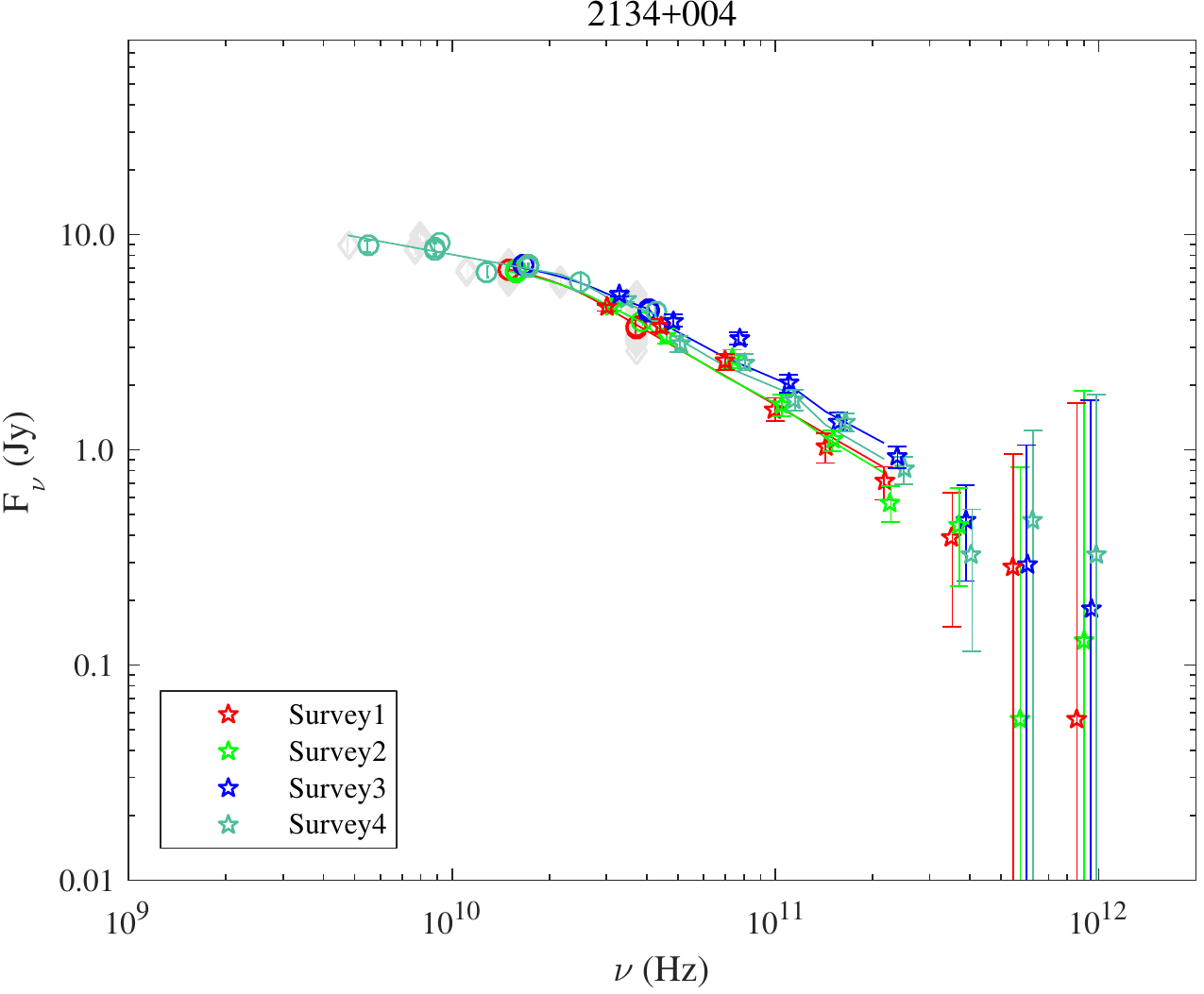}
	\caption{2134+004.}
	\label{2134+004_spectra}
	\end{minipage}\qquad
	\begin{minipage}[b]{.47\textwidth}
	\includegraphics[width=\columnwidth]{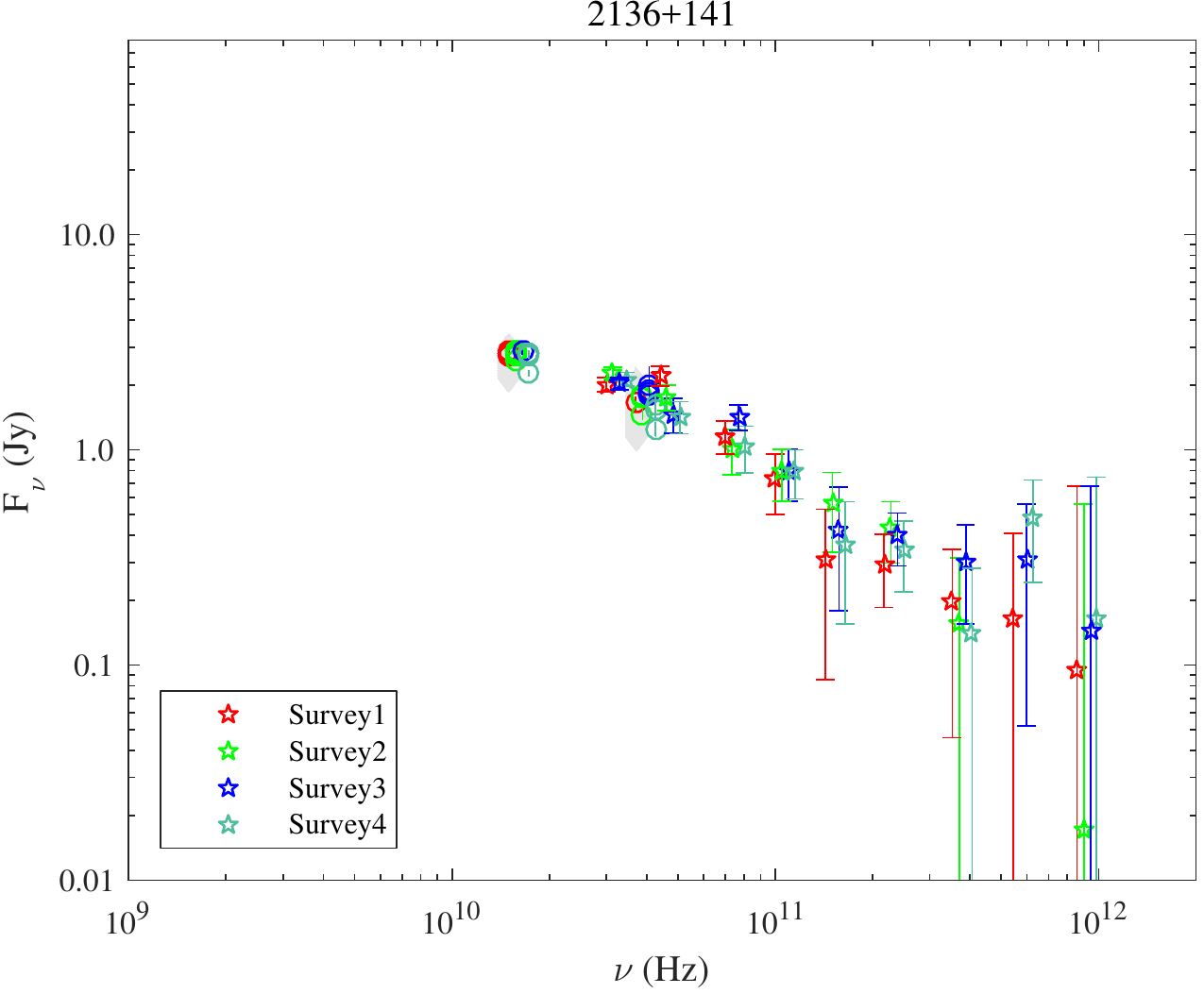}
	\caption{2136+141.}
	\label{2136+141_spectra}
	\end{minipage}
\end{figure*}

\begin{figure*}
	\centering
	\begin{minipage}[b]{.47\textwidth}
	\includegraphics[width=\columnwidth]{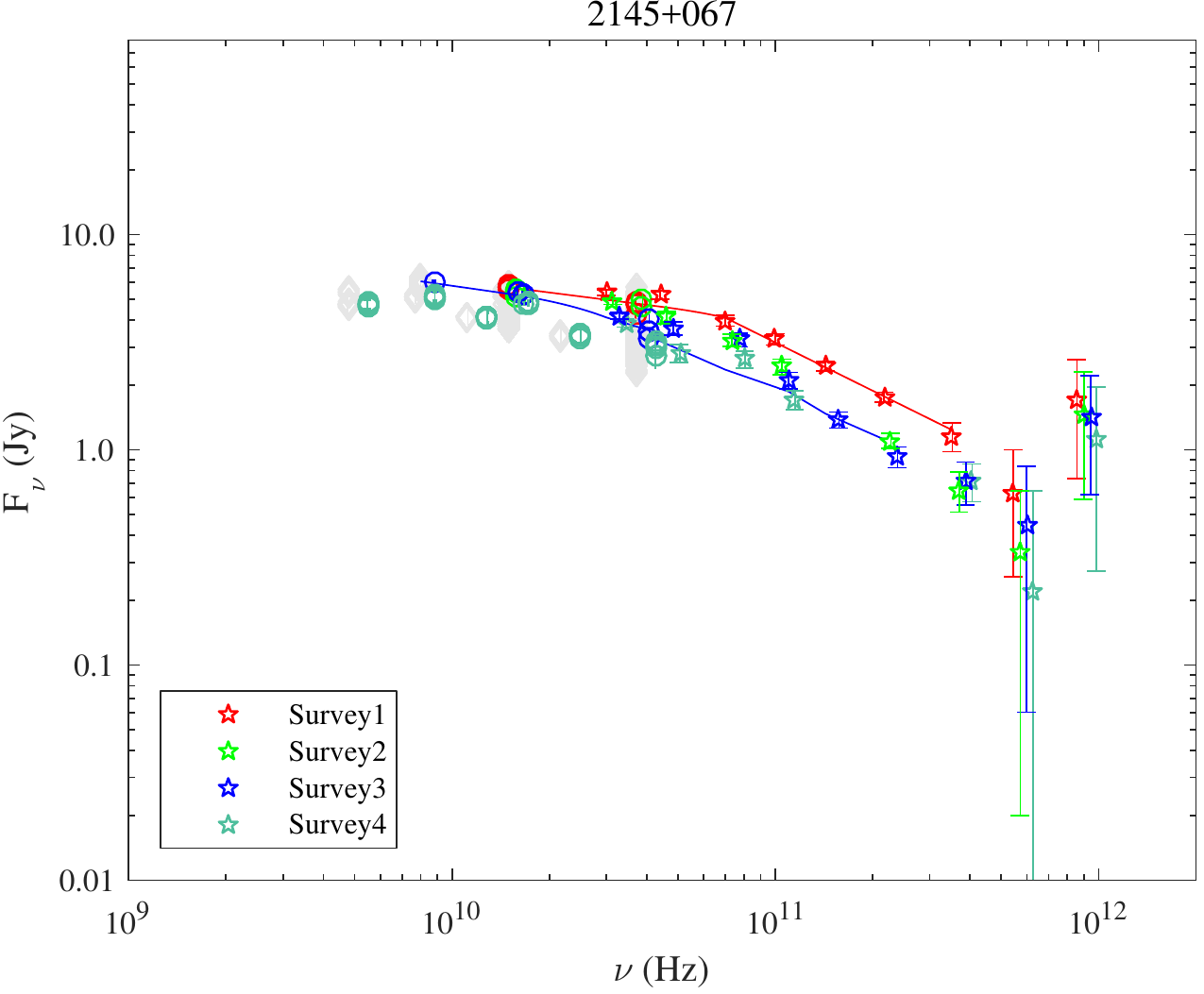}
	\caption{2145+067.}
	\label{2145+067_spectra}
	\end{minipage}\qquad
	\begin{minipage}[b]{.47\textwidth}
	\includegraphics[width=\columnwidth]{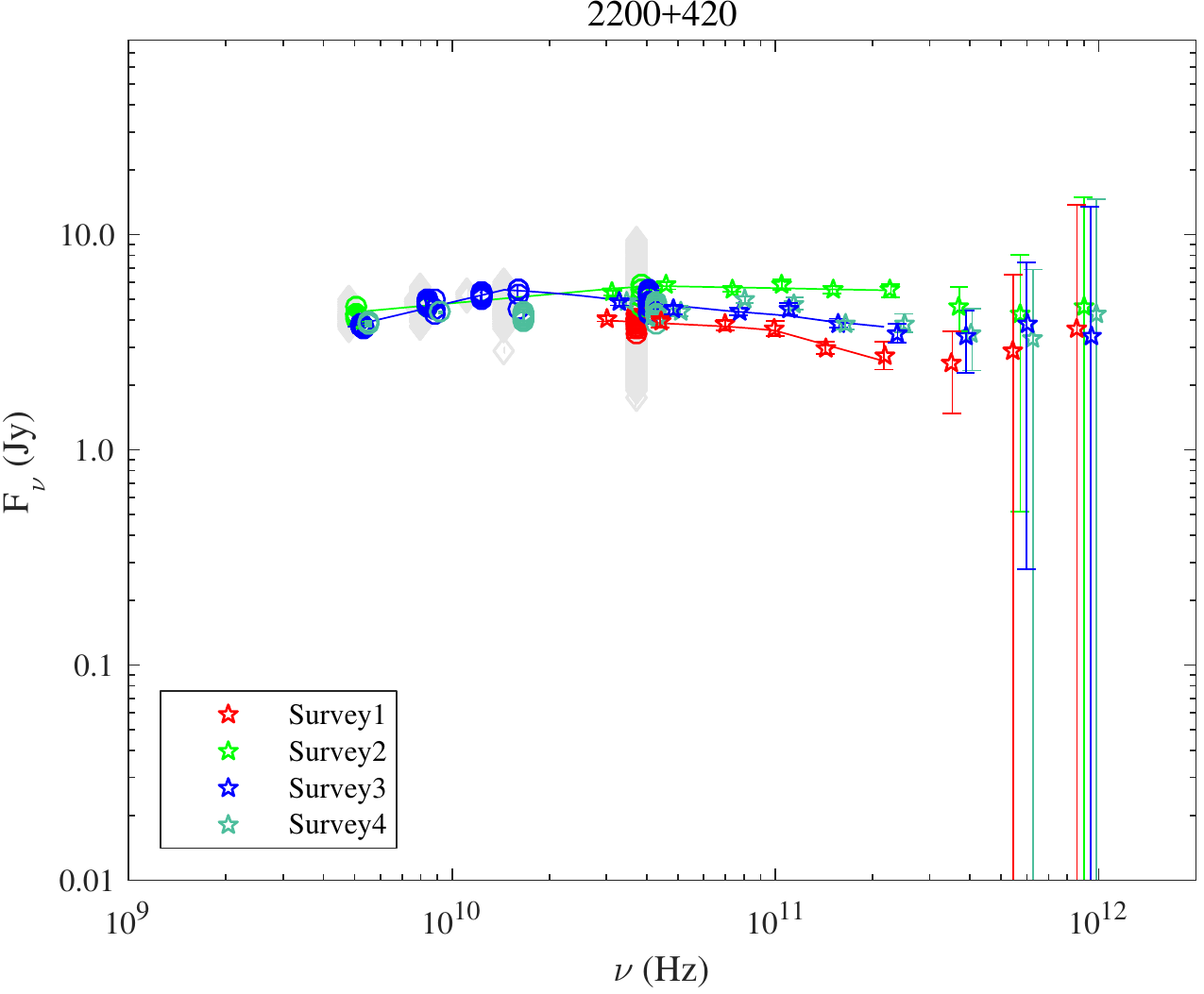}
	\caption{2200+420.}
	\label{2200+420_spectra}
	\end{minipage}
\end{figure*}

\begin{figure*}
	\centering
	\begin{minipage}[b]{.47\textwidth}
	\includegraphics[width=\columnwidth]{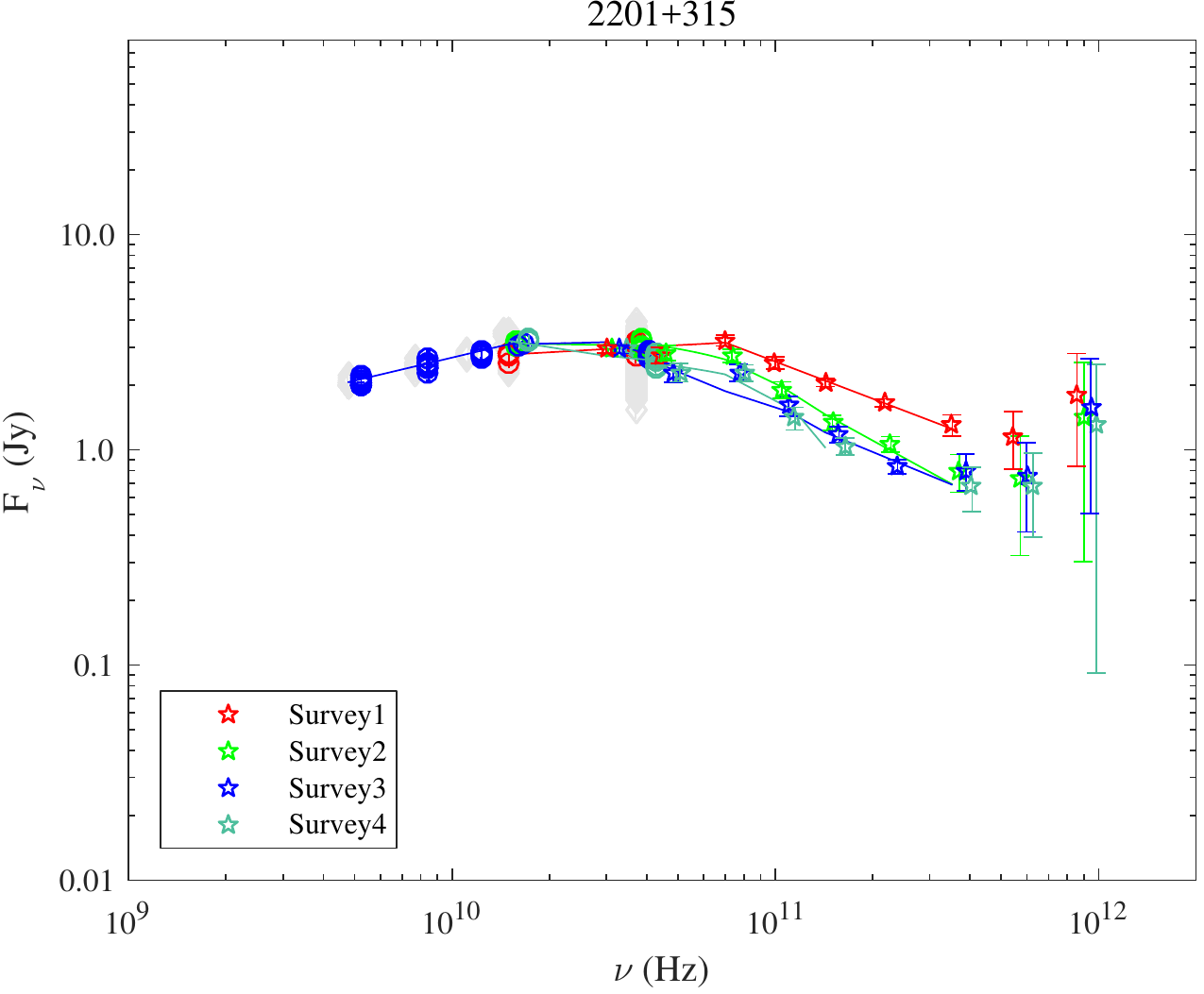}
	\caption{2201+315.}
	\label{2201+315_spectra}
	\end{minipage}\qquad
	\begin{minipage}[b]{.47\textwidth}
	\includegraphics[width=\columnwidth]{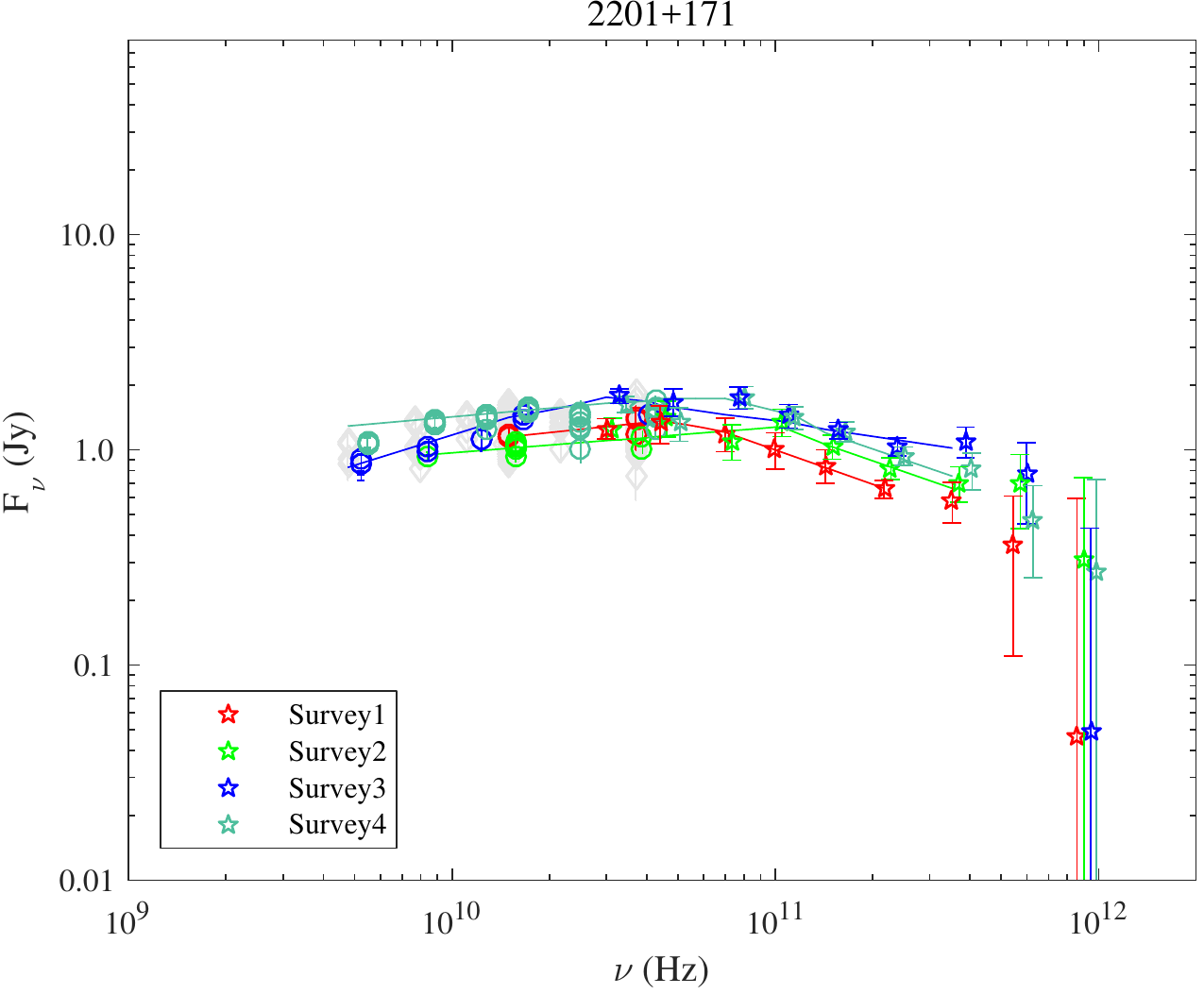}
	\caption{2201+171.}
	\label{2201+171_spectra}
	\end{minipage}
\end{figure*}

\clearpage

\begin{figure*}
	\centering
	\begin{minipage}[b]{.47\textwidth}
	\includegraphics[width=\columnwidth]{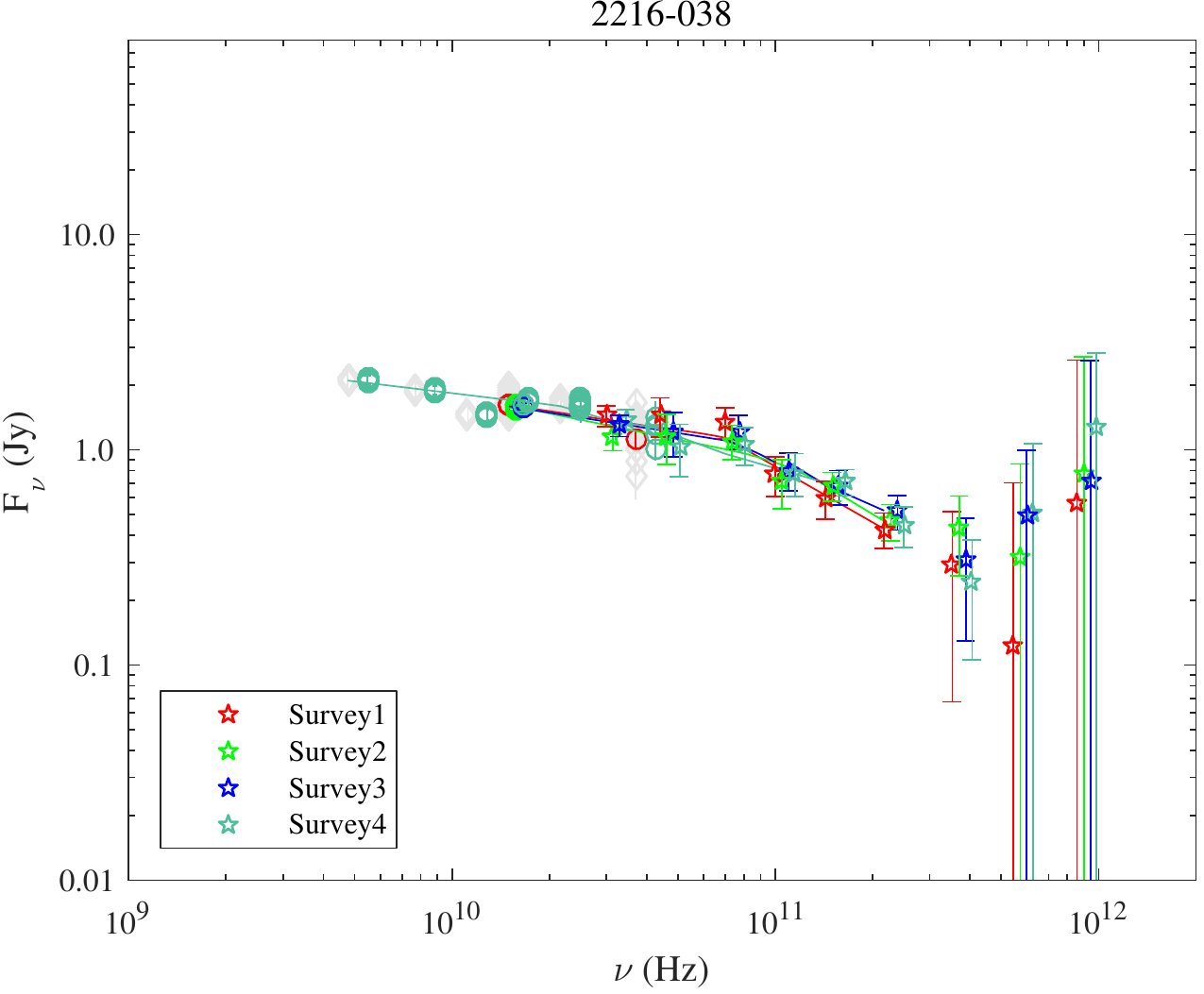}
	\caption{2216$-$038.}
	\label{2216-038_spectra}
	\end{minipage}\qquad
	\begin{minipage}[b]{.47\textwidth}
	\includegraphics[width=\columnwidth]{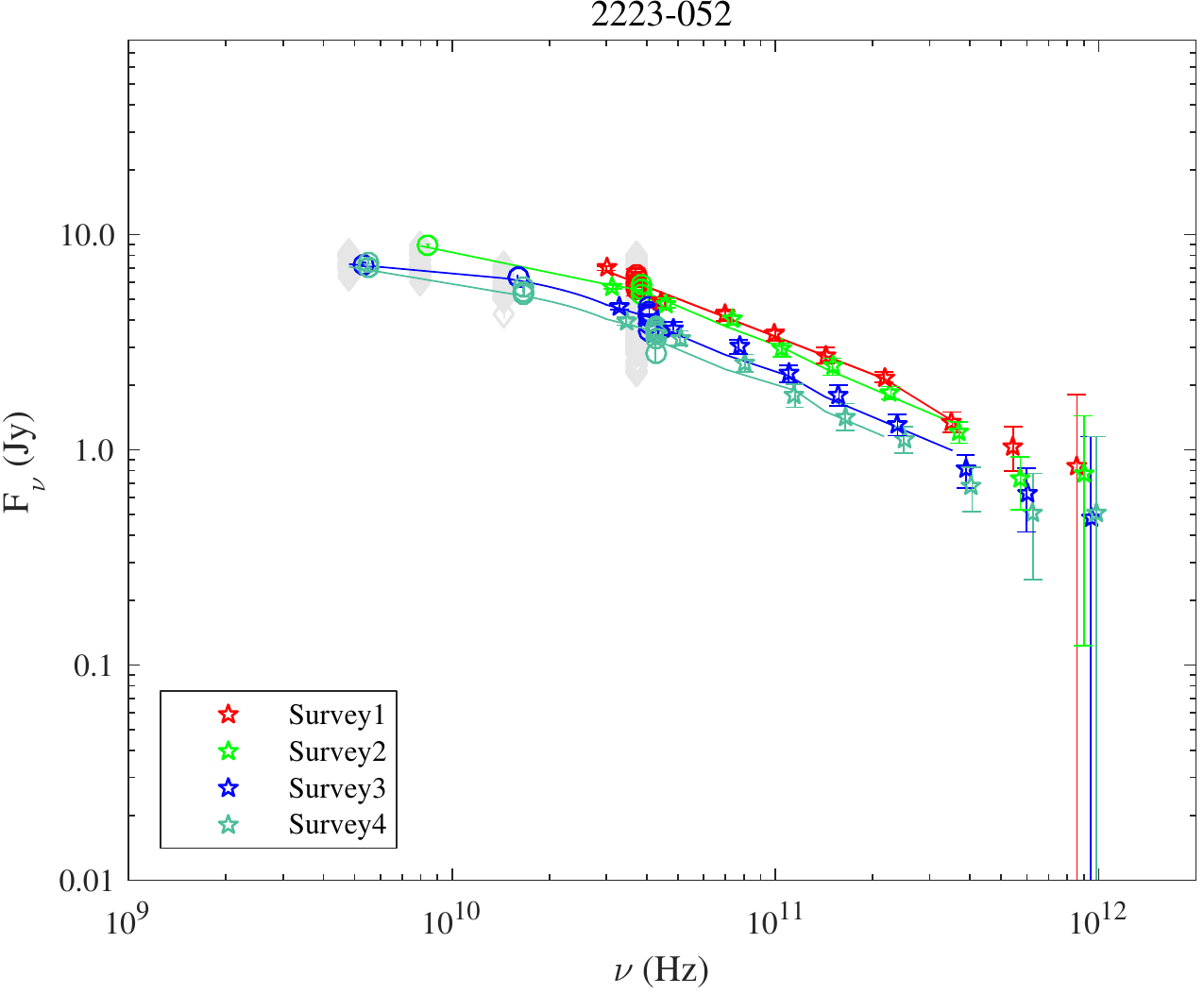}
	\caption{2223$-$052.}
	\label{2223-052_spectra}
	\end{minipage}
\end{figure*}

\begin{figure*}
	\centering
	\begin{minipage}[b]{.47\textwidth}
	\includegraphics[width=\columnwidth]{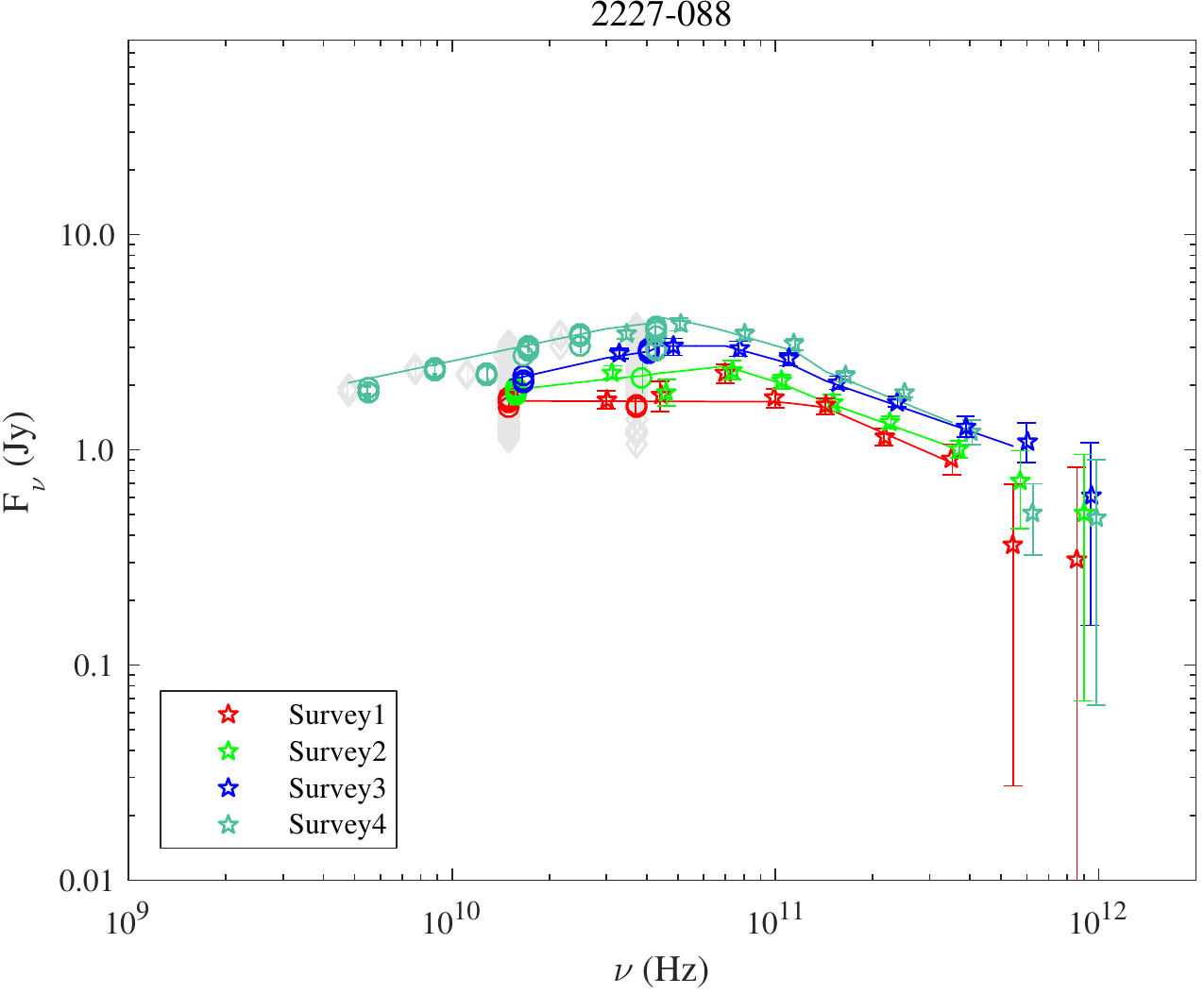}
	\caption{2227$-$088.}
	\label{2227-088_spectra}
	\end{minipage}\qquad
	\begin{minipage}[b]{.47\textwidth}
	\includegraphics[width=\columnwidth]{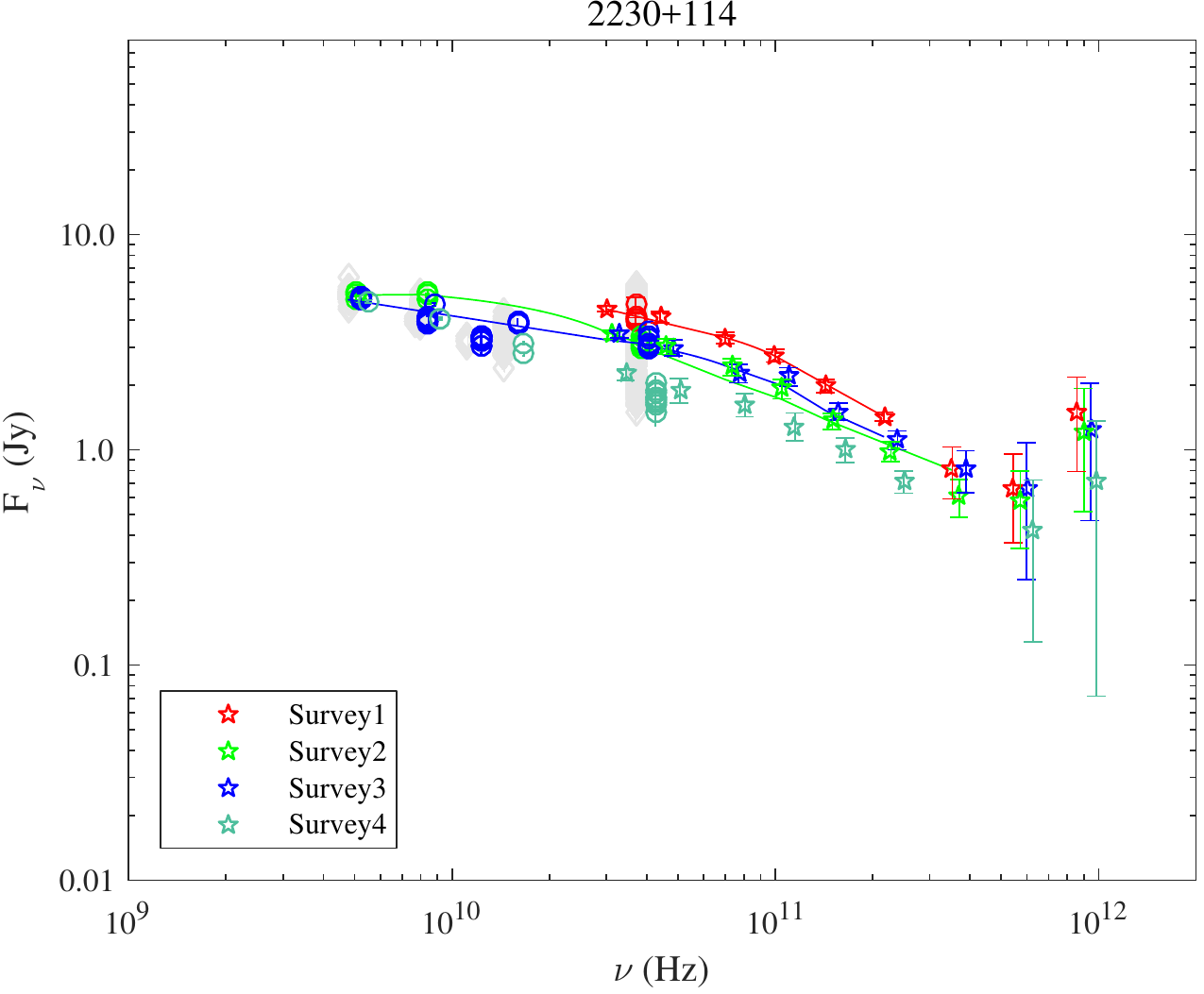}
	\caption{2230+114.}
	\label{2230+114_spectra}
	\end{minipage}
\end{figure*}

\begin{figure*}
	\centering
	\begin{minipage}[b]{.47\textwidth}
	\includegraphics[width=\columnwidth]{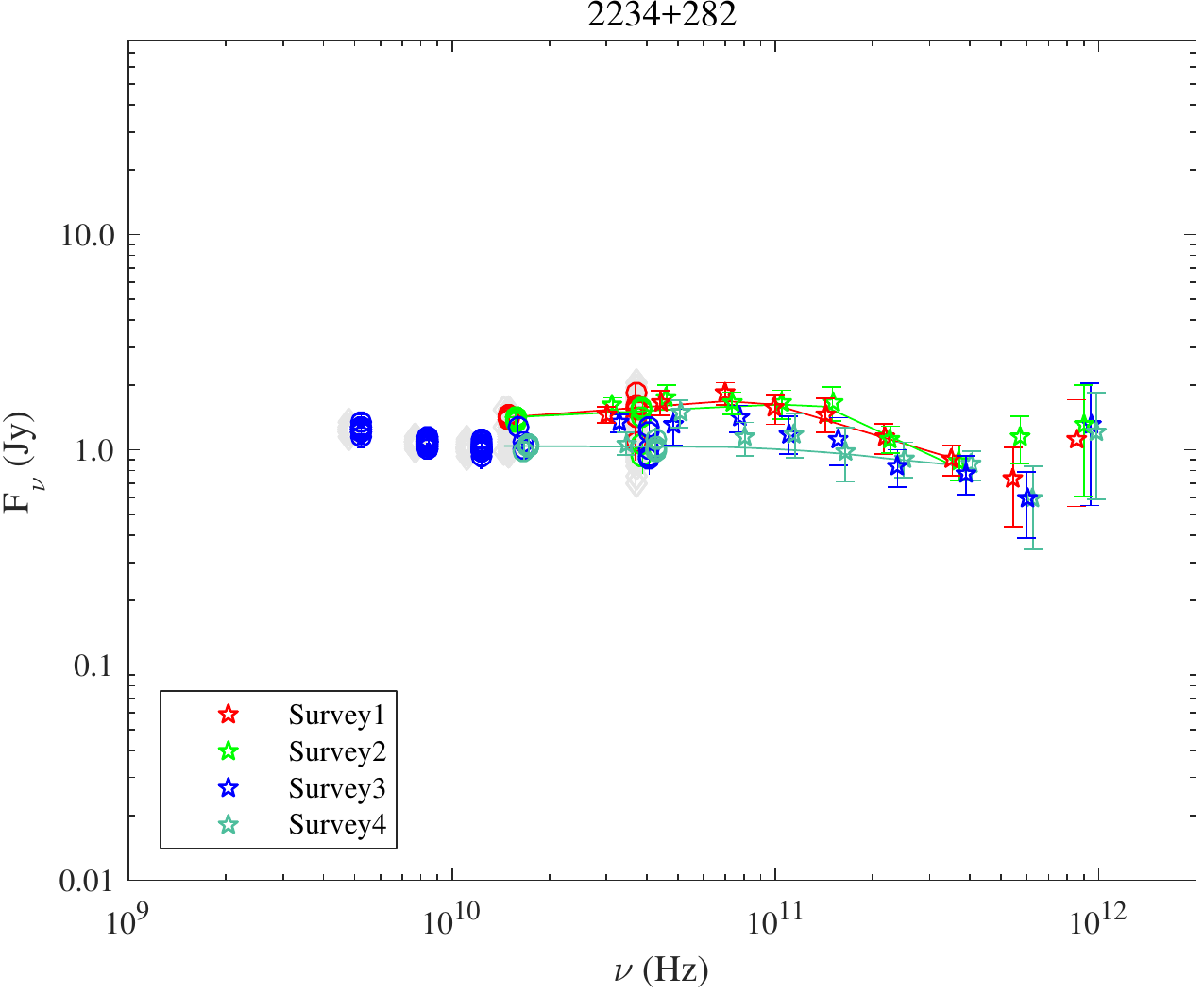}
	\caption{2234+282.}
	\label{2234+282_spectra}
	\end{minipage}\qquad
	\begin{minipage}[b]{.47\textwidth}
	\includegraphics[width=\columnwidth]{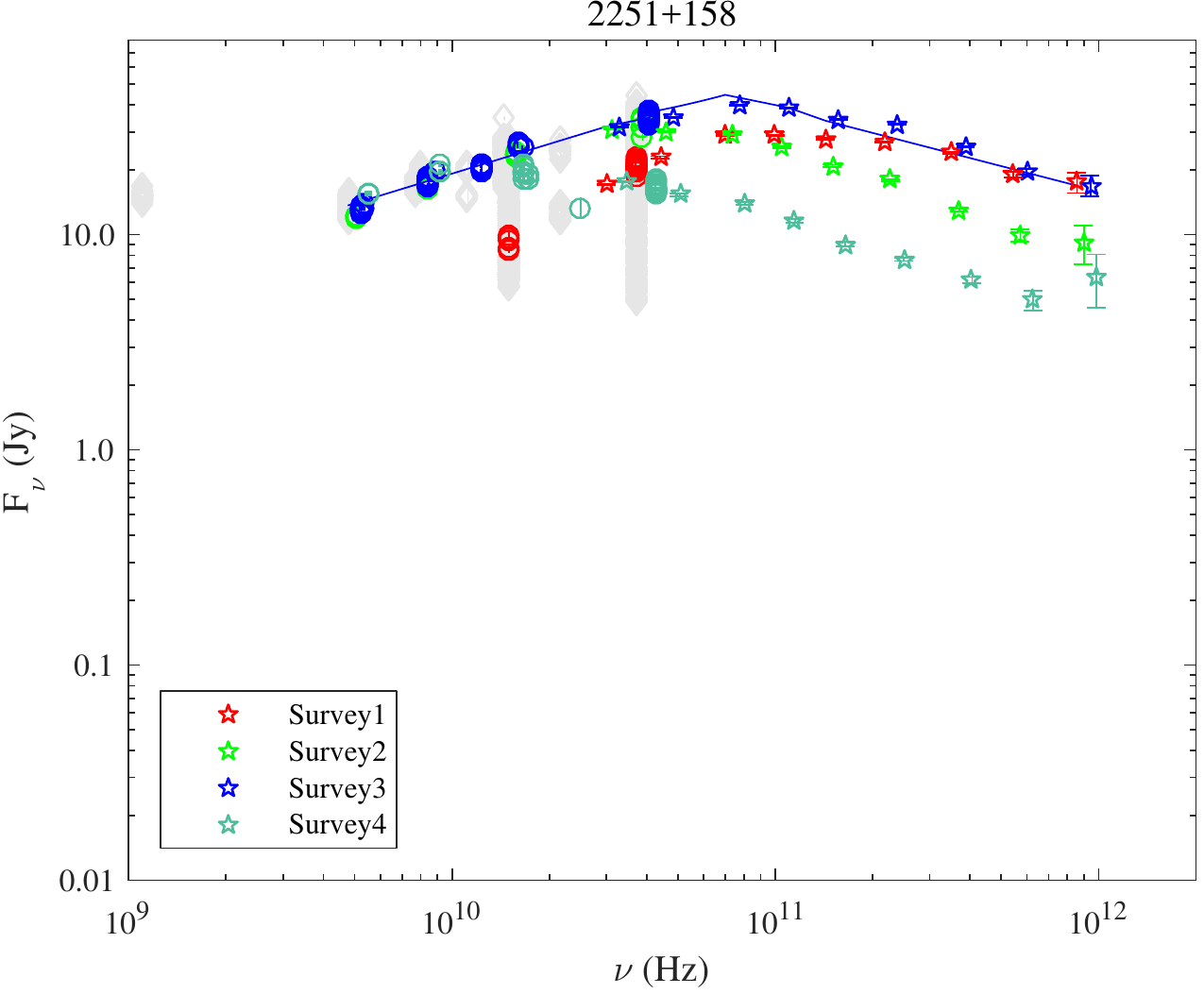}
	\caption{2251+158.}
	\label{2251+158_spectra}
	\end{minipage}
\end{figure*}

\clearpage

\begin{figure*}
	\centering
	\begin{minipage}[b]{.47\textwidth}
	\includegraphics[width=\columnwidth]{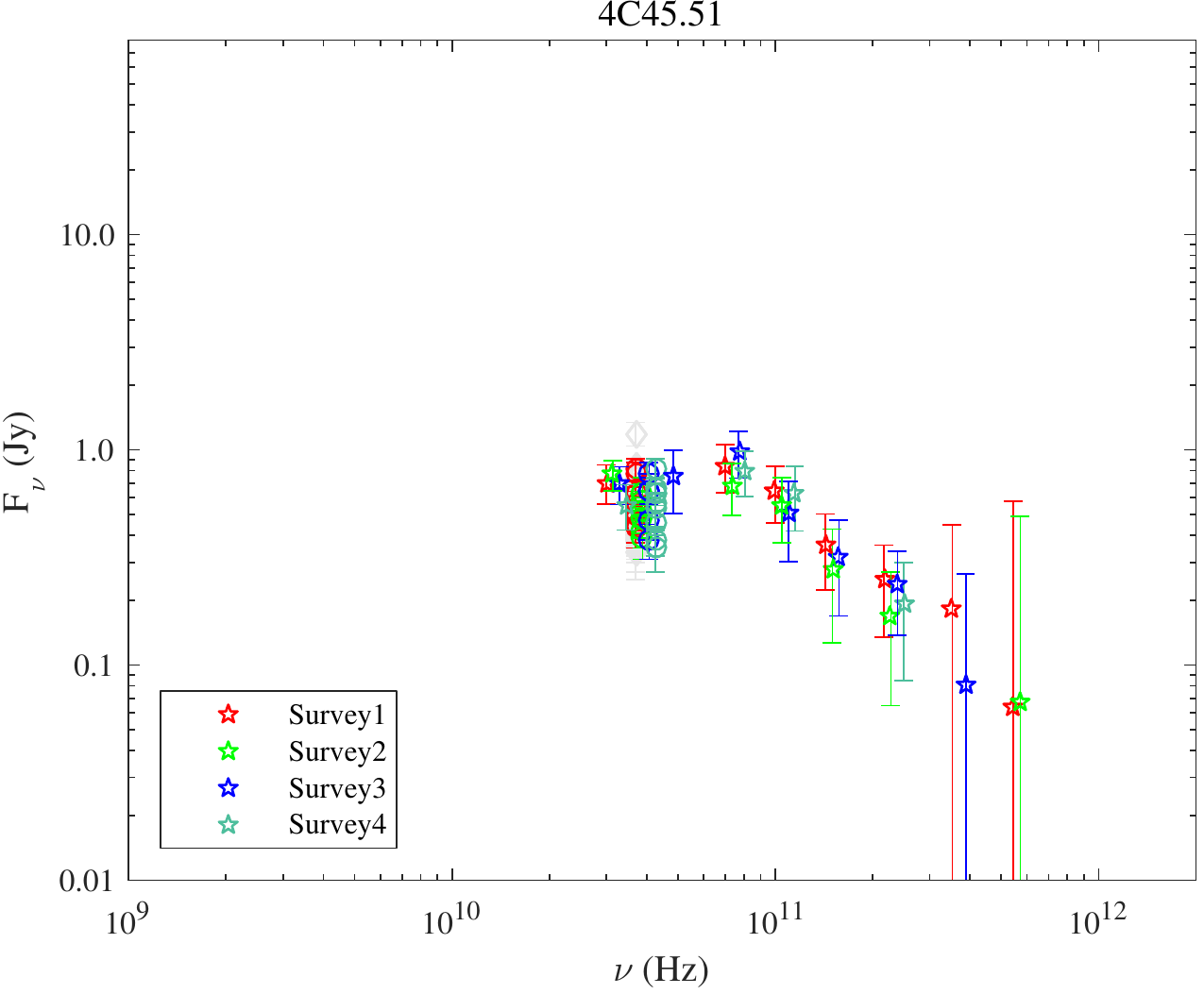}
	\caption{4C45.51.}
	\label{4C4551_spectra}
	\end{minipage}\qquad
	\begin{minipage}[b]{.47\textwidth}
	\includegraphics[width=\columnwidth]{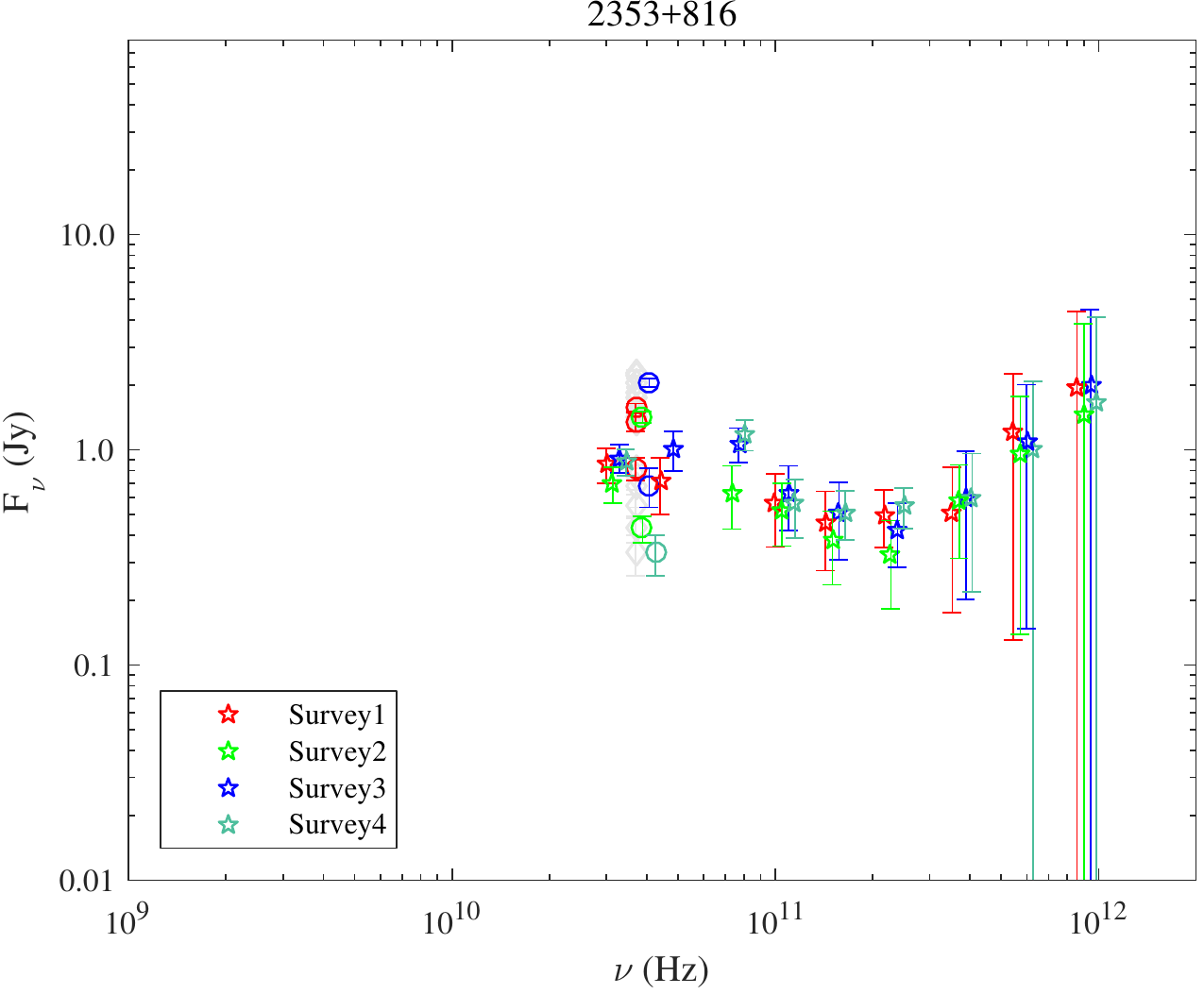}
	\caption{2353+816.}
	\label{2353+816_spectra}
	\end{minipage}
\end{figure*}

}


\begin{acknowledgements}

The Planck Collaboration acknowledges the support of: ESA; CNES, and CNRS/INSU-IN2P3-INP (France); ASI, CNR, and INAF (Italy); NASA and DoE (USA); STFC and UKSA (UK); CSIC, MINECO, JA, and RES (Spain); Tekes, AoF, and CSC (Finland); DLR and MPG (Germany); CSA (Canada); DTU Space (Denmark); SER/SSO (Switzerland); RCN (Norway); SFI (Ireland); FCT/MCTES (Portugal); ERC and PRACE (EU). A description of the Planck Collaboration and a list of its members, indicating which technical or scientific activities they have been involved in, can be found at \url{http://www.cosmos.esa.int/web/planck/planck-collaboration}. The Mets\"ahovi team acknowledges the support from the Academy of Finland to our observing projects (numbers 212656, 210338, 121148, and others). The Submillimeter Array is a joint project between the Smithsonian Astrophysical Observatory and the Academia Sinica Institute of Astronomy and Astrophysics and is funded by the Smithsonian Institution and the Academia Sinica. The OVRO 40-m monitoring programme is supported in part by NASA grants NNX08AW31G and NNX11A043G, and NSF grants AST-0808050 and AST-1109911. UMRAO has been supported by a series of grants from the NSF and NASA, and by the University of Michigan. We also acknowledge support through the Russian Government Programme of Competitive Growth of Kazan Federal University.
\end{acknowledgements}


\bibliographystyle{aa}
\bibliography{enbib_141016,Planck_bib}

\clearpage

\longtab{
\setcounter{table}{0}
\begin{table}[ht!]
\caption{Complete 1\,Jy northern AGN sample. The columns and source class types are defined in Sect.~\ref{section_sample}.}
\label{sample}
\vskip -6mm
\scriptsize
\setbox\tablebox=\vbox{
 \newdimen\digitwidth
 \setbox0=\hbox{\rm 0}
 \digitwidth=\wd0
 \catcode`*=\active
 \def*{\kern\digitwidth}
  \newdimen\signwidth
  \setbox0=\hbox{+}
  \signwidth=\wd0
  \catcode`!=\active
  \def!{\kern\signwidth}
\halign{\hbox to 2.4cm{#\leaderfil}\tabskip 1.0em&
    #\hfil\tabskip 1em&
    \hfil#\hfil&
    \hfil#\hfil&
    \hfil#\hfil&
    \hfil#\hfil\tabskip=0pt\cr
\noalign{\doubleline}
\omit\hfil Source\hfil&\omit\hfil Alias\hfil&Class&RA (J2000)&Dec (J2000)&$z$\cr
\noalign{\vskip 3pt\hrule\vskip 5pt}
0003$-$066& NRAO~5&        BLO& 00:06:13.90& $-$06:23:13.90& 0.347\cr
0007+106&   IIIZW2&        GAL& 00:10:31.01&   !10:58:31.01& 0.089\cr
0048$-$097& PKS~0048$-$097&BLO& 00:50:41.20& $-$09:29:41.20& 0.300\cr
0059+581&   TXS~0059+581&  QSO& 01:02:45.76&   !58:24:45.76& 0.644\cr
0106+013&   OC~012&        HPQ& 01:08:38.77&   !01:35:38.77& 2.099\cr
J0125$-$0005&&             QSO& 01:25:28.84& $-$00:05:28.84& 1.075\cr
0133+476&   DA~55&         HPQ& 01:36:58.59&   !47:51:58.59& 0.859\cr
0149+218&&                 HPQ& 01:52:18.06&   !22:07:18.06& 1.320\cr
0202+149&   4C~15.05&      HPQ& 02:04:50.41&   !15:14:50.41& 0.405\cr
0212+735&&                 BLO& 02:17:30.81&   !73:49:30.81& 2.367\cr
0224+671&&                 QSO& 02:28:50.05&   !67:21:50.05& 0.523\cr
0234+285&   4C~28.07&      HPQ& 02:37:52.41&   !28:48:52.41& 1.207\cr
0235+164&&                 BLO& 02:38:38.80&   !16:36:38.80& 0.940\cr
0238$-$084&&               GAL& 02:41:04.80& $-$08:15:04.80& 0.005\cr
0306+102&   PKS~0306+102&  BLO& 03:09:03.60&   !10:29:03.60& 0.863\cr
0316+413&   3C~84&         GAL& 03:19:48.16&   !41:30:48.16& 0.018\cr
0333+321&   NRAO~140&      HPQ& 03:36:30.11&   !32:18:30.11& 1.258\cr
0336$-$019& CTA~026&       HPQ& 03:39:30.94& $-$01:46:30.94& 0.852\cr
0355+508&   NRAO~150&      LPQ& 03:59:29.75&   !50:57:29.75& 1.510\cr
0415+379&   3C~111&        GAL& 04:18:21.28&   !38:01:21.28& 0.049\cr
0420$-$014& OA~129&        HPQ& 04:23:15.80& $-$01:20:15.80& 0.915\cr
0430+052&   3C~120&        GAL& 04:33:11.10&   !05:21:11.10& 0.033\cr
0446+112&   PKS~0446+112&  GAL& 04:49:07.67&   !11:21:07.67& 1.207\cr
0458$-$020& PKS~0458$-$020&HPQ& 05:01:12.81& $-$01:59:12.81& 2.286\cr
0507+179&&                 HPQ& 05:10:02.37&   !18:00:02.37& 0.416\cr
0528+134&   PKS~0528+134&  HPQ& 05:30:56.42&   !13:31:56.42& 2.070\cr
0552+398&   DA~193&        LPQ& 05:55:30.81&   !39:48:30.81& 2.363\cr
0605$-$085& PKS~0605$-$085&HPQ& 06:07:59.70& $-$08:34:59.70& 0.872\cr
0642+449&   OH~471&        LPQ& 06:46:32.03&   !44:51:32.03& 3.408\cr
0716+714&&                 BLO& 07:21:53.30&   !71:20:53.30& 0.300\cr
0723$-$008& PKS~0723$-$008&BLO& 07:25:50.70& $-$00:54:50.70& 0.127\cr
0735+178&   PKS~0735+178&  BLO& 07:38:07.40&   !17:42:07.40& 0.424\cr
0736+017&&                 HPQ& 07:39:18.03&   !01:37:18.03& 0.191\cr
0748+126&&                 LPQ& 07:50:52.05&   !12:31:52.05& 0.889\cr
0754+100&   OI~090.4&      BLO& 07:57:06.64&   !09:56:06.64& 0.266\cr
0805$-$077&&               QSO& 08:08:15.54& $-$07:51:15.54& 1.837\cr
0804+499&&                 HPQ& 08:08:39.67&   !49:50:39.67& 1.432\cr
0823+033&   PKS~0823+033&  BLO& 08:25:50.30&   !03:09:50.30& 0.506\cr
0827+243&   OJ~248&        LPQ& 08:30:52.08&   !24:10:52.08& 0.941\cr
0836+710&   4C~71.07&      LPQ& 08:41:24.37&   !70:53:24.37& 2.172\cr
0851+202&   OJ~287&        BLO& 08:54:48.80&   !20:06:48.80& 0.306\cr
0906+430&   3C~216&        HPQ& 09:09:33.50&   !42:53:33.50& 0.670\cr
0917+449&&                 QSO& 09:20:58.46&   !44:41:58.46& 2.188\cr
0923+392&   4C~39.25&      LPQ& 09:27:03.01&   !39:02:03.01& 0.698\cr
0945+408&   4C~40.24&      LPQ& 09:48:55.34&   !40:39:55.34& 1.252\cr
0953+254&&                 LPQ& 09:56:49.88&   !25:15:49.88& 0.712\cr
0954+658&   S4~0954+65&    BLO& 09:58:47.20&   !65:33:47.20& 0.367\cr
1036+054&&                 QSO& 10:38:46.78&   !05:12:46.78& 0.473\cr
TEX~1040+244&&             BLO& 10:43:09.00&   !24:08:09.00& 0.560\cr
1055+018&   OL~093&        HPQ& 10:58:29.61&   !01:33:29.61& 0.888\cr
J1130+3815& QSO~B1128+385& QSO& 11:30:53.28&   !38:15:53.28& 1.741\cr
1150+812&&                 QSO& 11:53:12.50&   !80:58:12.50& 1.250\cr
1150+497&&                 HPQ& 11:53:24.47&   !49:31:24.47& 0.334\cr
1156+295&   4C~29.45&      HPQ& 11:59:31.83&   !29:14:31.83& 0.729\cr
1219+044&&                 QSO& 12:22:22.55&   !04:13:22.55& 0.965\cr
1222+216&   PKS~1222+216&  HPQ& 12:24:54.51&   !21:22:54.51& 0.435\cr
1226+023&   3C~273&        HPQ& 12:29:06.69&   !02:03:06.69& 0.158\cr
1228+126&   3C~274&        GAL& 12:30:49.42&   !12:23:49.42& 0.004\cr
1253$-$055& 3C~279&        HPQ& 12:56:11.17& $-$05:47:11.17& 0.536\cr
1308+326&   AUCVn&         BLO& 13:10:28.66&   !32:20:28.66& 0.997\cr
1324+224&&                 QSO& 13:27:00.86&   !22:10:00.86& 1.400\cr
1413+135&&                 BLO& 14:15:58.80&   !13:20:58.80& 0.247\cr
1418+546&   OQ~530&        BLO& 14:19:56.60&   !54:23:56.60& 0.151\cr
1502+106&   OR~103&        HPQ& 15:04:24.98&   !10:29:24.98& 1.839\cr
1510$-$089& PKS~1510$-$089&HPQ& 15:12:50.53& $-$09:05:50.53& 0.360\cr
1546+027&&                 HPQ& 15:49:29.44&   !02:37:29.44& 0.414\cr
1548+056&&                 HPQ& 15:50:35.27&   !05:27:35.27& 1.417\cr
1606+106&   4C~10.45&      LPQ& 16:08:46.20&   !10:29:46.20& 1.226\cr
1611+343&   DA~406&        HPQ& 16:13:41.00&   !34:12:41.00& 1.401\cr
1633+382&   4C~38.41&      HPQ& 16:35:15.49&   !38:08:15.49& 1.814\cr
1638+398&&                 QSO& 16:40:29.63&   !39:46:29.63& 1.660\cr
1642+690&&                 HPQ& 16:42:07.85&   !68:56:07.85& 0.751\cr
1641+399&   3C~345&        HPQ& 16:42:58.81&   !39:48:58.81& 0.593\cr
1652+398&   MARK~501&      BLO& 16:53:52.20&   !39:45:52.20& 0.034\cr
1739+522&   S4~1739+52&    HPQ& 17:40:36.98&   !52:11:36.98& 1.375\cr
1741$-$038& PKS~1741$-$038&HPQ& 17:43:58.86& $-$03:50:58.86& 1.054\cr
1749+096&   PKS~1749+096&  BLO& 17:51:32.70&   !09:39:32.70& 0.322\cr
1803+784&   S5~1803+784&   BLO& 18:00:45.40&   !78:28:45.40& 0.684\cr
1807+698&   3C~371.0&      BLO& 18:06:50.70&   !69:49:50.70& 0.051\cr
1823+568&   4C~56.27&      BLO& 18:24:07.07&   !56:51:07.07& 0.663\cr
1828+487&   3C~380&        HPQ& 18:29:31.80&   !48:44:31.80& 0.692\cr
J184915+67064&&            UNK& 18:49:15.89&   !67:06:15.89& 0.657\cr
1928+738&   4C~73.18&      HPQ& 19:27:48.50&   !73:58:48.50& 0.302\cr
\noalign{\vskip 3pt\hrule\vskip 5pt}}}
\endPlancktable
\end{table}

\addtocounter{table}{-1}
\begin{table}[ht!]
\caption{Continued.}
\label{sample2}
\vskip -6mm
\scriptsize
\setbox\tablebox=\vbox{
 \newdimen\digitwidth
 \setbox0=\hbox{\rm 0}
 \digitwidth=\wd0
 \catcode`*=\active
 \def*{\kern\digitwidth}
  \newdimen\signwidth
  \setbox0=\hbox{+}
  \signwidth=\wd0
  \catcode`!=\active
  \def!{\kern\signwidth}
\halign{\hbox to 2.4cm{#\leaderfil}\tabskip 1.0em&
    #\hfil\tabskip 1em&
    \hfil#\hfil&
    \hfil#\hfil&
    \hfil#\hfil&
    \hfil#\hfil\tabskip=0pt\cr
\noalign{\doubleline}
\omit\hfil Source\hfil&\omit\hfil Alias\hfil&Class&RA (J2000)&Dec (J2000)&$z$\cr
\noalign{\vskip 3pt\hrule\vskip 5pt}
1954+513&&                 LPQ& 19:55:42.74&   !51:31:42.74& 1.223\cr
2007+776&   S5~2007+77&    BLO& 20:05:31.10&   !77:52:31.10& 0.342\cr
2005+403&&                 QSO& 20:07:44.94&   !40:29:44.94& 1.736\cr
2021+614&   OW~637&        GAL& 20:22:06.68&   !61:36:06.68& 0.227\cr
2037+511&&                 QSO& 20:38:37.04&   !51:19:37.04& 1.686\cr
2121+053&&                 HPQ& 21:23:44.52&   !05:35:44.52& 1.941\cr
2131$-$021& PKS~2131$-$021&HPQ& 21:34:10.31& $-$01:53:10.31& 1.285\cr
2134+004&   OX~057&        LPQ& 21:36:38.59&   !00:41:38.59& 1.932\cr
2136+141&&                 LPQ& 21:39:01.31&   !14:23:01.31& 2.427\cr
2145+067&&                 LPQ& 21:48:05.46&   !06:57:05.46& 0.990\cr
2200+420&   BL~LAC&        BLO& 22:02:43.30&   !42:16:43.30& 0.069\cr
2201+315&   4C~31.63&      HPQ& 22:03:14.98&   !31:45:14.98& 0.298\cr
2201+171&&                 QSO& 22:03:26.89&   !17:25:26.89& 1.075\cr
2216$-$038&&               QSO& 22:18:52.04& $-$03:35:52.04& 0.901\cr
2223$-$052& 3C~446&        BLO& 22:25:47.26& $-$04:57:47.26& 1.404\cr
2227$-$088&&               HPQ& 22:29:40.09& $-$08:32:40.09& 1.562\cr
2230+114&   CTA~102&       HPQ& 22:32:36.41&   !11:43:36.41& 1.037\cr
2234+282&&                 HPQ& 22:36:22.47&   !28:28:22.47& 0.795\cr
2251+158&   3C~454.3&      HPQ& 22:53:57.75&   !16:08:57.75& 0.859\cr
4C~45.51&&                 QSO& 23:54:21.68&   !45:53:21.68& 1.986\cr
2353+816&&                 QSO& 23:56:22.79&   !81:52:22.79& 1.340\cr
\noalign{\vskip 3pt\hrule\vskip 5pt}}}
\endPlancktable
\end{table}

\setcounter{table}{2}
\begin{table*}[ht!]
\caption{Start and end times [yyyy-mm-dd] for the \Planck\ Surveys, calculated with {\tt POFF}.}
\label{obstime}\vskip -6mm
\scriptsize
\setbox\tablebox=\vbox{
 \newdimen\digitwidth
 \setbox0=\hbox{\rm 0}
 \digitwidth=\wd0
 \catcode`*=\active
 \def*{\kern\digitwidth}
  \newdimen\signwidth
  \setbox0=\hbox{+}
  \signwidth=\wd0
  \catcode`!=\active
  \def!{\kern\signwidth}
\halign{\hbox to 1.8cm{#\leaderfil}\tabskip 2.0em&
    \hfil#\hfil\tabskip 1.0em&
    \hfil#\hfil\tabskip 2.0em&
    \hfil#\hfil\tabskip 1.0em&
    \hfil#\hfil\tabskip 2.0em&
    \hfil#\hfil\tabskip 1.0em&
    \hfil#\hfil\tabskip 2.0em&
    \hfil#\hfil\tabskip 1.0em&
    \hfil#\hfil\tabskip 0em\cr
\noalign{\doubleline}
\omit&\multispan2\hfil Survey 1\hfil&\multispan2\hfil Survey 2\hfil&\multispan2\hfil Survey 3\hfil& 
             \multispan2\hfil Survey 4\hfil\cr
\noalign{\vskip -3pt}
\omit&\multispan2\hrulefill&\multispan2\hrulefill&\multispan2\hrulefill&\multispan2\hrulefill\cr
\omit\hfil Class\hfil&Start&End&Start&End&Start&End&Start&End\cr
\noalign{\vskip 3pt\hrule\vskip 4pt}
  0003$-$066& 2009-12-12& 2009-12-21& 2010-06-25& 2010-07-04& 2010-12-12& 2010-12-21& 2011-06-25& 2011-07-04\cr
    0007+106& 2009-12-25& 2010-01-02& 2010-07-04& 2010-07-13& 2010-12-25& 2011-01-02& 2011-07-04& 2011-07-13\cr
  0048$-$097& 2009-12-23& 2010-01-01& 2010-07-08& 2010-07-16& 2010-12-23& 2011-01-01& 2011-07-08& 2011-07-16\cr
    0059+581& 2009-08-17& 2009-08-28& 2010-01-28& 2010-02-05& 2010-08-18& 2010-08-28& 2011-01-28& 2011-02-05\cr
    0106+013& 2010-01-04& 2010-01-12& 2010-07-17& 2010-07-24& 2011-01-04& 2011-01-12& 2011-07-17& 2011-07-24\cr
J0125$-$0005& 2010-01-08& 2010-01-16& 2010-07-21& 2010-07-28& 2011-01-08& 2011-01-16& 2011-07-21& 2011-07-28\cr
    0133+476& 2009-08-14& 2009-08-23& 2010-01-29& 2010-02-05& 2010-08-15& 2010-08-23& 2011-01-29& 2011-02-05\cr
    0149+218& 2010-01-24& 2010-01-30& 2010-08-05& 2010-08-11& 2011-01-24& 2011-01-30& 2011-08-04& 2011-08-11\cr
    0202+149& 2010-01-24& 2010-01-31& 2010-08-05& 2010-08-11& 2011-01-24& 2011-01-31& 2011-08-05& 2011-08-11\cr
    0212+735& 2009-09-11& 2009-09-21& 2010-02-11& 2010-02-20& 2010-09-11& 2010-09-21& 2011-02-11& 2011-02-20\cr
    0224+671& 2009-09-06& 2009-09-15& 2010-02-10& 2010-02-18& 2010-09-06& 2010-09-15& 2011-02-10& 2011-02-18\cr
    0234+285& 2009-08-16& 2009-08-23& 2010-02-03& 2010-02-09& 2010-08-17& 2010-08-23& 2011-02-03& 2011-02-09\cr
    0235+164& 2009-08-12& 2009-08-17& 2010-02-01& 2010-02-07& 2010-08-12& 2010-08-19& 2011-02-01& 2011-02-07\cr
  0238$-$084& 2010-01-24& 2010-02-01& 2010-08-05& 2010-08-11& 2011-01-24& 2011-02-01& 2011-08-05& 2011-08-11\cr
    0306+102& 2009-08-15& 2009-08-22& 2010-02-06& 2010-02-12& 2010-08-17& 2010-08-22& 2011-02-06& 2011-02-12\cr
    0316+413& 2009-08-29& 2009-09-04& 2010-02-13& 2010-02-19& 2010-08-29& 2010-09-04& 2011-02-13& 2011-02-19\cr
    0333+321& 2009-08-28& 2009-09-03& 2010-02-15& 2010-02-20& 2010-08-28& 2010-09-03& 2011-02-15& 2011-02-20\cr
  0336$-$019& 2009-08-18& 2009-08-25& 2010-02-10& 2010-02-17& 2010-08-19& 2010-08-25& 2011-02-10& 2011-02-17\cr
    0355+508& 2009-09-07& 2009-09-14& 2010-02-19& 2010-02-26& 2010-09-07& 2010-09-14& 2011-02-19& 2011-02-26\cr
    0415+379& 2009-09-06& 2009-09-12& 2010-02-22& 2010-02-28& 2010-09-06& 2010-09-12& 2011-02-22& 2011-02-28\cr
  0420$-$014& 2009-08-28& 2009-09-03& 2010-02-21& 2010-02-27& 2010-08-28& 2010-09-03& 2011-02-21& 2011-02-27\cr
    0430+052& 2009-09-01& 2009-09-07& 2010-02-24& 2010-03-01& 2010-09-01& 2010-09-07& 2011-02-24& 2011-03-01\cr
    0446+112& 2009-09-05& 2009-09-11& 2010-02-27& 2010-03-05& 2010-09-05& 2010-09-11& 2011-02-27& 2011-03-03\cr
  0458$-$020& 2009-09-05& 2009-09-11& 2010-03-01& 2010-03-07& 2010-09-05& 2010-09-11& 2011-03-01& 2011-03-05\cr
    0507+179& 2009-09-11& 2009-09-16& 2010-03-03& 2010-03-09& 2010-09-11& 2010-09-16& 2011-03-02& 2011-03-07\cr
    0528+134& 2009-09-14& 2009-09-20& 2010-03-07& 2010-03-13& 2010-09-14& 2010-09-20& 2011-03-05& 2011-03-11\cr
    0552+398& 2009-09-22& 2009-09-28& 2010-03-09& 2010-03-15& 2010-09-22& 2010-09-28& 2011-03-07& 2011-03-14\cr
  0605$-$085& 2009-09-19& 2009-09-26& 2010-03-17& 2010-03-23& 2010-09-19& 2010-09-26& 2011-03-16& 2011-03-23\cr
    0642+449& 2009-09-30& 2009-10-06& 2010-03-16& 2010-03-22& 2010-09-30& 2010-10-06& 2011-03-15& 2011-03-22\cr
    0716+714& 2009-10-04& 2009-10-12& 2010-03-10& 2010-03-18& 2010-10-04& 2010-10-12& 2011-03-08& 2011-03-18\cr
  0723$-$008& 2009-10-09& 2009-10-16& 2010-04-02& 2010-04-08& 2010-10-09& 2010-10-16& 2011-04-01& 2011-04-08\cr
    0735+178& 2009-10-10& 2009-10-17& 2010-03-31& 2010-04-06& 2010-10-10& 2010-10-17& 2011-03-31& 2011-04-06\cr
    0736+017& 2009-10-12& 2009-10-19& 2010-04-04& 2010-04-10& 2010-10-12& 2010-10-19& 2011-04-04& 2011-04-10\cr
    0748+126& 2009-10-14& 2009-10-20& 2010-04-04& 2010-04-10& 2010-10-14& 2010-10-20& 2011-04-04& 2011-04-10\cr
    0754+100& 2009-10-15& 2009-10-22& 2010-04-06& 2010-04-12& 2010-10-15& 2010-10-22& 2011-04-06& 2011-04-12\cr
  0805$-$077& 2009-10-22& 2009-10-31& 2010-04-12& 2010-04-19& 2010-10-22& 2010-10-31& 2011-04-12& 2011-04-19\cr
    0804+499& 2009-10-12& 2009-10-19& 2010-03-27& 2010-04-03& 2010-10-12& 2010-10-19& 2011-03-27& 2011-04-03\cr
    0823+033& 2009-10-24& 2009-11-01& 2010-04-14& 2010-04-20& 2010-10-24& 2010-11-01& 2011-04-14& 2011-04-20\cr
    0827+243& 2009-10-21& 2009-10-27& 2010-04-09& 2010-04-16& 2010-10-21& 2010-10-27& 2011-04-09& 2011-04-16\cr
    0836+710& 2009-10-12& 2009-10-19& 2010-03-17& 2010-03-27& 2010-10-12& 2010-10-19& 2011-03-16& 2011-03-27\cr
    0851+202& 2009-10-27& 2009-11-03& 2010-04-15& 2010-04-22& 2010-10-27& 2010-11-02& 2011-04-15& 2011-04-22\cr
    0906+430& 2009-10-23& 2009-10-30& 2010-04-09& 2010-04-17& 2010-10-23& 2010-10-30& 2011-04-09& 2011-04-17\cr
    0917+449& 2009-10-25& 2009-11-01& 2010-04-11& 2010-04-19& 2010-10-25& 2010-11-01& 2011-04-11& 2011-04-19\cr
    0923+392& 2009-10-28& 2009-11-03& 2010-04-15& 2010-04-23& 2010-10-28& 2010-11-03& 2011-04-15& 2011-04-23\cr
    0945+408& 2009-10-31& 2009-11-07& 2010-04-18& 2010-04-27& 2010-10-31& 2010-11-07& 2011-04-18& 2011-04-27\cr
    0953+254& 2009-11-08& 2009-11-15& 2010-04-27& 2010-05-06& 2010-11-08& 2010-11-15& 2011-04-27& 2011-05-06\cr
    0954+658& 2009-10-21& 2009-10-29& 2010-03-30& 2010-04-11& 2010-10-21& 2010-10-29& 2011-03-30& 2011-04-11\cr
    1036+054& 2009-11-29& 2009-12-08& 2010-05-16& 2010-05-25& 2010-11-29& 2010-12-08& 2011-05-16& 2011-05-25\cr
 TEX1040+244& 2009-11-19& 2009-11-27& 2010-05-10& 2010-05-19& 2010-11-19& 2010-11-27& 2011-05-10& 2011-05-19\cr
    1055+018& 2009-12-07& 2009-12-16& 2010-05-24& 2010-06-02& 2010-12-07& 2010-12-16& 2011-05-24& 2011-06-02\cr
  J1130+3815& 2009-09-05& 2009-09-11& 2010-03-18& 2010-03-24& 2010-09-05& 2010-09-11& 2011-03-17& 2011-03-24\cr
    1150+812& 2009-10-16& 2009-10-26& 2010-03-07& 2010-03-22& 2010-10-16& 2010-10-26& 2011-03-05& 2011-03-22\cr
    1150+497& 2009-11-16& 2009-11-26& 2010-05-11& 2010-05-26& 2010-11-16& 2010-11-26& 2011-05-11& 2011-05-26\cr
    1156+295& 2009-12-04& 2009-12-14& 2010-05-31& 2010-06-11& 2010-12-04& 2010-12-14& 2011-05-31& 2011-06-11\cr
    1219+044& 2009-12-29& 2010-01-07& 2010-06-20& 2010-06-29& 2010-12-29& 2011-01-07& 2011-06-20& 2011-06-29\cr
    1222+216& 2009-12-18& 2009-12-28& 2010-06-13& 2010-06-23& 2010-12-18& 2010-12-28& 2011-06-13& 2011-06-23\cr
    1226+023& 2010-01-02& 2010-01-10& 2010-06-23& 2010-07-02& 2011-01-02& 2011-01-10& 2011-06-23& 2011-07-02\cr
    1228+126& 2009-12-26& 2010-01-04& 2010-06-19& 2010-06-29& 2010-12-26& 2011-01-04& 2011-06-19& 2011-06-29\cr
  1253$-$055& 2010-01-13& 2010-01-20& 2010-07-05& 2010-07-13& 2011-01-13& 2011-01-20& 2011-07-05& 2011-07-13\cr
    1308+326& 2009-12-21& 2010-01-03& 2010-06-21& 2010-07-01& 2010-12-21& 2011-01-03& 2011-06-21& 2011-07-01\cr
    1324+224& 2010-01-05& 2010-01-15& 2010-07-02& 2010-07-11& 2011-01-05& 2011-01-15& 2011-07-02& 2011-07-11\cr
    1413+135& 2010-01-25& 2010-02-02& 2010-07-19& 2010-07-26& 2011-01-25& 2011-02-02& 2011-07-19& 2011-07-26\cr
    1418+546& 2009-12-06& 2009-12-28& 2010-06-18& 2010-07-03& 2010-12-06& 2010-12-27& 2011-06-18& 2011-07-03\cr
    1502+106& 2010-02-08& 2010-02-15& 2010-08-01& 2010-08-07& 2011-02-08& 2011-02-15& 2011-08-01& 2011-08-07\cr
  1510$-$089& 2009-08-12& 2009-08-14& 2010-02-14& 2010-02-20& 2010-08-09& 2010-08-16& 2011-02-14& 2011-02-20\cr
    1546+027& 2009-08-12& 2009-08-18& 2010-02-21& 2010-02-27& 2010-08-14& 2010-08-20& 2011-02-21& 2011-02-27\cr
    1548+056& 2009-08-12& 2009-08-18& 2010-02-21& 2010-02-27& 2010-08-13& 2010-08-19& 2011-02-20& 2011-02-27\cr
    1606+106& 2009-08-14& 2009-08-21& 2010-02-24& 2010-03-03& 2010-08-15& 2010-08-21& 2011-02-24& 2011-03-02\cr
    1611+343& 2010-02-23& 2010-03-06& 2010-08-05& 2010-08-13& 2011-02-23& 2011-03-04& 2011-08-05& 2011-08-13\cr
    1633+382& 2009-08-12& 2009-08-14& 2010-03-02& 2010-03-14& 2010-08-07& 2010-08-16& 2011-03-02& 2011-03-12\cr
    1638+398& 2009-08-12& 2009-08-14& 2010-03-04& 2010-03-16& 2010-08-07& 2010-08-16& 2011-03-03& 2011-03-15\cr
    1642+690& 2009-11-06& 2009-11-29& 2010-01-17& 2010-02-27& 2010-04-08& 2010-05-23& 2010-06-05& 2010-07-05\cr
    1641+399& 2009-08-12& 2009-08-15& 2010-03-05& 2010-03-17& 2010-08-07& 2010-08-17& 2011-03-04& 2011-03-17\cr
    1652+398& 2009-08-12& 2009-08-17& 2010-03-09& 2010-03-21& 2010-08-10& 2010-08-20& 2011-03-07& 2011-03-21\cr
    1739+522& 2009-08-12& 2009-08-18& 2010-03-30& 2010-04-15& 2010-08-06& 2010-08-21& 2011-03-30& 2011-04-15\cr
  1741$-$038& 2009-09-08& 2009-09-14& 2010-03-18& 2010-03-23& 2010-09-08& 2010-09-14& 2011-03-17& 2011-03-23\cr
    1749+096& 2009-09-07& 2009-09-14& 2010-03-21& 2010-03-27& 2010-09-07& 2010-09-14& 2011-03-21& 2011-03-27\cr
    1803+784& 2009-10-11& 2009-10-27& 2010-02-02& 2010-02-19& 2010-10-11& 2010-10-27& 2011-02-02& 2011-02-19\cr
    1807+698& 2009-10-21& 2009-11-15& 2010-01-07& 2010-01-31& 2010-05-11& 2010-06-05& 2010-06-16& 2010-07-12\cr
    1823+568& 2009-08-12& 2009-08-29& 2010-04-17& 2010-05-04& 2010-08-05& 2010-08-29& 2011-04-17& 2011-05-04\cr
    1828+487& 2009-08-25& 2009-09-12& 2010-04-11& 2010-04-24& 2010-08-25& 2010-09-12& 2011-04-11& 2011-04-24\cr
 J184915+670& 2009-10-12& 2009-11-14& 2009-12-25& 2010-01-19& 2010-05-12& 2010-06-10& 2010-06-26& 2010-07-31\cr
    1928+738& 2009-09-28& 2009-10-20& 2010-01-18& 2010-02-03& 2010-09-28& 2010-10-20& 2011-01-18& 2011-02-03\cr
    1954+513& 2009-10-09& 2009-12-06& 2010-05-06& 2010-05-20& 2010-10-09& 2010-12-06& 2011-05-06& 2011-05-20\cr
\noalign{\vskip 4pt\hrule\vskip 5pt}}}
\endPlancktablewide
\end{table*}

\addtocounter{table}{-1}
\begin{table*}[ht!]
\caption{Continued.}
\label{obstime}\vskip -6mm
\scriptsize
\setbox\tablebox=\vbox{
 \newdimen\digitwidth
 \setbox0=\hbox{\rm 0}
 \digitwidth=\wd0
 \catcode`*=\active
 \def*{\kern\digitwidth}
  \newdimen\signwidth
  \setbox0=\hbox{+}
  \signwidth=\wd0
  \catcode`!=\active
  \def!{\kern\signwidth}
\halign{\hbox to 1.8cm{#\leaderfil}\tabskip 2.0em&
    \hfil#\hfil\tabskip 1.0em&
    \hfil#\hfil\tabskip 2.0em&
    \hfil#\hfil\tabskip 1.0em&
    \hfil#\hfil\tabskip 2.0em&
    \hfil#\hfil\tabskip 1.0em&
    \hfil#\hfil\tabskip 2.0em&
    \hfil#\hfil\tabskip 1.0em&
    \hfil#\hfil\tabskip 0em\cr
\noalign{\doubleline}
\omit&\multispan2\hfil Survey 1\hfil&\multispan2\hfil Survey 2\hfil&\multispan2\hfil Survey 3\hfil& 
             \multispan2\hfil Survey 4\hfil\cr
\noalign{\vskip -3pt}
\omit&\multispan2\hrulefill&\multispan2\hrulefill&\multispan2\hrulefill&\multispan2\hrulefill\cr
\omit\hfil Class\hfil&Start&End&Start&End&Start&End&Start&End\cr
\noalign{\vskip 3pt\hrule\vskip 4pt}

    2007+776& 2009-09-25& 2009-10-12& 2010-01-27& 2010-02-10& 2010-09-25& 2010-10-12& 2011-01-27& 2011-02-10\cr
    2005+403& 2009-10-17& 2009-11-09& 2010-05-02& 2010-05-12& 2010-10-17& 2010-11-09& 2011-05-02& 2011-05-12\cr
    2021+614& 2009-08-15& 2009-08-21& 2009-12-20& 2010-01-08& 2010-05-26& 2010-07-12& 2010-07-28& 2010-08-22\cr
    2037+511& 2009-11-25& 2009-12-19& 2010-05-18& 2010-06-03& 2010-11-25& 2010-12-19& 2011-05-18& 2011-06-03\cr
    2121+053& 2009-11-02& 2009-11-11& 2010-05-08& 2010-05-16& 2010-11-02& 2010-11-11& 2011-05-08& 2011-05-16\cr
  2131$-$021& 2009-11-02& 2009-11-10& 2010-05-09& 2010-05-17& 2010-11-02& 2010-11-10& 2011-05-09& 2011-05-17\cr
    2134+004& 2009-11-04& 2009-11-12& 2010-05-10& 2010-05-19& 2010-11-04& 2010-11-12& 2011-05-10& 2011-05-19\cr
    2136+141& 2009-11-11& 2009-11-21& 2010-05-16& 2010-05-25& 2010-11-11& 2010-11-21& 2011-05-16& 2011-05-25\cr
    2145+067& 2009-11-10& 2009-11-19& 2010-05-16& 2010-05-24& 2010-11-10& 2010-11-19& 2011-05-16& 2011-05-24\cr
    2200+420& 2009-12-13& 2009-12-26& 2010-06-06& 2010-06-22& 2010-12-13& 2010-12-26& 2011-06-06& 2011-06-22\cr
    2201+315& 2009-12-02& 2009-12-15& 2010-05-31& 2010-06-12& 2010-12-02& 2010-12-15& 2011-05-30& 2011-06-12\cr
    2201+171& 2009-11-21& 2009-12-02& 2010-05-24& 2010-06-03& 2010-11-21& 2010-12-02& 2011-05-24& 2011-06-03\cr
  2216$-$038& 2009-11-14& 2009-11-22& 2010-05-21& 2010-05-31& 2010-11-14& 2010-11-22& 2011-05-21& 2011-05-31\cr
  2223$-$052& 2009-11-15& 2009-11-23& 2010-05-23& 2010-06-01& 2010-11-15& 2010-11-23& 2011-05-23& 2011-06-01\cr
  2227$-$088& 2009-11-14& 2009-11-22& 2010-05-23& 2010-06-01& 2010-11-14& 2010-11-22& 2011-05-23& 2011-06-01\cr
    2230+114& 2009-11-27& 2009-12-07& 2010-05-31& 2010-06-10& 2010-11-27& 2010-12-07& 2011-05-31& 2011-06-10\cr
    2234+282& 2009-12-10& 2009-12-21& 2010-06-09& 2010-06-21& 2010-12-10& 2010-12-21& 2011-06-09& 2011-06-21\cr
    2251+158& 2009-12-06& 2009-12-16& 2010-06-10& 2010-06-20& 2010-12-06& 2010-12-16& 2011-06-10& 2011-06-20\cr
   4C\,45.51& 2010-01-11& 2010-01-19& 2010-07-20& 2010-08-02& 2011-01-11& 2011-01-19& 2011-07-20& 2011-08-02\cr
    2353+816& 2009-09-19& 2009-10-01& 2010-02-08& 2010-02-19& 2010-09-19& 2010-10-01& 2011-02-08& 2011-02-19\cr
\noalign{\vskip 4pt\hrule\vskip 5pt}}}
\endPlancktablewide
\end{table*}

\setcounter{table}{3}
\begin{table*}
\caption{Number of \Planck\ pointings for each source, Survey, and frequency. Because the visibility periods calculated with {\tt POFF} may overlap in time, for some single-pointing sources there can be zero to three pointings per Survey.}          
\label{pointings}
\vskip -6mm
\scriptsize
\setbox\tablebox=\vbox{
\newdimen\digitwidth
\setbox0=\hbox{\rm 0}
\digitwidth=\wd0
\catcode`*=\active
\def*{\kern\digitwidth}
\newdimen\signwidth
\setbox0=\hbox{+}
\signwidth=\wd0
\catcode`!=\active
\def!{\kern\signwidth}
\newdimen\decimalwidth
\setbox0=\hbox{.}
\decimalwidth=\wd0
\catcode`@=\active
\def@{\kern\decimalwidth}
\halign{\hbox to 2.2cm{#\leaderfil}\tabskip 1.0em&
    \hfil#\hfil\tabskip 0.3em&
    \hfil#\hfil&
    \hfil#\hfil&
    \hfil#\hfil\tabskip1.0em&
    \hfil#\hfil\tabskip 0.3em&
    \hfil#\hfil&
    \hfil#\hfil&
    \hfil#\hfil\tabskip1.0em&
    \hfil#\hfil\tabskip 0.3em&
    \hfil#\hfil&
    \hfil#\hfil&
    \hfil#\hfil\tabskip1.0em&
    \hfil#\hfil\tabskip 0.3em&
    \hfil#\hfil&
    \hfil#\hfil&
    \hfil#\hfil\tabskip1.0em&
    \hfil#\hfil\tabskip 0.3em&
    \hfil#\hfil&
    \hfil#\hfil&
    \hfil#\hfil\tabskip1.0em&
    \hfil#\hfil\tabskip 0.3em&
    \hfil#\hfil&
    \hfil#\hfil&
    \hfil#\hfil\tabskip1.0em&
    \hfil#\hfil\tabskip 0.3em&
    \hfil#\hfil&
    \hfil#\hfil&
    \hfil#\hfil\tabskip1.0em&
    \hfil#\hfil\tabskip 0.3em&
    \hfil#\hfil&
    \hfil#\hfil&
    \hfil#\hfil\tabskip1.0em&
    \hfil#\hfil\tabskip 0.3em&
    \hfil#\hfil&
    \hfil#\hfil&
    \hfil#\hfil\tabskip 0em\cr
\noalign{\doubleline}
\omit&\multispan4\hfil 30\,GHz\hfil&\multispan4\hfil 44\,GHz\hfil&\multispan4\hfil 70\,GHz\hfil&\multispan4\hfil 100\,GHz\hfil&
      \multispan4\hfil 143\,GHz\hfil&\multispan4\hfil 217\,GHz\hfil&\multispan4\hfil 353\,GHz\hfil&\multispan4\hfil 545\,GHz\hfil&
      \multispan4\hfil 857\,GHz\hfil\cr
\noalign{\vskip -3pt}
\omit&\multispan4\hrulefill&\multispan4\hrulefill&\multispan4\hrulefill&\multispan4\hrulefill&\multispan4\hrulefill&\multispan4\hrulefill&
      \multispan4\hrulefill&\multispan4\hrulefill&\multispan4\hrulefill\cr
\omit\hfil Source\hfil& S1& S2& S3& S4& S1& S2& S3& S4& S1& S2& S3& S4& S1& S2& S3& S4& S1& S2& S3& S4& S1& S2& S3& S4& S1& S2& S3& S4& 
                        S1& S2& S3& S4& S1& S2& S3& S4\cr
\noalign{\vskip 3pt\hrule\vskip 4pt}
   0003$-$066& 1& 1& 1& 1& 1& 1& 1& 1& 1& 1& 1& 1& 1& 1& 1& 1& 1& 1& 1& 1& 1& 1& 1& 1& 1& 1& 1& 1& 1& 1& 1& 1& 1& 1& 1& 1\cr
    0007+106& 1& 1& 1& 1& 1& 1& 1& 1& 1& 1& 1& 1& 1& 1& 1& 1& 1& 1& 1& 1& 1& 1& 1& 1& 1& 1& 1& 1& 1& 1& 1& 1& 1& 1& 1& 1\cr
    0048$-$097& 1& 1& 1& 1& 1& 1& 1& 1& 1& 1& 1& 1& 1& 1& 1& 1& 1& 1& 1& 1& 1& 1& 1& 1& 1& 1& 1& 1& 1& 1& 1& 1& 1& 1& 1& 1\cr
    0059+581& 2& 0& 2& 0& 2& 0& 2& 0& 2& 0& 2& 0& 2& 0& 2& 0& 2& 0& 2& 0& 2& 0& 2& 0& 2& 0& 2& 0& 2& 0& 2& 0& 2& 0& 2& 0\cr
    0106+013& 1& 1& 1& 1& 1& 1& 1& 1& 1& 1& 1& 1& 1& 1& 1& 1& 1& 1& 1& 1& 1& 1& 1& 1& 1& 1& 1& 1& 1& 1& 1& 1& 1& 1& 1& 1\cr
    J0125$-$0005& 1& 1& 1& 1& 1& 1& 1& 1& 1& 1& 1& 1& 1& 1& 1& 1& 1& 1& 1& 1& 1& 1& 1& 1& 1& 1& 1& 1& 1& 1& 1& 1& 1& 1& 1& 1\cr
    0133+476& 2& 0& 2& 0& 2& 0& 2& 0& 2& 0& 2& 0& 2& 0& 2& 0& 2& 0& 2& 0& 2& 0& 2& 0& 2& 0& 2& 0& 2& 0& 2& 0& 2& 0& 2& 0\cr
    0149+218& 1& 1& 1& 0& 1& 1& 1& 0& 1& 1& 1& 0& 1& 1& 1& 0& 1& 1& 1& 0& 1& 1& 1& 0& 1& 1& 1& 0& 1& 1& 1& 0& 1& 1& 1& 0\cr
    0202+149& 1& 1& 1& 0& 1& 1& 1& 0& 1& 1& 1& 0& 1& 1& 1& 0& 1& 1& 1& 0& 1& 1& 1& 0& 1& 1& 1& 0& 1& 1& 1& 0& 1& 1& 1& 0\cr
    0212+735& 1& 1& 1& 1& 1& 1& 1& 1& 1& 1& 1& 1& 1& 1& 1& 1& 1& 1& 1& 1& 1& 1& 1& 1& 1& 1& 1& 1& 1& 1& 1& 1& 1& 1& 1& 1\cr
    0224+671& 1& 1& 1& 1& 1& 1& 1& 1& 1& 1& 1& 1& 1& 1& 1& 1& 1& 1& 1& 1& 1& 1& 1& 1& 1& 1& 1& 1& 1& 1& 1& 1& 1& 1& 1& 1\cr
    0234+285& 1& 1& 1& 1& 2& 1& 2& 1& 1& 1& 1& 1& 2& 0& 2& 0& 2& 0& 2& 0& 2& 0& 2& 0& 2& 0& 2& 0& 2& 0& 2& 0& 2& 0& 2& 0\cr
    0235+164& 2& 0& 2& 0& 2& 0& 2& 0& 2& 0& 2& 0& 2& 0& 2& 0& 2& 0& 2& 0& 2& 0& 2& 0& 2& 0& 2& 0& 2& 0& 2& 0& 2& 0& 2& 0\cr
    0238$-$084& 1& 1& 1& 0& 1& 1& 1& 0& 1& 1& 1& 0& 1& 1& 1& 0& 1& 1& 1& 0& 1& 1& 1& 0& 1& 1& 1& 0& 1& 1& 1& 0& 1& 1& 1& 0\cr
    0306+102& 1& 1& 1& 1& 2& 1& 2& 1& 1& 1& 1& 1& 1& 1& 1& 1& 2& 0& 2& 0& 1& 1& 1& 1& 1& 1& 1& 1& 1& 1& 1& 1& 1& 1& 1& 1\cr
    0316+413& 1& 1& 1& 1& 1& 1& 1& 1& 1& 1& 1& 1& 1& 1& 1& 1& 1& 1& 1& 1& 1& 1& 1& 1& 1& 1& 1& 1& 1& 1& 1& 1& 1& 1& 1& 1\cr
    0333+321& 1& 1& 1& 1& 1& 1& 1& 1& 1& 1& 1& 1& 1& 1& 1& 1& 1& 1& 1& 1& 1& 1& 1& 1& 1& 1& 1& 1& 1& 1& 1& 1& 1& 1& 1& 1\cr
    0336$-$019& 1& 1& 1& 1& 1& 1& 1& 1& 1& 1& 1& 1& 1& 1& 1& 1& 1& 1& 1& 1& 1& 1& 1& 1& 1& 1& 1& 1& 1& 1& 1& 1& 1& 1& 1& 1\cr
    0355+508& 1& 1& 1& 1& 1& 1& 1& 1& 1& 1& 1& 1& 1& 1& 1& 1& 1& 1& 1& 1& 1& 1& 1& 1& 1& 1& 1& 1& 1& 1& 1& 1& 1& 1& 1& 1\cr
    0415+379& 1& 1& 1& 1& 1& 1& 1& 1& 1& 1& 1& 1& 1& 1& 1& 1& 1& 1& 1& 1& 1& 1& 1& 1& 1& 1& 1& 1& 1& 1& 1& 1& 1& 1& 1& 1\cr
    0420$-$014& 1& 1& 1& 1& 1& 1& 1& 1& 1& 1& 1& 1& 1& 1& 1& 1& 1& 1& 1& 1& 1& 1& 1& 1& 1& 1& 1& 1& 1& 1& 1& 1& 1& 1& 1& 1\cr
    0430+052& 1& 1& 1& 1& 1& 1& 1& 1& 1& 1& 1& 1& 1& 1& 1& 1& 1& 1& 1& 1& 1& 1& 1& 1& 1& 1& 1& 1& 1& 1& 1& 1& 1& 1& 1& 1\cr
    0446+112& 1& 1& 1& 1& 1& 1& 1& 1& 1& 1& 1& 1& 1& 1& 1& 1& 1& 1& 1& 1& 1& 1& 1& 1& 1& 1& 1& 1& 1& 1& 1& 1& 1& 1& 1& 1\cr
    0458$-$020& 1& 1& 1& 1& 1& 1& 1& 1& 1& 1& 1& 1& 1& 1& 1& 1& 1& 1& 1& 1& 1& 1& 1& 1& 1& 1& 1& 1& 1& 1& 1& 1& 1& 1& 1& 1\cr
    0507+179& 1& 1& 1& 1& 1& 1& 1& 1& 1& 1& 1& 1& 1& 1& 1& 1& 1& 1& 1& 1& 1& 1& 1& 1& 1& 1& 1& 1& 1& 1& 1& 1& 1& 1& 1& 1\cr
    0528+134& 1& 1& 1& 1& 1& 1& 1& 1& 1& 1& 1& 1& 1& 1& 1& 1& 1& 1& 1& 1& 1& 1& 1& 1& 1& 1& 1& 1& 1& 1& 1& 1& 1& 1& 1& 1\cr
    0552+398& 1& 1& 1& 1& 1& 1& 1& 1& 1& 1& 1& 1& 1& 1& 1& 1& 1& 1& 1& 1& 1& 1& 1& 1& 1& 1& 1& 1& 1& 1& 1& 1& 1& 1& 1& 1\cr
    0605$-$085& 1& 1& 1& 1& 1& 1& 1& 1& 1& 1& 1& 1& 1& 1& 1& 1& 1& 1& 1& 1& 1& 1& 1& 1& 1& 1& 1& 1& 1& 1& 1& 1& 1& 1& 1& 1\cr
    0642+449& 1& 1& 1& 1& 1& 1& 1& 1& 1& 1& 1& 1& 1& 1& 1& 1& 1& 1& 1& 1& 1& 1& 1& 1& 1& 1& 1& 1& 1& 1& 1& 1& 1& 1& 1& 1\cr
    0716+714& 1& 1& 1& 1& 1& 1& 1& 1& 1& 1& 1& 1& 1& 1& 1& 1& 1& 1& 1& 1& 1& 1& 1& 1& 1& 1& 1& 1& 1& 1& 1& 1& 1& 1& 1& 1\cr
    0723$-$008& 1& 1& 1& 1& 1& 1& 1& 1& 1& 1& 1& 1& 1& 1& 1& 1& 1& 1& 1& 1& 1& 1& 1& 1& 1& 1& 1& 1& 1& 1& 1& 1& 1& 1& 1& 1\cr
    0735+178& 1& 1& 1& 1& 1& 1& 1& 1& 1& 1& 1& 1& 1& 1& 1& 1& 1& 1& 1& 1& 1& 1& 1& 1& 1& 1& 1& 1& 1& 1& 1& 1& 1& 1& 1& 1\cr
    0736+017& 1& 1& 1& 1& 1& 1& 1& 1& 1& 1& 1& 1& 1& 1& 1& 1& 1& 1& 1& 1& 1& 1& 1& 1& 1& 1& 1& 1& 1& 1& 1& 1& 1& 1& 1& 1\cr
    0748+126& 1& 1& 1& 1& 1& 1& 1& 1& 1& 1& 1& 1& 1& 1& 1& 1& 1& 1& 1& 1& 1& 1& 1& 1& 1& 1& 1& 1& 1& 1& 1& 1& 1& 1& 1& 1\cr
    0754+100& 1& 1& 1& 1& 1& 1& 1& 1& 1& 1& 1& 1& 1& 1& 1& 1& 1& 1& 1& 1& 1& 1& 1& 1& 1& 1& 1& 1& 1& 1& 1& 1& 1& 1& 1& 1\cr
    0805$-$077& 1& 1& 1& 1& 1& 1& 1& 1& 1& 1& 1& 1& 1& 1& 1& 1& 1& 1& 1& 1& 1& 1& 1& 1& 1& 1& 1& 1& 1& 1& 1& 1& 1& 1& 1& 1\cr
    0804+499& 1& 1& 1& 1& 1& 1& 1& 1& 1& 1& 1& 1& 1& 1& 1& 1& 1& 1& 1& 1& 1& 1& 1& 1& 1& 1& 1& 1& 1& 1& 1& 1& 1& 1& 1& 1\cr
    0823+033& 1& 1& 1& 1& 1& 1& 1& 1& 1& 1& 1& 1& 1& 1& 1& 1& 1& 1& 1& 1& 1& 1& 1& 1& 1& 1& 1& 1& 1& 1& 1& 1& 1& 1& 1& 1\cr
    0827+243& 1& 1& 1& 1& 1& 1& 1& 1& 1& 1& 1& 1& 1& 1& 1& 1& 1& 1& 1& 1& 1& 1& 1& 1& 1& 1& 1& 1& 1& 1& 1& 1& 1& 1& 1& 1\cr
    0836+710& 1& 1& 1& 1& 1& 1& 1& 1& 1& 1& 1& 1& 1& 1& 1& 1& 1& 1& 1& 1& 1& 1& 1& 1& 1& 1& 1& 1& 1& 1& 1& 1& 1& 1& 1& 1\cr
    0851+202& 1& 1& 1& 1& 1& 1& 1& 1& 1& 1& 1& 1& 1& 1& 1& 1& 1& 1& 1& 1& 1& 1& 1& 1& 1& 1& 1& 1& 1& 1& 1& 1& 1& 1& 1& 1\cr
    0906+430& 1& 1& 1& 1& 1& 1& 1& 1& 1& 1& 1& 1& 1& 1& 1& 1& 1& 1& 1& 1& 1& 1& 1& 1& 1& 1& 1& 1& 1& 1& 1& 1& 1& 1& 1& 1\cr
    0917+449& 1& 1& 1& 1& 1& 1& 1& 1& 1& 1& 1& 1& 1& 1& 1& 1& 1& 1& 1& 1& 1& 1& 1& 1& 1& 1& 1& 1& 1& 1& 1& 1& 1& 1& 1& 1\cr
    0923+392& 1& 1& 1& 1& 1& 1& 1& 1& 1& 1& 1& 1& 1& 1& 1& 1& 1& 1& 1& 1& 1& 1& 1& 1& 1& 1& 1& 1& 1& 1& 1& 1& 1& 1& 1& 1\cr
    0945+408& 1& 1& 1& 1& 1& 1& 1& 1& 1& 1& 1& 1& 1& 1& 1& 1& 1& 1& 1& 1& 1& 1& 1& 1& 1& 1& 1& 1& 1& 1& 1& 1& 1& 1& 1& 1\cr
    0953+254& 1& 1& 1& 1& 1& 1& 1& 1& 1& 1& 1& 1& 1& 1& 1& 1& 1& 1& 1& 1& 1& 1& 1& 1& 1& 1& 1& 1& 1& 1& 1& 1& 1& 1& 1& 1\cr
    0954+658& 1& 1& 1& 1& 1& 1& 1& 1& 1& 1& 1& 1& 1& 1& 1& 1& 1& 1& 1& 1& 1& 1& 1& 1& 1& 1& 1& 1& 1& 1& 1& 1& 1& 1& 1& 1\cr
    1036+054& 1& 1& 1& 1& 1& 1& 1& 1& 1& 1& 1& 1& 1& 1& 1& 1& 1& 1& 1& 1& 1& 1& 1& 1& 1& 1& 1& 1& 1& 1& 1& 1& 1& 1& 1& 1\cr
    TEX1040+244& 1& 1& 1& 1& 1& 1& 1& 1& 1& 1& 1& 1& 1& 1& 1& 1& 1& 1& 1& 1& 1& 1& 1& 1& 1& 1& 1& 1& 1& 1& 1& 1& 1& 1& 1& 1\cr
    1055+018& 1& 1& 1& 1& 1& 1& 1& 1& 1& 1& 1& 1& 1& 1& 1& 1& 1& 1& 1& 1& 1& 1& 1& 1& 1& 1& 1& 1& 1& 1& 1& 1& 1& 1& 1& 1\cr
    J1130+3815& 1& 1& 1& 1& 1& 1& 1& 1& 1& 1& 1& 1& 1& 1& 1& 1& 1& 1& 1& 1& 1& 1& 1& 1& 1& 1& 1& 1& 1& 1& 1& 1& 1& 1& 1& 1\cr
    1150+812& 1& 1& 1& 1& 1& 2& 1& 2& 1& 1& 1& 1& 1& 1& 1& 1& 1& 1& 1& 1& 1& 1& 1& 1& 1& 1& 1& 1& 1& 1& 1& 1& 1& 1& 1& 1\cr
    1150+497& 1& 1& 1& 1& 1& 2& 1& 2& 1& 1& 1& 1& 1& 1& 1& 1& 1& 1& 1& 1& 1& 1& 1& 1& 1& 1& 1& 1& 1& 1& 1& 1& 1& 1& 1& 1\cr
    1156+295& 1& 1& 1& 1& 1& 1& 1& 1& 1& 1& 1& 1& 1& 1& 1& 1& 1& 1& 1& 1& 1& 1& 1& 1& 1& 1& 1& 1& 1& 1& 1& 1& 1& 1& 1& 1\cr
    1219+044& 1& 1& 1& 1& 1& 1& 1& 1& 1& 1& 1& 1& 1& 1& 1& 1& 1& 1& 1& 1& 1& 1& 1& 1& 1& 1& 1& 1& 1& 1& 1& 1& 1& 1& 1& 1\cr
    1222+216& 1& 1& 1& 1& 1& 1& 1& 1& 1& 1& 1& 1& 1& 1& 1& 1& 1& 1& 1& 1& 1& 1& 1& 1& 1& 1& 1& 1& 1& 1& 1& 1& 1& 1& 1& 1\cr
    1226+023& 1& 1& 1& 1& 1& 1& 1& 1& 1& 1& 1& 1& 1& 1& 1& 1& 1& 1& 1& 1& 1& 1& 1& 1& 1& 1& 1& 1& 1& 1& 1& 1& 1& 1& 1& 1\cr
    1228+126& 1& 1& 1& 1& 1& 1& 1& 1& 1& 1& 1& 1& 1& 1& 1& 1& 1& 1& 1& 1& 1& 1& 1& 1& 1& 1& 1& 1& 1& 1& 1& 1& 1& 1& 1& 1\cr
    1253$-$055& 1& 1& 1& 1& 1& 1& 1& 1& 1& 1& 1& 1& 1& 1& 1& 1& 1& 1& 1& 1& 1& 1& 1& 1& 1& 1& 1& 1& 1& 1& 1& 1& 1& 1& 1& 1\cr
    1308+326& 1& 1& 1& 1& 1& 1& 1& 1& 1& 1& 1& 1& 1& 1& 1& 1& 1& 1& 1& 1& 1& 1& 1& 1& 1& 1& 1& 1& 1& 1& 1& 1& 1& 1& 1& 1\cr
    1324+224& 1& 1& 1& 1& 1& 1& 1& 1& 1& 1& 1& 1& 1& 1& 1& 1& 1& 1& 1& 1& 1& 1& 1& 1& 1& 1& 1& 1& 1& 1& 1& 1& 1& 1& 1& 1\cr
    1413+135& 1& 1& 1& 1& 1& 1& 1& 1& 1& 1& 1& 1& 1& 1& 1& 1& 1& 1& 1& 1& 1& 1& 1& 1& 1& 1& 1& 1& 1& 1& 1& 1& 1& 1& 1& 1\cr
    1418+546& 1& 1& 1& 1& 2& 2& 2& 2& 1& 1& 1& 1& 1& 1& 1& 1& 1& 1& 1& 1& 1& 1& 1& 1& 1& 1& 1& 1& 1& 1& 1& 1& 1& 1& 1& 1\cr
    1502+106& 0& 2& 0& 1& 0& 2& 0& 1& 0& 2& 0& 1& 0& 2& 0& 1& 0& 2& 0& 1& 0& 2& 0& 1& 0& 2& 0& 1& 0& 2& 0& 1& 0& 2& 0& 1\cr
    1510$-$089& 1& 1& 1& 1& 1& 2& 1& 1& 1& 1& 1& 1& 1& 1& 1& 1& 0& 2& 0& 1& 1& 1& 1& 1& 0& 1& 1& 1& 0& 2& 0& 1& 0& 2& 0& 1\cr
    1546+027& 1& 1& 1& 1& 1& 1& 1& 1& 1& 1& 1& 1& 1& 1& 1& 1& 1& 1& 1& 1& 1& 1& 1& 1& 1& 1& 1& 1& 1& 1& 1& 1& 1& 1& 1& 1\cr
    1548+056& 1& 1& 1& 1& 1& 1& 1& 1& 1& 1& 1& 1& 1& 1& 1& 1& 1& 1& 1& 1& 1& 1& 1& 1& 1& 1& 1& 1& 1& 1& 1& 1& 1& 1& 1& 1\cr
    1606+106& 1& 1& 1& 1& 1& 1& 1& 1& 1& 1& 1& 1& 1& 1& 1& 1& 1& 1& 1& 1& 1& 1& 1& 1& 1& 1& 1& 1& 1& 1& 1& 1& 1& 1& 1& 1\cr
    1611+343& 0& 1& 1& 1& 0& 2& 1& 1& 0& 2& 1& 1& 0& 2& 0& 1& 0& 2& 0& 1& 0& 2& 0& 1& 0& 2& 0& 1& 0& 2& 0& 1& 0& 2& 0& 1\cr
    1633+382& 1& 1& 1& 1& 1& 2& 1& 1& 1& 1& 1& 1& 0& 1& 1& 1& 0& 2& 0& 1& 0& 2& 1& 1& 0& 2& 0& 1& 0& 2& 0& 1& 0& 2& 0& 1\cr
    1638+398& 1& 1& 1& 1& 1& 2& 1& 1& 1& 1& 1& 1& 0& 1& 1& 1& 0& 2& 0& 1& 0& 2& 1& 1& 0& 2& 0& 1& 0& 2& 0& 1& 0& 2& 0& 1\cr
    1642+690& 1& 3& 1& 3& 3& 3& 3& 3& 1& 3& 1& 3& 1& 3& 1& 3& 2& 0& 2& 0& 2& 2& 2& 2& 2& 2& 2& 2& 2& 2& 2& 2& 2& 2& 2& 2\cr
    1641+399& 1& 1& 1& 1& 1& 2& 1& 1& 1& 1& 1& 1& 0& 1& 1& 1& 0& 2& 0& 1& 0& 2& 1& 1& 0& 2& 0& 1& 0& 2& 0& 1& 0& 2& 0& 1\cr
    1652+398& 1& 1& 1& 1& 1& 2& 1& 1& 1& 1& 1& 1& 1& 1& 1& 1& 0& 2& 1& 1& 1& 1& 1& 1& 0& 1& 1& 1& 0& 1& 1& 1& 0& 1& 1& 1\cr
    1739+522& 1& 1& 1& 1& 1& 2& 1& 2& 1& 1& 1& 1& 1& 1& 1& 1& 0& 2& 0& 1& 0& 1& 1& 1& 0& 2& 1& 1& 0& 2& 0& 1& 0& 2& 0& 1\cr
    1741$-$038& 1& 1& 1& 1& 1& 1& 1& 1& 1& 1& 1& 1& 1& 1& 1& 1& 1& 1& 1& 1& 1& 1& 1& 1& 1& 1& 1& 1& 1& 1& 1& 1& 1& 1& 1& 1\cr
    1749+096& 1& 1& 1& 1& 1& 1& 1& 1& 1& 1& 1& 1& 1& 1& 1& 1& 1& 1& 1& 1& 1& 1& 1& 1& 1& 1& 1& 1& 1& 1& 1& 1& 1& 1& 1& 1\cr
    1803+784& 1& 1& 1& 1& 3& 1& 3& 1& 1& 1& 1& 1& 1& 1& 1& 1& 2& 0& 2& 0& 1& 1& 1& 1& 1& 1& 1& 1& 2& 0& 2& 0& 2& 0& 2& 0\cr
    1807+698& 2& 2& 2& 2& 3& 2& 3& 2& 2& 2& 2& 2& 2& 2& 2& 2& 2& 0& 2& 0& 2& 2& 2& 2& 2& 0& 2& 0& 2& 0& 2& 0& 2& 0& 2& 0\cr
    1823+568& 1& 1& 1& 1& 1& 3& 1& 2& 1& 1& 1& 1& 1& 1& 1& 1& 0& 2& 0& 1& 1& 1& 1& 1& 0& 1& 1& 1& 0& 1& 1& 1& 0& 1& 1& 1\cr
    1828+487& 1& 1& 1& 1& 2& 1& 2& 1& 1& 1& 1& 1& 1& 1& 1& 1& 1& 1& 1& 1& 1& 1& 1& 1& 1& 1& 1& 1& 1& 1& 1& 1& 1& 1& 1& 1\cr
\noalign{\vskip 4pt\hrule\vskip 5pt}}}
\endPlancktablewide
\end{table*}

\addtocounter{table}{-1}
\begin{table*}
\caption{Continued.}          
\label{pointings}
\vskip -6mm
\scriptsize
\setbox\tablebox=\vbox{
\newdimen\digitwidth
\setbox0=\hbox{\rm 0}
\digitwidth=\wd0
\catcode`*=\active
\def*{\kern\digitwidth}
\newdimen\signwidth
\setbox0=\hbox{+}
\signwidth=\wd0
\catcode`!=\active
\def!{\kern\signwidth}
\newdimen\decimalwidth
\setbox0=\hbox{.}
\decimalwidth=\wd0
\catcode`@=\active
\def@{\kern\decimalwidth}
\halign{\hbox to 2.2cm{#\leaderfil}\tabskip 1.0em&
    \hfil#\hfil\tabskip 0.3em&
    \hfil#\hfil&
    \hfil#\hfil&
    \hfil#\hfil\tabskip1.0em&
    \hfil#\hfil\tabskip 0.3em&
    \hfil#\hfil&
    \hfil#\hfil&
    \hfil#\hfil\tabskip1.0em&
    \hfil#\hfil\tabskip 0.3em&
    \hfil#\hfil&
    \hfil#\hfil&
    \hfil#\hfil\tabskip1.0em&
    \hfil#\hfil\tabskip 0.3em&
    \hfil#\hfil&
    \hfil#\hfil&
    \hfil#\hfil\tabskip1.0em&
    \hfil#\hfil\tabskip 0.3em&
    \hfil#\hfil&
    \hfil#\hfil&
    \hfil#\hfil\tabskip1.0em&
    \hfil#\hfil\tabskip 0.3em&
    \hfil#\hfil&
    \hfil#\hfil&
    \hfil#\hfil\tabskip1.0em&
    \hfil#\hfil\tabskip 0.3em&
    \hfil#\hfil&
    \hfil#\hfil&
    \hfil#\hfil\tabskip1.0em&
    \hfil#\hfil\tabskip 0.3em&
    \hfil#\hfil&
    \hfil#\hfil&
    \hfil#\hfil\tabskip1.0em&
    \hfil#\hfil\tabskip 0.3em&
    \hfil#\hfil&
    \hfil#\hfil&
    \hfil#\hfil\tabskip 0em\cr
\noalign{\doubleline}
\omit&\multispan4\hfil 30\,GHz\hfil&\multispan4\hfil 44\,GHz\hfil&\multispan4\hfil 70\,GHz\hfil&\multispan4\hfil 100\,GHz\hfil&
      \multispan4\hfil 143\,GHz\hfil&\multispan4\hfil 217\,GHz\hfil&\multispan4\hfil 353\,GHz\hfil&\multispan4\hfil 545\,GHz\hfil&
      \multispan4\hfil 857\,GHz\hfil\cr
\noalign{\vskip -3pt}
\omit&\multispan4\hrulefill&\multispan4\hrulefill&\multispan4\hrulefill&\multispan4\hrulefill&\multispan4\hrulefill&\multispan4\hrulefill&
      \multispan4\hrulefill&\multispan4\hrulefill&\multispan4\hrulefill\cr
\omit\hfil Source\hfil& S1& S2& S3& S4& S1& S2& S3& S4& S1& S2& S3& S4& S1& S2& S3& S4& S1& S2& S3& S4& S1& S2& S3& S4& S1& S2& S3& S4& 
                        S1& S2& S3& S4& S1& S2& S3& S4\cr
\noalign{\vskip 3pt\hrule\vskip 4pt}
    J184915+670& 2& 2& 2& 2& 3& 2& 3& 1& 2& 2& 2& 2& 2& 2& 2& 2& 2& 0& 2& 0& 2& 2& 2& 2& 2& 2& 2& 2& 2& 2& 2& 2& 2& 2& 2& 2\cr
    1928+738& 2& 0& 2& 0& 3& 0& 3& 0& 2& 0& 2& 0& 2& 0& 2& 0& 2& 0& 2& 0& 2& 0& 2& 0& 2& 0& 2& 0& 2& 0& 2& 0& 2& 0& 2& 0\cr
    1954+513& 1& 1& 1& 1& 2& 1& 2& 1& 1& 1& 1& 1& 1& 1& 1& 1& 1& 1& 1& 1& 1& 1& 1& 1& 1& 1& 1& 1& 1& 1& 1& 1& 1& 1& 1& 1\cr
    2007+776& 1& 1& 1& 1& 3& 1& 3& 1& 2& 1& 2& 1& 2& 0& 2& 0& 2& 0& 2& 0& 2& 0& 2& 0& 2& 0& 2& 0& 2& 0& 2& 0& 2& 0& 2& 0\cr
    2005+403& 1& 1& 1& 1& 2& 1& 2& 1& 1& 1& 1& 1& 1& 1& 1& 1& 1& 1& 1& 1& 1& 1& 1& 1& 1& 1& 1& 1& 1& 1& 1& 1& 1& 1& 1& 1\cr
    2021+614& 1& 1& 1& 1& 3& 3& 3& 2& 1& 1& 1& 1& 1& 1& 1& 1& 2& 1& 1& 1& 1& 1& 1& 1& 1& 1& 1& 1& 1& 1& 1& 1& 1& 1& 1& 1\cr
    2037+511& 1& 1& 1& 1& 2& 2& 2& 2& 1& 1& 1& 1& 1& 1& 1& 1& 1& 1& 1& 1& 1& 1& 1& 1& 1& 1& 1& 1& 1& 1& 1& 1& 1& 1& 1& 1\cr
    2121+053& 1& 1& 1& 1& 1& 1& 1& 1& 1& 1& 1& 1& 1& 1& 1& 1& 1& 1& 1& 1& 1& 1& 1& 1& 1& 1& 1& 1& 1& 1& 1& 1& 1& 1& 1& 1\cr
    2131$-$021& 1& 1& 1& 1& 1& 1& 1& 1& 1& 1& 1& 1& 1& 1& 1& 1& 1& 1& 1& 1& 1& 1& 1& 1& 1& 1& 1& 1& 1& 1& 1& 1& 1& 1& 1& 1\cr
    2134+004& 1& 1& 1& 1& 1& 1& 1& 1& 1& 1& 1& 1& 1& 1& 1& 1& 1& 1& 1& 1& 1& 1& 1& 1& 1& 1& 1& 1& 1& 1& 1& 1& 1& 1& 1& 1\cr
    2136+141& 1& 1& 1& 1& 1& 1& 1& 1& 1& 1& 1& 1& 1& 1& 1& 1& 1& 1& 1& 1& 1& 1& 1& 1& 1& 1& 1& 1& 1& 1& 1& 1& 1& 1& 1& 1\cr
    2145+067& 1& 1& 1& 1& 1& 1& 1& 1& 1& 1& 1& 1& 1& 1& 1& 1& 1& 1& 1& 1& 1& 1& 1& 1& 1& 1& 1& 1& 1& 1& 1& 1& 1& 1& 1& 1\cr
    2200+420& 1& 1& 1& 1& 1& 2& 1& 2& 1& 1& 1& 1& 1& 1& 1& 1& 1& 1& 1& 1& 1& 1& 1& 1& 1& 1& 1& 1& 1& 1& 1& 1& 1& 1& 1& 1\cr
    2201+315& 1& 1& 1& 1& 1& 1& 1& 1& 1& 1& 1& 1& 1& 1& 1& 1& 1& 1& 1& 1& 1& 1& 1& 1& 1& 1& 1& 1& 1& 1& 1& 1& 1& 1& 1& 1\cr
    2201+171& 1& 1& 1& 1& 1& 1& 1& 1& 1& 1& 1& 1& 1& 1& 1& 1& 1& 1& 1& 1& 1& 1& 1& 1& 1& 1& 1& 1& 1& 1& 1& 1& 1& 1& 1& 1\cr
    2216$-$038& 1& 1& 1& 1& 1& 1& 1& 1& 1& 1& 1& 1& 1& 1& 1& 1& 1& 1& 1& 1& 1& 1& 1& 1& 1& 1& 1& 1& 1& 1& 1& 1& 1& 1& 1& 1\cr
    2223$-$052& 1& 1& 1& 1& 1& 1& 1& 1& 1& 1& 1& 1& 1& 1& 1& 1& 1& 1& 1& 1& 1& 1& 1& 1& 1& 1& 1& 1& 1& 1& 1& 1& 1& 1& 1& 1\cr
    2227$-$088& 1& 1& 1& 1& 1& 1& 1& 1& 1& 1& 1& 1& 1& 1& 1& 1& 1& 1& 1& 1& 1& 1& 1& 1& 1& 1& 1& 1& 1& 1& 1& 1& 1& 1& 1& 1\cr
    2230+114& 1& 1& 1& 1& 1& 1& 1& 1& 1& 1& 1& 1& 1& 1& 1& 1& 1& 1& 1& 1& 1& 1& 1& 1& 1& 1& 1& 1& 1& 1& 1& 1& 1& 1& 1& 1\cr
    2234+282& 1& 1& 1& 1& 1& 1& 1& 1& 1& 1& 1& 1& 1& 1& 1& 1& 1& 1& 1& 1& 1& 1& 1& 1& 1& 1& 1& 1& 1& 1& 1& 1& 1& 1& 1& 1\cr
    2251+158& 1& 1& 1& 1& 1& 1& 1& 1& 1& 1& 1& 1& 1& 1& 1& 1& 1& 1& 1& 1& 1& 1& 1& 1& 1& 1& 1& 1& 1& 1& 1& 1& 1& 1& 1& 1\cr
    4C~45.51& 1& 1& 1& 1& 1& 1& 1& 1& 1& 1& 1& 1& 1& 1& 1& 1& 1& 1& 1& 0& 1& 1& 1& 1& 1& 1& 1& 1& 1& 1& 1& 0& 1& 1& 1& 0\cr
    2353+816& 1& 1& 1& 1& 1& 1& 1& 1& 1& 1& 1& 1& 1& 1& 1& 1& 1& 1& 1& 1& 1& 1& 1& 1& 1& 1& 1& 1& 1& 1& 1& 1& 1& 1& 1& 1\cr
\noalign{\vskip 4pt\hrule\vskip 5pt}}}
\endPlancktablewide
\end{table*}

\setcounter{table}{5}
\begin{table*}
\caption{Broken power-law fits to the radio spectrum.}            
\label{fit_par}    
\vskip -6mm
\scriptsize
\setbox\tablebox=\vbox{
\newdimen\digitwidth
\setbox0=\hbox{\rm 0}
\digitwidth=\wd0
\catcode`*=\active
\def*{\kern\digitwidth}
\newdimen\signwidth
\setbox0=\hbox{+}
\signwidth=\wd0
\catcode`!=\active
\def!{\kern\signwidth}
\newdimen\decimalwidth
\setbox0=\hbox{.}
\decimalwidth=\wd0
\catcode`@=\active
\def@{\kern\decimalwidth}

\halign{\hbox to 2.2cm{#\leaderfil}\tabskip 2.0em&
    \hfil#\hfil\tabskip 1.0em&
    \hfil$#$\hfil\tabskip 1.0em&
    \hfil$#$\hfil&
    \hfil$#$\hfil\tabskip 2.0em&
    \hfil$#$\hfil\tabskip 1.0em&
    \hfil$#$\hfil&
    \hfil$#$\hfil\tabskip 2.0em&
    \hfil$#$\hfil\tabskip 1.0em&
    \hfil$#$\hfil&
    \hfil$#$\hfil\tabskip 2.0em&
    \hfil$#$\hfil\tabskip 1.0em&
    \hfil$#$\hfil&
    \hfil$#$\hfil\tabskip 0em\cr
\noalign{\doubleline}
\omit&\omit&\multispan3\hfil Survey 1\hfil&\multispan3\hfil Survey 2\hfil&\multispan3\hfil Survey 3\hfil& 
             \multispan3\hfil Survey 4\hfil\cr
\noalign{\vskip -3pt}
\omit&\omit&\multispan3\hrulefill&\multispan3\hrulefill&\multispan3\hrulefill&\multispan3\hrulefill\cr
\omit\hfil Source\hfil&\omit Class&\alpha_{\rm LF}& \alpha_{\rm HF}& \nu_{\rm break}& \alpha_{\rm LF}& \alpha_{\rm HF}& \nu_{\rm break}& \alpha_{\rm LF}& \alpha_{\rm HF}& \nu_{\rm break}& \alpha_{\rm LF}& \alpha_{\rm HF}& \nu_{\rm break}\cr
\noalign{\vskip 3pt\hrule\vskip 4pt}
0003$-$066&  BLO&-0.07& -0.33& *23.18& -0.26& -0.39& 241.31& -0.15& -0.36& *42.78& -0.26& -0.4*& *70@**\cr
  0007+106&  GAL&!0.59& -0.45& *37.56& !0.14& -0.81& 100@**& \dots& \dots&  \dots& \dots& \dots&  \dots\cr
0048$-$097&  BLO&-0.02& -0.52& 120.25& \dots& \dots&  \dots& \dots& \dots&  \dots& \dots& \dots&  \dots\cr
  0059+581&  QSO&!0.36& -0.7*& *46.18& \dots& \dots&  \dots& \dots& \dots&  \dots& \dots& \dots&  \dots\cr
  0106+013&  HPQ&!0.13& -0.44& *23.6*& -0.15& -0.67& *56.79& -0.38& -0.67& 100@**& \dots& \dots&  \dots\cr
J0125$-$0005&QSO&\dots& \dots&  \dots& \dots& \dots&  \dots& \dots& \dots&  \dots& \dots& \dots&  \dots\cr
  0133+476&  HPQ&!0.03& -0.4*& *57.49& \dots& \dots&  \dots& -0.18& -0.39& *37.15& \dots& \dots&  \dots\cr
  0149+218&  HPQ&\dots& \dots&  \dots& !0.13& -0.58& *81.23& \dots& \dots&  \dots& \dots& \dots&  \dots\cr
  0202+149&  HPQ&\dots& \dots&  \dots& \dots& \dots&  \dots& \dots& \dots&  \dots& \dots& \dots&  \dots\cr
  0212+735&  BLO&\dots& \dots&  \dots& !0.32& -0.6*& *18.15& \dots& \dots&  \dots& !0.09& -0.46& **9.31\cr
  0224+671&  QSO&!0.84& -0.24& *39.61& -0.09& -0.31& *70@**& \dots& \dots&  \dots& \dots& \dots&  \dots\cr
  0234+285&  HPQ&-0.15& -0.61& *57.91& \dots& \dots&  \dots& !0.07& -0.5*& *59.13& \dots& \dots&  \dots\cr
  0235+164&  BLO&!0.01& -0.5*& 123.65& \dots& \dots&  \dots& !0.04& -0.34& 100@**& \dots& \dots&  \dots\cr
0238$-$084&  GAL&-0.12& !0.23& 181.71& -0.12& -0.02& 147@**& \dots& \dots&  \dots& \dots& \dots&  \dots\cr
  0306+102&  BLO&!0.37& -0.75& *70@**& \dots& \dots&  \dots& !0.28& -0.59& *70@**& \dots& \dots&  \dots\cr
  0316+413&  GAL&!0.21& -0.54& *19.56& !0.18& -0.57& *21.17& !0.36& -0.54& *19.54& !0.37& -0.57& *19.88\cr
  0333+321&  HPQ&!0.19& -1.01& *53.9*& -0.18& -0.75& *65.16& !0.62& -0.63& *20.19& -0.21& -0.63& *37.81\cr
0336$-$019&  HPQ&\dots& \dots&  \dots& \dots& \dots&  \dots& -0.02& -0.7*& *88.92& -0.1*& -0.71& 106.44\cr
  0355+508&  LPQ&-0.41& -0.69& *77.18& !0.1*& -0.57& *37.47& !0.22& -0.59& *19.68& \dots& \dots&  \dots\cr
  0415+379&  GAL&-0.44& -0.34& *37@**& !0.69& -0.38& *44.88& -0.58& -0.07& *28.41& -0.56& !0@**& *15.98\cr
0420$-$014&  HPQ&!0.12& -0.46& *58.72& !0.07& -0.59& *54.35& \dots& \dots&  \dots& \dots& \dots&  \dots\cr
  0430+052&  GAL&!0.02& -0.45& *22.88& -0.16& -0.19& *19.63& \dots& \dots&  \dots& \dots& \dots&  \dots\cr
  0446+112&  GAL&!0.2*& -0.67& *70.19& !0.02& -0.71& *82.85& !0.21& -0.48& *41.72& \dots& \dots&  \dots\cr
0458$-$020&  HPQ&\dots& \dots&  \dots& !0.12& -0.72& *68.27& !0.24& -0.64& *70@**& -0.05& -0.31& *70@**\cr
  0507+179&  HPQ&\dots& \dots&  \dots& \dots& \dots&  \dots& \dots& \dots&  \dots& \dots& \dots&  \dots\cr
  0528+134&  HPQ&\dots& \dots&  \dots& \dots& \dots&  \dots& \dots& \dots&  \dots& \dots& \dots&  \dots\cr
  0552+398&  LPQ&\dots& \dots&  \dots& \dots& \dots&  \dots& \dots& \dots&  \dots& -0.13& -0.56& *10.4*\cr
0605$-$085&  HPQ&!0.32& -0.43& *46.9*& !0.12& -0.83& *66.2*& \dots& \dots&  \dots& !0.18& -0.78& *52.11\cr
  0642+449&  LPQ&!0.37& -0.78& *22.89& !0.18& -0.61& *19.85& !0.31& -0.65& *20.41& -0.01& -0.57& *21.75\cr
  0716+714&  BLO&\dots& \dots&  \dots& !0.12& -0.17& *70@**& \dots& \dots&  \dots& \dots& \dots&  \dots\cr
0723$-$008&  BLO&!0.01& -0.44& *84.7*& !0.36& -0.46& *67.93& !0.37& -0.39& *73.62& -0.01& -0.42& *55.18\cr
  0735+178&  BLO&\dots& \dots&  \dots& \dots& \dots&  \dots& \dots& \dots&  \dots& \dots& \dots&  \dots\cr
  0736+017&  HPQ&!0.26& -0.41& *57.65& \dots& \dots&  \dots& -0.39& !0.04& **4.76& !0.18& !0.07& *82.27\cr
  0748+126&  LPQ&-0.15& -0.78& *77.58& -0.25& -0.65& *85.86& !0.05& -0.64& *37.21& -0.04& -0.54& *31.49\cr
  0754+100&  BLO&-0.04& -0.53& *70@**& !0.03& -0.31& *68.99& -0.04& -0.22& *52.08& \dots& \dots&  \dots\cr
0805$-$077&  QSO&-0.11& -0.31& 100@**& !0.02& -0.43& *62.54& -0.03& -0.35& *41.48& !0.01& -0.65& *81.94\cr
  0804+499&  HPQ&\dots& \dots&  \dots& \dots& \dots&  \dots& \dots& \dots&  \dots& \dots& \dots&  \dots\cr
  0823+033&  BLO&\dots& \dots&  \dots& \dots& \dots&  \dots& !0.24& -0.83& *81.59& \dots& \dots&  \dots\cr
  0827+243&  LPQ&\dots& \dots&  \dots& !0.44& -0.52& *55.54& !0.52& -0.41& *81.36& !0.1*& -0.81& *63.38\cr
  0836+710&  LPQ&!0.44& -0.7*& *50.95& !0.5*& -0.9*& *67.64& -0.37& -0.98& *70@**& !0.06& -0.62& *59.59\cr
  0851+202&  BLO&!0.28& -0.26& *53.19& \dots& \dots&  \dots& !0.14& -0.36& *52.99& \dots& \dots&  \dots\cr
  0906+430&  HPQ&!0.09& -0.35& *44@**& \dots& \dots&  \dots& -0.45& !0.45& 127.38& -0.25& -0.5*& *77.75\cr
  0917+449&  QSO&!0.3*& -0.33& *32.12& !0.17& -0.52& *44@**& !0.29& -0.39& *33.3*& !0.47& -0.53& *26.08\cr
  0923+392&  LPQ&!0.33& -0.6*& *19.9*& -0.24& -0.62& *35.59& \dots& \dots&  \dots& -0.04& -0.6*& *23.97\cr
  0945+408&  LPQ&-0.12& -1.11& 100.18& -0.15& -0.78& *70@**& -0.3*& -0.38& *64.78& -0.3*& -0.85& 111.51\cr
  0953+254&  LPQ&\dots& \dots&  \dots& \dots& \dots&  \dots& \dots& \dots&  \dots& \dots& \dots&  \dots\cr
  0954+658&  BLO&\dots& \dots&  \dots& !0.24& -0.08& *17.28& !0.07& -0.15& *44@**& -0.02& !0.11& *15@**\cr
  1036+054&  QSO&\dots& \dots&  \dots& \dots& \dots&  \dots& \dots& \dots&  \dots& \dots& \dots&  \dots\cr
TEX1040+244& BLO&\dots&\dots&  \dots& \dots& \dots&  \dots& \dots& \dots&  \dots& -0.01& -0.4*& *37@**\cr
  1055+018&  HPQ&!0.13& -0.66& *58.33& \dots& \dots&  \dots& !0.24& -0.49& *50.02& !0.27& -0.41& *55.53\cr
J1130+3815&  QSO&\dots& \dots&  \dots& -0.02& -0.54& *52.48& !0.42& -0.44& *44@**& !0.31& -0.37& *35.17\cr
  1150+812&  QSO&\dots& \dots&  \dots& \dots& \dots&  \dots& \dots& \dots&  \dots& \dots& \dots&  \dots\cr
  1150+497&  HPQ&\dots& \dots&  \dots& !0.53& -0.44& *82.51& -0.05& -0.39& 171.43& \dots& \dots&  \dots\cr
  1156+295&  HPQ&-0.02& -0.22& *44@**& -0.02& -0.3*& *37@**& !0.06& -0.35& *28@**& -0.15& -0.53& *23.74\cr
  1219+044&  QSO&!0.39& -0.56& *46.31& !0.02& -0.51& *54.02& !0.43& -0.34& *67.44& !0.29& -0.54& *45.63\cr
  1222+216&  HPQ&-0.07& -0.62& *78.2*& !0.34& -0.63& *59.63& !0.19& -0.77& *61.4*& !0.27& -0.47& *23.1*\cr
  1226+023&  HPQ&\dots& \dots&  \dots& -0.24& -0.85& *58.59& -0.35& -0.61& *31.99& \dots& \dots&  \dots\cr
  1228+126&  GAL&!0.37& -0.72& *38.34& !0.34& -0.73& *37.96& \dots& \dots&  \dots& !0.39& -0.74& *39.21\cr
1253$-$055&  HPQ&\dots& \dots&  \dots& \dots& \dots&  \dots& \dots& \dots&  \dots& \dots& \dots&  \dots\cr
  1308+326&  BLO&!0.08& -0.57& *44@**& !0.39& -0.63& *24.83& !0.1*& -0.57& *21.7*& -0.04& -0.5*& *27.37\cr
  1324+224&  QSO&\dots& \dots&  \dots& \dots& \dots&  \dots& \dots& \dots&  \dots& \dots& \dots&  \dots\cr
  1413+135&  BLO&\dots& \dots&  \dots& \dots& \dots&  \dots& \dots& \dots&  \dots& \dots& \dots&  \dots\cr
  1418+546&  BLO&!0.11& -0.21& *56.71& !0.19& -0.13& *17.61& -0.18& -0.1*& *86.05& \dots& \dots&  \dots\cr
  1502+106&  HPQ&\dots& \dots&  \dots& \dots& \dots&  \dots& \dots& \dots&  \dots& -0.07& -1.14& *93.1*\cr
1510$-$089&  HPQ&\dots& \dots&  \dots& !0.02& -0.73& *72.75& \dots& \dots&  \dots& !0.48& -0.41& *54@**\cr
  1546+027&  HPQ&!0.09& -0.59& *67.18& -0.07& -0.42& *45.38& !0.22& -0.58& *52.44& !0.21& -0.41& *70@**\cr
  1548+056&  HPQ&!0.22& -0.53& *22.39& -0.32& -0.47& *64.45& -0.15& -0.64& *56.46& -0.19& -0.58& *41.89\cr
  1606+106&  LPQ&\dots& \dots&  \dots& \dots& \dots&  \dots& \dots& \dots&  \dots& \dots& \dots&  \dots\cr
  1611+343&  HPQ&\dots& \dots&  \dots& !0.39& -0.64& *37@**& \dots& \dots&  \dots& !0.14& -0.67& *27.3*\cr
  1633+382&  HPQ&\dots& \dots&  \dots& !0.07& -0.71& *68.84& \dots& \dots&  \dots& \dots& \dots&  \dots\cr
  1638+398&  QSO&\dots& \dots&  \dots& \dots& \dots&  \dots& \dots& \dots&  \dots& -0.11& -0.8*& *59.42\cr
  1642+690&  HPQ&-0.17& -0.64& *33.91& \dots& \dots&  \dots& \dots& \dots&  \dots& \dots& \dots&  \dots\cr
  1641+399&  HPQ&\dots& \dots&  \dots& -0.12& -0.64& *68.03& \dots& \dots&  \dots& -0.14& -0.48& *27.51\cr
  1652+398&  BLO&\dots& \dots&  \dots& \dots& \dots&  \dots& \dots& \dots&  \dots& -0.24& -0.46& *75.56\cr
  1739+522&  HPQ&\dots& \dots&  \dots& -0.27& -0.86& *92.73& \dots& \dots&  \dots& \dots& \dots&  \dots\cr
1741$-$038&  HPQ&-0.22& -0.77& *76.3*& !0.37& -0.96& *58.92& -0.02& -0.6*& *39.19& !0.2*& -0.6*& *50.06\cr
  1749+096&  BLO&!0.04& -0.45& *59.36& !0@**& -0.52& *70@**& !0.26& -0.44& *61.43& \dots& \dots&  \dots\cr
  1803+784&  BLO&-0.18& -0.27& *53.61& \dots& \dots&  \dots& -0.09& -0.33& *56.7*& \dots& \dots&  \dots\cr
  1807+698&  BLO&!0.12& -0.31& *20.33& \dots& \dots&  \dots& -0.1*& -0.3*& *37@**& \dots& \dots&  \dots\cr
  1823+568&  BLO&!0.02& -0.31& *70@**& -0.07& -0.47& *60.42& -0.07& -0.59& *67.18& -0.08& -0.19& *22.48\cr
  1828+487&  HPQ&-0.43& -0.96& *59.86& -0.39& -0.74& *80@**& \dots& \dots&  \dots& -0.23& -0.64& *43.7*\cr
J184915+670& UNK&0.07&-0.43& *57.05& !0.23& -0.5*& *90.79& !0.27& -0.47& *30.88& \dots& \dots&  \dots\cr
  1928+738&  HPQ&!0.15& -0.68& *55.06& \dots& \dots&  \dots& !0.25& -0.58& *30.33& \dots& \dots&  \dots\cr
  1954+513&  LPQ&\dots& \dots&  \dots& \dots& \dots&  \dots& \dots& \dots&  \dots& \dots& \dots&  \dots\cr
\noalign{\vskip 4pt\hrule\vskip 5pt}}}
\endPlancktablewide
\end{table*}

\addtocounter{table}{-1}
\begin{table*}
\caption{Continued.}
\vskip -6mm
\scriptsize
\setbox\tablebox=\vbox{
\newdimen\digitwidth
\setbox0=\hbox{\rm 0}
\digitwidth=\wd0
\catcode`*=\active
\def*{\kern\digitwidth}
\newdimen\signwidth
\setbox0=\hbox{+}
\signwidth=\wd0
\catcode`!=\active
\def!{\kern\signwidth}
\newdimen\decimalwidth
\setbox0=\hbox{.}
\decimalwidth=\wd0
\catcode`@=\active
\def@{\kern\decimalwidth}

\halign{\hbox to 2.2cm{#\leaderfil}\tabskip 2.0em&
    \hfil#\hfil\tabskip 1.0em&
    \hfil$#$\hfil\tabskip 1.0em&
    \hfil$#$\hfil&
    \hfil$#$\hfil\tabskip 2.0em&
    \hfil$#$\hfil\tabskip 1.0em&
    \hfil$#$\hfil&
    \hfil$#$\hfil\tabskip 2.0em&
    \hfil$#$\hfil\tabskip 1.0em&
    \hfil$#$\hfil&
    \hfil$#$\hfil\tabskip 2.0em&
    \hfil$#$\hfil\tabskip 1.0em&
    \hfil$#$\hfil&
    \hfil$#$\hfil\tabskip 0em\cr
\noalign{\doubleline}
\omit&\omit&\multispan3\hfil Survey 1\hfil&\multispan3\hfil Survey 2\hfil&\multispan3\hfil Survey 3\hfil& 
             \multispan3\hfil Survey 4\hfil\cr
\noalign{\vskip -3pt}
\omit&\omit&\multispan3\hrulefill&\multispan3\hrulefill&\multispan3\hrulefill&\multispan3\hrulefill\cr
\omit\hfil Source\hfil&\omit Class& \alpha_{\rm LF}& \alpha_{\rm HF}& \nu_{\rm break}& \alpha_{\rm LF}& \alpha_{\rm HF}& \nu_{\rm break}& \alpha_{\rm LF}& \alpha_{\rm HF}& \nu_{\rm break}& \alpha_{\rm LF}& \alpha_{\rm HF}& \nu_{\rm break}\cr
\noalign{\vskip 3pt\hrule\vskip 4pt}
  2007+776&  BLO&    0& -0.47& 143.86& \dots& \dots&  \dots& !0.31& -0.6*& *89.07& \dots& \dots&  \dots\cr
  2005+403&  QSO&\dots& \dots&  \dots& \dots& \dots&  \dots& \dots& \dots&  \dots& !0.28& -0.5*& *14.74\cr
  2021+614&  GAL&\dots& \dots&  \dots& \dots& \dots&  \dots& \dots& \dots&  \dots& \dots& \dots&  \dots\cr
  2037+511&  QSO&-0.4*& -0.68& *70.67& !0.27& -0.62& *33.4*& -0.09& -0.6*& *37@**& !0.33& -0.61& *34.45\cr
  2121+053&  HPQ&\dots& \dots&  \dots& \dots& \dots&  \dots& \dots& \dots&  \dots& -0.25& -0.48& *22.31\cr
2131$-$021&  HPQ&-0.14& -0.48& *70@**& -0.12& -0.45& *70@**& -0.21& -0.48& *70@**& !0.01& -0.32& *16.24\cr
  2134+004&  LPQ&!0.38& -0.87& *17.51& !0.38& -0.91& *18.55& !0.37& -0.81& *18.71& -0.28& -0.88& *23.36\cr
  2136+141&  LPQ&\dots& \dots&  \dots& \dots& \dots&  \dots& \dots& \dots&  \dots& \dots& \dots&  \dots\cr
  2145+067&  LPQ&-0.19& -0.74& *66.34& \dots& \dots&  \dots& -0.21& -0.66& *23.8*& \dots& \dots&  \dots\cr
  2200+420&  BLO&-0.08& -0.42& *95.04& !0.14& -0.03& *39.12& !0.36& -0.15& *14.99& \dots& \dots&  \dots\cr
  2201+315&  HPQ&!0.09& -0.57& *69.64& -0.02& -0.83& *59.19& !0.36& -0.62& *23.82& -0.23& -1.1*& *69.91\cr
  2201+171&  QSO&!0.16& -0.53& *52.76& !0.12& -0.52& 100@**& !0.47& -0.22& *25.62& !0.13& -0.51& *63.63\cr
2216$-$038&  QSO&-0.23& -0.86& *70@**& -0.31& -0.91& 107.29& -0.24& -0.66& *70@**& -0.18& -0.43& *21.7*\cr
2223$-$052&  BLO&-0.58& -0.93& 217@**& -0.31& -0.64& *39.78& -0.14& -0.63& *20.72& -0.25& -0.63& *23.14\cr
2227$-$088&  HPQ&    0& -0.66& 130.13& !0.17& -0.54& *69.61& !0.34& -0.52& *58.31& !0.31& -0.61& *51.01\cr
  2230+114&  HPQ&-0.36& -0.84& *86.28& !0.01& -0.6*& *15.17& -0.23& -0.67& *58.5*& \dots& \dots&  \dots\cr
  2234+282&  HPQ&!0.11& -0.45& *86.98& !0.07& -0.69& 134.01& \dots& \dots&  \dots& -0.01& -0.15& *96.81\cr
  2251+158&  HPQ&\dots& \dots&  \dots& \dots& \dots&  \dots& !0.46& -0.39& *66.03& \dots& \dots&  \dots\cr
 4C\,45.51&  QSO&\dots& \dots&  \dots& \dots& \dots&  \dots& \dots& \dots&  \dots& \dots& \dots&  \dots\cr
  2353+816&  QSO&\dots& \dots&  \dots& \dots& \dots&  \dots& \dots& \dots&  \dots& \dots& \dots&  \dots\cr
\noalign{\vskip 4pt\hrule\vskip 5pt}}}
\endPlancktablewide
\end{table*}

}

\end{document}